\documentclass[english,3p,floatfix]{elsarticle}
\usepackage{float}
\usepackage{amsmath}
\usepackage{graphicx}
\usepackage{amssymb}
\usepackage{esint}

\makeatletter

\providecommand{\tabularnewline}{\\}

\usepackage{mathrsfs}\usepackage{amsfonts}\@ifundefined{definecolor}
 {\@ifundefined{definecolor}
 {\@ifundefined{definecolor}
 {\@ifundefined{definecolor}
 {\usepackage{color}}{}
}{} }{} }{}
\usepackage[colorlinks,linkcolor=blue]{hyperref}

\makeatother

\usepackage{babel}

\makeatother

\usepackage{babel}

\makeatother

\usepackage{babel}

\makeatother

\usepackage{babel}

\begin{document}

\title{Quantum spin squeezing}

\author[RIKEN,ZJU]{Jian Ma}

\author[RIKEN,ZJU]{Xiaoguang Wang}

\ead{xgwang@zimp.zju.edu.cn }

\author[RIKEN,ITP]{C. P. Sun}

\author[RIKEN,UM]{Franco Nori}

\address[RIKEN]{Advanced Science Institute, RIKEN, Wako-shi, Saitama 351-0198, Japan}

\address[ZJU]{Zhejiang Institute of Modern Physics, Department of Physics, Zhejiang
University, Hangzhou 310027, China}

\address[ITP]{Institute of Theoretical Physics, Chinese Academy of Sciences, Beijing
100190, China}

\address[UM]{Physics Department, The University of Michigan, Ann Arbor, Michigan
48109-1040, USA}
\begin{abstract}
This paper reviews quantum spin squeezing, which characterizes the
sensitivity of a state with respect to SU(2) rotations, and is
significant for both entanglement detection and high-precision
metrology. We first present various definitions of spin squeezing
parameters, explain their origin and properties for typical states,
and then discuss spin-squeezed states produced with nonlinear
twisting Hamiltonians. Afterwards, we explain pairwise correlations
and entanglement in spin-squeezed states, as well as the relations
between spin squeezing and quantum Fisher information, where the
latter plays a central role in quantum metrology. We also review the
applications of spin squeezing for detecting quantum chaos and
quantum phase transitions, as well as the influence of decoherence
on spin squeezing. Finally, we review several experimental
realizations of spin squeezing, as well as their corresponding
theoretical backgrounds, including: producing spin-squeezed states
via particle collisions in Bose-Einstein condensates, transferring
photon squeezing to atomic ensembles, and generating spin squeezing
via quantum non-demolition measurements.
\end{abstract}
\maketitle
\tableofcontents{}

\section{Introduction}

In the past two decades, spin squeezing
\cite{KITAGAWA1993,WINELAND1992,WINELAND1994} has attracted
considerable attention, both theoretically and experimentally. The
notion of spin squeezing has arisen mainly from two considerations:
The study of particle correlations and entanglement
\cite{Sorensen2001,Bigelow2001,Guehne2009}, and the improvement of
measurement precision in experiments
\cite{WINELAND1992,WINELAND1994,Polzik2008,Cronin2009}. However, the
definition of spin-squeezing parameter is not unique, and we will
here review several definitions that have been used in the past. The
most widely studied squeezing parameters were proposed by Kitagawa
and Ueda in Ref.~\cite{KITAGAWA1993}, denoted by $\xi_{S}^{2}$, and
by Wineland \textit{et al.} in
Ref.~\cite{WINELAND1992,WINELAND1994}, denoted by $\xi_{R}^{2}$. The
spin squeezing parameter $\xi_{S}^{2}$~\cite{KITAGAWA1993} was
inspired by the well-known photon squeezing; while the parameter
$\xi_{R}^{2}$~\cite{WINELAND1992,WINELAND1994} was introduced
naturally in the standard Ramsey spectroscopy experiment, where the
squeezing parameter is the ratio of the phase resolution when using
a correlated state versus that when using a coherent spin state
(CSS).

One application of spin squeezing is to detect quantum
entanglement~\cite{Guehne2009,Amico2008,Horodecki2009}, which plays
a key role in both the foundations of quantum physics and
quantum-information processing~\cite{Nielsen2000,Stolze2008}.
Parameter $\xi_{S}^{2}$, found to be related to negative pairwise
correlations~\cite{Ulam-Orgikh2001} and concurrence
~\cite{Wootters1998,Wang2003}, is able to characterize pairwise
entanglement for a class of many-body spin-1/2 states. Reference
\cite{Sorensen2001} proved that a many-body spin-1/2 state is
entangled if it is spin squeezed corresponding to $\xi_{R}^{2}<1$.
Indeed, spin-squeezing parameters are multipartite entanglement
witnesses \cite{Guehne2009}. Another important reason for choosing
spin-squeezing parameters as measures of multipartite correlations
is that spin squeezing is relatively easy to be generated and
measured experimentally
~\cite{Hald1999,Orzel2001,Esteve2008,Gross2010,Riedel2010,Julsgaard2001,Fernholz2008,Appel2009,Louchet-Chauvet2010,Takano2009}.
Spin-squeezing parameters only involve the first and second moments
of the collective angular momentum operators, and in many practical
cases, e.g., in Bose-Einstein condensations (BECs), particles cannot
be addressed individually, and only the collective operators can be
measured. To detect entanglement for realistic experiments, a class
of spin squeezing inequalities were proposed
~\cite{Korbicz2005,Korbicz2006,Toth2009a}, based the first, second,
and even the third moments of collective spin operators. The use of
spin-squeezing parameters as entanglement detectors has already been
discussed in a recent review~\cite{Guehne2009}. Besides, spin
squeezing could also be useful for quantum computation (see, e.g.,
the reviews in \cite{Buluta2010,You2005,Shevchenko2010}), as well as
for quantum simulations (see, e.g., the reviews in
\cite{Buluta2009}).

Another application of spin squeezing, especially in experiments, is
to improve the precision of measurements, e.g., in the Ramsey
spectroscopy~\cite{WINELAND1992,WINELAND1994,Bollinger1996,Agarwal1996,Berman1997,Xu1999,Meyer2001,Smith2004,Cronin2009,Doering2010},
and in making more precise atom clocks
\cite{WINELAND1994,Bigelow2001,Polzik2008,Appel2009,Oblak2005,Sorensen1999,Meiser2008,Andre2004,Leroux2010a}
and gravitational-wave
interferometers~\cite{WALLS1981,Dunningham2004,Goda2008}. Therefore,
many efforts have been devoted to the generation of squeezing in
atomic systems. Basically, these works can be sorted by two
categories.

(i) Generating spin squeezing in atomic ensembles via atom-photon
interactions \cite{Hammerer2010}. Within this category, a very
natural idea is to transfer squeezing from light to atoms
\cite{WINELAND1994,Hald1999,Julsgaard2001,Palma1989,Banerjee1996,Kuzmich1997,Sorensen1998,Molmer1999,Hald2000,Vernac2000,Vernac2001,Hald2001,Vernac2002,Dantan2003a,Dantan2005b,Dantan2006,Civitarese2009},
and many proposals focused on transferring squeezing via
electromagnetically induced
transparency~\cite{Duan2000a,Fleischhauer2000,Lukin2000,Dantan2003,Akamatsu2004,Dantan2004,Dantan2005,Dantan2006a,Honda2008,Gong2010}.
When considering light-atom interactions, the detuning between the
light and the atoms are very important. In the large detuning
regime, an effective Hamiltonian is obtained, consisting of a
dispersive interaction term and a nonlinear interaction term, and
the magnitude of the two terms can be adjusted. (a) The dispersive
interaction between the light and atoms, results in a Faraday
rotation of the polarization of the light \cite{Smith2003}. Then by
performing a quantum nondemolition (QND) measurement of the output
light, the atomic ensembles can be squeezed conditioned to the
measurement results
\cite{Kuzmich1998,Takahashi1999,Kuzmich1999,Kuzmich2000,Thomsen2002a,Bouchoule2002,Zhang2003,Smith2003,Saito2003,Auzinsh2004,Smith2004,Madsen2004,Kuzmich2004,Echaniz2005,Geremia2005,Oblak2005,Petersen2005,Genes2006,Smith2006,Cviklinski2007a,Meiser2008,Nielsen2008,Teper2008,Windpassinger2008,Windpassinger2009,Chase2009,Appel2009,Louchet-Chauvet2010,Saffman2009,KubasikPR2009,Takano2009,Shah2010,Kurucz2010,Marek2010,Wasilewski2010},
and this method was realized in experiments
\cite{Kuzmich1998,Duan2000a,Julsgaard2001,Appel2009,Louchet-Chauvet2010,Schleier-Smith2010a,KoschorreckPRL2010a,KoschorreckPRL2010b}.
(b) The nonlinear term, which describes interactions between atoms,
is a one-axis twisting
Hamiltonian~\cite{KITAGAWA1993,Klimov2005,Pathak2008}, and can be
used to generate spin squeezing
\cite{Agarwal1997,Klimov1998,Shindo2004,Deb2006,Chaudhury2007,Fernholz2008}.
There are also some other proposals, including placing atoms in a
high-Q cavity such that the atoms interact with a single field mode
(not squeezed) repeatedly \cite{Ueda1996}, illuminating bichromatic
light on the atoms in a bad cavity \cite{Sorensen2002a} without
requiring strong atom-cavity coupling. The intrinsic spin squeezing
in a large atomic radiating system was studied in
Ref.~\cite{Yukalov2004}, where spin-squeezed states are generated
due to photon-exchanging induced strong interatomic correlations.
Spin squeezing can also be produced via squeezing exchange between
the motional and internal degrees of freedom\cite{Ramon1999}. In
addition, squeezing can also be transferred from the atomic
ensembles to photons \cite{Saito1997,Saito1999,Poulsen2001c}.

(ii) Generating spin squeezing in the BEC via atomic collisions. In
the last decade, generating spin-squeezed states in a BEC attracted
many interests. The nonlinear atom-atom collisions in a
two-component BEC can be described by a one-axis twisting
Hamiltonian
\cite{Sorensen2001,Orzel2001,Law2001,Poulsen2001a,Raghavan2001,Jenkins2002,Jaksch2002,Jing2002,Sorensen2002,Micheli2003,Molmer2003,Jaaskelainen2004,Choi2005,Jaaskelainen2005,Jin2007,Jin2007a,Thanvanthri2007,Esteve2008,Jin2008,Grond2009,Grond2009a},
which can generate spin-squeezed states. Besides, two-axis twisting
Hamiltonian can be realized via Raman processes
\cite{Helmerson2001,Zhang2003a}, or via two-component condensates in
a double-well potential \cite{Ng2003}. Bragg scattering induced spin
squeezing was studied in Refs.~\cite{Deb2003,Gasenzer2002a}. Spin
squeezing can also be created in a spinor-1 BEC
\cite{Yi2002,Mutecaplioglu2007}, via spin-exchange interactions
\cite{Duan2000,Pu2000}, or free dynamical evolution
\cite{Duan2002,Mustecaplioglu2002}. As studied in
Ref.~\cite{Yi2006}, the ground state of a $F=1$ dipolar spinor BEC
with transverse external magnetic field is spin squeezed. The
squeezing of the ground state of spin-1 condensates in an optical
lattice has been studied in \cite{Gerbier2006,Oztop2009}, where spin
squeezing occurs in the Mott insulator phase. The particle
collisions induced decoherence in the generation of spin squeezing
was studied in \cite{Gasenzer2002}, and spin-squeezed states
generated in BEC are robust to dissipation \cite{Dunningham2002a},
and particle losses \cite{Micheli2003,Li2008,Li2009}.
Photodissociation induced spin squeezing was studied in
Refs.~\cite{Poulsen2001,Yi2001a}. The spin-squeezed states in BEC
have been used to perform sub-shot-noise measurements
\cite{Dunningham2004,Lee2006}, detect weak forces
\cite{Jaaskelainen2006}. Experimental realizations of spin-squeezed
states in BEC were reported in
Refs.~\cite{Orzel2001,Jo2007,Li2007,Vengalattore2007,Esteve2008,Gross2010,Riedel2010}.
Recently, spin squeezing in BEC was re-examined by considering the
effects of indistinguishable of particles
\cite{Benatti2010,Benatti2010a}, where spin squeezing is neither
necessary to achieve sub-shot noise accuracies in quantum metrology,
nor in general an entanglement witness. Below, we only consider
distinguishable cases. Besides the above two categories, other
proposals such as using Ising Hamiltonian to generate squeezed
states was studied in
Ref.~\cite{Sorensen1999,Wang2001b,Wang2004,Reboiro2007}.

This review is organized as follows. First, in Sec.~\ref{definition},
we present a full list of definitions of spin squeezing, which were
proposed for different tasks, and give the relations between the spin
squeezing and the bosonic squeezing. We make a comparison between
bosonic squeezing and spin squeezing. We demonstrate their relations
under the large-$N$ limit and small excitations. In this case spin
squeezing will reduce to bosonic squeezing.

In Sec.~\ref{states}, we review the generation of spin-squeezed
states via the one-axis twisting and two-axis twisting Hamiltonians,
which have both fixed parity and particle-exchange symmetry. The
proposal of using nonlinear Hamiltonians to generate spin-squeezed
states originates from quantum optics, where the nonlinear Kerr
effect is employed to prepare squeezed light. The dynamics
controlled by the one-axis twisting Hamiltonian can be solve
analytically. This type of Hamiltonian is widely studied in the
regime of spin squeezing, and could be implemented in BEC
\cite{Sorensen2001,Orzel2001,Esteve2008,Gross2010,Riedel2010}, and
large detuning atom-field interaction models
\cite{AGARWAL1994,Agarwal1997,Fernholz2008}, etc. The two-axis
twisting has advantages in generating spin squeezing as compared to
the one-axis twisting, however, it is not easy to implemented in
experiments, and analytical results cannot be obtained for arbitrary
system size.

Since the nonlinear Hamiltonians involve two-body interactions, one
can infer that spin-squeezed states may be pairwise correlated or
entangled. This is reviewed in Sec.~\ref{entanglement}. We study the
relations between spin squeezing and negative pairwise correlation.
Then we review the relations between spin squeezing and
entanglement, and this is one of the most important branches in the
study of spin squeezing. The spin-squeezing parameter $\xi_{S}^{2}$
is closely related to pairwise entanglement. Moreover, the squeezing
parameter proposed in~\cite{Toth2009a,Wang2003} has been found to be
qualitatively equivalent to the concurrence for systems with
exchange symmetry and parity. The parameter $\xi_{R}^{2}$ can
provide a criterion for multipartite entanglement. Thus, spin
squeezing parameters can also be used to detect entanglement beyond
pairwise entanglement. Spin-squeezing parameters can also be used to
detect entanglement in a system of spin-$j$ particles. In the end of
this section, we also discuss the concept of two-mode spin
squeezing, of which the bipartite entanglement is a valuable quantum
information resource.

In Sec.~\ref{Fisher}, we discuss spin squeezing and quantum Fisher
information (QFI) in quantum metrology. The spin squeezing parameter
$\xi_{R}^{2}$ arises in the Ramsey interferometer, and the QFI lies
at the heart of parameter estimation theory. Both quantities
characterize the entanglement enhanced quantum metrology. Their
relations are reviewed and compared in the Ramsey interferometry. In
Sec.~\ref{qpt}, we discuss the applications of spin squeezing in
quantum phase transitions (QPTs) and quantum chaos. In this section,
we first employ the Lipkin-Meshkov-Glick model, which has a
second-order QPT, to show that spin squeezing can be used to
identify the critical point. Then, we review spin squeezing in
quantum chaos. Two typical chaotic models are described: the quantum
kicked-top model and the Dicke model. Some quantities, like entropy
and concurrence, have been studied in these two models, and
spin-squeezing parameters can also reflect their chaotic behaviors.
The lifetime of a spin-squeezed state is much shorter when the
system is chaotic than when the system is regular.

In Sec.~\ref{decoherence}, we consider the effects of decoherence.
Firstly, in Sec.~7.1, we discuss how decoherence affects the generation
of spin squeezing. Then in Sec.~7.2 and 7.3, we discuss the robustness
and lifetime of spin squeezing in the presence of decoherence. Finally
in Sec.~7.4, we discuss how decoherence affects the ability of spin-squeezed
states to the improvement of phase resolution in Ramsey processes.

Finally, in Sec.~\ref{experiment}, we review the generation of spin
squeezing in experiments, including BEC, light-matter interaction,
QND measurements. The closely related theoretical backgrounds are
provided as well. For each physical system, we give a brief yet comprehensive
summary of a series experimental results, as well as detailed discussions
of several notable experimental improvements. We first discuss experiments
on BEC, which involves Mott insulator transitions and the one-axis
twisting. Then we review the transfer of squeezing, from light to
atomic ensemble, via absorption of squeezed vacuum. Atomic ensembles
can be viewed as quantum memories, and light is a very promising information
carrier. Spin-squeezed states have been prepared in an optically thick
Cesium atomic ensemble by using a squeezed vacuum. We also discuss
a QND-measurement scheme, which generates conditional spin-squeezed
states; and by utilizing feedback techniques, one can produce unconditional
spin-squeezed states. Finally, we summarize some useful results and
derivations in the appendices.

\section{Definitions of spin squeezing\label{definition}}

The definition of spin squeezing is not unique, and it depends on
the context where squeezing is considered. The most popular
definitions are proposed by\textbf{ }Kitagawa and
Ueda~\cite{KITAGAWA1993}, in analogy to photon squeezing, and by
Wineland \textit{et al.}~\cite{WINELAND1992,WINELAND1994}, in Ramsey
experiments.\textbf{\emph{ }}The latter one is directly associated
with quantum metrology. Besides these two widely studied
definitions, there are some other definitions of spin squeezing,
which were introduced for certain considerations. Before we discuss
spin squeezing, we first give a brief review on bosonic squeezing,
widely studied in quantum optics. These will give some initial
indications on the possible definitions and the generation of spin
squeezing.

\subsection{Coherent state and bosonic squeezing \label{bcs}}

Let us first review some basic concepts of bosonic squeezing, which
is widely studied in quantum optics
\cite{Yuen1976,Walls1994,Scully1997,Dodonov2002}, and has
applications in practical precision
measurements~\cite{Caves1981,Bondurant1984}. Squeezed states have
also been studied in other contexts, including squeezed phonons in
condensed matter physics~\cite{Hu1996,Hu1996a,Hu1997,Hu1999} and
squeezed states of microwave radiation in a superconducting resonant
circuit~\cite{Zagoskin2008}.

Consider the bosonic creation and annihilation operators $a^{\dagger}$
and $a$, with the corresponding commutation relation\begin{equation}
\left[a,a^{\dagger}\right]=1,\label{bose_commu}\end{equation}
 satisfying $a|n\rangle=\sqrt{n}|n-1\rangle$ and $a^{\dagger}|n\rangle=\sqrt{n+1}|n+1\rangle$,
where $|n\rangle$ is the number state, i.e., $a^{\dagger}a|n\rangle=n|n\rangle$.
Below we consider two dimensionless operators $X$ and $P$ given
by\begin{align}
X & \equiv a+a^{\dagger},\text{ }\quad P\equiv\frac{a-a^{\dagger}}{i},\label{eq:xp}\end{align}
 which are the position and momentum amplitudes, respectively. The
commutator\begin{equation}
\left[X,P\right]=2i\end{equation}
 results in the Heisenberg uncertainty relation\begin{equation}
\Delta X\Delta P\geq1,\label{xp_uncertain}\end{equation}
 where $\Delta A=\sqrt{\left\langle A^{2}\right\rangle -\left\langle A\right\rangle ^{2}}$
is the standard deviation, and $\left(\Delta A\right)^{2}$ is the
variance.

\subsubsection{Coherent state}

The coherent state~\cite{Barut1971,Gilmore1972,Perelomov1972},
usually called the minimum uncertainty state, satisfies the
following simple condition\begin{equation} \Delta X=\Delta
P=1.\label{CS_criterion}\end{equation}
 The coherent state and its applications have been reviewed in \cite{Perelomov,Zhang1990}.
From the above criterion (\ref{CS_criterion}), the simplest coherent
state is the vacuum state $|0\rangle$. A coherent state is defined
as\begin{equation}
a|\alpha\rangle=\alpha|\alpha\rangle,\label{cs3}\end{equation}
 from which the first moments of $X$ and $P$ for the coherent state
are directly obtained as Eq.~(\ref{CS_criterion}).

The uncertainty relation (\ref{xp_uncertain}) cannot be violated.
However, we can choose a state to make $\Delta X$ (or $\Delta P$)
smaller than $1$, and this type of state is called coherent squeezed
state. Generally, squeezing occurs when variance is less than 1 in
any direction of the $X$-$P$ plane. We introduce an operator in
the $X$-$P$ plane as\begin{align}
X_{\theta} & =e^{i\theta a^{\dagger}a}Xe^{-i\theta a^{\dagger}a}\notag\\
 & =ae^{-i\theta}+a^{\dagger}e^{i\theta},\label{xtheta}\end{align}
 with $X=X_{0}$ and $P=X_{\pi/2}$ being special cases. We see that
the generalized operator $X_{\theta}$ can be obtained by rotating
operator $X$ in the phase space. The so-called principal quadrature
squeezing~\cite{LUKS1988,LUKS1988a} is characterized by a
parameter\begin{equation}
\zeta_{B}^{2}=\min_{\theta\in(0,2\pi)}(\Delta
X_{\theta})^{2},\label{principle_squ}\end{equation}
 which is the minimum value of $(\Delta X_{\theta})^{2}$ with respect
to $\theta$, and $\zeta_{B}^{2}<1$ indicates bosonic principal squeezing.
The minimization in the above definition can be easily performed (See
\ref{principle})\begin{equation}
\zeta_{B}^{2}=1+2\left(\langle a^{\dagger}a\rangle-|\langle a\rangle|^{2}\right)-2|\langle a^{2}\rangle-\langle a\rangle^{2}|,\label{xibb}\end{equation}
 which only contains expectation values $\langle a\rangle$, $\langle a^{2}\rangle$,
and $\langle a^{\dagger}a\rangle$.

\subsubsection{Generation of bosonic squeezed state}

A bosonic squeezed state can be generated by applying the following
nonlinear Hamiltonian\begin{equation}
H=i\left(ga^{\dagger2}-g^{\ast}a^{2}\right)\label{bose_squ_Ham}\end{equation}
 on a coherent states as\begin{equation}
|\alpha,\eta\rangle=S\left(\eta\right)|\alpha\rangle,\end{equation}
 with squeezing operator $S$: \begin{equation}
S\left(\eta\right)=\exp\left(-iHt\right)=\exp\left(\frac{1}{2}\eta^{\ast}a^{2}-\frac{1}{2}\eta a^{\dagger2}\right)\end{equation}
 and complex number $\eta=r\exp\left(i\theta\right),$ where $r=-2\left\vert g\right\vert t$.

In Sec.~\ref{taws}, we will see that the two-axis twisting
Hamiltonian that generates spin-squeezed states is inspired from
this squeezing operator $S$. The nonlinear operator $S$ mixes the
$X$ and $P$ components and thus performs not only a rotation, but
also squeezing. As shown in Ref.~\cite{Scully1997}, the standard
deviations for the rotated operators are derived as\begin{equation}
\Delta X_{\theta/2}=e^{-r},\text{ }\Delta
P_{\theta/2}=e^{r},\end{equation} The variance in $X_{\theta/2}$ is
reduced with a factor $e^{-r}$, and the squeezing is determined by
$r$, which is usually called the squeezing parameter. Computer
animations showing various types of squeezed states, including
squeezed wave packets are available online at~\cite{nori_website}.

Bosonic squeezed states can also be generated from the Kerr
interaction Hamiltonian~\cite{Kitagawa1986} \begin{equation}
H=\kappa(a^{\dagger}a)^{2}.\end{equation}
 Choosing an initial state be the coherent state, we have the state
at time $t,$\begin{equation}
|\psi(t)\rangle_{\text{Kerr}}=\exp\left[{-i\kappa t(a^{\dagger}a)^{2}}\right]|\alpha\rangle.\label{bose_squ_Ham2}\end{equation}

\subsection{Coherent spin state}

Now we turn to spin systems. Hereafter, we will mainly consider the
system consisting of $N$ spin-1/2 particles, since in many practical
cases we deal with two-level systems or qubits. A two-level atom
interacting with the radiation field can be treated as a spin-1/2
particle in a magnetic filed. The angular momentum operators for
this ensemble of spin-1/2 particles are given by \begin{equation}
J_{\alpha}=\frac{1}{2}\sum_{l=1}^{N}\sigma_{l\alpha},\text{ \
}\alpha=x,y,z\end{equation}
 where $\sigma_{l\alpha}$ is the Pauli matrix for the $l$-th particle.
Before we discuss spin squeezing, we introduce the CSS.

\begin{figure}[H]
\begin{centering}
\includegraphics[width=15cm]{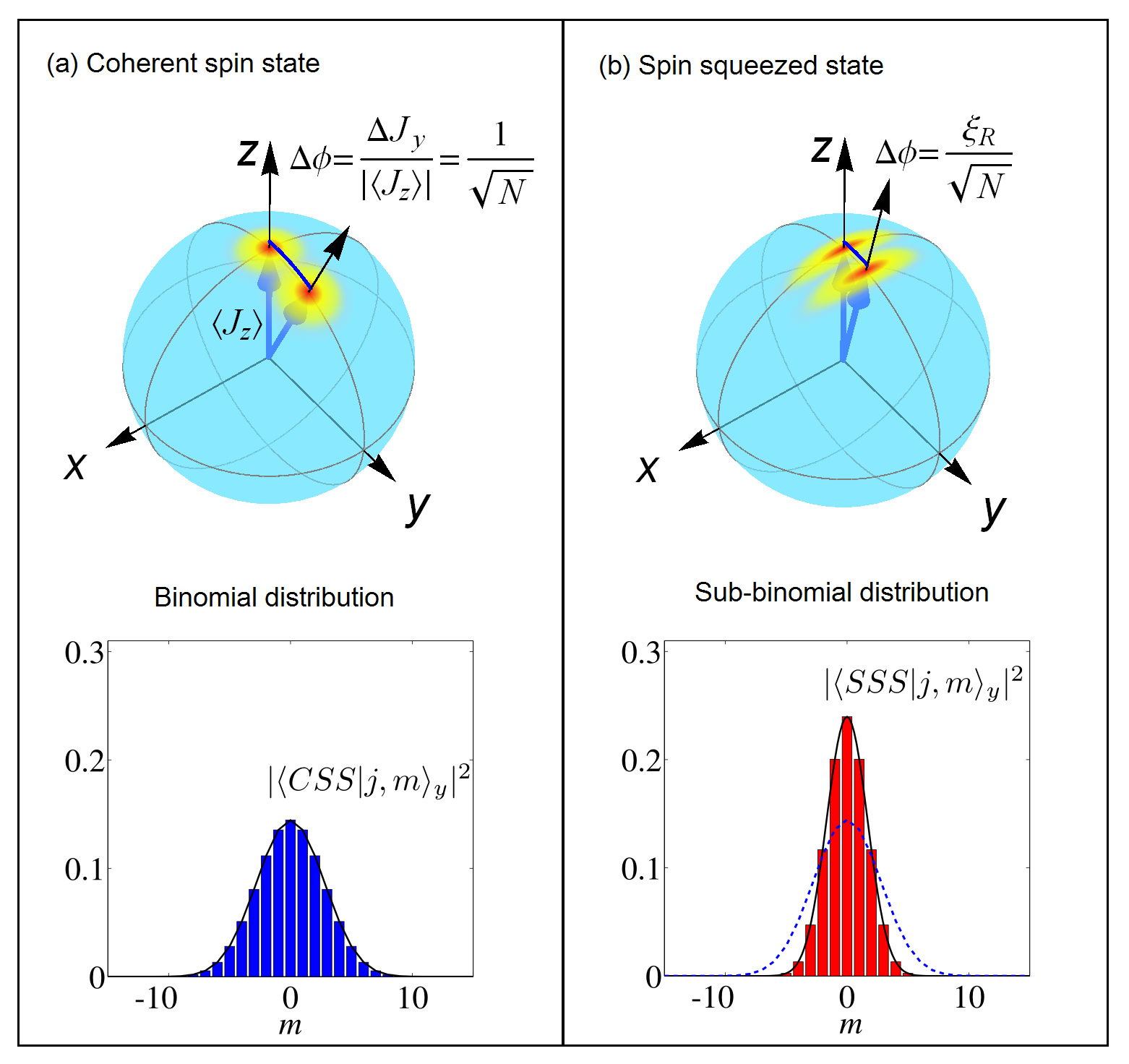}
\par\end{centering}

\caption{(Color online) Bloch sphere representations for a coherent spin state
$|CSS\rangle$ shown in (a), and a spin-squeezed state $|SSS\rangle$
in (b). The radius of the Bloch sphere is $|\langle\vec{J}\rangle|$.
The fluctuations of the spin components are represented by the circular
multi-color disks in (a) and the multi-color elliptical disks in (b).
The phase variance $\Delta\phi$, determined by spin fluctuation disks
and spin lengths, is the resolution of the spin state with respect
to rotations, which characterizes the frequency precision in Ramsey
spectroscopy. For a CSS, the projection noise is characterized by
a binomial distribution, as shown in (a), while for a SSS, the projection
noise is not binomial, and may be a sub-binomial distribution (b).}

\label{fig_spin_bloch}
\end{figure}

The CSS is defined as a direct product of single spin states\begin{equation}
|\theta,\phi\rangle={\bigotimes\limits _{l=1}^{N}}\left[\cos\frac{\theta}{2}|0\rangle_{l}+e^{i\phi}\sin\frac{\theta}{2}|1\rangle_{l}\right],\label{CSS}\end{equation}
 where $|0\rangle_{l}$ and $|1\rangle_{l}$ are the eigenstates of
$\sigma_{lz}$ with eigenvalues $1$ and $-1$, respectively. The above
definition (\ref{CSS}) gives an intuitive geometric description as
shown in Fig.~\ref{fig_spin_bloch}, i.e., all the spins point in the
same direction.

The CSS can also be written as\begin{equation}
|\eta\rangle\equiv|\theta,\phi\rangle=(1+|\eta|^{2})^{-j}\sum_{m=-j}^{j}{\binom{2j}{j+m}}^{1/2}\eta^{j+m}|j,m\rangle,\text{ }\;\eta\in\mathbb{C},\label{CSS1}\end{equation}
 in terms of the Dicke states $|j,m\rangle,$ which are the eigenstates
of $J_{z}$ with eigenvalue $m$, and\begin{equation}
\eta=-\tan\frac{\theta}{2}\exp({-i\phi}).\label{eta_theta}\end{equation}
 Note that, when $\theta=\pi$, $\eta$ is divergent, and only the
coefficient before $|j,j\rangle$ is nonzero, thus
$|\theta=\pi,\phi\rangle=|j,j\rangle$. The above form is expressed
as the projection of a coherent state onto Dicke states
$|j,m\rangle$, and $\left\vert \left\langle j,m|\eta\right\rangle
\right\vert ^{2}$ are related to the projection noise
\cite{Itano1993}, obeying a binomial distribution, shown in
Fig.~\ref{fig_spin_bloch}.

Below, we shall derive Eq.~(\ref{CSS1}). By using the identity
\begin{equation}
\exp[{i(\xi\sigma_{+}+\eta\sigma_{-})}]=\cos\sqrt{\xi\eta}+i\frac{\sin\sqrt{\xi\eta}}{\sqrt{\xi\eta}}(\xi\sigma_{+}+\eta\sigma_{-}),\end{equation}
 one can write the CSS in the following form \begin{equation}
|\theta,\phi\rangle={\bigotimes\limits _{l=1}^{N}}R_{l}(\theta,\phi)|0\rangle_{l}={\bigotimes\limits _{l=1}^{N}}\exp\left(\zeta\sigma_{l+}-\zeta^{\ast}\sigma_{l-}\right)|0\rangle_{l},\end{equation}
 where \begin{equation}
\zeta=-\frac{\theta}{2}\exp({-i\phi}).\label{zeta_theta}\end{equation}
 It can be further written as\begin{equation}
|\theta,\phi\rangle=R(\theta,\phi)|j,j\rangle=\exp\left(\zeta J_{+}-\zeta^{\ast}J_{-}\right)|j,j\rangle,\label{CSS_temp}\end{equation}
 where $|j,j\rangle\equiv{\bigotimes\limits _{l=1}^{N}}$ $|0\rangle_{l}$
is the eigenstate of $J_{z}$ with eigenvalue $j=N/2$, which represents
all the spins polarize to the $z$-direction. Here \begin{equation}
J_{\pm}=J_{x}\pm iJ_{y}\label{ladder_op}\end{equation}
 are the ladder operators and $R(\theta,\phi)$ is the rotation operator,
which can be also given by\begin{equation}
R(\theta,\phi)=\exp\left(-i\theta
J_{\vec{n}}\right)=\exp\left[i\theta\left(J_{x}\sin\phi-J_{y}\cos\phi\right)\right],\end{equation}
 where $\vec{n}=\left(-\sin\phi,\cos\phi,0\right)$.

Now we write Eq.~(\ref{CSS_temp}) in a more explicit form by using
the formula~\cite{Arecchi1972,Zhang1990}\begin{equation}
\exp\left(\zeta J_{+}-\zeta^{\ast}J_{-}\right)=\exp\left(\eta
J_{+}\right)\exp\left[\ln\left(1+\left\vert \eta\right\vert
^{2}\right)J_{z}\right]\exp\left(-\eta^{\ast}J_{-}\right),\label{rotation_relation}\end{equation}
 where $\eta$ is defined as Eq.~(\ref{eta_theta}). The above relations
can be verified as below. The rotation operator $R\left(\theta,\phi\right)$
can be expressed in a product form as\begin{equation}
R\left(\theta,\phi\right)={\bigotimes\limits _{l=1}^{N}}R_{l}(\theta,\phi)=\bigotimes\limits _{l=1}^{N}\exp\left(\zeta\sigma_{l+}-\zeta^{\ast}\sigma_{l-}\right),\end{equation}
 and $R_{l}\left(\theta,\phi\right)$ can be evaluated readily as\begin{equation}
R_{l}(\theta,\phi)=\begin{pmatrix}\cos\left\vert \zeta\right\vert  & -e^{-i\phi}\sin\left\vert \zeta\right\vert \\
e^{i\phi}\sin\left\vert \zeta\right\vert  & \cos\left\vert \zeta\right\vert \end{pmatrix},\end{equation}
 similarly, we find\begin{equation}
\exp\left(\eta\sigma_{+}\right)\exp\left[\ln\sqrt{1+\left\vert \eta\right\vert ^{2}}\sigma_{z}\right]\exp\left(-\eta^{\ast}\sigma_{-}\right)=\begin{pmatrix}\left(1+\left\vert \eta\right\vert ^{2}\right)^{-1/2} & \eta\left(1+\left\vert \eta\right\vert ^{2}\right)^{-1/2}\\
-\eta^{\ast}\left(1+\left\vert \eta\right\vert ^{2}\right)^{-1/2} & \left(1+\left\vert \eta\right\vert ^{2}\right)^{-1/2}\end{pmatrix}.\end{equation}
 Then, taking Eqs.~(\ref{zeta_theta}) and (\ref{eta_theta}) into
the above two matrices, the relation (\ref{rotation_relation}) is
obtained.

\subsection{Spin-squeezing parameters based on the Heisenberg uncertainty relation}

The definition of spin squeezing is not unique. When talking about
spin squeezing, we should specify a certain spin-squeezing parameter.
The uncertainty relation for angular momentum operators results from
the commutation relation \begin{equation}
\left[J_{\alpha},J_{\beta}\right]=i\varepsilon_{\alpha\beta\gamma}J_{\gamma},\label{angular_commu_rel}\end{equation}
 where $\alpha$, $\beta$, $\gamma$ denote the components in any
three orthogonal directions, and $\varepsilon_{\alpha\beta\gamma}$
is the Levi-Civita symbol. The uncertainty relation is \begin{equation}
\left(\Delta J_{\alpha}\right)^{2}\left(\Delta J_{\beta}\right)^{2}\geq\left\vert \left\langle J_{\gamma}\right\rangle \right\vert ^{2}/4.\label{spin_comm_rel}\end{equation}
 In analogy to bosonic squeezing, spin squeezing could be defined
when one of the fluctuations in the left-hand side of
Eq.~(\ref{spin_comm_rel}) satisfies~\cite{WALLS1981,Walls1981b}
$\left(\Delta J_{\alpha}\right)^{2}\leq\left\vert \left\langle
J_{\gamma}\right\rangle \right\vert /2$, and consequently, a
squeezing parameter $\xi_{H}^{2}$ is given by\begin{equation}
\xi_{H}^{2}=\frac{2\left(\Delta J_{\alpha}\right)^{2}}{\left\vert
\left\langle J_{\gamma}\right\rangle \right\vert }\text{,
}\alpha\neq\gamma\in\left(x,y,z\right),\label{xi_h}\end{equation}
 and if $\xi_{H}^{2}<1$, the state is squeezed. The subscript $H$
refers to the Heisenberg uncertainty relations. More generally, one
can define\begin{equation}
\xi_{H}^{2}=\frac{2\left(\Delta J_{\vec{n}_{1}}\right)^{2}}{\left\vert \left\langle J_{\vec{n}_{2}}\right\rangle \right\vert },\end{equation}
 where $\vec{n}_{1}$ and $\vec{n}_{2}$ are two orthogonal unit vectors.

Below, we calculate $\xi_{H}^{2}$ for a CSS $|\theta,\phi\rangle$.
According to the definition (\ref{xi_h}), the value of $\xi_{H}^{2}$
depends on the choice of the directions $\vec{n}_{1}$ and $\vec{n}_{2}$,
and also on the parameter $\theta$ and $\phi$ of the CSS. From Eq.~(\ref{expvar}),
one can find that\begin{equation}
\xi_{H}^{2}=\frac{1-(\vec{n}_{0}\cdot\vec{n}_{1})^{2}}{\left\vert \vec{n}_{0}\cdot\vec{n}_{2}\right\vert }.\end{equation}
 More specifically, let $\vec{n}_{0}$ be the $z$-direction and $\vec{n}_{1}$
the $x$-direction.The squeezing parameter then becomes\begin{equation}
\xi_{H}^{2}=\left\vert \frac{\sin\theta}{\cos\phi}\right\vert ,\end{equation}
 where we have used Eq.~(\ref{expectation}). It is found that $\xi_{H}^{2}$
may be less than 1 for a CSS, and this is not expected, since a CSS
should not be spin squeezed.

Now we discuss a generalized spin-squeezing parameter, introduced in
Refs.~\cite{Prakash2005,Prakash2007}, for which spin squeezing could
occur in two orthogonal directions simultaneously. The subscript
$H^{\prime}$ indicates that this parameter is also related to the
Heisenberg uncertainty relation. We first discuss spin squeezing in
the $x$-axis, considering two orthogonal angular momentum operators
in the $y$-$z$ plane\begin{align}
J_{\theta} & \equiv J_{y}\cos\theta+J_{z}\sin\theta,\notag\\
J_{\theta+\pi/2} & \equiv-J_{y}\sin\theta+J_{z}\cos\theta,\end{align}
 and the commutation relations $\left[J_{x},J_{\theta}\right]=iJ_{\theta+\pi/2}\text{, }\left[J_{x},J_{\theta+\pi/2}\right]=-iJ_{\theta}$.
The state is squeezed in the $x$-axis if\begin{equation}
\left(\Delta J_{x}\right)^{2}<\frac{1}{2}\left\vert \left\langle J_{\theta+\pi/2}\right\rangle \right\vert \text{, }\ \text{and/or }\left(\Delta J_{x}\right)^{2}<\frac{1}{2}\left\vert \left\langle J_{\theta}\right\rangle \right\vert \text{.}\end{equation}
 Since the maximum value of $\left\langle J_{\theta}\right\rangle $
is $\sqrt{\left\langle J_{z}\right\rangle ^{2}+\left\langle J_{y}\right\rangle ^{2}}$,
we can say that the state is $J_{x}$-squeezed if $\left(\Delta J_{x}\right)^{2}<\frac{1}{2}\sqrt{\left\langle J_{z}\right\rangle ^{2}+\left\langle J_{y}\right\rangle ^{2}}$,
otherwise, the state is $J_{y}$-squeezed if $\left(\Delta J_{y}\right)^{2}<\frac{1}{2}\sqrt{\left\langle J_{z}\right\rangle ^{2}+\left\langle J_{x}\right\rangle ^{2}}$.
The above two inequalities could hold simultaneously.

More generally, we could write the above squeezing parameter as follows\begin{equation}
\xi_{H^{\prime}}^{2}=\frac{2\left(\Delta J_{\vec{n}_{1}}\right)^{2}}{\sqrt{\langle J_{\vec{n}_{2}}\rangle^{2}+\langle J_{\vec{n}_{3}}\rangle^{2}}}.\end{equation}
 For the CSS $|\theta,\phi\rangle$, from Eq.~(\ref{expvar}), one
can find that $\xi_{H^{\prime}}^{2}=\sqrt{1-(\vec{n}_{0}\cdot\vec{n}_{1})^{2}}$.
More specifically, let $\vec{n}_{0}$ be the $z$-direction, we have\begin{equation}
\xi_{H^{\prime}}^{2}=|\sin\theta|.\end{equation}
 The above equation indicates that the CSS can be squeezed, implying
that $\xi_{H^{\prime}}^{2}$ is not a desirable definition for spin
squeezing.

\subsection{Squeezing parameter $\xi_{S}^{2}$ given by Kitagawa and Ueda}

Unlike the bosonic systems where the variance is equal in any direction
for a bosonic coherent state, for a CSS the variance of spin operators
depends on $\vec{n}$, and there exists \textit{a prior} direction:
the mean-spin direction (MSD) \begin{equation}
\vec{n}_{0}=\frac{\langle\vec{J}\rangle}{|\langle\vec{J}\rangle|}=\frac{\langle\vec{\sigma}_{1}\rangle}{|\langle\vec{\sigma}_{1}\rangle|},\label{msd}\end{equation}
 where the second inequality results from the exchange symmetry.

Below, we use $\vec{n}_{\perp}$ to denote the direction
perpendicular to the MSD. For a CSS, we have $\left(\Delta
J_{\vec{n}_{\perp}}\right)^{2}=j/2$ (See \ref{varn}), thus a state
is spin squeezed if the variance of $J_{\vec{n}_{\perp}}$ is less
than $j/2$. Compared with the principal squeezing
(\ref{principle_squ}), we arrive at the spin-squeezing
parameter~\cite{KITAGAWA1993}\begin{equation}
\xi_{S}^{2}=\frac{\min\left(\Delta
J_{\vec{n}_{\perp}}^{2}\right)}{j/2}=\frac{4\min\left(\Delta
J_{\vec{n}_{\perp}}^{2}\right)}{N},\end{equation}
 where $j=N/2$, and $\vec{n}_{\perp}$ refers to an axis perpendicular
to the MSD and the minimization is over all directions $\vec{n}_{\perp}$.
It is desirable that the spin-squeezing parameter $\xi_{S}^{2}$ is
equal to 1 for the CSS.

To calculate the parameter $\xi_{S}^{2},$ the first step is to compute
the MSD determined by the expectation values $\langle J_{\alpha}\rangle$,
with $\alpha\in\{x,y,z\}$. The MSD $\vec{n}_{0}$ can be written
in spherical coordinates as \begin{equation}
\vec{n}_{0}=(\sin{\theta}\cos{\phi},\sin{\theta}\sin{\phi},\cos{\theta}),\end{equation}
 where $\theta$ and $\phi$ are polar and azimuthal angles, respectively.
The angles $\theta$ and $\phi$ are given by \cite{Song2006}
\begin{align}
\theta= & \arccos\left(\frac{\langle J_{z}\rangle}{|{\vec{J}}|}\right),\notag\\
\phi= & \left\{ \begin{array}{ll}
\arccos\Big(\frac{\langle J_{x}\rangle}{|{\vec{J}}|\sin(\theta)}\Big) & \text{if}~~\langle J_{y}\rangle>0,\\
2\pi-\arccos\Big(\frac{\langle J_{x}\rangle}{|{\vec{J}}|\sin(\theta)}\Big) & \text{if}~~\langle J_{y}\rangle\leq0,\end{array}\right.\end{align}
 where $|{\vec{J}}|=\sqrt{\langle J_{x}\rangle^{2}+\langle J_{y}\rangle^{2}+\langle J_{z}\rangle^{2}}$
is the magnitude of the mean spin. With respect to $\vec{n}_{0}$,
the other two orthogonal bases are given as \begin{align}
\vec{n}_{1} & =(-\sin{\phi},\cos{\phi},0),\\
\vec{n}_{2} & =(\cos{\theta}\cos{\phi},\cos{\theta}\sin{\phi},-\sin{\theta}).\end{align}
 The above expressions are valid for $\theta\neq0,\pi$. For $\theta=0,\pi$,
the mean spin is along the $\pm z$ direction, and the possible choices
of $\phi$ can be $0$ or $\pi$.

The second step now is to find the minimal variance of $J_{\vec{n}_{\perp}}=\vec{J}\cdot\vec{n}_{\perp}$.
The direction $\vec{n}_{\perp}$ can be represented as\begin{equation}
\vec{n}_{\perp}=\vec{n}_{1}~O^{T}=\vec{n}_{1}\cos\varphi+\vec{n}_{2}\sin\varphi,\label{phi}\end{equation}
 where $O$ is a $2\times2$ orthogonal matrix that performs rotations
in the normal plane. The variance $(\Delta J_{\vec{n}_{\perp}})^{2}$
can be written as\begin{equation}
\left(\Delta J_{\vec{n}_{\perp}}\right)^{2}=\left\langle J_{\vec{n}_{\perp}}^{2}\right\rangle =\vec{n}_{\perp}\Gamma\vec{n}_{\perp}^{T},\end{equation}
 where the symmetric matrix\begin{equation}
\Gamma=\begin{pmatrix}\left\langle J_{\vec{n}_{1}}^{2}\right\rangle  & \text{Cov}\left(J_{\vec{n}_{1}},J_{\vec{n}_{2}}\right)\\
\text{Cov}\left(J_{\vec{n}_{1}},J_{\vec{n}_{2}}\right) & \left\langle J_{\vec{n}_{2}}^{2}\right\rangle \end{pmatrix},\end{equation}
 in which \begin{eqnarray}
\text{Cov}\left(J_{\vec{n}_{1}},J_{\vec{n}_{2}}\right) & = & \frac{1}{2}\left\langle \left[J_{\vec{n}_{1}},J_{\vec{n}_{2}}\right]_{+}\right\rangle -\left\langle J_{\vec{n}_{1}}\right\rangle \left\langle J_{\vec{n}_{2}}\right\rangle \notag\\
 & = & \frac{1}{2}\left\langle \left[J_{\vec{n}_{1}},J_{\vec{n}_{2}}\right]_{+}\right\rangle ,\end{eqnarray}
 is the covariance between $J_{\vec{n}_{1}}$ and $J_{\vec{n}_{2}}$,
and $\left[X,Y\right]_{+}=XY+YX$ is the anti-commutator. In the above
equation, $\left\langle J_{\vec{n}_{1}}\right\rangle =\left\langle J_{\vec{n}_{2}}\right\rangle =0$
since $\vec{n}_{1}$ and $\vec{n}_{2}$ are perpendicular to the MSD.
The variance can be written as \begin{equation}
\left(\Delta J_{\vec{n}_{\perp}}\right)^{2}=\vec{n}_{1}O^{T}\Gamma O\vec{n}_{1}^{T},\end{equation}
 and the matrix $O$ can be chosen such that \begin{equation}
O^{T}\Gamma O=\text{diag}\{\lambda_{-},~\lambda_{+}\},\end{equation}
 where the eigenvalues\begin{equation}
\lambda_{\pm}=\frac{1}{2}\left[\left\langle J_{\vec{n}_{1}}^{2}+J_{\vec{n}_{2}}^{2}\right\rangle \pm\sqrt{\left(\left\langle J_{\vec{n}_{1}}^{2}-J_{\vec{n}_{2}}^{2}\right\rangle \right)^{2}+4\text{Cov}\left(J_{\vec{n}_{1}},J_{\vec{n}_{2}}\right)^{2}}\right],\end{equation}
 and $\min\left(\Delta J_{\vec{n}_{\perp}}\right)^{2}=\lambda_{-}$
(See \ref{nGamman}), thus the squeezing parameters becomes\begin{equation}
\xi_{S}^{2}=\frac{4}{N}\lambda_{-}=\frac{2}{N}\left[\left\langle J_{\vec{n}_{1}}^{2}+J_{\vec{n}_{2}}^{2}\right\rangle \pm\sqrt{\left(\left\langle J_{\vec{n}_{1}}^{2}-J_{\vec{n}_{2}}^{2}\right\rangle \right)^{2}+4\text{Cov}\left(J_{\vec{n}_{1}},J_{\vec{n}_{2}}\right)^{2}}\right].\label{KU}\end{equation}
 The optimal squeezing angle in Eq.~(\ref{phi}) is given as\begin{equation}
\varphi=\left\{ \begin{array}{ll}
\frac{1}{2}\arccos\Big(\frac{-A}{\sqrt{A^{2}+B^{2}}}\Big) & \text{if}~~B\leq0,\\
\pi-\frac{1}{2}\arccos\Big(\frac{-A}{\sqrt{A^{2}+B^{2}}}\Big) & \text{if}~~B>0,\end{array}\right.\label{optsquangle}\end{equation}
 where we define\begin{equation}
A\equiv\langle J_{\vec{n}_{1}}^{2}-J_{\vec{n}_{2}}^{2}\rangle,\text{ \ }\; B\equiv2\text{Cov}\left(J_{\vec{n}_{1}},J_{\vec{n}_{2}}\right).\label{AB}\end{equation}

We know that the parameter $\xi_{S}^{2}=1$ for the uncorrelated pure
CSS $|\theta,\phi\rangle$ in Eq.~(\ref{CSS}). Thus, if there are
certain quantum correlations among the elementary spins, we may have
$\xi_{S}^{2}<1$, i.e., the fluctuation in one direction is reduced,
as shown in Fig.~\ref{fig_spin_bloch}. Therefore, the squeezing
parameter $\xi_{S}^{2}$ has natural connections with quantum
correlations (entanglement). Indeed, it has been found that
$\xi_{S}^{2}$ has a very close relation with quantities such as
negative correlations~\cite{Ulam-Orgikh2001} and concurrence
\cite{Wang2003}, and we will discuss them in
Sec.~\ref{entanglement}.

\subsection{Squeezing parameter $\xi_{R}^{2}$ given by Wineland \textit{et al.}}

Now, we discuss the spin-squeezing parameter proposed by Wineland
\textit{et al}.~\cite{WINELAND1992,WINELAND1994} in the study of
Ramsey spectroscopy. The squeezing parameter $\xi_{R}^{2}$ is the
ratio of the fluctuations between a general state and a CSS in the
determination of the resonance frequency in Ramsey spectroscopy. The
CSS here acts as a noise reference state. In contrast with
$\xi_{S}^{2}$, which is the analog of bosonic squeezing, the
parameter $\xi_{R}^{2}$ is substantially connected to the
improvement of the sensitivity of angular-momentum states to
rotations, and thus is attractive for experiments.

Here we use a simple graphical way to describe this type of
parameter, as shown in Fig.~\ref{fig_spin_bloch}, while the
mathematical form and essential physics are the same as for Ramsey
spectroscopy and the Mach-Zender interferometer employed in
Ref.~\cite{WINELAND1994}. More details of the Ramsey process will be
discussed in Sec.~\ref{sub:metrology}.

Consider now a spin state $|\psi\rangle$. Without loss of
generality, we assume the MSD to be along the $z$ direction, and
thus $\left\langle J_{x}\right\rangle =\left\langle
J_{y}\right\rangle =0$. This spin state is shown in
Fig.~\ref{fig_spin_bloch}, the circular or elliptical disks
represent the variance $\Delta J_{\vec{n}_{\perp}}$, which is also
called the projection noise. This state can be represented by a cone
ending in a Bloch sphere.

Let us now rotate the state around the $x$-axis. Then, in the Heisenberg
picture, we have\begin{equation}
J_{y}^{\text{out}}=\exp(i\phi J_{x})J_{y}\exp(-i\phi J_{x})=\cos\phi J_{y}-\sin\phi J_{z}.\end{equation}
 From the above equation, we can immediately obtain\begin{align}
\langle J_{y}^{\text{out}}\rangle & =-\sin\phi\langle J_{z}\rangle,\\
(\Delta J_{y}^{\text{out}})^{2} & =\cos^{2}\phi(\Delta J_{y})^{2}+\sin^{2}\phi(\Delta J_{z})^{2}-\frac{1}{2}\sin(2\phi)\langle\lbrack J_{y},J_{z}]_{+}\rangle.\end{align}
 According to the error propagation formula, $\Delta x=\Delta f(x)/\left|\partial\left\langle f(x)\right\rangle /\partial x\right|$,
the phase sensitivity $\Delta\phi$ can be calculated as

\begin{equation}
\Delta\phi=\frac{\Delta J_{y}^{\text{out}}}{\left\vert \partial\langle J_{y}^{\text{out}}\rangle/\partial\phi\right\vert }=\frac{\Delta J_{y}^{\text{out}}}{\left\vert \cos\phi\langle J_{z}\rangle\right\vert }.\end{equation}
 For the rotation angle $\phi\sim0,$ $(\Delta J_{y}^{\text{out}})^{2}\sim(\Delta J_{y})^{2}$
and $\cos\phi\sim1.$ Thus, the above equation reduces to\begin{equation}
\Delta\phi=\frac{\Delta J_{y}}{\left\vert \langle J_{z}\rangle\right\vert },\end{equation}
 which allows us to know the phase sensitivity from the expectations
and variances of the collective spin operators.

From the above procedure, for the more general case that the MSD is
not along the $z$ direction, we can obtain the phase sensitivity
as\begin{equation} \Delta\phi=\frac{\Delta
J_{\vec{n}_{\perp}}}{\left\vert \langle\vec{J}\rangle\right\vert
}.\label{xi_r_phase}\end{equation}
 For the CSS, we can obtain the phase sensitivity from Eqs.~(\ref{expectation})
and ~(\ref{variances}) as\begin{equation}
\left(\Delta\phi\right)_{\text{CSS}}=\frac{1}{\sqrt{N}},\end{equation}
 which is the so-called standard quantum limit (SQL) or shot-noise
limit. This is the limit of precision in atomic interferometry experiments
when we use uncorrelated atoms.

The squeezing parameter proposed by Wineland \textit{et al}. is
defined as \cite{WINELAND1992,WINELAND1994}\begin{equation}
\xi_{R}^{2}=\frac{\left(\Delta\phi\right)^{2}}{\left(\Delta\phi\right)_{\text{CSS}}^{2}}=\frac{N\left(\Delta
J_{\vec{n}_{\perp}}\right)^{2}}{\left\vert
\langle\vec{J}\rangle\right\vert
^{2}}.\label{ss_wineland}\end{equation}
 This is the ratio of the phase sensitivity of a general state versus
the CSS. Here, we choose the direction $\vec{n}_{\perp}$ where
$\Delta J_{\vec{n}_{\perp}}$ is minimized. This definition is
related to the $\xi_{S}^{2}$ by Kitagawa and Ueda via
\begin{equation}
\xi_{R}^{2}=\left(\frac{j}{|\langle\vec{J}\rangle|}\right)^{2}\xi_{S}^{2}.\label{ss_w_k}\end{equation}
 Since $j=\frac{N}{2}\geq|\langle\vec{J}\rangle|,$ we have $\xi_{S}^{2}\leq\xi_{R}^{2}$.
Even though these two parameters are similar (when $j=|\langle\vec{J}\rangle|$,
$\xi_{R}^{2}=\xi_{S}^{2}$), their physical meanings are different.
When $\xi_{R}^{2}<1$, the state is spin squeezed, and its phase sensitivity
to rotation is improved over the shot-noise limit. According to Eq.~(\ref{xi_r_phase}),
the phase sensitivity can be written as \begin{equation}
\Delta\phi=\frac{\xi_{R}}{\sqrt{N}}.\label{phase_xi_r}\end{equation}
 If $\xi_{R}^{2}<1$, $\Delta\phi<\left(\Delta\phi\right)_{\text{CSS}}$
beats the shot-noise limit.

The lower bound of the phase sensitivity is given by the Heisenberg
uncertainty relation $(\Delta J_{\vec{n}_{\perp}})^{2}(\Delta
J_{\vec{n}_{\perp}^{\prime}})^{2}\geq\frac{1}{4}\vert\langle
J_{\vec{n}}\rangle \vert ^{2}$, from which we have
$\xi_{R}^{2}\,4(\Delta J_{\vec{n}_{\perp}^{\prime}})^{2}/N\geq1$.
Using the relations $N^{2}/4=j^{2}\geq\langle
J_{\vec{n}}^{2}\rangle\geq\left(\Delta J_{\vec{n}}\right)^{2}$ and
the fact that the largest eigenvalue of $J_{\vec{n}}^{2}$ is
$j^{2},$ we obtain\begin{equation}
\xi_{R}^{2}\geq\frac{1}{N}.\end{equation}
 By using Eq.~(\ref{phase_xi_r}), we further have \begin{equation}
\Delta\phi\geq\left(\Delta\phi\right)_{\text{HL}}=\frac{1}{N},\end{equation}
 where $\left(\Delta\phi\right)_{\text{HL}}$ is the Heisenberg limit
(HL)~\cite{Giovannetti2004}.

Reference~\cite{WINELAND1992} proposed another definition,
\begin{equation}
\xi_{H^{\prime\prime}}^{2}=\frac{j}{|\langle\vec{J}\rangle|}\xi_{S}^{2}=\frac{2\left(\Delta
J_{\vec{n}_{\perp}}\right)^{2}}{|\langle\vec{J}\rangle|}.\label{xir_tilde}\end{equation}
 We may choose another set of three orthogonal directions $\{\vec{n}_{\perp},\vec{n}_{\perp}^{\prime},\vec{n}_{0}\}$.
Then the uncertainty relation can be written as \begin{equation}
(\Delta J_{\vec{n}_{\perp}})^{2}(\Delta J_{\vec{n}_{\perp}^{\prime}})^{2}\geq\left\vert \left\langle J_{\vec{n}_{0}}\right\rangle \right\vert ^{2}/4.\label{spin_comm_relll}\end{equation}
 Thus, the parameter $\xi_{H^{\prime\prime}}^{2}$ is naturally defined
in this coordinate system. We will show that $\xi_{H^{\prime\prime}}^{2}=1$
for the CSS, indicating that this parameter is appropriate for characterizing
spin squeezing. For a CSS, the variance of $J_{\vec{n}_{0}}$ is zero,
i.e., no fluctuations along the MSD (See \ref{varn}), $\left(\Delta J_{\vec{n}_{0}}\right)^{2}=0$.
The variances of the angular momenta in the plane perpendicular to
$\vec{n}_{0}$ behave similar to bosonic operators, for which we obtain
(See \ref{varn})\begin{equation}
\left(\Delta J_{\alpha}\right)^{2}=\left(\Delta J_{\beta}\right)^{2}=j/2,\end{equation}
 where the subscripts $\alpha$ and $\beta$ denote two orthogonal
axes perpendicular to the mean-spin direction $\vec{n}_{0}$. So,
we have $\xi_{H^{\prime\prime}}^{2}=1$.

The three squeezing definitions $\xi_{S}^{2}$, $\xi_{H^{\prime\prime}}^{2}$,
and $\xi_{R}^{2}$ discussed above are not equivalent, however, they
satisfy the following relation\begin{equation}
\xi_{S}^{2}\leq\xi_{H^{\prime\prime}}^{2}\leq\xi_{R}^{2},\label{cond}\end{equation}
 since $j\geq|\langle\vec{J}\rangle|$ always holds. Therefore, if
a state is spin squeezed according to criterion $\xi_{R}^{2}<1$,
it is definitely squeezed according to parameter $\xi_{H^{\prime\prime}}^{2}$
and $\xi_{R}^{2}$, and thus $\xi_{R}^{2}<1$ is the most stringent
condition of squeezing among these three parameters in Eq.~(\ref{cond}).

We also note that, the projection noise can be schematically
visualized in Fig.~\ref{fig_spin_bloch}, where we show the
projection noise distributions for a CSS and a SSS. For CSS, take
$|j,j\rangle$ for example, and the MSD is along the $z$ direction.
If we measure $J_{y}$, the expectation value $\left\langle
J_{y}\right\rangle =0$, while the variance $\left(\Delta
J_{y}\right)^{2}=j/2$. The probability of the outcome states
$|j,m\rangle_{y}$ obeys the binomial distribution\begin{equation}
P\left(m\right)=\left\vert \left\langle j,j|j,m\right\rangle
_{y}\right\vert ^{2}=\left\vert \left\langle
j,m\left|\exp\left({-i\frac{\pi}{2}J_{x}}\right)\right|j,j\right\rangle
\right\vert ^{2}=\frac{1}{2^{N}}\left(\begin{array}{c}
N\\
N-m\end{array}\right),\end{equation}
 where $|j,m\rangle_{y}=\exp({i\frac{\pi}{2}J_{x}})|j,m\rangle$ is
the eigenstate of $J_{y}$ with eigenvalue $m$. The phase sensitivity
$\Delta\phi$ is determined by the width of the binomial
distribution, and a spin-squeezed state may thus have a sub-binomial
distribution~\cite{Takahashi1999}.

\subsection{Other spin-squeezing parameters}

In this section, we review some other definitions of spin squeezing,
and discuss their applications. Reference \cite{Sorensen2001}
proposed a spin-squeezing parameter $\xi_{R^{\prime}}^{2}$, which is
a criterion for multipartite entanglement. The subscript
$R^{\prime}$ indicates the close relationship between the spin
squeezing parameters $\xi_{R^{\prime}}^{2}$ and $\xi_{R}^{2}$. The
parameter $\xi_{R^{\prime}}^{2}$ is defined as \begin{equation}
\xi_{R^{\prime}}^{2}=\frac{N\left(\Delta
J_{\vec{n}_{1}}\right)^{2}}{\left\langle
J_{\vec{n}_{2}}\right\rangle ^{2}+\left\langle
J_{\vec{n}_{3}}\right\rangle ^{2}}.\label{defxir}\end{equation}
 For spin-1/2 many-body systems, it has been proved \cite{Sorensen2001}
that, if $\xi_{R^{\prime}}^{2}<1$ the state is entangled. However,
an entangled state may not be necessarily squeezed. Based on this
parameter, spin squeezing is directly connected to multipartite
entanglement. Inspired by this finding, many works studied the
relations between spin squeezing and entanglement (see,
e.g.,~\cite{Guehne2009}). The choices of the directions
$\vec{n}_{i}$ are arbitrary, and if $\vec{n}_{2}$ is chosen to be
the MSD, while $\vec{n}_{1}$ is chosen to minimize the variance
$\Delta J_{\vec{n}_{1}}$, then $\xi_{R^{\prime}}^{2}$ reduces to
$\xi_{R}^{2}$.

Below, we show that for the CSS, $\xi_{R^{\prime}}^{2}=1$. Substituting
Eq.~(\ref{expvar}) into Eq.~(\ref{defxir}), we obtain\begin{equation}
\xi_{R^{\prime}}^{2}=\frac{1-\left(\vec{n}_{0}\cdot\vec{n}_{1}\right)^{2}}{\left(\vec{n}_{0}\cdot\vec{n}_{2}\right)^{2}+\left(\vec{n}_{0}\cdot\vec{n}_{3}\right)^{2}}=1.\end{equation}
 In summary, $\xi_{R^{\prime}}^{2}$ can be viewed as a generalization
of $\xi_{R}^{2}$, and also provides a useful criterion for many-body
entanglement.

In the study of Dicke states $|j,m\rangle$ in BEC, Raghavan
\textit{et al}. \cite{Raghavan2001} proposed a new kind of squeezing
parameter\begin{equation} \xi_{D}^{2}=\frac{N\left(\Delta
J_{\vec{n}}\right)^{2}}{N^{2}/4-\left\langle
J_{\vec{n}}\right\rangle ^{2}},\label{spsqu_Raghavan}\end{equation}
 where the subscript $D$ indicates that this parameter can detect
entanglement in Dicke states. If $\xi_{D}^{2}<1$, the state is
squeezed along $\vec{n}$. Actually, in Ref.~\cite{Raghavan2001}, the
authors only consider the variance $\left(\Delta J_{z}\right)^{2}$
along the $z$-axis. For a CSS $|\theta,\phi\rangle$,
$\xi_{D}^{2}=1$, which can be directly proved by substituting Eq.
(\ref{expvar}) into Eq.~(\ref{spsqu_Raghavan}). For all entangled
symmetric Dicke states $|j,m\rangle$, we will show that this
parameter can detect entanglement in Dicke states in
Sec.~\ref{states}.

As discussed previously, an important application of the spin
squeezing parameter is to detect entanglement. This kind of
entanglement criterion is based on collective-spin inequalities, and
it is attractive for experiments, because in practice we cannot
always address individual particles, while the expectation values
and variances for collective spin operators are easier to measure,
such as in population spectroscopies. Therefore, Tóth \textit{et
al.}~\cite{Toth2009a} generalized the spin squeezing definitions and
gave a set of spin inequalities. We find that one of the spin
inequalities is suitable to be rewritten as a new type of
spin-squeezing parameter, and this inequality reads\begin{equation}
(N-1)\left(\Delta J_{\vec{n}_{1}}\right)^{2}\geq\left\langle
J_{\vec{n}_{2}}^{2}\right\rangle +\left\langle
J_{\vec{n}_{3}}^{2}\right\rangle
-N/2,\label{toth_ineq}\end{equation}
 which holds for any separable states, and the violation of this inequality
indicates entanglement. This inequality can be further written in
the following form\begin{equation}
N\left(\Delta J_{\vec{n}_{1}}\right)^{2}\geq\left\langle \vec{J}\,^{2}\right\rangle -\left\langle J_{\vec{n}_{1}}\right\rangle ^{2}-N/2,\end{equation}
 and one can define a spin squeezing parameter related to entanglement\begin{equation}
\xi_{E}^{2}=\frac{N\left(\Delta J_{\vec{n}_{1}}\right)^{2}}{\left\langle \vec{J}\,^{2}\right\rangle -N/2-\left\langle J_{\vec{n}_{1}}\right\rangle ^{2}},\end{equation}
 whose form is similar to $\xi_{D}^{2}.$ In fact, for symmetric states
with only Dicke states populated, $\left\langle \vec{J}\,^{2}\right\rangle =N/2(N/2+1)$,
and then $\xi_{E}^{2}$ reduces to $\xi_{D}^{2}.$ Thus, for the CSS,
$\xi_{E}^{2}=\xi_{D}^{2}=1.$ In Table \ref{tab_definition}, we give
a summary of the spin squeezing parameters discussed above, and we
also show their values for the CSS state $|\theta,\phi\rangle$ in
Eq.~(\ref{CSS}).

\begin{table}[H]
 \caption{This table shows the definitions for different spin-squeezing parameters.
Except for $\xi_{H}^{2}$ and $\xi_{H^{\prime}}^{2}$, the other spin-squeezing
parameters all equal to 1 for the CSS.}

\label{tab_definition} 

\centering{}\begin{tabular}{c||c|c|c}
\hline
Squeezing parameters  & Definitions  & Coherent spin state  & References \tabularnewline
\hline
\hline
\parbox[c]{2cm}{%
$\xi_{H}^{2}$%
}  & %
\parbox[c]{3.7cm}{%
\vspace{2mm}
 ${\displaystyle \frac{2(\Delta J_{\vec{n}_{1}})^{2}}{|\langle J_{\vec{n}_{2}}\rangle|}}$
\vspace{2mm}
}  & %
\parbox[c]{3.7cm}{%
${\displaystyle \frac{1-(\vec{n}_{0}\cdot\vec{n}_{1})^{2}}{\left\vert \left\langle \vec{n}_{0}\cdot\vec{n}_{2}\right\rangle \right\vert }}$%
}  & ~\cite{WALLS1981} \tabularnewline \hline
\parbox[c]{2cm}{%
$\xi_{H'}^{2}$%
}  & %
\parbox[c]{3.7cm}{%
\vspace{2mm}
 ${\displaystyle \frac{2(\Delta J_{\vec{n}_{1}})^{2}}{\sqrt{\langle J_{\vec{n}_{2}}\rangle^{2}+\langle J_{\vec{n}_{3}}\rangle^{2}}}}$
\vspace{2mm}
}  & %
\parbox[c]{3.7cm}{%
{${\displaystyle \sqrt{1-(\vec{n}_{0}\cdot\vec{n}_{1})^{2}}}$}%
}  & \cite{Prakash2005} \tabularnewline \hline
\parbox[c]{2cm}{%
$\xi_{{H''}}^{2}$%
}  & %
\parbox[c]{3.7cm}{%
\vspace{2mm}
 ${\displaystyle \frac{2(\Delta J_{\vec{n}_{\perp}})_{\min}^{2}}{|\langle\vec{J}\rangle|}}$
\vspace{2mm}
}  & %
\parbox[c]{1cm}{%
$1$ %
}  & ~\cite{WINELAND1992} \tabularnewline \hline
\parbox[c]{2cm}{%
$\xi_{S}^{2}$%
}  & %
\parbox[c]{3.7cm}{%
\vspace{2mm}
 ${\displaystyle \frac{4(\Delta J_{\vec{n}_{\perp}})_{\min}^{2}}{N}}$
\vspace{2mm}
}  & %
\parbox[c]{1cm}{%
${\displaystyle 1}$%
}  & \cite{KITAGAWA1993} \tabularnewline \hline
\parbox[c]{2cm}{%
$\xi_{R}^{2}$%
}  & %
\parbox[c]{3.7cm}{%
\vspace{2mm}
 ${\displaystyle \frac{N^{2}}{4\langle\vec{J}\rangle^{2}}\xi_{S}^{2}}$
\vspace{2mm}
}  & %
\parbox[c]{1cm}{%
${\displaystyle 1}$ %
}  & ~\cite{WINELAND1994} \tabularnewline \hline
\parbox[c]{2cm}{%
$\xi_{R'}^{2}$%
}  & %
\parbox[c]{3.7cm}{%
\vspace{2mm}
 ${\displaystyle \frac{N(\Delta J_{\vec{n}_{1}})^{2}}{\langle J_{\vec{n}_{2}}\rangle^{2}+\langle J_{\vec{n}_{3}}\rangle^{2}}}$
\vspace{2mm}
}  & %
\parbox[c]{1cm}{%
${\displaystyle 1}$%
}  & ~\cite{Sorensen2001} \tabularnewline \hline
\parbox[c]{2cm}{%
$\xi_{D}^{2}$%
}  & %
\parbox[c]{3.7cm}{%
\vspace{2mm}
 ${\displaystyle \frac{N(\Delta J_{\vec{n}})^{2}}{N^{2}/4-\langle J_{\vec{n}}\rangle^{2}}}$
\vspace{2mm}
}  & %
\parbox[c]{1cm}{%
$1$%
}  & ~\cite{Raghavan2001} \tabularnewline \hline
\parbox[c]{2cm}{%
$\xi_{E}^{2}$%
}  & %
\parbox[c]{3.9cm}{%
\vspace{2mm}
 \ ${\displaystyle \frac{N\left(\Delta J_{\vec{n}_{1}}\right)^{2}}{\left\langle \vec{J}^{2}\right\rangle -N/2-\left\langle J_{\vec{n}_{1}}\right\rangle ^{2}}}$
\vspace{2mm}
}  & %
\parbox[c]{1cm}{%
${\displaystyle 1}$%
}  & ~\cite{Toth2009a} \tabularnewline \hline
\end{tabular}
\end{table}

\subsection{Rotationally invariant extensions of squeezing parameters}

As discussed above, for a given state, the denominators of
parameters $\xi_{R^{\prime}}^{2}$, $\xi_{D}^{2}$, and $\xi_{E}^{2}$
depend on the choice of the directions. Unlike parameters
$\xi_{S}^{2}$ and $\xi_{R}^{2}$, for which the denominators are
constants. Thus it is difficult to determine the minima of
$\xi_{R^{\prime}}^{2}$, $\xi_{D}^{2}$, and $\xi_{E}^{2}$. Inspired
by T\"{o}h's discussions ~\cite{Toth2009a,Rivas2008,Rivas2008a}, we
will give slightly different definitions for $\xi_{R^{\prime}}^{2}$
and $\xi_{D}^{2}$, and the new definitions provide us a simple way
to determine whether a state is spin squeezed.

Spin squeezing with respect to $\xi_{R^{\prime}}^{2},$
$\xi_{D}^{2},$ and $\xi_{E}^{2}$ are equivalent to the following
three inequalities\begin{align}
N\left(\Delta J_{\vec{n}_{1}}\right)^{2} & <\left\langle J_{\vec{n}_{2}}\right\rangle ^{2}+\left\langle J_{\vec{n}_{3}}\right\rangle ^{2},\notag\\
N\left(\Delta J_{\vec{n}}\right)^{2} & <\frac{N^{2}}{4}-\left\langle J_{\vec{n}}\right\rangle ^{2},\notag\\
N\left(\Delta J_{\vec{n}}\right)^{2} & <\left\langle \vec{J}\,\right\rangle ^{2}-\frac{N}{2}-\left\langle J_{\vec{n}}\right\rangle ^{2},\end{align}
 The right-hand sides of the above three inequalities can be written
in a rotation-invariant form by adding $\left\langle J_{\vec{n}_{1}}\right\rangle ^{2}$
to both sides of the first inequality, and $\left\langle J_{\vec{n}}\right\rangle ^{2}$
to the second and third ones. Then we obtain\begin{align}
\left(N-1\right)\left(\Delta J_{\vec{n}}\right)^{2}+\left\langle J_{\vec{n}}^{2}\right\rangle  & <\langle\vec{J}\,\rangle^{2},\notag\\
\left(N-1\right)\left(\Delta J_{\vec{n}}\right)^{2}+\left\langle J_{\vec{n}}^{2}\right\rangle  & <\frac{N^{2}}{4},\notag\\
\left(N-1\right)\left(\Delta J_{\vec{n}}\right)^{2}+\left\langle J_{\vec{n}}^{2}\right\rangle  & <\left\langle \vec{J}\,\right\rangle ^{2}-\frac{N}{2},\end{align}
 which are equivalent to the original inequalities, and we replace
$\vec{n}_{1}$ with $\vec{n}$ in the first inequality. Note that
the right-hand side of the inequality is invariant under rotations.
Thus, to detect entanglement, we shall find the minimum value of the
left-hand side by rotating the direction $\vec{n}$. So, the spin
squeezing parameters can be defined as\begin{align}
\tilde{\xi}_{R^{\prime}}^{2} & =\frac{\min_{_{\vec{n}}}[\left(N-1\right)\left(\Delta J_{\vec{n}}\right)^{2}+\left\langle J_{\vec{n}}^{2}\right\rangle ]}{\langle\vec{J}\,\rangle^{2}},\text{ \ }\\
\tilde{\xi}_{D}^{2} & =\frac{4}{N^{2}}\min_{_{\vec{n}}}[\left(N-1\right)\left(\Delta J_{\vec{n}}\right)^{2}+\left\langle J_{\vec{n}}^{2}\right\rangle ],\\
\tilde{\xi}_{E}^{2} & =\frac{\min_{_{\vec{n}}}[\left(N-1\right)\left(\Delta J_{\vec{n}}\right)^{2}+\left\langle J_{\vec{n}}^{2}\right\rangle ]}{\left\langle \vec{J}\,\right\rangle ^{2}-{N}/{2}}.\label{xiexie}\end{align}
 These parameters are rotationally invariant.

The minimization procedure works as below. First the term
$\left(N-1\right)\left(\Delta J_{\vec{n}}\right)^{2}+\left\langle
J_{\vec{n}}^{2}\right\rangle $ can be written as~\cite{Toth2009a}
\begin{equation} \left(N-1\right)\left(\Delta
J_{\vec{n}}\right)^{2}+\left\langle J_{\vec{n}}^{2}\right\rangle
=\vec{n}\Gamma\vec{n}^{T},\end{equation}
 where the superscript $T$ denotes the transpose, and $\Gamma$ is
a $3\times3$ matrix, which is defined as \begin{equation}
\Gamma=(N-1)\gamma+\mathbf{C},\label{gamma}\end{equation}
 where the covariance matrix $\gamma$ is given as\begin{equation}
\gamma_{kl}=\mathbf{C}_{kl}-\langle J_{k}\rangle\langle J_{l}\rangle\;\;\text{for}\;\; k,l\in\{x,y,z\}=\{1,2,3\},\label{comatrix}\end{equation}
 with a correlation matrix\begin{equation}
\mathbf{C}_{kl}=\frac{1}{2}\langle J_{l}J_{k}+J_{k}J_{l}\rangle.\label{cmatrix}\end{equation}
 The minimum value of $\vec{n}\Gamma\vec{n}^{T}$ is the minimum eigenvalue
of $\Gamma$ (See Appendix B), and thus the spin-squeezing parameter
based on these inequality can be defined
as~\cite{Wang2010}\begin{equation}
\tilde{\xi}_{R^{\prime}}^{2}=\frac{\lambda_{\min}}{\langle\vec{J}\rangle^{2}},\quad\text{
\
}\tilde{\xi}_{D}^{2}=\frac{4\lambda_{\min}}{N^{2}},\quad\tilde{\xi}_{E}^{2}=\frac{\lambda_{\min}}{\langle\vec{J}^{2}\rangle-N/2}.\end{equation}
 If $\tilde{\xi}_{E}^{2}<1$, the state is spin squeezed and entangled.
In the case when $\big{\langle} \vec{J}^{2}\,big{\rangle}
=j\left(j+1\right)$ and when $\lambda_{\min}$ is obtained in the
$\vec{n}_{\perp}$-direction, we find
$\tilde{\xi}_{E}^{2}=\xi_{S}^{2}$, and thus $\tilde{\xi}_{E}^{2}$
can be regarded as a generalization of $\xi_{S}^{2}$. For a CSS,
$\lambda_{\min}=j^{2}$ and $\tilde{\xi}_{E}^{2}=1$.

It is interesting to note that the left-hand sides of the above inequalities
are equal to $\vec{n}\Gamma\vec{n}^{T}$, and thus the minimum values
are just $\lambda_{\min}$. Therefore, the new definitions of squeezing
parameters become proportional to $\lambda_{\text{min}}$. Although
these new parameters are not quantitatively equal to their original
ones, they are qualitatively equivalent to the original ones in the
sense that they all can detect whether a state is spin squeezed or
not.

In the above discussion, spin squeezing was defined for many-qubit
states belonging to the maximum multiplicity subspace of the
collective angular momentum operator $\vec{J}$. This means that
these states exhibit particle exchange symmetry. The concept of spin
squeezing is therefore restricted to symmetric many-body systems
that are accessible to collective operations alone. There is no
\emph{a} \emph{priori} reason for this restriction in real
experiments, thus Ref.~\cite{Devi2003a} explored the possibility of
extending the concept of spin squeezing to multi-qubit systems,
where individual qubits are controllable in the sense that they are
accessible to local operations. This requires a criterion of spin
squeezing that exhibits invariance under local unitary operations on
the qubits.

Now we introduce a local unitary invariant spin squeezing criterion
for $N$ qubits. We denote the mean-spin direction for a local qubit
$i$ by\begin{equation}
\hat{n}_{i0}=\frac{\left\langle \vec{\sigma}_{i}\right\rangle }{\left\vert \left\langle \vec{\sigma}_{i}\right\rangle \right\vert }.\end{equation}
 Associating a mutually-orthogonal set \{$\hat{n}_{i\perp}$, $\hat{n}_{i\perp^{\prime}}$,
$\hat{n}_{i0}$\} of unit vectors with each qubit, we may define the
collective operators\begin{equation}
\mathcal{J}_{\perp}=\frac{1}{2}\sum_{i=1}^{N}\vec{\sigma}_{i}\cdot\hat{n}_{i\perp},\quad\text{
}\mathcal{J}_{\perp^{\prime}}=\frac{1}{2}\sum_{i=1}^{N}\vec{\sigma}_{i}\cdot\hat{n}_{i\perp},\text{
}\quad\mathcal{J}_{0}=\frac{1}{2}\sum_{i=1}^{N}\vec{\sigma}_{i}\cdot\hat{n}_{i0},\label{locally_invar}\end{equation}
 which satisfy the usual angular momentum commutation relations as
Eq.~(\ref{angular_commu_rel}), and this leads to the uncertainty
relation\begin{equation}
\left(\Delta\mathcal{J}_{\perp}\right)\left(\Delta\mathcal{J}_{\perp^{\prime}}\right)\geq\frac{1}{2}\left\langle \mathcal{J}_{0}\right\rangle .\end{equation}
 Analogous to the above spin definitions, like $\xi_{S}^{2}$ and
$\xi_{R}^{2}$, we may define the corresponding squeezing
parameters~\cite{Devi2003a} as\begin{equation}
\bar{\xi}_{S}^{2}=\frac{4\left(\Delta\mathcal{J}_{\perp}\right)_{\min}^{2}}{N},\text{
}\quad\bar{\xi}_{R}^{2}=\frac{N\left(\Delta\mathcal{J}_{\perp}\right)_{\min}^{2}}{\left\langle
\mathcal{J}_{0}\right\rangle ^{2}}.\label{SU2_invar}\end{equation}
 One important advantage of these squeezing parameters~\cite{Devi2003a}
is that they are locally unitary invariant due to the definition of
$\mathcal{J}$ in Eq.~(\ref{locally_invar}). Immediately, one can
verify that for all $N$-body separable states, the above two parameters
in Eq.~(\ref{SU2_invar}) equal to one, since the single spin-1/2
particle is never spin squeezed. This is in contrast with the original
squeezing parameters $\xi_{S}^{2}$ and $\xi_{R}^{2}$, which are
equal to 1 only for CSS.

\subsection{Spin-squeezing parameters for states with parity}

In preceding discussions, we reviewed several different spin-squeezing
parameters presented in the literature. In the following, we mainly
focus on two typical spin-squeezing parameters, $\xi_{S}^{2}$ and
$\xi_{R}^{2}$, which have wide applications in detecting entanglement,
and in quantum metrology, etc. Besides these two parameters, we also
consider parameter $\tilde{\xi}_{E}^{2}$ in some parts of the paper,
since $\tilde{\xi}_{E}^{2}$ has close relations with $\xi_{S}^{2}$
and can be viewed as a generalization of $\xi_{S}^{2}$. Furthermore,
$\tilde{\xi}_{E}^{2}$ is significant in discussing spin squeezing
and entanglement.

We shall see that most typical spin-squeezed states are of a fixed
parity. In this subsection we will study spin-squeezing parameters
$\xi_{S}^{2}$, $\xi_{R}^{2}$ and $\tilde{\xi}_{E}^{2}$ for states
with even (odd) parity, and give relations among them. For general
states, it is hard to find explicit relations of these three parameters.
At first we explain the parity of a spin state and restrict ourselves
to states with exchange symmetry and with only Dicke states being
populated. For these states, we may define the parity operator\begin{equation}
P=\left(-1\right)^{J_{z}+j},\label{parity_op}\end{equation}
 with eigenvalues $p=1$ and $-1$ corresponding to even and odd parity,
respectively. Thus the state $|j,-j\rangle$, which means each spin
pointing to the $-z$ direction, has even parity, and for a Dicke
state $|j,-j+n\rangle$, if $n$ is even (odd), the state has even
(odd) parity. In general, if a spin state is spanned only by Dicke
states of even (odd) parity, this state is of even (odd) parity. States
with parity are very common, e.g., the state generated by the one-axis
twisting model, which we will discuss in Sec.~\ref{states}.

\subsubsection{Spin-squeezing parameters $\xi_{S}^{2}$ and $\xi_{R}^{2}$}

Here, we consider spin-squeezing parameters $\xi_{S}^{2}$ and $\xi_{R}^{2}$.
For states with parity, we have the following relations,\begin{equation}
\left\langle J_{\alpha}\right\rangle =\left\langle J_{\alpha}J_{z}\right\rangle =\left\langle J_{z}J_{\alpha}\right\rangle =0,\text{ \ }\alpha=x,y,\end{equation}
 which mean the MSD is along the $z$ direction. The above equation
leads to the following zero covariance \begin{equation}
\text{Cov}\big(J_{z},J_{\vec{n}_{\perp}}\big)=\frac{1}{2}\left\langle \left[J_{z},J_{\vec{n}_{\perp}}\right]_{+}\right\rangle -\left\langle J_{z}\right\rangle \left\langle J_{\vec{n}_{\perp}}\right\rangle =0,\label{cov_Jz_Jn}\end{equation}
 which implies that there are no spin correlations between the longitudinal
($z$) and transverse directions ($x$-$y$ plane). The MSD is along
the $z$-axis, then the general expression for the spin-squeezing
parameter $\xi_{S}^{2}$ in Eq.~(\ref{KU}) reduces to \cite{Wang2003}
\begin{equation}
\xi_{S}^{2}=\frac{2}{N}\left(\left\langle J_{x}^{2}+J_{y}^{2}\right\rangle -\left\vert \left\langle J_{-}^{2}\right\rangle \right\vert \right).\label{min_spsqu_s}\end{equation}

Due to the exchange symmetry, we find\begin{align}
\left\langle J_{\alpha}\right\rangle  & =\frac{N}{2}\left\langle \sigma_{1\alpha}\right\rangle ,\text{ }\label{exp}\\
\left\langle J_{\alpha}^{2}\right\rangle  & =\frac{N}{4}+\frac{N\left(N-1\right)}{4}\left\langle \sigma_{1\alpha}\sigma_{2\alpha}\right\rangle ,\text{ }\alpha=x,y,z,\label{global_local1}\\
\left\langle \vec{J}^{2}\right\rangle  & =\frac{3N}{4}+\frac{N\left(N-1\right)}{4}\left\langle \vec{\sigma}_{1}\cdot\vec{\sigma}_{2}\right\rangle ,\label{jjj}\\
(\Delta J_{\alpha})^{2} & =\frac{N}{4}\left[1+(N-1)\left\langle \sigma_{1\alpha}\sigma_{2\alpha}\right\rangle -N\left\langle \sigma_{1\alpha}\right\rangle ^{2}\right]\notag\\
 & =\frac{N}{4}\left[1+N\mathcal{C}_{\alpha\alpha}-\left\langle \sigma_{1\alpha}\sigma_{2\alpha}\right\rangle \right],\label{var}\end{align}
 where\begin{equation}
\mathcal{C}_{\alpha\alpha}=\left\langle \sigma_{1\alpha}\sigma_{2\alpha}\right\rangle -\left\langle \sigma_{1\alpha}\right\rangle \left\langle \sigma_{2\alpha}\right\rangle \label{eq:czz}\end{equation}
 is the correlation function along the $\alpha$-direction. Furthermore,
we obtain \begin{align}
\left\langle J_{-}^{2}\right\rangle  & =N\left(N-1\right)\left\langle \sigma_{1-}\sigma_{2-}\right\rangle ,\label{jm}\\
\left\langle J_{x}^{2}+J_{y}^{2}\right\rangle  & =\frac{N}{2}+\frac{N\left(N-1\right)}{4}\left\langle \sigma_{1x}\sigma_{2x}+\sigma_{1y}\sigma_{2y}\right\rangle \notag\\
 & =\frac{N}{2}+\frac{N\left(N-1\right)}{2}\left\langle \sigma_{1+}\sigma_{2-}+\sigma_{1-}\sigma_{2+}\right\rangle .\label{global_local2}\end{align}
 By substituting Eqs.~(\ref{jm}) and (\ref{global_local2}) into
Eq.~(\ref{min_spsqu_s}), we obtain\begin{equation}
\xi_{S}^{2}=1-2\left(N-1\right)\left(\left\vert \left\langle \sigma_{1-}\sigma_{2-}\right\rangle \right\vert -\left\langle \sigma_{1+}\sigma_{2-}\right\rangle \right),\text{ \ }\label{spsqp}\end{equation}
 where we have used $\left\langle \sigma_{1+}\sigma_{2-}\right\rangle =\left\langle \sigma_{1-}\sigma_{2+}\right\rangle ,$
which results from the exchange symmetry. From Eq.~(\ref{exp}) and
the relation between the parameters $\xi_{R}^{2}$ and $\xi_{S}^{2},$
we find\begin{equation}
\xi_{R}^{2}=\left(\frac{N}{2|J_{z}|}\right)^{2}\xi_{S}^{2}=\frac{\xi_{S}^{2}}{\left\langle \sigma_{1z}\right\rangle ^{2}}.\label{ss_w_k_1}\end{equation}
 The above expressions for the two spin-squeezing parameters establish
explicit relations between spin squeezing and local two-spin correlations,
and we will further discuss these relations in Sec.~\ref{entanglement}.
\begin{table}[H]
 \caption{Comparison between different squeezing parameters for states with
parity (extended from Ref.~\protect\cite{Wang2010}). In the third
column, simplified expressions are displayed for squeezing
parameters for states with parity. The squeezing parameters are also
expressed in terms of local expectations (fourth column).}

\label{tab_comp}

\centering{}\begin{tabular}{c||c|c|c}
\hline
Parameters  & Definitions  & States with parity  & In terms of local expectations \tabularnewline
\hline
\hline
\parbox[c]{2cm}{%
$\xi_{S}^{2}$%
}  & %
\parbox[c]{3cm}{%
\vspace{2mm}
 ${\displaystyle \frac{4(\Delta J_{\vec{n}_{\perp}})_{\min}^{2}}{N}}$\vspace{2mm}
}  & %
\parbox[c]{4cm}{%
\vspace{2mm}
 ${\displaystyle \frac{2}{N}(\langle J_{x}^{2}+J_{y}^{2}\rangle-|\langle J_{-}^{2}\rangle|)}$\vspace{2mm}
}  & %
\parbox[c]{6cm}{%
\vspace{2mm}
 ${\displaystyle 1-2(N-1)\left(|\langle\sigma_{1-}\sigma_{2-}\rangle|-\langle\sigma_{1+}\sigma_{2-}\rangle\right)}$\vspace{2mm}
} \tabularnewline
\hline
\parbox[c]{2cm}{%
$\xi_{R}^{2}$%
}  & %
\parbox[c]{3cm}{%
\vspace{2mm}
 ${\displaystyle \frac{N^{2}}{4\langle\vec{J}\rangle^{2}}\xi_{S}^{2}}$\vspace{2mm}
}  & %
\parbox[c]{4cm}{%
{\vspace{2mm}
 ${\displaystyle \frac{\xi_{S}^{2}}{4\langle J_{z}\rangle^{2}/N^{2}}}$\vspace{2mm}
 } %
}  & %
\parbox[c]{6cm}{%
\vspace{2mm}
 ${\displaystyle \frac{\xi_{S}^{2}}{\langle\sigma_{1z}\rangle^{2}}}$\vspace{2mm}
} \tabularnewline
\hline
\parbox[c]{2cm}{%
$\tilde{\xi}_{R'}^{2}$%
}  & %
\parbox[c]{3cm}{%
\vspace{2mm}
 ${\displaystyle \frac{\lambda_{\min}}{\langle\vec{J}\rangle^{2}}}$\vspace{2mm}
}  & %
\parbox[c]{4cm}{%
\vspace{2mm}
 ${\displaystyle \frac{\min\left\{ \xi_{S}^{2},\varsigma^{2}\right\} }{4\langle J_{z}\rangle^{2}/N^{2}}}$\vspace{2mm}
}  & %
\parbox[c]{6cm}{%
\vspace{2mm}
 ${\displaystyle \frac{\min\{\xi_{S}^{2},1+(N-1)\mathcal{C}_{zz}\}}{\langle\sigma_{1z}\rangle^{2}}}$\vspace{2mm}
} \tabularnewline
\hline
\parbox[c]{2cm}{%
$\tilde{\xi}_{D}^{2}$%
}  & %
\parbox[c]{3cm}{%
\vspace{2mm}
 ${\displaystyle \frac{4\lambda_{\min}}{N^{2}}}$\vspace{2mm}
}  & %
\parbox[c]{4cm}{%
\vspace{2mm}
 ${\displaystyle \min\left\{ \xi_{S}^{2},\varsigma^{2}\right\} }$\vspace{2mm}
}  & %
\parbox[c]{6cm}{%
\vspace{2mm}
 ${\displaystyle \min\{\xi_{S}^{2},1+(N-1)\mathcal{C}_{zz}\}}$\vspace{2mm}
} \tabularnewline
\hline
\parbox[c]{2cm}{%
$\tilde{\xi}_{E}^{2}$%
}  & %
\parbox[c]{3cm}{%
\vspace{2mm}
 ${\displaystyle \frac{\lambda_{\min}}{\langle\vec{J}^{2}\rangle-N/2}}$\vspace{2mm}
}  & %
\parbox[c]{4cm}{%
\vspace{2mm}
 ${\displaystyle \frac{\min\left\{ \xi_{S}^{2},\varsigma^{2}\right\} }{4\langle\vec{J}^{2}\rangle/N^{2}-2/N}}$\vspace{2mm}
}  & %
\parbox[c]{6cm}{%
\vspace{2mm}
 ${\displaystyle \frac{\min\{\xi_{S}^{2},1+(N-1)\mathcal{C}_{zz}\}}{(1-N^{-1})\langle\vec{\sigma}_{1}\cdot\vec{\sigma}_{2}\rangle+N^{-1}}}$\vspace{2mm}
} \tabularnewline
\hline
\end{tabular}
\end{table}

\subsubsection{Spin-squeezing parameter $\tilde{\xi}_{E}^{2}$}

Here, we derive explicit expression of $\tilde{\xi}_{E}^{2}$ for
states with parity. By using Eq.~(\ref{cov_Jz_Jn}), the correlation
matrix (\ref{cmatrix}) is simplified to the following form\begin{equation}
\mathbf{C}=\begin{pmatrix}\left\langle J_{x}^{2}\right\rangle  & \mathbf{C}_{xy} & 0\\
\mathbf{C}_{xy} & \left\langle J_{y}^{2}\right\rangle  & 0\\
0 & 0 & \left\langle J_{z}^{2}\right\rangle
\end{pmatrix},\end{equation} where $\mathbf{C}_{xy}=\big{\langle}
\left[J_{x},J_{y}\right]_{+}\big{\rangle} /2$. From the correlation
matrix $\mathbf{C}$ and the definition of covariance matrix $\gamma$
given by Eq.~(\ref{comatrix}), one finds\begin{equation}
\mathbf{\Gamma}=\begin{pmatrix}N\left\langle J_{x}^{2}\right\rangle  & N\mathbf{C}_{xy} & 0\\
N\mathbf{C}_{xy} & N\left\langle J_{y}^{2}\right\rangle  & 0\\
0 & 0 & N\left(\Delta J_{z}\right)^{2}+\left\langle J_{z}\right\rangle ^{2}\end{pmatrix}.\end{equation}
 This matrix has a block-diagonal form and the eigenvalues of the
$2\times2$ block are obtained as\begin{equation}
\lambda_{\pm}=\frac{N}{2}\Big(\left\langle J_{x}^{2}+J_{y}^{2}\right\rangle \pm\left\vert \left\langle J_{-}^{2}\right\rangle \right\vert \Big).\end{equation}
 Therefore, the smallest eigenvalue $\lambda_{\min}$ of $\Gamma$
is obtained as\begin{equation}
\lambda_{\min}=\min\left\{ \lambda_{-},N\left(\Delta J_{z}\right)^{2}+\left\langle J_{z}\right\rangle ^{2}\right\} .\end{equation}
 It is interesting to see that the eigenvalue $\lambda_{-}$ is simply
related to the spin squeezing parameter $\xi_{S}^{2}$ in Eq.~(\ref{min_spsqu_s})
via\begin{equation}
\xi_{S}^{2}=\frac{4}{N^{2}}\lambda_{-}.\end{equation}
 Thus, from the definition of $\tilde{\xi}_{E}^{2}$ given by Eq.~(\ref{xiexie}),
we finally find\begin{equation}
\tilde{\xi}_{E}^{2}=\frac{\min\left\{ \xi_{S}^{2},\varsigma^{2}\right\} }{4\left\langle \vec{J}^{2}\right\rangle /N^{2}-2/N},\label{xi_t_local}\end{equation}
 where \begin{equation}
\varsigma^{2}=\frac{4}{N^{2}}\left(N\left(\Delta J_{z}\right)^{2}+\left\langle J_{z}\right\rangle ^{2}\right).\end{equation}

The meaning of $\varsigma^{2}$ will be clear by substituting Eqs.~(\ref{exp})
and (\ref{var}) into the above equation. Then, we obtain \begin{equation}
\varsigma^{2}=N\left(\Delta J_{z}\right)^{2}+\left\langle J_{z}^{2}\right\rangle =1+\left(N-1\right)\mathcal{C}_{zz}.\end{equation}
 Parameter $\varsigma^{2}$ is just a linear function of the correlation
function $\mathcal{C}_{zz}$ given in Eq.~(\ref{eq:czz}). From Eq.~(\ref{jjj})
and the above expression, we write parameter $\tilde{\xi}_{E}^{2}$
$\,$in terms of expectations of local operators as \begin{equation}
\tilde{\xi}_{E}^{2}=\frac{\min\left\{ \xi_{S}^{2},\varsigma^{2}\right\} }{\left(1-N^{-1}\right)\left\langle \vec{\sigma}_{1}\cdot\vec{\sigma}_{2}\right\rangle +N^{-1}}.\end{equation}
 In the special case when only spin $j=N/2$ is populated, i.e., $\left\langle \vec{J}\,^{2}\right\rangle =N/2(N/2+1)$,
we have $\left\langle \vec{\sigma}_{1}\cdot\vec{\sigma}_{2}\right\rangle =1,$
and therefore, the above equation reduces to \begin{equation}
\tilde{\xi}_{E}^{2}=\min\left\{ \xi_{S}^{2},\varsigma^{2}\right\} .\label{Toth_pc}\end{equation}
 Thus, the relations among the spin-squeezing parameters $\xi_{S}^{2}$,
$\xi_{R}^{2}$, and $\tilde{\xi}_{E}^{2}$ are clear for states with
parity, and in this case, the spin-squeezing parameters are
determined by pairwise correlations in the $z$-axis and the $x$-$y$
plane. The above results are summarized in Table~\ref{tab_comp}
extended from Ref.~\cite{Wang2001}.

\subsubsection{Dicke States}

Below, we first consider Dicke states and simple superpositions of
two Dicke states. The Dicke states $|j,m\rangle$ are entangled state
except for $m=\pm j$. Since we mainly consider an ensemble of spin-1/2
particles, the state $|j,m\rangle$ can be written as\begin{align}
|j,m\rangle & =\sqrt{\frac{\left(j-m\right)!}{\left(j+m\right)!\left(2j\right)!}}\, J_{+}^{j+m}|j,-j\rangle\notag\\
 & =\sqrt{\frac{\left(j-m\right)!}{\left(j+m\right)!\left(2j\right)!}}\left(\sum_{i=1}^{N}\sigma_{i+}\right)^{j+m}\!\!\!\!\!|1\rangle^{\otimes N}.\end{align}
 Since the Dicke states $|j,m\rangle$ are eigenstates of $J_{z}$
with eigenvalue $m$, the MSD is along the $z$-axis. A single Dicke
state has either even or odd parity. Thus, we can use formulas shown
in Table~\ref{tab_comp}, and we only need to calculate $\xi_{S}^{2}$
and $\varsigma^{2}$.

From Eq.~(\ref{min_spsqu_s}), the squeezing parameter $\xi_{S}^{2}$
can be written as\begin{equation}
\xi_{S}^{2}=1+j-\frac{1}{j}\Big(\langle J_{z}^{2}\rangle+|\langle J_{-}^{2}\rangle|\Big),\label{xiss}\end{equation}
 which is determined by two expectation values: $\langle J_{z}^{2}\rangle$
and $\langle J_{-}^{2}\rangle.$ For the Dicke states, we obtain \begin{equation}
\xi_{S}^{2}=1+\frac{j^{2}-m^{2}}{j}\geq1,\end{equation}
 since $j\geq m$. The equal sign holds for $m=\pm j$, when the Dicke
state becomes a CSS. The quantity $\varsigma^{2}$ is obtained as
\begin{align}
\varsigma^{2}=\frac{1}{j^{2}}\left[N\left(\Delta J_{z}\right)^{2}+\left\langle J_{z}\right\rangle ^{2}\right]=\frac{m^{2}}{j^{2}}\leq1.\end{align}
 Thus, from Table \ref{tab_comp}, one finds\begin{equation}
\tilde{\xi}_{D}^{2}\,=\,\tilde{\xi}_{E}^{2}\,=\,\min\left(\xi_{S}^{2},\varsigma^{2}\right)\,=\,\varsigma^{2}\,\leq\,1,\end{equation}
 indicating that the state is squeezed and entangled for $m\neq\pm j.$
The parameters $\xi_{R}^{2}$ and $\tilde{\xi}_{R^{\prime}}^{2}$
are given by \begin{align}
\xi_{R}^{2} & ~=~\left(\frac{j}{m}\right)^{2}\xi_{S}^{2}\,\geq\,1\text{,}\\
\tilde{\xi}_{R^{\prime}}^{2} &
~=~\frac{\varsigma^{2}}{m^{2}/j^{2}}\,=\,1,\end{align} thus spin
squeezing according to the three parameters $\xi_{S}^{2},$
$\xi_{R}^{2},$ and $\tilde{\xi}_{R^{\prime}}^{2}$ cannot reflect the
underlying entanglement in the Dicke states.

\subsection{Relations between spin squeezing and bosonic squeezing}

Above, we reviewed some basic concepts about bosonic and spin
squeezing, and now we will demonstrate the relationship between
these two mathematically distinct, yet intuitively connected
squeezing~\cite{Liu1996,Wang2003a}. It has been shown in
Ref.~\cite{Wang2003a} that the spin-squeezing parameter
$\xi_{S}^{2}$ reduces to the bosonic squeezing in the limit of large
number of atoms and small excitations. For this purpose, we consider
the principal quadrature squeezing~defined as (\ref{principle_squ}).
The definition of $\zeta_{B}^{2}$ provides an atomic squeezing
counterpart to bosonic squeezing, and is similar to the definition
of $\xi_{S}^{2}$, both of them searching for minimum squeezing.

It is well-known that the Heisenberg-Weyl algebra describing the
bosonic mode can be obtained by contraction from the SU(2) algebra
describing the ensemble of atoms~\cite{Chumakov1999}. To see this,
we define $b\equiv J_{-}/\sqrt{2j}$ and $b^{\dagger}\equiv
J_{+}/\sqrt{2j}.$ From the commutation relation (\ref{bose_commu}),
we have \begin{equation}
\lbrack\mathcal{N},b^{\dagger}]=b^{\dagger},\quad[\mathcal{N},b]=-b,\quad[b,b^{\dagger}]=1-\frac{\mathcal{N}}{j},\end{equation}
 where $\mathcal{N}=J_{z}+j$ is the `number operator', and its eigenvalues
vary from 0 to $N$, counting the number of excited atoms. In the
limit of $j\rightarrow\infty$ and small $\langle\mathcal{N}\rangle,$
the operators $\mathcal{N},$ $b,$ and $b^{\dagger}$ satisfy the
commutation relations of the Heisenberg-Weyl algebra. Note that, when
we take this limit, the average number of excited atoms $\langle\mathcal{N}\rangle$
should be much less than the total number of atoms $N$.

We can also use the usual Holstein-Primakoff
transformation~\cite{Holstein1940}:
\begin{equation}
J_{+}=a^{\dagger}\sqrt{2j-a^{\dagger}a},\quad J_{-}=\sqrt{2j-a^{\dagger}a}~a,\quad J_{z}=a^{\dagger}a-j.\end{equation}
 In the limit of $j\rightarrow\infty,$ we have \begin{equation}
\frac{J_{+}}{\sqrt{2j}}\rightarrow a^{\dagger},\quad\frac{J_{-}}{\sqrt{2j}}\rightarrow a,\quad-\frac{J_{z}}{j}\rightarrow1,\label{limit2}\end{equation}
 by expanding the square root and neglecting terms of $O(1/j)$. We
see that the bosonic system and the atomic spin system are connected
by the large-$j$ limit from an algebraic point of view.

To display this connection, we consider even (odd) states. These states
refer to those being a superposition of even (odd) Fock states for
bosonic systems, and those being a superposition of Dicke states $|n\rangle_{j}\equiv|j,-j+n\rangle$
with the even (odd) excitations for the atomic systems. The Dicke
states $|n\rangle_{j}$ satisfy $\mathcal{N}|n\rangle_{j}=n|n\rangle_{j}$.
Specifically, even and odd bosonic coherent states have been realized
experimentally in various physical systems. The even (odd) states
serve as examples for demonstrating connections between bosonic and
atomic squeezing. For even (odd) states, $\langle a\rangle=0$; thus,
from Eq.~(\ref{principle_squ}), we obtain \begin{equation}
\zeta_{B}^{2}=1+2\langle a^{\dagger}a\rangle-2|\langle a^{2}\rangle|.\label{squu1}\end{equation}
 Obviously, one necessary condition for squeezing is that $|\langle a^{2}\rangle|\neq0$.

For even (odd) atomic states, the squeezing parameter $\xi_{S}^{2}$
has already been given in Eq.~(\ref{min_spsqu_s}), and we rewrite
it as\begin{align}
\xi_{S}^{2} & =1+2\langle\mathcal{N}\rangle-\frac{\langle\mathcal{N}^{2}\rangle}{j}-\frac{|\langle J_{-}^{2}\rangle|}{j}.\label{squuu}\end{align}
 Using Eq.~(\ref{limit2}), in the limit of $j\rightarrow\infty$,
we find that Eq.~(\ref{squuu}) reduces to Eq.~(\ref{squu1}) for even
(odd) states. Also, we may find that the squeezing parameter
$\xi_{R}^{2}$ also reduce to $\zeta_{B}^{2}$ in this limit. This
result displays a direct connection between bosonic squeezing and
atomic squeezing. From an experimental point of view, the number of
atoms is typically large enough, so the observed atomic squeezing is
expected to approximate the bosonic quadrature squeezing. As a
remark, Eqs.~(\ref{squu1}) and (\ref{squuu}) obtained for even and
odd states are also applicable to arbitrary states, which is
discussed in Ref.~\cite{Wang2003a}.

\section{Generation of spin squeezing with nonlinear twisting Hamiltonians\label{states}}

In this section, we discuss generating spin-squeezed states with the
one-axis twisting and two-axis twisting Hamiltonians. The one-axis
twisting Hamiltonian is one of the most important models studied in
generating spin squeezing, both theoretically and experimentally. It
also describes a nonlinear rotator, and was studied in
Ref.~\cite{Sanders1989} before its applications in spin squeezing.
The proposal of using these two types of twisting Hamiltonians to
generate spin-squeezed states is directly inspired by using the
nonlinear Hamiltonian (\ref{bose_squ_Ham}) to produce bosonic
squeezing.

In Sec.~3.1, we first present the analytical results of the
evolution of the one-axis twisting. Then we discuss how to implement
this Hamiltonian in a two-component BEC, and by using large-detuned
light-atom interactions. The experimental progresses are reviewed in
Sec.~\ref{experiment-BEC}. Then in Sec.~3.2, we discuss the two-axis
twisted state. As compared with the one-axis twisting, one can
obtain higher degree of squeezing by using the two-axis twisting
Hamiltonian. However, this type of Hamiltonian is not easy to be
implemented in experiments, and analytical results are not available
for arbitrary system size.

\subsection{One-axis twisting Hamiltonian}

Here we discuss the generation of spin squeezing by using the
one-axis twisting Hamiltonian~\cite{KITAGAWA1993,Sanders1989}. The
one-axis twisting model is very simple, and is one of the most
widely studied models in generating spin-squeezed state
\cite{AGARWAL1994,Agarwal1997,Fernholz2008,Law2001,Raghavan2001,Jaksch2002,Jenkins2002,Wesenberg2002,Micheli2003,Molmer2003,Ocak2003,Choi2005,Jaaskelainen2005,Takeuchi2005,Deb2006,Lee2006,Chaudhury2007,Jin2007,Jin2007a,Esteve2008,Fernholz2008,Grond2009,Jin2009,Leroux2010,Schleier-Smith2010}.
It works in analogy to the squeezing operator,
Eq.~(\ref{bose_squ_Ham2}), in photon system and has been implemented
in BEC via atomic collisions
\cite{Sorensen2001,Orzel2001,Esteve2008,Gross2010,Riedel2010} and in
atomic
ensembles~\cite{AGARWAL1994,Agarwal1997,Chaudhury2007,Fernholz2008,Schleier-Smith2010,Leroux2010}.
Moreover, it allows simple derivations of various analytical
results.

\subsubsection{One-axis twisted states}

Consider now an ensemble of $N$ spin-1/2 particles with exchange
symmetry, and assume that its dynamical properties can be described
by collective operators $J_{\alpha}$, $\alpha=x,y,z$. The one-axis
twisting Hamiltonian reads\begin{equation}
H_{\text{OAT}}~=~\chi J_{x}^{2}~=~\frac{\chi}{4}\sum_{k,l=1}^{N}\sigma_{kx}\sigma_{lx},\label{OAT_Ham}\end{equation}
 which is a nonlinear operator with coupling constant $\chi$ and
involves all pairwise interactions, which indicates that the spin-squeezed
states generated by this Hamiltonian may exhibit pairwise correlations.
The most commonly used twisting Hamiltonian is along the $z$-axis,
with the initial state being a CSS pointing along the $x$-axis. Since
the twisting is along the $x$-axis, we choose the initial CSS along
the $z$-axis to make our analysis consistent with the $z$-axis twisting
version. Here we prefer the $x$-axis twisting Hamiltonian satisfying
\begin{equation}
\left[H_{\text{OAT}},P\right]=0,\end{equation}
 where $P$ is the parity operator given by Eq.~(\ref{parity_op}).
We choose the initial state as $|j,-j\rangle=|1\rangle^{\otimes N}$.
Considering its dynamic evolution, the spin-squeezed state at time
$t$ is formally written as\begin{equation}
|\Psi(t)\rangle=\exp\left({-i\theta J_{x}^{2}/2}\right)|1\rangle^{\otimes N},\label{initial}\end{equation}
 where \begin{equation}
\theta=2\chi t\label{angle}\end{equation}
 is the one-axis twisting angle. This state is the one-axis twisted
state with even parity, and the MSD is along the $z$ direction, thus
the results derived in Sec.~\ref{definition} can be directly used
here.

\begin{table}[tbph]
 \caption{Expectation values of local observables for the one-axis twisted state.}

\label{Exp_val_OAT}

\centering{}\begin{tabular}{c||c}
\hline
\parbox[c]{2cm}{%
$\langle\sigma_{1z}\rangle$%
}  & %
\parbox[c]{3cm}{%
\vspace{2mm}
 ${\displaystyle -\cos^{N-1}\left(\theta/2\right)}$\vspace{2mm}
} \tabularnewline
\hline
\parbox[c]{2cm}{%
$\langle\sigma_{1z}\sigma_{2z}\rangle$%
}  & %
\parbox[c]{3cm}{%
\vspace{2mm}
 ${\displaystyle \frac{1}{2}\left(1+\cos^{N-2}\theta\right)}$\vspace{2mm}
} \tabularnewline
\hline
\parbox[c]{2cm}{%
$\langle\sigma_{1+}\sigma_{2-}\rangle$%
}  & %
\parbox[c]{3cm}{%
\vspace{2mm}
 ${\displaystyle \frac{1}{8}\left(1-\cos^{N-2}\theta\right)}$\vspace{2mm}
} \tabularnewline
\hline
\parbox[c]{2cm}{%
$\langle\sigma_{1-}\sigma_{2-}\rangle$%
}  & %
\parbox[c]{7cm}{%
\vspace{1mm}
 ${\displaystyle -\frac{1}{8}\left(1-\cos^{N-2}\theta\right)-\frac{i}{2}\sin\left(\theta/2\right)\cos^{N-2}\left(\theta/2\right)}$\vspace{1mm}
} \tabularnewline
\hline
\end{tabular}
\end{table}

The expectation values needed for calculating spin squeezing parameters
are derived in \ref{oat}, and are summarized in Table \ref{Exp_val_OAT}.
By substituting expressions $\langle\sigma_{1+}\sigma_{2-}\rangle$
and $\langle\sigma_{1-}\sigma_{2-}\rangle$ into Eq.~(\ref{spsqp})$,$
we obtain \begin{equation}
\xi_{S}^{2}=1-C_{r}=1-(N-1)C,\label{xis_conc}\end{equation}
 where\begin{equation}
C=\frac{1}{4}\left\{ \left[\Big(1-\cos^{N-2}\theta\Big)^{2}+16\sin^{2}\left(\frac{\theta}{2}\right)\cos^{2N-4}\left(\frac{\theta}{2}\right)\right]^{1/2}-\Big(1-\cos^{N-2}\theta\Big)\right\} .\label{C2}\end{equation}
 It will be clear in Sec.~\ref{entanglement} that the quantity $C$
is the concurrence~\cite{Wootters1998}, measuring entanglement of
two spin $s=1/2$ particles.

In the case that $N\gg1$ and $\left\vert \theta\right\vert \ll1$,
while $N\left\vert \theta\right\vert \gg1$ and $N\left\vert
\theta\right\vert ^{2}\ll1$, we can expand Eq.~(\ref{C2}) and find
that the spin-squeezing parameter scales
as~\cite{KITAGAWA1993}\begin{equation}
\xi_{S}^{2}~\sim~N^{-2/3},\end{equation}
 at $\left\vert \theta\right\vert =\theta_{0}=12^{1/6}\left(N/2\right)^{-2/3}$.
Since $\theta$ is very small and $N$ is large, at $\theta_{0}$
we find\begin{equation}
\left\langle \sigma_{1z}\right\rangle ~\sim~-\exp\left[-\left(\frac{4}{3}N\right)^{-1/3}\right],\end{equation}
 and then, for large enough $N$, we have\begin{equation}
\xi_{R}^{2}~\sim~N^{-2/3}.\end{equation}
 Therefore, the projection noise is reduced by a factor of $N^{-2/3}$.
The optimal squeezing angle for the one-axis twisting
is~\cite{KITAGAWA1993}
\begin{equation}
\delta\sim\frac{1}{2}\arctan\left(N^{-1/3}\right),\end{equation}
 which varies with the particle number $N$; although if $N$ is large
enough, $\delta$ is close to $0$.

\begin{figure}[H]

\begin{centering}
\includegraphics[width=11cm]{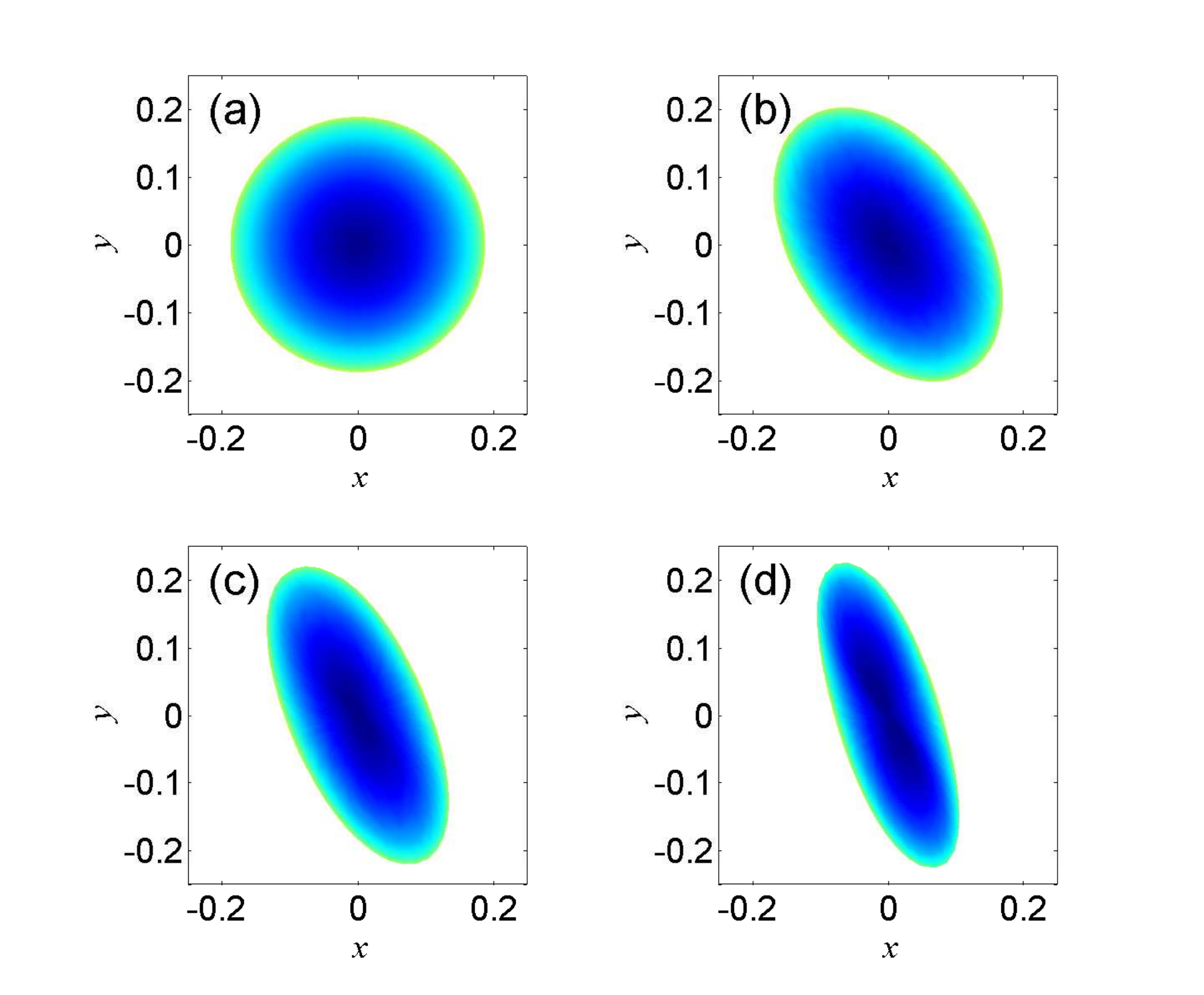}
\par\end{centering}

\caption{(Color online) Husimi $Q$ function of spin-squeezed states generated
by the one-axis twisting model for various times: (a) $\chi t=0$,
(b) $\chi t=0.1$, (c) $\chi t=0.2$ and (d) $\chi t=0.3$. The system
size is $N=60$, and at the beginning $\chi t=0$, the state is a
CSS and thus the $Q$ function is a circle. The optimal squeezing
angle rotates with time. }

\label{fig_qpd_oat_a}
\end{figure}

The squeezing and the dynamic evolution of the state can also be
illustrated by calculating the Husimi-$Q$ function\begin{equation}
Q\left(\theta_{0},\phi_{0}\right)=\left\vert \left\langle
\theta_{0},\phi_{0}|\Psi(t)\right\rangle \right\vert
^{2},\end{equation}
 which represents the quasiprobability distribution of $|\Psi\left(t\right)\rangle$,
and $|\theta_{0},\phi_{0}\rangle$ is the CSS given in Eq.~(\ref{CSS}).
The Husimi-$Q$ function is shown in Fig.~\ref{fig_qpd_oat_a}, where
the coordinate we used is determined by\begin{align}
x & =Q\cos\theta_{0}\cos\phi_{0},\notag\\
y & =Q\cos\theta_{0}\sin\phi_{0}.\end{align}
 As we can see that, the initial CSS is a circle given by the Husimi
function, and during the evolution, the Husimi-$Q$ function becomes
squeezed and elliptical, while the squeezing angle rotates.

\begin{table}[ptb]
 \caption{Spin-squeezing parameters in terms of the rescaled concurrence $C_{r}$
given by Eqs.~(\ref{xis_conc}) and (\protect\ref{C2}) for the
one-axis twisted state.}

\label{Tab_SP_OAT}

\centering{}\begin{tabular}{c||c}
\hline
\parbox[c]{1cm}{%
$\xi_{S}^{2}$%
}  & %
\parbox[c]{2cm}{%
\vspace{2mm}
 ${\displaystyle 1-C_{r}}$\vspace{2mm}
} \tabularnewline
\hline
\parbox[c]{1cm}{%
$\xi_{R}^{2}$%
}  & %
\parbox[c]{2cm}{%
\vspace{2mm}
 ${\displaystyle \frac{1-C_{r}}{\langle\sigma_{1z}\rangle^{2}}}$\vspace{2mm}
} \tabularnewline
\hline
\parbox[c]{1cm}{%
$\tilde{\xi}_{E}^{2}$%
}  & %
\parbox[c]{2cm}{%
\vspace{2mm}
 ${\displaystyle 1-C_{r}}$\vspace{2mm}
} \tabularnewline
\hline
\end{tabular}
\end{table}

Now we discuss another two parameters $\xi_{R}^{2}$ and $\tilde{\xi}_{E}^{2}$.
Parameter $\xi_{R}^{2}$ is easily obtained, as we know both $\xi_{S}^{2}$
and $\langle\sigma_{1z}\rangle^{2}$. To obtain $\tilde{\xi}_{E}^{2}$,
we also need to derive $\mathcal{C}_{zz}$ and $\left\langle \vec{\sigma}_{1}\cdot\vec{\sigma}_{2}\right\rangle $
as seen from Table \ref{tab_comp}. For this state, $\langle\vec{\sigma}_{1}\cdot\vec{\sigma}_{2}\rangle=1$,
thus the expression of $\tilde{\xi}_{E}^{2}$ reduces to \begin{align}
\tilde{\xi}_{E}^{2} & =\min\left\{ \xi_{S}^{2},\varsigma^{2}\right\} \notag\\
 & =\min\left\{ 1-C_{r},1+(N-1)\mathcal{C}_{zz}\right\} .\label{thirdd}\end{align}
 By substituting expressions of $\langle\sigma_{1z}\sigma_{2z}\rangle^{2}$
and $\langle\sigma_{1z}\rangle^{2}$ (Table \ref{Exp_val_OAT}) into
definition of the correlation function, we obtain, \begin{equation}
\mathcal{C}_{zz}=\frac{1}{2}\left(1+\cos^{N-2}\theta\right)-\cos^{2N-2}{(\theta/2)}\geq0.\label{c5}\end{equation}
 The proof of the above inequality is given in Appendix C of Ref.~\cite{Wang2010}.

As the correlation function $\mathcal{C}_{zz}$ and the rescaled concurrence
$C_{r}$ are always larger than zero, Eq.~(\ref{thirdd}) reduces
to \begin{equation}
\tilde{\xi}_{E}^{2}=\xi_{S}^{2}=1-C_{r}.\end{equation}
 So, for the state given by Eq.~(\ref{initial}), the spin-squeezing
parameters $\tilde{\xi}_{E}^{2}$ and $\xi_{S}^{2}$ are equal , and
we summarize these results in Table \ref{Tab_SP_OAT}.

\subsubsection{One-axis twisting with a transverse field}

In Ref.~\cite{Law2001}, it was found that a one-axis twisting
Hamiltonian with a transverse control field \begin{equation} H=\chi
J_{x}^{2}+BJ_{z},\label{driven_OAT}\end{equation}
 is more effective in generating squeezed states, where $B$ is the
strength of the external field. Reference \cite{Rojo2003} proved
that the optimally spin-squeezed states that maximize the
sensitivity of the Ramsey spectroscopy are eigensolutions of this
Hamiltonian. This type of Hamiltonian was considered in BEC
\cite{Jaksch2002,Jenkins2002,Choi2005,Jaaskelainen2005,Jin2007a,Grond2009},
and atomic ensembles \cite{Chaudhury2007}. The initial state is also
a CSS $|j,-j\rangle$, while the dynamic evolution cannot be solved
analytically, except for $N\leq3$, and numerical results show that
the external field leads to an improvement of the degrees of
squeezing in an extended period of time.

\subsubsection{One-axis twisting in Bose-Einstein condensates\label{sub:Mapping-a-two-component}}

Below, we discuss how to derive the one-axis twisting Hamiltonian
from a two-component BEC, which can be regarded as a BEC with atoms
in two internal states, or similarly, a BEC in a double-well
potential. We consider the case of $N$ atoms in two internal states
$|A\rangle$ and $|B\rangle$. The Hamiltonian of the system is given
by \cite{Sorensen2001,Sorensen2002}
$\left(\hbar\equiv1\right)$\begin{align}
H= & \int d\vec{r}\sum_{k=A,B}\left[\hat{\psi}_{k}^{\dagger}\left(\vec{r}\right)\hat{h}_{k}\hat{\psi}_{k}\left(\vec{r}\right)+\frac{g_{kk}}{2}~\hat{\psi}_{k}^{\dagger}\left(\vec{r}\right)\hat{\psi}_{k}^{\dagger}\left(\vec{r}\right)\hat{\psi}_{k}\left(\vec{r}\right)\hat{\psi}_{k}\left(\vec{r}\right)\right]\notag\\
 & +\int d\vec{r}~g_{AB}~\hat{\psi}_{A}^{\dagger}\left(\vec{r}\right)\hat{\psi}_{B}^{\dagger}\left(\vec{r}\right)\hat{\psi}_{A}\left(\vec{r}\right)\hat{\psi}_{B}\left(\vec{r}\right)\label{BEC_Ham1}\end{align}
 where the single-particle Hamiltonian \begin{equation}
\hat{h}_{k}=-\frac{\nabla^{2}}{2m}+V_{k}\left(\vec{r}\right)\end{equation}
 governs atoms in the internal state $k$, with atom mass $m$ and
trapping potential $V_{k}\left(\vec{r}\right)$. The bosonic field
operator $\hat{\psi}_{k}\left(\vec{r}\right)$ annihilates an atom
at position $\vec{r}$ in the internal state $k$, which obeys\begin{equation}
\left[\hat{\psi}_{k}\left(\vec{r}\right),\hat{\psi}_{k}^{\dagger}\left(\vec{r}^{\prime}\right)\right]=\delta\left(\vec{r}-\vec{r}^{\prime}\right).\end{equation}
 The interaction strengths \begin{equation}
g_{kl}=\frac{4\pi a_{kl}}{m},\label{eq:BEC_g}\end{equation}
 with $a_{kl}$ the $s$-wave scattering length and $g_{AA}$, $g_{BB}$,
and $g_{AB}$ are for collisions between atoms in states $A$, $B$,
and interspecies collisions, respectively.

Now we employ the single-mode approximation for each of the components,\begin{equation}
\hat{\psi}_{A}\left(\vec{r}\right)=\phi_{A}\left(\vec{r}\right)a,\text{ \ \ }\hat{\psi}_{B}\left(\vec{r}\right)=\phi_{B}\left(\vec{r}\right)b,\label{single_mode}\end{equation}
 where $\phi_{A}\left(\vec{r}\right)$ and $\phi_{B}\left(\vec{r}\right)$
are assumed to be real, $a$ and $b$ are the bosonic annihilation
operators that satisfy $\left[a,a^{\dagger}\right]=\left[b,b^{\dagger}\right]=1$
and $\left[a,b\right]=0$. In the single-mode approximation, the Hamiltonian
is rewritten as\begin{align}
H= & \left(\omega_{A}-U_{AA}\right)a^{\dagger}a+\left(\omega_{B}-U_{BB}\right)b^{\dagger}b\notag\\
 & +U_{AA}\left(a^{\dagger}a\right)^{2}+U_{BB}\left(b^{\dagger}b\right)^{2}+2U_{AB}a^{\dagger}ab^{\dagger}b,\label{BEC_Ham2}\end{align}
 where \begin{align}
\omega_{k} & =\int d\vec{r}~\phi_{k}^{\ast}\left(\vec{r}\right)~\hat{h}_{k}~\phi_{k}\left(\vec{r}\right),\notag\\
U_{kl} & =\frac{g_{kl}}{2}\int d\vec{r}~\left\vert \phi_{k}\left(\vec{r}\right)\right\vert ^{2}\left\vert \phi_{l}\left(\vec{r}\right)\right\vert ^{2}.\end{align}
 Note that, $\phi_{k}\left(\vec{r}\right)$ obeys the coupled Gross-Pitaevskii
equation, thus $\omega_{k}$ and $U_{kl}$ depend on the time $t$.
Now the effective Hamiltonian is rewritten by using angular momentum
operators via the following Schwinger representation\begin{align}
J_{z} & =\frac{1}{2}\left(a^{\dagger}a-b^{\dagger}b\right),\notag\\
J_{+} & =a^{\dagger}b,\text{ }J_{-}=ab^{\dagger},\label{Schwinger_Rep}\end{align}
 where the operator $2J_{z}$ measures the population difference between
states $|A\rangle$ and $|B\rangle$, and $J_{\pm}$ describe the
atomic tunneling between the two internal states. The total atom number
operator $\hat{N}=a^{\dagger}a+b^{\dagger}b$ is a conserved quantity
here. Using the Schwinger representation (\ref{Schwinger_Rep}) the
Hamiltonian (\ref{BEC_Ham2}) can be written as\begin{equation}
H=e\left(t\right)\hat{N}+E\left(t\right)\hat{N}^{2}+\delta\left(t\right)J_{z}+\chi\left(t\right)J_{z}^{2},\label{BEC_Ham3}\end{equation}
 where the nonlinear interaction coefficient is\begin{equation}
\chi\left(t\right)=U_{AA}+U_{BB}-2U_{AB},\label{eq:BEC_chi}\end{equation}
 and the other coefficients are\begin{align}
e\left(t\right) & =\left(\omega_{A}+\omega_{B}-U_{AA}-U_{BB}\right)/2,\nonumber \\
E\left(t\right) & =\left(U_{AA}+U_{BB}+2U_{AB}\right)/4,\nonumber \\
\delta\left(t\right) & =\omega_{A}-\omega_{B}+\left(U_{AA}-U_{BB}\right)\big(\hat{N}-1\big).\end{align}
 As the number operator $\hat{N}$ is a conserved quantity, if $\chi\left(t\right)\neq0$
as presented in Sec.~8.1, spin-squeezed states are generated as
discussed previously. The validity of using the single-mode
assumption to calculate spin squeezing was verified in
Refs.~\cite{Poulsen2001a,Sorensen2002}, based on both Bogoliubov
theory and positive-$P$ simulations. Note that, the original
Hamiltonian (\ref{BEC_Ham1}) entangles the internal and motional
states of the atoms, which is a source of decoherence for the spin
squeezing. With direct numerical simulation
\cite{Sorensen2001,Li2009}, spin squeezing produced via the original
Hamiltonian (\ref{BEC_Ham1}) is roughly in agreement with the
one-axis twisting Hamiltonian (\ref{BEC_Ham3}).

We now consider that a driving microwave field is applied
\cite{Choi2005}, and the Hamiltonian becomes\begin{equation}
H_{1}=H+\frac{1}{2}\int
d\vec{r}\left[\hat{\psi}_{A}^{\dagger}\left(\vec{r}\right)\hat{\psi}_{B}\left(\vec{r}\right)\Omega_{R}e^{-i\Delta
t}+\text{h.c.}\right],\end{equation}
 where $\Delta$ is the detuning of the field from resonance and $\Omega_{R}$
is the effective Rabi frequency assumed to be positive. Following
the same steps, we effectively obtain the Hamiltonian in terms of
the angular-momentum operators as \begin{equation}
H_{1}=\delta\left(t\right)J_{z}+\chi\left(t\right)J_{z}^{2}+\Omega~\left[J_{+}e^{-i\delta t}+J_{-}e^{i\delta t}\right],\end{equation}
 where \begin{equation}
\Omega=\Omega_{R}\int d\vec{r}~\phi_{A}^{\ast}\left(\vec{r}\right)\phi_{B}\left(\vec{r}\right).\end{equation}
 Usually, we can assume $\tilde{\delta}\left(t\right)=0$. As presented
in Ref.~\cite{Law2001}, by using the control field to assist the
one-axis twisting, spin squeezing can be maintained for an extended
period of time. In Ref.~\cite{Jin2007}, $\chi\left(t\right)$ is
assumed to be independent of time $t$, and the external field is
turned off rapidly at a time $t_{M}$, so that
$\Omega\left(t\right)=\Omega_{R}~\Theta\!\left(t_{M}-t\right)$,
where $\Theta\left(t\right)$ is the usual step function, and the
maximal-squeezing time $t_{M}$ is obtained
analytically\begin{equation}
\chi~t_{M}\simeq\frac{\pi}{4}\sqrt{\frac{\chi}{\Omega_{R}N}},\end{equation}
 which is valid for large $N$ $\left(\geq10^{3}\right)$. The time-dependent
field $\Omega(t)$ provides the control for storing the spin squeezing.

\subsubsection{One-axis twisting from large-detuned atom-field interaction}

Next we discuss the derivation of the one-axis twisting Hamiltonian
from a collection of two-level atoms interacting with a large
detuned field
\cite{Agarwal1997,Klimov1998,Shindo2004,Deb2006,Chaudhury2007,Fernholz2008}.
Consider an ideal model, a collection of $N$ two-level atoms laid in
a cavity, the Hamiltonian for the whole system is \begin{equation}
H=H_{0}+H_{I},\end{equation}
 where \begin{align}
H_{0} & =\omega_{0}J_{z}+\omega_{c}a^{\dagger}a,\nonumber \\
H_{I} & =g\left(J_{+}a+J_{-}a^{\dagger}\right).\end{align}
 $H_{0}$ describes the free dynamics of the atoms and field, and $H_{I}$
is the interaction term under the rotating wave approximation. The
spin operators $J_{z,\pm}$ describe the atomic system, $a$,
$a^{\dagger}$ describe the cavity field, and $g$ is the atom-field
interaction strength. We can see from the Hamiltonian that, atoms
interact with the cavity field, while no direct interaction exists
between atoms. However, under the large detuning condition, i.e. the
atom-field detuning $\Delta=\omega_{0}-\omega_{c}$ is very large as
compared to $g$ such that $|\Delta|\gg g\sqrt{N}$, and we can
perform the Fr\"{o}ich-Nakajima transform to obtain an effective
Hamiltonian describing nonlinear atom-atom interaction. The
Fr\"{o}ich-Nakajima transform is performed as, \begin{equation}
H_{S}=e^{-S}He^{S},\label{eq:Nakajima}\end{equation}
 where $S$ is of the same order as the interaction term $H_{I}$.
Expand the above transform to the second order of $S$, we have\begin{align}
H_{S} & =H+\left[H,\, S\right]+\frac{1}{2}\left[\left[H,\, S\right],\, S\right]\nonumber \\
 & =H_{0}+\left(H_{I}+\left[H_{0},\, S\right]\right)+\frac{1}{2}\left[\left(H_{I}+\left[H_{0},\, S\right]\right),\, S\right]+\frac{1}{2}\left[H_{I},\, S\right],\end{align}
 and let $H_{I}+\left[H_{0},\, S\right]=0$, which gives\begin{equation}
H_{S}=H_{0}+\frac{1}{2}\left[H_{I},\, S\right],\label{eq:Hs}\end{equation}
 where\begin{equation}
S=\frac{g}{\Delta}\left(a^{\dagger}J_{-}-aJ_{+}\right).\label{eq:TransS}\end{equation}
 Inserting Eq.~(\ref{eq:TransS}) into Eq.~(\ref{eq:Hs}), we obtain
the effective Hamiltonian\begin{align}
H_{S} & =H_{0}-\eta\left[J_{z}^{2}-\left(2a^{\dagger}a+1\right)J_{z}\right],\label{eq:effHam}\end{align}
 and the factor $\eta=g^{2}/\Delta$. Note that, the effective Hamiltonian
(\ref{eq:effHam}) contains a one-axis twisting term $J_{z}^{2}$,
and a dispersive interaction term proportional to $a^{\dagger}aJ_{z}$,
which is employed in the QND measurement and shall be discussed in
Sec.~8.

If photon loss is taken into account, the effective Hamiltonian is
of the same form as Eq.~(\ref{eq:effHam}), while the interaction
strength is modified to be \cite{Agarwal1997}\begin{equation}
\eta=\frac{g^{2}\Delta}{\text{\ensuremath{\Delta}}^{2}+\gamma^{2}/4},\label{eq:detune_fac}\end{equation}
 where $\gamma$ is the decay rate of the cavity field and satisfies
$\gamma\ll\Delta$. The derivation of Eq.~(\ref{eq:detune_fac}) can
refer to Ref.~\cite{Agarwal1997}, where the atoms are laid in a
cavity that is highly detuned from the atomic transition frequency,
while the atomic dissipation induced by spontaneous emission was
considered to be negligibly small as compared to the time scale of
interaction. Spin squeezing of atoms in cavity was also studied in
\cite{Klimov1998,Shindo2004,Deb2006}. In experiments, the effective
one-axis twisting Hamiltonian was demonstrated for squeezing
individual high spin ($F=4$ \cite{Fernholz2008}, and $F=3$
\cite{Smith2006,Chaudhury2007}) Cs atoms that interacts with
off-resonant light field.

References~\cite{Takeuchi2005,Schleier-Smith2010,Leroux2010}
proposed an interesting method to generate effective one-axis
twisting Hamiltonian via cavity feedback. The large detuned
atom-field interaction in cavity induces an effective QND-type
Hamiltonian,\begin{equation} H=\alpha
a^{\dagger}aJ_{z},\end{equation}
 while the backaction of the cavity light causes the photon number
operator $a^{\dagger}a$ to be linearly proportional to $J_{z}$, thus
the QND Hamiltonian becomes a one-axis twisting Hamiltonian. In a
recent experiment\cite{Leroux2010}, an effective one-axis twisting
Hamiltonian was generated by cavity feedback, and achieved $5.6(6)$
improvement in signal-to-noise ratio for $|F=1,\,
m_{F}=0\rangle\leftrightarrow|F=2,\, m_{F}=0\rangle$ hyperfine clock
transition in $^{87}$Rb atoms.

\subsection{Two-axis twisted states\label{taws}}

Although spin squeezing can be produced by the one-axis twisting model
effectively, the optimal squeezing angle depends on the system size
and evolution time. This problem is solved if the twisting is performed
simultaneously clockwise and counterclockwise about two orthogonal
axes in the plane normal to the MSD. The initial state is also $|j,-j\rangle$,
and the twisting is about two axes in the $\theta=\pi/2$, $\phi=\pm\pi/4$
directions. The relevant two spin operators are written as

\begin{align}
J_{\frac{\pi}{2},\frac{\pi}{4}} & =\cos\left(\frac{\pi}{4}\right)J_{x}+\sin\left(\frac{\pi}{4}\right)J_{y}=\frac{1}{\sqrt{2}}\Big(J_{x}+J_{y}\Big),\\
J_{\frac{\pi}{2},-\frac{\pi}{4}} &
=\cos\left(\frac{\pi}{4}\right)J_{x}-\sin\left(\frac{\pi}{4}\right)J_{y}=\frac{1}{\sqrt{2}}\Big(J_{x}-J_{y}\Big).\end{align}
 The two-axis twisting Hamiltonian is written as~\cite{KITAGAWA1993}\begin{equation}
H_{\text{TAT}}=J_{\frac{\pi}{2},\frac{\pi}{4}}^{2}-J_{\frac{\pi}{2},-\frac{\pi}{4}}^{2}=\chi\Big(J_{x}J_{y}+J_{y}J_{x}\Big)=\frac{\chi}{2i}\Big(J_{+}^{2}-J_{-}^{2}\Big),\label{TAT}\end{equation}
 which is analogous to the Hamiltonian (\ref{bose_squ_Ham}) for producing
squeezed light that creates and annihilates photons in pairs. By
replacing $a$ and $a^{\dagger}$ with $J_{-}/\sqrt{N}$ and
$J_{+}/\sqrt{N}$, respectively, we will obtain the two-axis twisting
Hamiltonian shown in Eq.~(\ref{TAT}). Various approaches for
implementing this Hamiltonian were studied in
Refs.~\cite{Helmerson2001,Andre2002a,Bouchoule2002a,Wesenberg2002,Zhang2003a,Jafarpour2008}.
The MSD is also along the $z$-axis. Unfortunately, the two-axis
twisting model cannot be solved analytically for arbitrary $N$,
except for $N\le3$. Below, we list two advantages of the two-axis
twisting compared with the one-axis twisting case.

\begin{table}[tbh]
 \caption{Comparison of squeezing parameters when using either one-axis twisting
or two-axis twisting Hamiltonian.}

\label{tab_comp_oat}

\centering{}\begin{tabular}{c||c|c}
\hline
 & One-axis Twisting  & Two-axis Twisting \tabularnewline
\hline
\hline
\parbox[c]{4cm}{%
Minimum $\xi_{S}^{2}$ and $\xi_{R}^{2}$%
}  & %
\parbox[c]{3.9cm}{%
\vspace{1mm}
 ${\displaystyle \propto~~\frac{1}{N^{2/3}}}$\vspace{1mm}
}  & %
\parbox[c]{3.9cm}{%
\vspace{2mm}
 ${\displaystyle \propto~~\frac{1}{N}}$\vspace{2mm}
} \tabularnewline
\hline
\parbox[c]{4cm}{%
Optimal squeezing angle%
}  & %
\parbox[c]{3.9cm}{%
\vspace{2mm}
 ${\displaystyle \delta~\sim~\frac{1}{2}\arctan\left({N^{-1/3}}\right)}$\vspace{2mm}
}  & %
\parbox[c]{3.9cm}{%
\vspace{2mm}
 Unchanged \vspace{2mm}
} \tabularnewline
\hline
\parbox[c]{4cm}{%
Physical implementations%
}  & %
\parbox[c]{3.9cm}{%
\vspace{6mm}
 (i) Bose-Einstein condensation~\cite{Orzel2001,Esteve2008,Gross2010,Riedel2010};
(ii) Large detuning atom-field
interaction~\cite{Agarwal1997,Chaudhury2007}.\vspace{6mm}
}  & %
\parbox[c]{3.9cm}{%
\vspace{2mm}
 Effective atom-atom interaction via photon exchange~\cite{Andre2002,Andre2002a}.
\vspace{2mm}
} \tabularnewline
\hline
\end{tabular}
\end{table}

\begin{figure}[H]
\begin{centering}
\includegraphics[width=11cm]{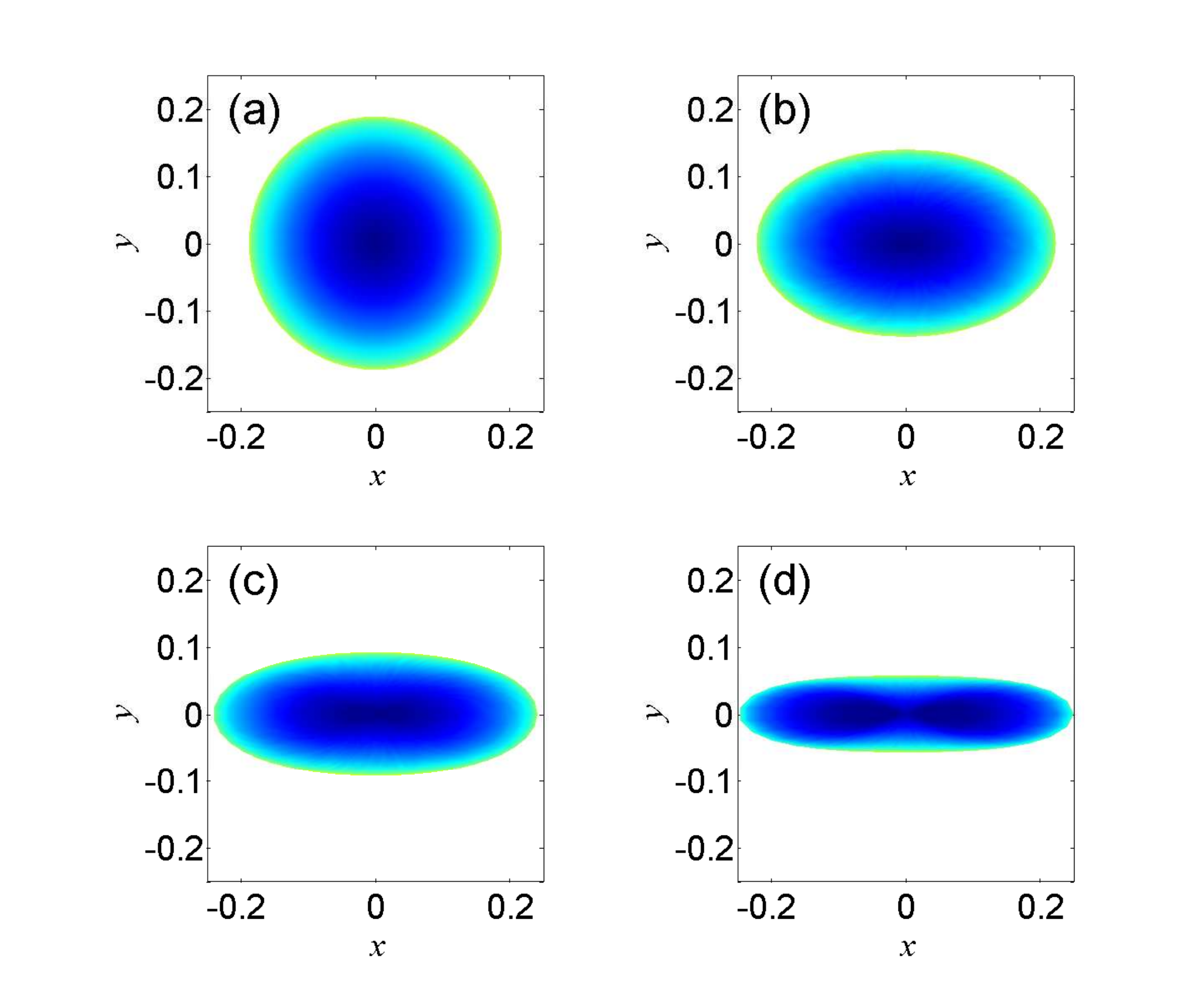}
\par\end{centering}

\caption{(Color online) Husimi $Q$ function of spin-squeezed states generated
by the two-axis twisting model for various times: (a) $\chi t=0$,
(b) $\chi t=0.1$, (c) $\chi t=0.2$, and (d) $\chi t=0.3$. The system
size is $N=60$. Compared with the one-axis twisting, the optimal
squeezing angle here is invariant during the time evolution.}

\label{fig_QPD_TAT_a}
\end{figure}

(i) The optimal squeezing angle is invariant during the evolution.
Since the MSD for the two-axis twisting Hamiltonian is along the $z$-axis,
according to Eq.~(\ref{optsquangle}), the optimal squeezing angle
is determined by two quantities, $\left\langle J_{x}^{2}-J_{y}^{2}\right\rangle $,
and $\left\langle J_{x}J_{y}+J_{y}J_{x}\right\rangle $. Due to \begin{equation}
\left[J_{x}J_{y}+J_{y}J_{x},\, H_{\text{TAT}}\right]=0,\end{equation}
 $\left\langle J_{x}J_{y}+J_{y}J_{x}\right\rangle $ is invariant
during the time evolution. Here, the initial state is the CSS $|j,-j\rangle$,
thus $\left\langle J_{x}J_{y}+J_{y}J_{x}\right\rangle =0$. Then,
the optimal squeezing direction is $\varphi=0,\pi/2$ during the evolution.
The Husimi function is shown in Fig.~\ref{fig_QPD_TAT_a}, and it
is clear that the optimal squeezing angle is invariant.

(ii) The degree of squeezing is high. By numerical calculations, the
spin squeezing parameters scales
as~\cite{KITAGAWA1993}\begin{equation}
\xi_{R}^{2}\propto\frac{1}{N},\quad\xi_{S}^{2}\propto\frac{1}{N},\end{equation}
 in the two-axis twisting model. Thus, according to Eq.~(\ref{phase_xi_r}),
the phase noise approaches the Heisenberg limit. Comparisons between
the one-axis twisting and two-axis twisting Hamiltonians are displayed
in Table \ref{tab_comp_oat}.

\section{Spin squeezing, negative pairwise correlations, and entanglement
\label{entanglement}}

In this and the following sections, we mainly concentrate on spin
squeezing, quantum correlations, and entanglement. Here, we first
consider spin squeezing and pairwise correlations in a system with
exchange symmetry, and consider the squeezing parameters
$\xi_{S}^{2}$ and $\tilde{\xi}_{E}^{2}$. Note that for a CSS, which
has no pairwise correlation, the variance of $J_{\vec{n}_{\perp}}$
is evenly distributed on individual spin components in the
$\vec{n}_{\perp}$-direction, thus if a state has negative pairwise
correlations in the $\vec{n}_{\perp}$-direction, the variance
$J_{\vec{n}_{\perp}}$ could be reduced as compared with CSSs. Below,
we show that, as discussed in Ref.~\cite{Ulam-Orgikh2001}, spin
squeezing with respect to $\xi_{S}^{2}<1$ implies negative pairwise
correlation in the $\vec{n}_{\perp}$ direction. Moveover, the
minimum pairwise correlation is associated with parameter
$\tilde{\xi}_{E}^{2}$, and $\tilde{\xi}_{E}^{2}<1$ is equivalent to
the existence of negative pairwise correlation.

\subsection{Spin squeezing and pairwise correlations}

As discussed in Refs.~\cite{KITAGAWA1993,Ulam-Orgikh2001}, the
spin-squeezing parameter $\xi_{S}^{2}$ is well defined in the
$j=N/2$ subspace, where the states are of exchange symmetry. Thus,
the expectation values and variances of the angular momentum
operators $J_{\alpha}$ can be expressed in terms of the local
pairwise correlations, which helps us to obtain the relations
between spin squeezing parameters and the pairwise correlations. The
pairwise correlation function is defined as \begin{align}
G_{i\vec{n},j\vec{n}} & \equiv\left\langle
\sigma_{i\vec{n}}\sigma_{j\vec{n}}\right\rangle -\left\langle
\sigma_{i\vec{n}}\right\rangle \left\langle
\sigma_{j\vec{n}}\right\rangle =\left\langle
\sigma_{1\vec{n}}\sigma_{2\vec{n}}\right\rangle -\left\langle
\sigma_{1\vec{n}}\right\rangle ^{2},\end{align}
 with $i$, $j$ being particle indices, and the second equality holds
due to exchange symmetry. One can further define the minimum pairwise
correlation as \begin{equation}
G_{m}=\min_{\vec{n}}\; G_{1\vec{n},2\vec{n}},\end{equation}
 where the minimization is over an arbitrary direction $\vec{n}.$

\subsubsection{Spin-squeezing parameter $\xi_{S}^{2}$ and correlation $G_{\vec{n}_{\perp},\vec{n}_{\perp}}$}

First, we consider the spin-squeezing parameter $\xi_{S}^{2}$. From
the relation $J_{\vec{n}}=1/2\sum_{i=1}^{N}\sigma_{i\vec{n}},$ one
finds the variance of $J_{\vec{n}},$ which is given by\begin{align}
\left(\Delta J_{\vec{n}}\right)^{2} & =\frac{1}{4}\sum_{ij}\left(\langle\sigma_{i\vec{n}}\sigma_{j\vec{n}}\rangle-\langle\sigma_{i\vec{n}}\rangle\langle\sigma_{j\vec{n}}\rangle\right)\nonumber \\
 & =\frac{1}{4}\left[N\left(1-\langle\sigma_{1\vec{n}}\rangle^{2}\right)+N(N-1)G_{1\vec{n},2\vec{n}}\right],\end{align}
 where we used the symmetry property in deriving the last equality.
From this equation, the correlation function can be written as \begin{align}
G_{1\vec{n},2\vec{n}} & =\frac{4\left[N\left(\Delta J_{\vec{n}}\right)^{2}+\langle J_{\vec{n}}\rangle^{2}\right]}{N^{2}(N-1)}-\frac{1}{N-1}.\label{correlation_var}\end{align}

If the correlation function is along the direction $\vec{n}_{\perp}$,
Eq.~(\ref{correlation_var}) reduces to \begin{equation}
G_{1\vec{n}_{\perp},2\vec{n}_{\perp}}=\frac{4\left\langle J_{\vec{n}_{\perp}}^{2}\right\rangle -N}{\left(N-1\right)N},\label{g233}\end{equation}
 where we have used the fact $\langle J_{\vec{n}_{\perp}}\rangle=0.$
From the relation between the spin squeezing parameter $\xi_{S}^{2}$
and the expectation value $\left\langle J_{\vec{n}_{\perp}}^{2}\right\rangle ,$\begin{equation}
\min_{\vec{n}_{\perp}}\left\langle J_{\vec{n}_{\perp}}^{2}\right\rangle =\frac{N}{4}\xi_{S}^{2},\end{equation}
 one obtain the minimum pairwise correlation~\cite{Ulam-Orgikh2001}
\begin{equation}
\min_{\vec{n}_{\perp}}G_{\vec{n}_{\perp},\vec{n}_{\perp}}=\frac{\xi_{S}^{2}-1}{N-1},\label{kita_pc}\end{equation}
 where we have used Eq.~(\ref{g233}). This implies that $\min_{\vec{n}_{\perp}}G_{\vec{n}_{\perp},\vec{n}_{\perp}}<0$
is equivalent to $\xi_{S}^{2}<1$. Therefore, a spin-squeezed state
($\xi_{S}^{2}<1$) has negative pairwise correlation in the $\vec{n}_{\perp}$
direction ($G_{\vec{n}_{\perp},\vec{n}_{\perp}}<0$).

\subsubsection{Spin-squeezing parameter $\tilde{\xi}_{E}^{2}$ and correlation $G_{1\vec{n},2\vec{n}}$}

Now, we investigate the relation between the correlation function
$G_{1\vec{n},2\vec{n}}$ and the squeezing parameter $\tilde{\xi}_{E}^{2}.$
It is more convenient to rewrite the correlation function as\begin{align}
G_{1\vec{n},2\vec{n}} & =\langle\vec{n}^{T}\vec{\sigma}_{1}\vec{\sigma}_{2}^{T}\vec{n}\rangle-\langle\vec{n}^{T}\vec{\sigma}_{1}\rangle\langle\vec{\sigma}_{2}^{T}\vec{n}\rangle=\vec{n}^{T}\mathbf{G}\vec{n},\label{pairwise_matrix}\end{align}
 where the normalized direction $\vec{n}=\left(n_{x},n_{y},n_{z}\right)^{T}$,
and the pairwise correlation matrix $\mathbf{G}$ is given by\begin{equation}
\mathbf{G=}\langle\vec{\sigma}_{1}\vec{\sigma}_{2}^{T}\rangle-\langle\vec{\sigma}_{1}\rangle\langle\vec{\sigma}_{2}^{T}\rangle.\end{equation}
 The matrix elements of $\mathbf{G}$ are\begin{equation}
\mathbf{G}_{k,l}=\left\langle \sigma_{1k}\sigma_{2l}\right\rangle -\left\langle \sigma_{1k}\right\rangle \left\langle \sigma_{2l}\right\rangle ,\text{ \ }k,l=x,y,z.\end{equation}

From Eq.~(\ref{correlation_var}), we know that the correlation function
$G_{1\vec{n},2\vec{n}}$ is a linear function of the quantity $N\left(\Delta J_{\vec{n}}\right)^{2}+\langle J_{\vec{n}}\rangle^{2}$,
which can be written as\begin{equation}
N\left(\Delta J_{\vec{n}}\right)^{2}+\langle J_{\vec{n}}\rangle^{2}=\vec{n}^{T}\left[N\left(\frac{\langle\vec{J}\vec{J}^{T}\rangle+\langle\vec{J}\vec{J}^{T}\rangle^{T}}{2}-\langle\vec{J}\rangle\langle\vec{J}\rangle^{T}\right)+\langle\vec{J}\rangle\langle\vec{J}\rangle^{T}\right]\vec{n}.\label{paiwisem}\end{equation}
The matrix \begin{equation}
N\left(\frac{\langle\vec{J}\vec{J}^{T}\rangle+\langle\vec{J}\vec{J}^{T}\rangle^{T}}{2}-\langle\vec{J}\rangle\langle\vec{J}\rangle^{T}\right)+\langle\vec{J}\rangle\langle\vec{J}\rangle^{T}\end{equation}
 is just the matrix $\Gamma$ given by Eq.~(\ref{gamma}). Therefore,
we have\begin{equation}
N\left(\Delta J_{\vec{n}}\right)^{2}+\langle J_{\vec{n}}\rangle^{2}=\vec{n}^{T}\Gamma\vec{n}.\label{pairwisem}\end{equation}

Substituting Eqs.~(\ref{pairwise_matrix}) and (\ref{pairwisem})
into Eq.~(\ref{correlation_var}) leads to the following relation\begin{equation}
\mathbf{G}=\frac{4\Gamma}{N^{2}\left(N-1\right)}-\frac{\mathbb{I}}{\left(N-1\right)},\label{rel_G_Gamma}\end{equation}
 where $\mathbb{I}$ is a $3\times3$ identity matrix, and thus $\mathbf{G}$
and $\Gamma$ can be diagonalized simultaneously. This indicates that,
$G_{\vec{n},\vec{n}}$ can be expressed in terms of $\xi_{S}^{2}$
or $\tilde{\xi}_{E}^{2}$ by choosing a specific direction $\vec{n}$.

Now we look for the relation between the parameter $\tilde{\xi}_{E}^{2}$
and the minimum pairwise correlation. From Eq.~(\ref{pairwise_matrix}),
the minimum pairwise correlation\begin{equation}
G_{m}=\min_{\vec{n}}G_{\vec{n},\vec{n}}=\min_{\vec{n}}\left(\vec{n}\mathbf{G}\vec{n}^{T}\right)=g_{\min},\end{equation}
 where $g_{\min}$ is the minimum eigenvalue of the pairwise correlation
matrix $\mathbf{G}$. From Eq.~(\ref{rel_G_Gamma}), we find\begin{equation}
g_{\min}=\frac{4\lambda_{\min}-N^{2}}{N^{2}\left(N-1\right)}.\end{equation}
 In the symmetric case ($j=N/2)$, as we have assumed here, $\tilde{\xi}_{E}^{2}=4\lambda_{\min}/N^{2}$,
and thus\begin{equation}
G_{m}=\frac{\tilde{\xi}_{E}^{2}-1}{N-1}.\label{rel_pairwise_toth}\end{equation}
 If the minimal pairwise correlation is in the plane normal to the
MSD, the above relation will degenerate to Eq.~(\ref{kita_pc}).
In another form, we write\begin{equation}
\tilde{\xi}_{E}^{2}=1+\left(N-1\right)G_{m}.\end{equation}
 This exact result indicates that the spin squeezing defined by the
parameter $\tilde{\xi}_{E}^{2}$ is equivalent to the negative pairwise
correlation, i.e., $G_{m}<0$ implies $\tilde{\xi}_{E}^{2}<1$ and
vice versa.

\subsection{Spin squeezing and pairwise entanglement}

One of the most useful applications of spin squeezing is to detect
entanglement for many-qubit system. To determine whether a state is
entangled, we just need to measure the collective operators, which
are particle populations in many cases. Moreover, in many
experiments, such as BEC, particles are not accessed individually,
and the spin-squeezing parameter is easier to obtain than the
concurrence and the entanglement entropy. Different kinds of
spin-squeezing inequalities may be used to detect various types of
entanglement~\cite{Korbicz2005,Korbicz2006,Toth2009a,Guehne2009}. In
Refs.~\cite{Ulam-Orgikh2001,Wang2002,Wang2003,Yan2005,Zeng2005}, the
relationships between negative pairwise correlation, concurrence,
and $\xi_{S}^{2}$ for symmetric states and even (odd)-parity states
were found. References \cite{Messikh2003a,Messikh2003} showed that
for a two-qubit Dicke system, parameter $\xi_{S}^{2}$ is better than
$\xi_{R}^{2}$ to measure entanglement. A multipartite entanglement
criterion for spin-squeezing parameter was given in
Ref.~\cite{Sorensen2001}. Inspired by this work, some other
generalized spin squeezing inequalities were proposed. In
Refs.~\cite{Korbicz2005,Korbicz2006}, by employing a positive
partial transpose method, they found generalized spin squeezing
inequalities as criteria for two- and three-qubit entanglement. In
Refs.~\cite{Toth2007,Toth2009a}, optimal spin squeezing inequalities
were proposed.

At the time when squeezing parameters $\xi_{S}^{2}$ and
$\xi_{R}^{2}$ were proposed,
Refs.~\cite{KITAGAWA1993,WINELAND1992,WINELAND1994} noticed the
potential relationship between spin squeezing and entanglement.
Since $\xi_{S}^{2}=1$ for CSS, Ref.~\cite{KITAGAWA1993} expected
that for an appropriate correlated state, $\xi_{S}^{2}<1$. As shown
above, the one-axis twisting Hamiltonian can produce squeezed
states, and this Hamiltonian involves pairwise interactions. This
indicates that spin squeezing is associated with pairwise
entanglement.

To detect two-qubit entanglement, it was
proven~\cite{Korbicz2005,Korbicz2006,Vidal2006} that if the
inequality\begin{equation} \left[\left\langle
J_{{\vec{n}}_{1}}^{2}\right\rangle
+\frac{N\left(N-2\right)}{4}\right]^{2}\geq\left[\left\langle
J_{{\vec{n}}_{2}}^{2}\right\rangle +\left\langle
J_{{\vec{n}}_{3}}^{2}\right\rangle
-\frac{N}{2}\right]^{2}+\left(N-1\right)^{2}\left\langle
J_{{\vec{n}}_{1}}\right\rangle ^{2}\end{equation}
 is violated then the state is two-qubit entangled. For symmetric
states, the above inequality is simplified to\begin{equation}
1-\frac{4\left\langle J_{{\vec{n}}}\right\rangle ^{2}}{N^{2}}\geq\frac{4\left(\Delta J_{{\vec{n}}}\right)^{2}}{N},\end{equation}
 and a symmetric state is two-qubit entangled if and only if it violates
the above inequality.

Entanglement of two-qubit systems is characterized by the
concurrence~\cite{Wootters1998}. The concurrence $C$, quantifying
the entanglement of a pair of qubits, is defined
as~\cite{Wootters1998}\begin{equation}
C=\max\left(0,\lambda_{1}-\lambda_{2}-\lambda_{3}-\lambda_{4}\right),\label{Cdef}\end{equation}
 where the quantities $\lambda_{i}$'s are the square roots of the
eigenvalues, in descending order, of the matrix \begin{equation}
\varrho_{12}=\rho_{12}(\sigma_{1y}\otimes\sigma_{2y})\rho_{12}^{\ast}(\sigma_{1y}\otimes\sigma_{2y}),\label{varrho}\end{equation}
 where $\rho_{12}$ is the two-qubit density matrix, and $\rho_{12}^{\ast}$
is the complex conjugate of $\rho_{12}$. From its definition (\ref{Cdef}),
the concurrence $C\geq0$, and the existence of two-qubit entanglement
is equivalent to $C>0$.

For example, we calculate the concurrence of a pure state\begin{equation}
|\psi\rangle=a|00\rangle+b|01\rangle+c|10\rangle+d|11\rangle.\end{equation}
 Then density matrix of $|\psi\rangle$ is\begin{equation}
\rho=|\psi\rangle\langle\psi|,\end{equation}
 and its conjugate $\rho^{\ast}$ is a pure state. Since $\sigma_{1y}\otimes\sigma_{2y}$
is a unitary operation, \begin{equation}
\tilde{\rho}=\sigma_{1y}\otimes\sigma_{2y}\,\rho_{12}^{\ast}\,\sigma_{1y}\otimes\sigma_{2y},\end{equation}
 is still a pure state. Therefore, the rank of $\rho\tilde{\rho}$
is less than or equal to one, and it has at most one nonzero eigenvalue
$\lambda$, and the concurrence\begin{align}
C & =\sqrt{\lambda}=\sqrt{\text{Tr}\left(\rho\tilde{\rho}\right)}=\left\vert \langle\psi|\sigma_{1y}\otimes\sigma_{2y}\left(|\psi\rangle\right)^{\ast}\right\vert \notag\\
 & =2\left\vert ad-bc\right\vert .\end{align}
 If $a=d=1/\sqrt{2}$ and $b=c=0$, then $|\psi\rangle$ is the Bell
state and is entangled since $C=1$.

Following the previous discussion, we proceed to give a quantitative
relation between the squeezing parameter and the concurrence~for
states with even (odd) parity and nonzero mean spin, i.e., $\langle
J_{z}\rangle\neq0$. The two-spin reduced density matrix for a parity
state with exchange symmetry can be written in a block-diagonal
form~\cite{Wang2002}
\begin{equation}
\rho_{12}=\left(\begin{array}{cc}
v_{+} & u^{\ast}\\
u & v_{-}\end{array}\right)\oplus\left(\begin{array}{cc}
w & y\\
y & w\end{array}\right),\label{re}\end{equation}
 in the basis \{$|00\rangle,|11\rangle,|01\rangle,|10\rangle$\},
where\begin{align}
v_{\pm}= & \frac{N^{2}-2N+4\langle J_{z}^{2}\rangle\pm4\langle J_{z}\rangle(N-1)}{4N(N-1)},\notag\\
w= & \frac{N^{2}-4\langle J_{z}^{2}\rangle}{4N(N-1)},\text{ \ }u=\frac{\langle J_{+}^{2}\rangle}{N(N-1)},\notag\\
y= & \frac{4\langle J_{x}^{2}+J_{y}^{2}\rangle-2N}{4N(N-1)},\label{eq:bbbaaa}\end{align}
 which can be written in terms of local expectations as\begin{align}
v_{\pm} & =\frac{1}{4}\left(1\pm2\langle\sigma_{1z}\rangle+\langle\sigma_{1z}\sigma_{2z}\rangle\right),\label{r1}\\
w & =\frac{1}{4}\left(1-\langle\sigma_{1z}\sigma_{2z}\rangle\right),\label{r3}\\
u & =\langle\sigma_{1-}\sigma_{2-}\rangle,\label{rrr}\\
y & =\langle\sigma_{1+}\sigma_{2-}\rangle.\label{rr}\end{align}
 The concurrence is then given by~\cite{Wootters1998} \begin{equation}
C=2\max\left\{ 0,|u|-w,y-\sqrt{v_{+}v_{-}}\right\} .\label{conc}\end{equation}
From the above expressions we know that, the spin-squeezing parameters
and concurrence are determined by the expectation value $\left\langle \sigma_{1z}\right\rangle $,
correlations $\left\langle \sigma_{1+}\sigma_{2-}\right\rangle $,
$\left\langle \sigma_{1-}\sigma_{2-}\right\rangle $, and $\left\langle \sigma_{1z}\sigma_{2z}\right\rangle $.

Since $\left\langle J^{2}\right\rangle ={N}/{2}\left({N}/{2}+1\right)$,
from Eq.~(\ref{eq:bbbaaa}), one can straightforwardly verify that\begin{equation}
w=y.\label{r8}\end{equation}
 Thus the concurrence given by Eq.~(\ref{conc}) becomes \begin{equation}
C=2\max\left\{ 0,|u|-y,y-\sqrt{v_{+}v_{-}}\right\} .\label{concurrence}\end{equation}
 Since the density matrix should be positive, we find \begin{equation}
\sqrt{v_{+}v_{-}}\geq\left\vert u\right\vert .\end{equation}
 From this inequality, one may find that if $|u|-y>0,$ then $y-\sqrt{v_{+}v_{-}}<0\,$and
if $y-\sqrt{v_{+}v_{-}}>0\,,|u|-y<0.$ In other words, two quantities
$|u|-y$ and $y-\sqrt{v_{+}v_{-}}$ cannot be simultaneously larger
than zero.
\begin{table}[H]
 \caption{Spin squeezing parameters $\xi_{S}^{2}$ and $\tilde{\xi}_{E}^{2}$
as well as the concurrence $C$ for parity states. Symbols are
defined in the text}

\label{tab_squeezing_conc}

\centering{}\begin{tabular}{c||c|c|c} \hline
\parbox[c]{2cm}{%
\vspace{2mm}
 ~\vspace{2mm}
}  & \multicolumn{2}{c|}{Pairwise entangled ($C>0$)} & %
\parbox[c]{2cm}{%
\vspace{2mm}
 Unentangled\vspace{2mm}
} \tabularnewline \hline
\parbox[c]{2cm}{%
Concurrence%
}  & %
\parbox[c]{4cm}{%
\vspace{2mm}
 $C=2(\left\vert u\right\vert -y)>0$\vspace{2mm}
}  & $C=2(y-\sqrt{v_{+}v_{-}})>0$  & $C=0$ \tabularnewline \hline
\parbox[c]{2cm}{%
$\xi_{S}^{2}$%
}  & %
\parbox[c]{4cm}{%
\vspace{2mm}
 $\xi_{S}^{2}=1-\left(N-1\right)C<1$\vspace{2mm}
}  & $\xi_{S}^{2}>1$  & $\xi_{S}^{2}\geq1$ \tabularnewline \hline
\parbox[c]{2cm}{%
$\tilde{\xi}_{E}^{2}$%
}  & %
\parbox[c]{4cm}{%
\vspace{2mm}
 $\tilde{\xi}_{E}^{2}=1-\left(N-1\right)C<1$\vspace{2mm}
}  & %
\parbox[c]{6cm}{%
$\tilde{\xi}_{E}^{2}=1-2(N-1)(y+\sqrt{v_{+}v_{-}})C<1$%
}  & $\tilde{\xi}_{E}^{2}\geq1$ \tabularnewline \hline
\end{tabular}
\end{table}

From Table~\ref{tab_comp} and Eqs.~(\ref{rrr}) and (\ref{rr}),
we can write the spin-squeezing parameter $\xi_{S}^{2}$ in terms
of the reduced matrix elements $u$ and $y$ as\begin{equation}
\xi_{S}^{2}=1-2(N-1)\left(\left\vert u\right\vert -y\right).\label{xi_conc}\end{equation}
 Again from Table \ref{tab_comp}, $\tilde{\xi}_{E}^{2}$ contains
the quantity \begin{equation}
\varsigma^{2}=1+(N-1)\mathcal{C}_{zz}.\label{varsig}\end{equation}
 So, to write $\tilde{\xi}_{E}^{2}$ in terms of the matrix elements,
we consider the correlation function $\mathcal{C}_{zz}.$ From Eqs.~(\ref{r1}),
(\ref{r3}), and (\ref{r8}), we obtain \begin{equation}
y^{2}-v_{+}v_{-}=-\frac{1}{4}\mathcal{C}_{zz}.\end{equation}
 This is a key step. Thus, from Eq.~(\ref{varsig}), we have\begin{equation}
\varsigma^{2}=1-4(N-1)(y+\sqrt{v_{+}v_{-}})(y-\sqrt{v_{+}v_{-}}).\end{equation}
 From the above equation and Table~\ref{tab_comp}, we obtain\begin{equation}
\tilde{\xi}_{E}^{2}=\min\left\{ \xi_{S}^{2},1-4(N-1)(y+\sqrt{v_{+}v_{-}})(y-\sqrt{v_{+}v_{-}})\right\} .\end{equation}

The relations between spin squeezing and concurrence are displayed
in Table \ref{tab_squeezing_conc}. From it we can see that, for a
symmetric state, $\tilde{\xi}_{E}^{2}<1$ is qualitatively equivalent
to $C>0$, implying that spin squeezing according to
$\tilde{\xi}_{E}^{2}$ is equivalent to pairwise entanglement
\cite{Yin2009}. Although $\xi_{S}^{2}<1$ indicates $C>0$, when
$C=2(y-\sqrt{v_{+}v_{-}})>0$, we find $\xi_{S}^{2}>1$. Therefore, a
spin-squeezed state ($\xi_{S}^{2}<1$) is pairwise entangled, while a
pairwise entangled state may not be spin-squeezed according to the
squeezing parameter $\xi_{S}^{2}$~\cite{Wang2003}.

Below, we give a simple example to illustrate the above results.
Consider a simple superposition of Dicke states \begin{equation}
|\psi_{D}\rangle=\cos\theta|j,m\rangle+e^{i\varphi}\sin\theta|j,m+2\rangle,\text{
}n=-j,\ldots,j-2\label{Dicke State_e}\end{equation}
 with the angle $\theta\in\lbrack0,\pi)$ and the relative phase $\varphi\in\lbrack0,2\pi)$,
and $j=N/2$. This state is of even parity, and the MSD is along the
$z$ direction. The relevant spin-expectation values can be obtained
as\begin{align}
\left\langle J_{z}\right\rangle  & =m+2\sin^{2}\theta,\notag\\
\langle J_{z}^{2}\rangle & =m^{2}+4\left(m+1\right)\sin^{2}\theta,\notag\\
\langle J_{-}^{2}\rangle & =\frac{1}{2}e^{i\varphi}\sin2\theta\sqrt{\mu_{m}},\label{mean_of_J}\end{align}
 where $\mu_{m}=\left(j+m+1\right)\left(j+m+2\right)\left(j-m\right)\left(j-m-1\right)$.
With the above results we find\begin{align}
u= & \frac{e^{i\varphi}\sin2\theta}{2N(N-1)}\sqrt{\mu_{m}},\notag\\
y= & \frac{1}{N-1}\left\{ \frac{N}{4}-\frac{1}{N}\big{[}m^{2}+4(m+1)\sin^{2}\theta\big{]}\right\} ,\notag\\
\sqrt{v_{+}v_{-}} & =\frac{\sqrt{(N^{2}-2N+4\left\langle J_{z}^{2}\right\rangle )^{2}-16(N-1)^{2}\left\langle J_{z}\right\rangle ^{2}}}{4N(N-1)},\label{rho_results}\end{align}
 and thus the spin-squeezing parameters are obtained as\begin{align}
\xi_{S}^{2} & =1-\frac{1}{N}\left\{ \left\vert \sin2\theta\right\vert \sqrt{\mu_{m}}-2\left[m^{2}+4\left(m+1\right)\sin^{2}\theta\right]-\frac{N^{2}}{2}\right\} ,\label{xi-theta}\\
\varsigma^{2} & =\frac{4}{N}\left[m^{2}+4(m+1)\sin^{2}\theta\right]-\frac{4(N-1)}{N^{2}}\left[m+2\sin^{2}\theta\right]^{2}.\end{align}
\begin{figure}[H]
\begin{centering}
\includegraphics[width=10cm]{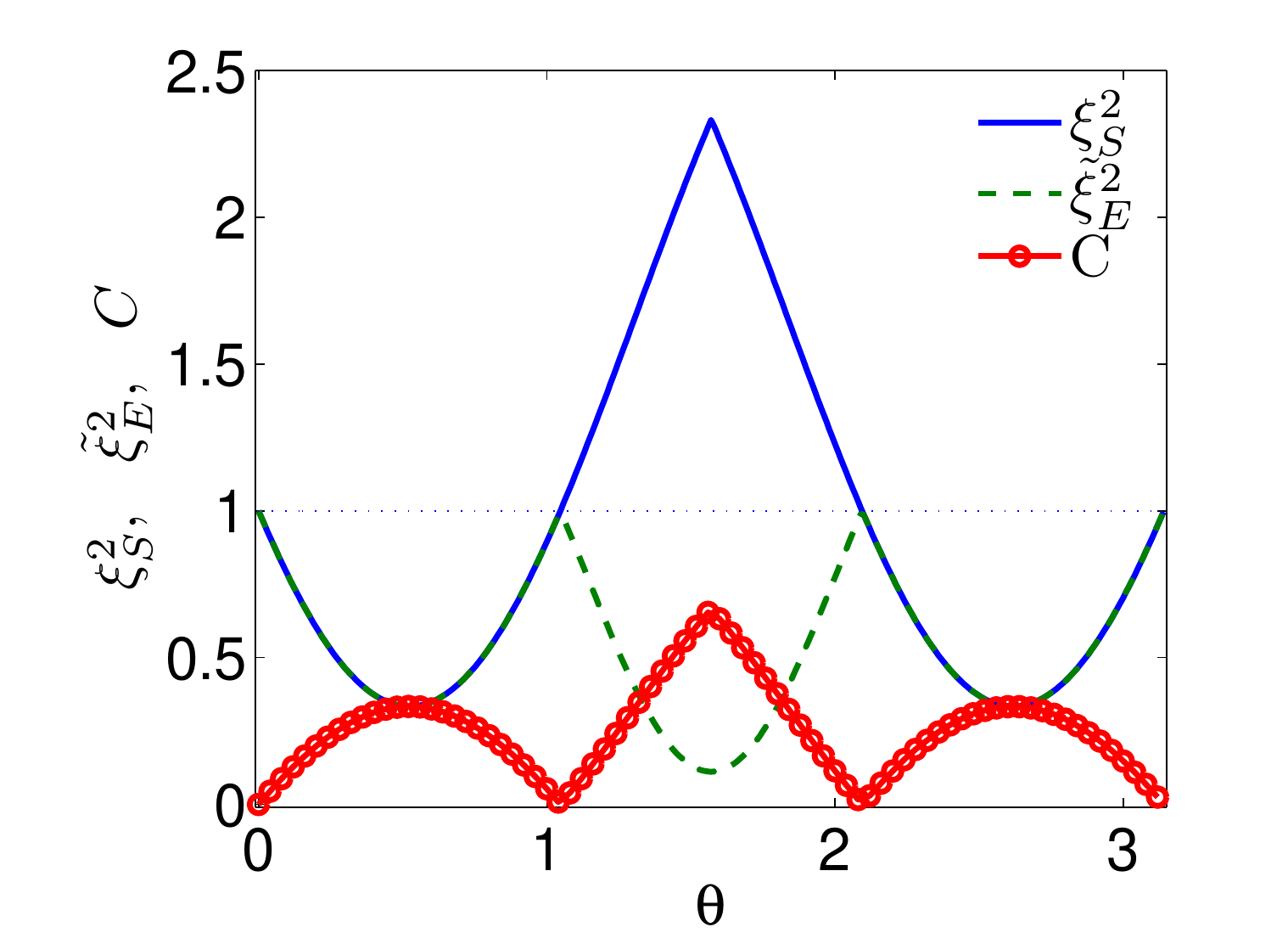}
\par\end{centering}

\caption{(Color online) Spin-squeezing parameters $\tilde{\xi}_{E}^{2}$, $\tilde{\xi}_{S}^{2}$
and concurrence $C$ as a function of $\theta$, with $N=3$ and $m=-N/2$.
The one-to-one correspondence between $\tilde{\xi}_{E}^{2}$ and the
concurrence $C$ is very clear. }

\label{xi_s_xi_t_conc}
\end{figure}

In Fig.~\ref{xi_s_xi_t_conc}, we plot these two spin-squeezing parameters
and concurrence versus $\theta$ within one period. The system size
$N=3$ and $m=-3/2$. We observe that for $\theta\in(0,\pi/3)\cup(2\pi/3,\pi)$,
$\tilde{\xi}_{E}^{2}=\xi_{S}^{2}<1$, therefore the state is spin-squeezed
in the $x$-$y$ plane. Moreover, as $C>0$, the state is pairwise
entangled. For $\theta\in(\pi/3,2\pi/3)$, it is obvious that the
state is also pairwise entangled, since $C>0,$ while spin squeezing
occurs in the $z$-axis since $\tilde{\xi}_{E}^{2}<1$ and $\xi_{S}^{2}>1$.
These results clearly show that $\tilde{\xi}_{E}^{2}<1$ is equivalent
to $C>0$.

\subsection{Spin squeezing and many-body entanglement}

To characterize and detect multipartite entanglement is one of the
most challenging open problems in quantum information
theory~\cite{Amico2008}. The simplest multipartite state is the
three-qubit pure state, for which there exists a good measure of
tripartite entanglement based on the concurrence~\cite{Cof00}. One
can also use the \emph{state preparation fidelity} $F$ for a
$N$-qubit state $\rho$ in order to investigate multipartite
entanglement. The state $\rho$ can be either pure or mixed. The
fidelity $F$ is defined as \cite{Sackett}
\begin{equation}
F(\rho)=\langle\Psi_{\text{GHZ}}|\rho|\Psi_{\text{GHZ}}\rangle,\label{eq:ff}\end{equation}
where
$|\Psi_{\text{GHZ}}\rangle=1/\sqrt{2}(|00...0\rangle+|11...1\rangle)$
is the $N$-particle Greenberger-Horne-Zeilinger (GHZ) state. The
sufficient condition for $N$-particle entanglement is given by
\cite{Sackett}
\begin{equation}
{F}(\rho)>1/2.\end{equation}
 We have the freedom to choose other GHZ states such as $|\Psi_{\text{GHZ}}\rangle=1/\sqrt{2}(|00...01\rangle\pm|11...10\rangle)$,
etc. By local unitary operations we can transfer these states to the
original GHZ state and these operations do not change the
entanglement. Detailed discussions on this sufficient condition can
be found in Refs. \cite{Sackett,Uffink}. They also discussed how to
use many-body Bell inequalities to detect multipartite entanglement.

Spin squeezing inequalities can act as entanglement criteria (see
also~Ref. \cite{Guehne2009}). To detect three-qubit entanglement, it
has been proven that it is necessary to measure the third-order
moments of the collective angular
momenta~\cite{Korbicz2005,Korbicz2006}. It was found that, for a
state $\rho$ and three orthogonal directions $\vec{n}_{1},$
$\vec{n}_{2}$, and $\vec{n}_{3}$, if the following
inequality~\cite{Korbicz2006} \begin{equation} -\frac{1}{3}\langle
J_{{\vec{n}}_{1}}^{3}\rangle+\langle
J_{{\vec{n}}_{2}}J_{{\vec{n}}_{1}}J_{{\vec{n}}_{2}}\rangle-\frac{N-2}{2}\langle
J_{{\vec{n}}_{3}}^{2}\rangle+\frac{1}{3}\langle
J_{{\vec{n}}_{1}}\rangle+\frac{N(N-1)(5N-2)}{24}<0\,,\label{ss1}\end{equation}
 is satisfied, the state $\rho$ possesses a genuine GHZ-type entanglement.
If one of the following inequalities \begin{align}
 & \langle J_{{\vec{n}}_{3}}^{3}\rangle-2\langle J_{{\vec{n}}_{2}}J_{{\vec{n}}_{3}}J_{{\vec{n}}_{2}}\rangle-2\langle J_{{\vec{n}}_{1}}J_{{\vec{n}}_{3}}J_{{\vec{n}}_{1}}\rangle\notag\\
 & -\frac{N-2}{2}\Big(2\langle J_{{\vec{n}}_{1}}^{2}\rangle+2\langle J_{{\vec{n}}_{2}}^{2}\rangle-\langle J_{{\vec{n}}_{3}}^{2}\rangle\Big)-\frac{N^{2}-4N+8}{4}\langle J_{{\vec{n}}_{3}}\rangle+\frac{N(N-2)(13N-4)}{24}<0,\label{ss2}\\
 & -\frac{1}{3}\langle J_{{\vec{n}}_{1}}^{3}\rangle+\langle J_{{\vec{n}}_{2}}J_{{\vec{n}}_{1}}J_{{\vec{n}}_{2}}\rangle-\frac{N-2}{2}\langle J_{{\vec{n}}_{3}}^{2}\rangle+\frac{1}{3}\langle J_{{\vec{n}}_{1}}\rangle+\frac{N^{2}(N-2)}{8}<0\,,\label{ss3}\end{align}
 are fulfilled, then the state $\rho$ possesses a genuine 3-qubit
entanglement.

\subsection{Spin-squeezing inequalities for higher spin-$j$ systems}

The entanglement criteria presented above are all suitable for
spin-1/2 systems. However, in most realistic experiments, atoms can
have larger spins. In general, we cannot straightforwardly
generalize the above criteria to spin-$j$ cases. To introduce the
spin-squeezing inequality for higher-spin systems, we first consider
the squeezing parameter $\xi_{R}^{2}$~\cite{Sorensen2001a}. Unlike
$\xi_{S}^{2}$, it is difficult to determine the minimum value of
$\xi_{R}^{2}$, since its denominator is not a constant. To find the
minimum value of $\xi_{R}^{2}$, which corresponds to the extreme
spin squeezing, one can use the Lagrange multiplier method, to find
the state that minimizes~\cite{Sorensen2001a}\begin{equation}
f=\mu\left\langle J_{z}\right\rangle +\left(\Delta
J_{x}\right)^{2}.\end{equation}
 Here the $z$-direction is assumed to be the MSD and $\Delta J_{x}$
is the minimum variance in the $x$-$y$ plane. As discussed in
Ref.~\cite{Sorensen2001a}, for integer spins the state that
minimizes $\Delta J_{x}$ for a given $\left\langle
J_{z}\right\rangle $ has vanishing $\left\langle J_{x}\right\rangle
$ and $\left\langle J_{y}\right\rangle $, thus $f=\mu\left\langle
J_{z}\right\rangle +\left\langle J_{x}^{2}\right\rangle $. To
minimizes $f$, one could find the ground state of $H=\mu
J_{z}+J_{x}^{2}$, which is just a transverse-field one-axis twisting
Hamiltonian. For half-integer spins, the problem becomes more
difficult, since the state that minimizes $\left\langle
J_{x}^{2}\right\rangle $ does not minimize $\Delta J_{x}$ for a
given $\left\langle J_{z}\right\rangle $, and then one cannot
formulate the problem as the diagonalization of an operator
containing a Lagrange multiplier term. A Monte Carlo variational
calculation that minimizes $f$ is presented in
Ref.~\cite{Sorensen2001a}.

Now, consider states of $N$ spin-$j$ particles. The collective spin
operator is $\mathcal{\vec{J}}\equiv\sum_{i}\vec{J}_{i}$, with $\vec{J}_{i}$
the spin operator for the $i$-th particle. A separable state of $N$
spin-$j$ particles could also be represented as \begin{equation}
\rho_{\text{sep}}=\sum_{k}p_{k}\rho_{k}^{(1)}\otimes\rho_{k}^{(2)}\otimes...\otimes\rho_{k}^{(N)},\end{equation}
 where $\rho_{k}^{\left(i\right)}$ is the density matrix of the $i$-th
spin-$j$ particle. For separable states, the variance of
$\mathcal{J}_{x}$ obeys the
inequality~\cite{Sorensen2001a}\begin{equation}
\left(\Delta\mathcal{J}_{x}\right)^{2}\geq\sum_{k}p_{k}\sum_{i=1}^{N}\left[\left(\Delta
J_{x}\right)^{2}\right]_{i}^{\left(k\right)}\geq\sum_{k}p_{k}\sum_{i=1}^{N}jF_{j}\left(\frac{\left\langle
J_{z}\right\rangle _{i}^{\left(k\right)}}{j}\right),\end{equation}
 where the first inequality comes from the concavity of the variance,
the function $F_{j}\left(x\right)$ is the minimum variance of $J_{x}$
divided by $j$ for a given $x$, $x=\left\langle J_{z}\right\rangle _{i}^{\left(k\right)}/j$,
and $F_{j}\left(x\right)$ is a convex function. Therefore, by considering
Jensen's inequality, we have\begin{align}
\left(\Delta\mathcal{J}_{x}\right)^{2} & \geq\sum_{k}p_{k}NjF_{j}\left(\sum_{i=1}^{N}\frac{\left\langle J_{z}\right\rangle _{i}^{\left(k\right)}}{Nj}\right)\geq NjF_{j}\left(\sum_{k}p_{k}\sum_{i=1}^{N}\frac{\left\langle J_{z}\right\rangle _{i}^{\left(k\right)}}{Nj}\right)\notag\\
 & =NjF_{j}\left(\frac{\left\langle \mathcal{J}_{z}\right\rangle }{Nj}\right),\end{align}
 and if the above inequality is violated, the spin-$j$ system is
entangled.

Another entanglement criterion is based on the following
inequality~\cite{T'oth2004a},
\begin{equation}
\left(\Delta J_{x}\right)^{2}+\left(\Delta J_{y}\right)^{2}+\left(\Delta J_{z}\right)^{2}\geq Nj,\label{singlet_critia}\end{equation}
 which holds for separable states of a multipartite $N$ spin-$j$
system. Based on this inequality, Ref.~\cite{T'oth2004a} gave a
spin-squeezing parameter\begin{equation}
\xi_{\text{singlet}}^{2}=\frac{\left(\Delta
J_{x}\right)^{2}+\left(\Delta J_{y}\right)^{2}+\left(\Delta
J_{z}\right)^{2}}{J},\end{equation}
 where $J=Nj$. The subscript `singlet' means that this parameter
can detect entanglement in singlet states. It has been proven in
Ref.~\cite{T'oth2004a} that, states satisfying \begin{equation}
\xi_{\text{singlet}}^{2}<1\end{equation}
 are entangled. This inequality can be used to detect entanglement
in the vicinity of many-qubit singlet states. These are pure states
that are invariant under a simultaneous unitary rotation on all qubits.
For example, for two qubits, the only state is the two-qubit singlet
state, which is invariant under rotation.

\subsection{Two-mode spin-squeezed states}

Spin squeezing discussed above is also called one-mode spin
squeezing, which describes the fluctuations of collective spin
operators $\hat{J}_{\alpha}$ ($\alpha=x,\, y,\, z$) of a total
system. The two-mode spin squeezing
\cite{Julsgaard2001,Kuzmich2000a,Berry2002} is analogous to the
definition of the two-mode squeezing of continuous observables
\cite{Duan2000}, and is proved to be a criterion of inseparability
between spin variables of two separated atomic samples.

A bipartite system is separable if and only if its state can be written
as \begin{equation}
\rho=\sum_{i}p_{i}\rho_{i}^{\left(1\right)}\otimes\rho_{i}^{\left(2\right)},\end{equation}
 where $p_{i}$ denotes the probability. In continuous case, it has
been proven that \cite{Duan2000b,Simon2000} the following
inequality\begin{equation}
\Delta\left(q^{\left(1\right)}+q^{\left(2\right)}\right)^{2}+\Delta\left(p^{\left(1\right)}-p^{\left(2\right)}\right)^{2}<2\label{eq:TM_S}\end{equation}
 is a sufficient condition for entanglement between subsystems 1 and
2. The continuous operators $q^{(i)}$ and $p^{(i)}$ belong to subsystems
$i$, satisfying $\left[q^{(k)},\, p^{(l)}\right]=i\delta_{k,l}$
($k,\, l=1,\,2$).

For spin systems, the observables of interest are spin operators\begin{equation}
\hat{J}_{\alpha}^{\left(\pm\right)}=\hat{J}_{\alpha}^{\left(1\right)}\pm\hat{J}_{\alpha}^{\left(2\right)}.\end{equation}
 The usual criterion for two-mode spin squeezing is \cite{Berry2005}\begin{equation}
\left(\Delta J_{z}^{(+)}\right)^{2}+\left(\Delta J_{y}^{(-)}\right)^{2}<\left\langle J_{x}^{\left(+\right)}\right\rangle .\label{eq:TM_SS}\end{equation}
 It has been shown that \cite{Julsgaard2001,Berry2002,Raymer2003},
two-mode spin squeezing implies entanglement between spin components
of the two subsystems. Furthermore, for pure states of two spin
systems of equal dimension, two-mode spin squeezing after
application of local unitaries is a necessary condition for
entanglement, except for a set of bipartite pure states of measure
zero \cite{Berry2005}. Account for the Heisenberg-Weyl algebra
discussed in Sec.~2.9, Eq.~(\ref{eq:TM_SS}) can reduce to
Eq.~(\ref{eq:TM_S}). In
Refs.~\cite{Julsgaard2001,Berry2002,Raymer2003}, two spatially
separated atomic ensembles, each containing about $10^{12}$ Cs
atoms, were entangled for 0.5~ms via interacting with a polarized
field. The entangled state generated in this experiment is similar
to a two-mode squeezing but not the maximally entangled state.
Besides, the two-mode spin-squeezed states can also be generated via
QND measurement with feedback \cite{Berry2002a}. Due to the
considerable long lifetime of entanglement, two-mode spin squeezing
was proposed to be a valuable resource of quantum information, and
could be used to perform atomic quantum state teleportation and
swapping \cite{Kuzmich2000a,Berry2002}.

\section{Spin squeezing, Fisher information, and quantum metrology\label{Fisher}}

As discussed previously, the squeezing parameter $\xi_{R}^{2}$
characterizes the sensitivity of a state with respect to SU(2)
rotations, and has been studied in quantum metrology. In this
section, we first introduce the
QFI~\cite{Helstrom1976,Caves1981,Holevo1982,Kok2010}, which
determines the precision of the parameter estimation. Then, we
discuss the relation between QFI and spin
squeezing~\cite{Pezze2009,Ma2009,Hyllus2010,Jin2010}. Finally, we
discuss the applications of spin squeezing and QFI in entanglement
enhanced quantum metrology, where the Ramsey and Mach-Zehnder
interferometers are discussed.

\subsection{Quantum Fisher information}

The Fisher information measures the amount of information of a
parameter that we can extract from a probability distribution. It
determines how precise we can attain when estimate a parameter, and
with a larger Fisher information we can estimate the parameter with
higher precision. Firstly, we begin with a brief discussion about
the Fisher information in probability theory, and for more
information please refer to Chapter 13 in Ref.~\cite{Kok2010}. In
the regime of the parameter estimation theory, the central problem
is to estimate the parameter $\lambda$ in a probability distribution
$p\left(x|\lambda\right)$, where $x$ is the random variable, and
below we only consider the single parameter case.

The minimum variance of our estimation is determined by the Fisher
information. Consider a general distribution $p\left(x|\lambda\right)$,
to estimate the parameter $\lambda$, we construct an estimator $E\left(x\right)$
which is a map from the experimental data $x$ to the parameter $\lambda$.
The expectation value of the estimator is \begin{equation}
\langle E\left(x\right)\rangle=\int dx\, p\left(x|\lambda\right)E\left(x\right).\end{equation}
 We consider the case $\left\langle E\left(x\right)\right\rangle =\lambda$,
i.e., so-called unbiased estimation, thus we have \begin{equation}
\int dx\, p\left(x|\lambda\right)\left[E\left(x\right)-\lambda\right]=0.\end{equation}
 Differentiating both sides with respect to $\lambda$ gives\begin{equation}
\int dx\, p\left(x|\lambda\right)L\left(x,\lambda\right)\left[E\left(x\right)-\lambda\right]=1,\label{eq:fisher-1}\end{equation}
 where \begin{equation}
L\left(x,\lambda\right)=\frac{\partial\ln p\left(x|\lambda\right)}{\partial\lambda}.\label{eq:LD}\end{equation}
 Now, square both sides of Eq.~(\ref{eq:fisher-1}) and use the Cauchy-Schwarz
inequality:\begin{equation}
\left|\langle f,\, g\rangle\right|^{2}\le\langle f,\, f\rangle\langle g,\, g\rangle,\end{equation}
 we obtain the Cram\'{e}r-Rao inequality: \begin{equation}
\left(\Delta E\left(x\right)\right)^{2}\ge\frac{1}{N\mathcal{I}_{\lambda}},\label{eq:CR_ineq}\end{equation}
 where \begin{equation}
\left(\Delta E\left(x\right)\right)^{2}=\int dx\, p\left(x|\lambda\right)\left[E\left(x\right)-\lambda\right]^{2}\end{equation}
 is the variance of $E\left(x\right)$, and \begin{equation}
\mathcal{I}_{\lambda}=\int dx\, p\left(x|\lambda\right)L\left(x,\lambda\right)^{2}\label{eq:FI}\end{equation}
is the Fisher information with respect to $\lambda$. $N$ is the
number of independent experiments (here $N=1$), i.e., by repeating
the experiment $N$ times, the precision of $E\left(x\right)$ is
improved by $1/N$. In practice, we need to maximize the precision
of $\lambda$, while this is fundamentally limited by the Fisher information
$\mathcal{I}_{\lambda}$, which can be regarded as the amount of information
of $\lambda$ that we can get from $p\left(x|\lambda\right)$, i.e.,
with more information, we can make more precise estimation.

Now we turn to discuss the
QFI~\cite{Helstrom1976,Holevo1982,Kok2010}, which is the extension
of the classical Fisher information in the quantum regime. Consider
an $n\times n$ density matrix $\rho\left(\lambda\right)$, by
performing the positive operator valued measure (POVM) $\left\{
P_{i}\right\} $, the parameter $\lambda$ resides in the outcomes
probability as \begin{equation}
p_{i}\left(\lambda\right)=\text{Tr}\left(\rho\left(\lambda\right)P_{i}\right).\end{equation}
 According to Eq.~(\ref{eq:FI}), we have\begin{equation}
\mathcal{I}_{\lambda}=\sum_{i}\frac{\left(\text{Re}\left\{ \text{Tr}\left[L_{\lambda}\rho\left(\lambda\right)P_{i}\right]\right\} \right)^{2}}{p_{i}\left(\lambda\right)},\end{equation}
 where $L_{\lambda}$ is the so-called symmetric logarithmic derivative
determined by the following equation \begin{equation}
\frac{\partial\rho\left(\lambda\right)}{\partial\lambda}=\frac{1}{2}\left[\rho\left(\lambda\right)L_{\lambda}+L_{\lambda}\rho\left(\lambda\right)\right].\end{equation}
 The operator $L_{\lambda}$ is a quantum analogy of $L\left(x,\lambda\right)$
in Eq.~(\ref{eq:LD}). Thus, a set of POVM yields a corresponding
probability distribution, which gives the Fisher information
$\mathcal{I}_{\lambda}$ of the parameter $\lambda$. It was proven
that\cite{Braunstein1994}
\begin{equation}
\mathcal{I}_{\lambda}\le F_{\lambda}=\text{Tr}\left[\rho\left(\lambda\right)L_{\lambda}^{2}\right],\end{equation}
 where $F_{\lambda}$ is the QFI, which is measurement-independent.
The explicit expression of the QFI is given by\begin{equation}
F_{\lambda}=\sum_{i=1}^{k}\frac{\left(\partial_{\lambda}p_{i}\right)^{2}}{p_{i}}+\sum_{i\neq j}^{k}\frac{2\left(p_{i}-p_{j}\right)^{2}}{p_{i}+p_{j}}\left\vert \left\langle \varphi_{i}|\partial_{\lambda}\varphi_{j}\right\rangle \right\vert ^{2},\text{ \ }(k\leq n),\label{FI1}\end{equation}
 where $p_{i}$ and $|\varphi_{i}\rangle$ are the $i$th nonzero
eigenvalue and eigenvector of $\rho\left(\lambda\right)$. Therefore,
they are all functions of the parameter $\lambda$. The probability
distribution that gives the maximum Fisher information, i.e. the QFI,
is produced by the optimal POVM built of the eigenprojectors of $L_{\lambda}$.
However, in some cases $L_{\lambda}$ is not a proper physical observable,
and the corresponding POVM cannot be realized in experiments, thus
the QFI is not always attainable.

The role of the QFI in parameter estimation is given by the quantum
Cram\'{e}r-Rao bound~\cite{Helstrom1976,Holevo1982},\begin{equation}
\left(\Delta\hat{\lambda}\right)^{2}\geq\left(\Delta\lambda\right)_{\text{QCB}}^{2}\equiv\frac{1}{N_{m}F_{\lambda}},\label{CRB}\end{equation}
 where $N_{m}$ is the number of independent experiments, $\hat{\lambda}$
is the so-called unbiased estimator of the parameter $\lambda$, i.e.
$\langle\hat{\lambda}\rangle=\lambda$. Indeed, $\hat{\lambda}$ is a
map from the experimental data to the parameter space. The
Cram\'{e}r-Rao bound gives the ultimate limit for the precision of
$\lambda$ that can be achieved. In a sense, parameter estimation is
equivalent to distinguishing neighboring states along the path in
parameter space. The QFI has a more intuitive geometric explanation,
and it is the geometric metric of the state
$\rho\left(\lambda\right)$ in parameter space, since
\begin{equation}
F_{\lambda}(d\lambda)^{2}=4(ds_{B})^{2},\end{equation}
 where $(ds_{B})^{2}$ is the Bures distance~\cite{Bures1969}.
Thus, the Fisher information is equivalent to the so-called fidelity
susceptibility~\cite{Gu2007}, that has been extensively studied for
characterizing QPT.

\subsection{Spin squeezing and quantum Fisher information}

Recently, it has been found that QFI gives a more stringent
criterion for entanglement than the spin-squeezing parameter
$\xi_{R}^{2}$ \cite{Pezze2009,Hyllus2010}. However, for identical
particles, it was shown that \cite{Benatti2010,Benatti2010a},
neither spin squeezing nor quantum Fisher information is an
entanglement witness. Below, we deal with distinguishable particles.
Consider an ensemble of $N$ spin-1/2 particles represented by a
density matrix $\rho$. The QFI with respect to $\theta$ is given by
\begin{equation}
F\left[\rho\left(\theta\right),J_{\vec{n}}\right]=\text{Tr}\left[\rho\left(\theta\right)L_{\theta}^{2}\right],\end{equation}
 where \begin{equation}
\rho\left(\theta\right)=\exp\left(-i\theta J_{\vec{n}}\right)\rho\exp\left(i\theta J_{\vec{n}}\right),\label{rho_theta}\end{equation}
 and $J_{\vec{n}}$ is the generator of the rotation along the direction
$\vec{n}.$ For pure states, the QFI
becomes~\cite{Pezze2009}\begin{equation}
F\left[\rho\left(\theta\right),J_{\vec{n}}\right]=4\left(\Delta
J_{\vec{n}}\right)^{2}\text{.}\label{eq:PureQFI}\end{equation}
 For mixed states\begin{equation}
F\left[\rho\left(\theta\right),J_{\vec{n}}\right]\equiv\sum_{i\neq j}\frac{2\left(p_{i}-p_{j}\right)^{2}}{p_{i}+p_{j}}\left\vert \left\langle \varphi_{i}|J_{\vec{n}}|\varphi_{j}\right\rangle \right\vert ^{2}.\label{FI2}\end{equation}
 Comparing Eq.~(\ref{FI1}) with Eq.~(\ref{FI2}), there are no
derivatives of $p_{i}$, since the transformation in Eq.~(\ref{rho_theta})
does not change the eigenvalues of $\rho$. Actually, as explained
in the previous section, $F\left[\rho\left(\theta\right),J_{\vec{n}}\right]$
characterizes the geometric properties of $\rho$ with respect to
rotation, and is the rotational sensitivity of the state, thus we
need not to perform a true rotation.

According to Eq.~(\ref{CRB}), the lower bound of the uncertainty
of $\theta$ is given by\begin{equation}
\left(\Delta\theta\right)_{\text{QCB}}^{2}=\frac{1}{F\left[\rho\left(\theta\right),J_{\vec{n}}\right]},\label{theta_bound1}\end{equation}
 where we set $N_{m}=1$. If we measure the angular momentum operator
$J$, the lower bound of $\Delta\theta$ becomes\begin{equation}
\left(\Delta\theta\right)_{\text{SS}}^{2}=\frac{\xi_{R}^{2}}{N},\label{theta_bound2}\end{equation}
 where $\xi_{R}^{2}$ is the spin-squeezing parameter. Since the Cram\'{e}r-Rao
bound~\cite{Helstrom1976,Holevo1982} gives the ultimate limit of the
precision of $\theta$, we must have\begin{equation}
\left(\Delta\theta\right)_{\text{QCB}}^{2}\leq\left(\Delta\theta\right)_{\text{SS}}^{2},\end{equation}
 and thus \begin{equation}
\chi^{2}\equiv\frac{N}{F\left[\rho\left(\theta\right),J_{\vec{n}}\right]}\leq\xi_{R}^{2},\label{ineq_chi_xi}\end{equation}
 which was proved in Ref.~\cite{Pezze2009}. They also proved that
if \begin{equation}
\chi^{2}<1,\end{equation}
 the state is entangled. It is known that $\xi_{R}^{2}<1$ also indicates
entanglement. Thus, $\chi^{2}<1$ is more stringent than $\xi_{R}^{2}<1$
in detecting multipartite entanglement, regarding the cases when $\chi^{2}<1$
and $\xi_{R}^{2}\geq1$.

The inequality (\ref{ineq_chi_xi})\ can be readily proven when $\rho\left(\theta\right)$
is a pure state, since in this case the QFI is just as Eq.~(\ref{eq:PureQFI}).
By using the Heisenberg uncertainty relation, we have\begin{equation}
F\left[\rho\left(\theta\right),J_{\vec{n}_{\bot}^{\prime}}\right]\left(\Delta J_{\vec{n}_{\perp}}\right)^{2}\geq|\langle J_{\vec{n}}\rangle|^{2},\end{equation}
 where the directions $\vec{n}_{\bot}$, $\vec{n}_{\bot}^{\prime}$,
and $\vec{n}$ are orthogonal to each other. Then, according to the
definitions of $\xi_{R}^{2}$ and $\chi^{2}$, we can obtain the inequality~(\ref{ineq_chi_xi}).
Both the spin-squeezing parameters and the QFI could be used as criteria
for entanglement, which is a resource for high-precision measurements.
Note that, generally, it is difficult for experiments to achieve the
precision given theoretically by the QFI due to practical difficulties
of the measurements.

\begin{figure}[H]
 \includegraphics[scale=0.6]{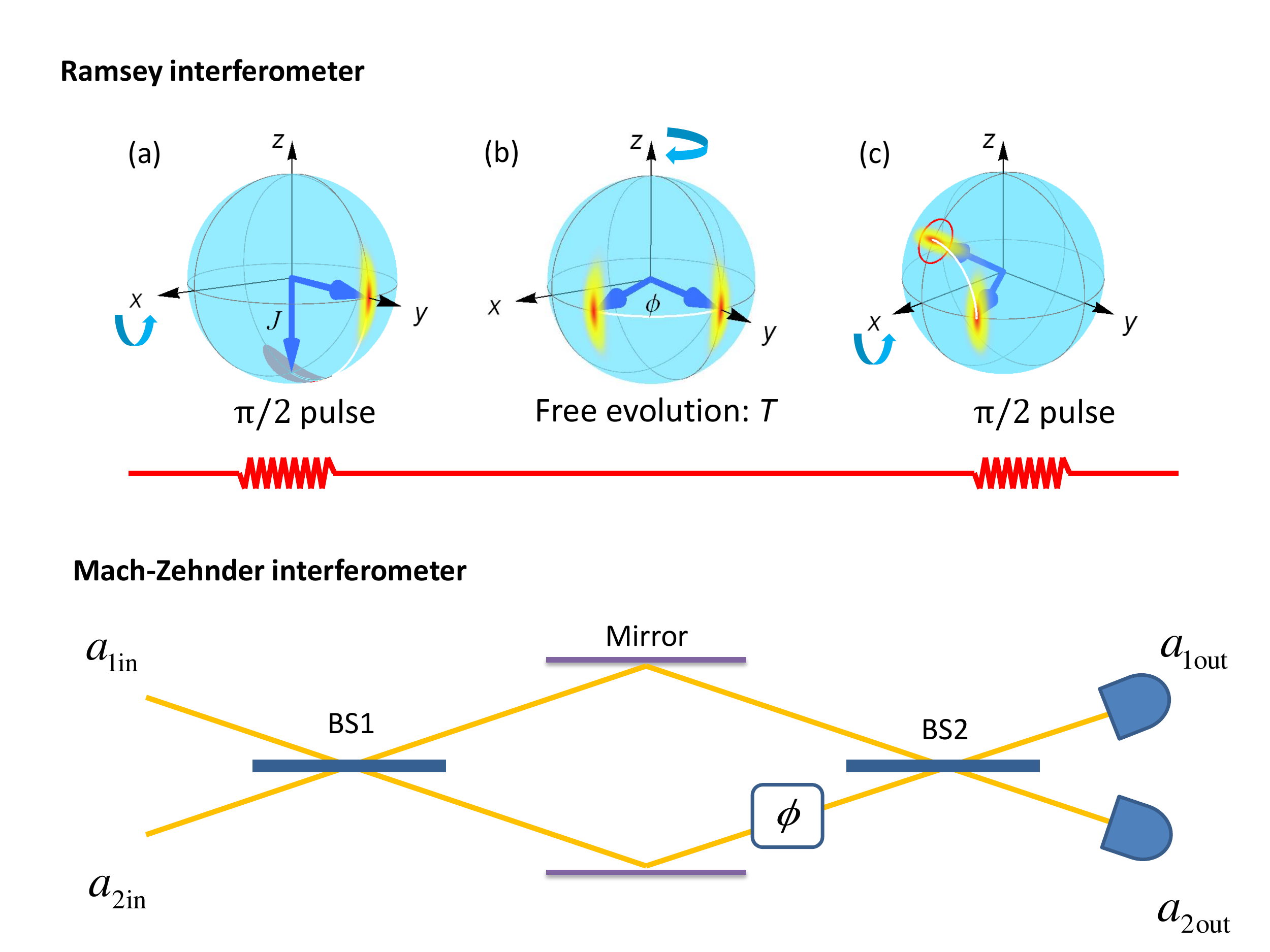}

\caption{(Color online) Schematic diagrams of the Ramsey (above) and Mach-Zehnder
(below) interferometers. For the Ramsey interferometer, the blue arrows
represent the states of the spin. The uncertainty of the spin components
are represented by the elliptical multi-color disks. The red circle
in (c) represent the uncertainty of a CSS. For the Mach-Zehnder interferometer,
two beams injected from both sides of the first beam spliter, BS1.
The parameter $\phi$ is the phase difference gained between the two
beam spliters BS1 and BS2. After the second beam spliter, BS2, photon
numbers are detected, and the number difference between the two detectors
is related to the phase difference $\phi$. The two beam spliters
correspond to the two $\pi/2$ pulses in the Ramsey process (a) and
(c).}

\label{fig_Mach_Zehnder}
\end{figure}

\subsection{Spin squeezing, Fisher information and metrology\label{sub:metrology}}

Here, we discuss spin squeezing and QFI in quantum
metrology~\cite{Caves1981,Sanders1995,Giovannetti2006,Boixo2008,Zanardi2008}.
In general, we cannot access parameters of a state or a Hamiltonian
directly, since they are not physical observables. But in many
cases, the parameter is related to an observable, like in the Ramsey
spectroscopy, where the phase $\phi$ gained in the free evolution is
related to the population differences, and the parameters can also
be obtained via Landau-Zener-Stückelberg
interferometry~\cite{Shevchenko2010}. This parameter estimation
scheme is illustrated below,\begin{gather}
\text{Prepare a state }\rho,\nonumber \\
\Downarrow\nonumber \\
\text{Quantum evolution }\rho_{\phi}=U\left(\phi\right)\rho U^{\dagger}\left(\phi\right),\nonumber \\
\Downarrow\nonumber \\
\text{Measurement \ }\langle\hat{O}\rangle_{\phi}=\text{Tr}\left[\hat{O}\rho_{\phi}\right],\label{param_est}\end{gather}
 where the information of the parameter $\phi$ is contained in the
measurement result $\langle\hat{O}\rangle_{\phi}$, and in some cases
the observable $\hat{O}$ cannot be accessed directly, and this was
studied in Ref.~\cite{Burgarth2009}. The fluctuation of the
observable $\hat{O}$ is unavoidable, and the variance of $\phi$
is\begin{equation}
\Delta\phi=\frac{(\Delta\hat{O})_{\phi}}{\left\vert
\partial\langle\hat{O}\rangle_{\phi}/\partial\phi\right\vert
}.\label{eq:error_prop}\end{equation} Since $\phi$ is periodic, its
mean and variance are not calculated as non-periodic variables, and
this should be declared. To calculate the mean of $\phi$, we first
obtain the mean of $\hat{O}$, that is
$\langle\hat{O}\rangle_{\phi}$. Then the mean of $\phi$ is the
inverse function of $\langle\hat{O}\rangle_{\phi}$. For more
details, please refer to Ref.~\cite{wiki_mean_circle}. Since in many
cases the evolution processes are fixed (or hard to change), we can
reduce the variance of $\phi$ by choosing an appropriate initial
state $\rho$ and operator $\hat{O}$. When we measure $J_{z}$, the
precision is given by squeezing parameter $\xi_{R}^{2}$ as shown in
Eq.~(\ref{phase_xi_r}).

We first give a brief summary of Ramsey processes, shown in Fig.~\ref{fig_Mach_Zehnder}.
Consider an ensemble of $N$ two-level particles interacting with
an applied magnetic field $\vec{B}$. The Hamiltonian of the two-level
particles\begin{equation}
H=-\vec{\mu}\cdot\vec{B},\end{equation}
 where $\vec{\mu}=\mu_{0}\vec{J}$ is the magnetic moment and $\vec{B}=B_{0}\vec{n}_{z}+\vec{B}_{1}$,
with a static magnetic field \begin{equation}
B_{0}=-\hbar\omega_{0}/\mu_{0}\text{, \ \ }\mu_{0}<0,\end{equation}
 and a time-dependent field\begin{equation}
\vec{B}_{1}=B_{1}\left[\vec{n}_{x}\cos(\omega t)+\vec{n}_{y}\sin(\omega t)\right].\end{equation}
 The Hamiltonian now becomes \begin{equation}
H=\hbar\omega_{0}J_{z}+\frac{\hbar\Omega_{R}}{2}\left(J_{+}e^{-i\omega t}+\text{h.c.}\right),\end{equation}
 where $\Omega_{R}=\left\vert \mu_{0}B_{1}\right\vert /\hbar$ is
the Rabi frequency. It is convenient to use a frame of reference rotating
around $B_{0}$ with frequency $\omega$,\begin{equation}
H_{R}=\hbar\left(\omega_{0}-\omega\right)J_{z}+\hbar\Omega_{R}J_{x},\label{rotating_Ham}\end{equation}
 where the second term in Eq.~(\ref{rotating_Ham}) acts as a pulse.
When this pulse is applied, the above Hamiltonian is approximated
as\begin{equation}
H_{R}\simeq\hbar\Omega_{R}J_{x},\label{pulse}\end{equation}
 since $\Omega_{R}\gg\left\vert \omega_{0}-\omega\right\vert $.

As shown in Fig.~(\ref{fig_Mach_Zehnder}), the Ramsey interferometry
consists of two $\pi/2$-pulses of length $t_{\pi/2}=\pi/\left(2\Omega_{R}\right)$
and a free evolution of length $t$. The first pulse plays as a $\pi/2$
rotation around the $x$-axis. Thereafter, during the free-evolution
period, when $B_{1}=0$, the state vector precesses about the $z$-axis
and acquires a phase $\phi=\left(\omega-\omega_{0}\right)t$. Assuming
$t\gg t_{\pi/2}$, the time of the entire process is \begin{equation}
t_{f}=2t_{\pi/2}+t\simeq t.\end{equation}
 After the free period, followed by a second Ramsey pulse, the spin
direction is rotated around the $x$-axis by $\pi/2$. The the initial
state evolves to\[
|\psi\left(t\right)\rangle=U|\psi\left(0\right)\rangle,\]
 where the unitary operator is\begin{align}
U & =\exp\left({-i\frac{\pi}{2}J_{x}}\right)\exp\left({i\phi J_{z}}\right)\exp\left({-i\frac{\pi}{2}J_{x}}\right)\nonumber \\
 & =\exp\left({-i\phi J_{y}}\right)\exp\left({-i\pi J_{x}}\right).\label{eq:ramsey_u}\end{align}
 Then we could measure the number of atoms in the excited energy level
$|0\rangle$ to estimate $\phi$. This is equivalent to measuring
$J_{z}$:

\begin{equation}
\left\langle J_{z}\right\rangle _{t}=\left\langle J_{x}\right\rangle _{t=0}\sin\phi-\left\langle J_{z}\right\rangle _{t=0}\cos\phi,\label{eq:Jzt}\end{equation}
 where the subscript $t$ denotes the time, and in the right-hand
side, the average $\left\langle \cdot\right\rangle _{t=0}$ is carried
out using the initial state. From the evolution operator $U$, we
can further find the variance \begin{align}
\left(\Delta J_{z}\right)_{t}^{2} & =\cos^{2}\phi\left(\Delta J_{z}^{2}\right)_{t=0}+\sin^{2}\phi\left(\Delta J_{x}^{2}\right)_{t=0}-\sin(2\phi)\text{Cov}(J_{x},J_{z})_{t=0}.\label{eq:dJzt}\end{align}
 Then $\omega_{0}$ is related to the measurement data of $\left\langle J_{z}\right\rangle $
via\begin{equation}
\omega_{0}=\omega+\frac{1}{t}\arccos\left(-\frac{\left\langle J_{z}\right\rangle _{t}}{\left\langle J_{z}\right\rangle _{t=0}}\right).\end{equation}
 In practice, we repeat the above procedure $N_{m}$ times, with a
total experimental time $T=N_{m}t$. Then, we can estimate $\phi$
with the uncertainty \begin{align}
\left(\Delta\phi\right)^{2} & =\frac{\left(\Delta J_{z}\right)_{t}^{2}}{N_{m}\left\vert \partial\left\langle J_{z}\right\rangle _{t}/\partial\phi\right\vert ^{2}},\label{eq:dphi}\end{align}
 which is obtained by using the propagation of the fluctuation (\ref{eq:error_prop}).
Below, we set $N_{m}=1$.

\begin{figure}[ptb]

\begin{centering}
\includegraphics[width=14cm]{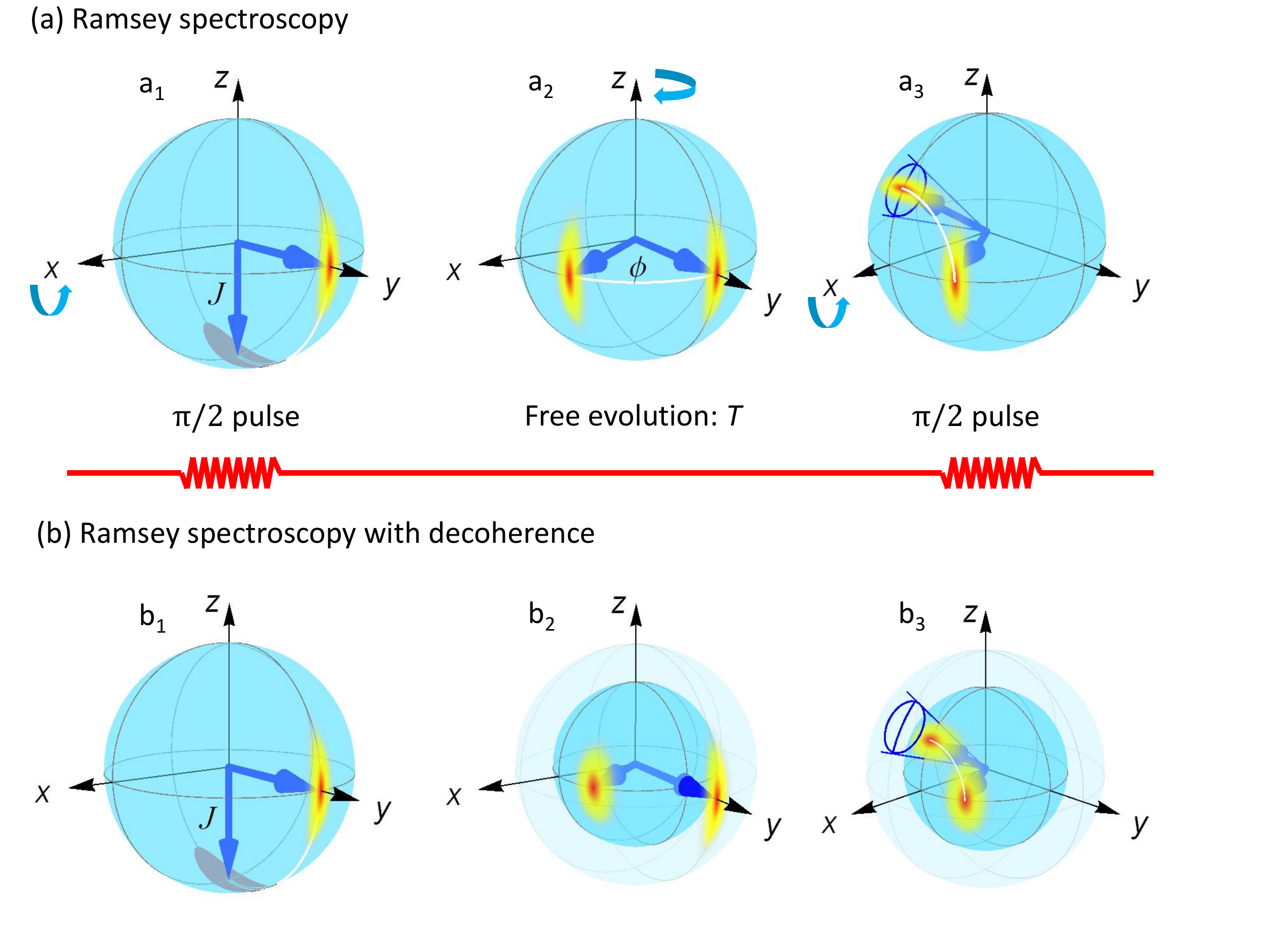}
\par\end{centering}

\caption{(Color online) Schematic diagram of the state evolution for Ramsey
spectroscopy. Here the initial state is spin-squeezed in the $x$-direction,
with spin length $J$. At each {}``snapshot\textquotedblright{},
the initial and final states are joined by a white {}``path\textquotedblright{}or
trajectory on the Bloch sphere. The phase $\phi$ gained in the free-evolution
period is determined by measuring the population differences, between
particles in states $|0\rangle$ and $|1\rangle$, which is equivalent
to measuring $J_{z}$. The precision of $\phi$ is thus determined
by the spin length and the fluctuation in $J_{z}$. In the absence
of decoherence (a), a squeezed state (fuzzy elliptic rainbow) performs
better (i.e., has smaller fluctuations in $J_{z}$) than a coherent
spin state (blue circle), shown in a$_{3}$. When decoherence is considered,
during the free-evolution period, the spin length is shorter and the
fluctuations become larger. Thus, in the final population measurement,
b$_{3}$, the precision of $\phi$ becomes lower.}

\label{fig_Ramsey}
\end{figure}

From Eqs.~(\ref{eq:Jzt}) (\ref{eq:dJzt}) and (\ref{eq:dphi}),
the phase uncertainty is related to the initial states and the phase.
Below, we discuss the uncertainty of $\phi$ for different initial
states. If the initial state is prepared as a CSS \begin{equation}
|\psi\left(0\right)\rangle_{\text{CSS}}=|j,-j\rangle=|1\rangle^{\otimes N},\label{ramsey_ini}\end{equation}
 i.e., all the atoms are prepared in their ground states and the spin
vector points in the $-z$ direction, as shown in Fig.~\ref{fig_Ramsey}.
For this initial state shown in Eq.~(\ref{ramsey_ini}), we can find
\begin{equation}
\left(\Delta\phi\right)_{\text{CSS}}^{2}=\frac{1}{N},\end{equation}
 which is the shot-noise limit, and we have already known in Sec.~2.5
since $\xi_{R}^{2}=1$ for a CSS. We can also obtain $F_{\phi}=1$
by using Eq.~(\ref{FI1}).

Next, we consider a spin-squeezed state\begin{equation}
|\psi\left(0\right)\rangle_{\text{SSS}}=\frac{1}{\sqrt{2}}\left[|j,0\rangle_{x}-\frac{1}{\sqrt{2}}\left(|j,+1\rangle_{x}+|j,-1\rangle_{x}\right)\right],\end{equation}
 where the subscript $x$ denotes the state is the eigenstate of $J_{x}$,
i.e., $J_{x}|j,m\rangle_{x}=m|j,m\rangle_{x}$. This state was
studied in Ref.~\cite{Andre2002a}, the expectation values
are\begin{align}
\left\langle J_{x}\right\rangle _{t=0} & =\left\langle J_{y}\right\rangle _{t=0}=\text{Cov}\left(J_{x},J_{z}\right)=0,\nonumber \\
\left\langle J_{z}\right\rangle _{t=0} & =\sqrt{\frac{j(j+1)}{2}},\end{align}
 and\begin{align}
\left(\Delta J_{x}\right)_{t=0}^{2} & =\frac{1}{2},\nonumber \\
\left(\Delta J_{y}\right)_{t=0}^{2} & =\frac{3}{8}j(j+1)-\frac{1}{4},\nonumber \\
\left(\Delta J_{z}\right)_{t=0}^{2} & =\frac{1}{8}j(j+1)-\frac{1}{4}.\end{align}
 This state is spin squeezed, since\begin{equation}
\xi_{R}^{2}=\frac{2}{N/2+1},\end{equation}
 and the phase uncertainty is optimized at $\phi\simeq\pi/2$,\begin{equation}
\left(\Delta\phi\right)_{\text{SSS}}^{2}=\frac{1}{j(j+1)}\propto\frac{1}{N^{2}},\end{equation}
 which attains the Heisenberg limit.

Now, we take the initial state to be a maximally entangled state\begin{equation}
|\psi\left(0\right)\rangle_{\text{GHZ}}=\frac{1}{\sqrt{2}}\left(|j,j\rangle_{y}+|j,-j\rangle_{y}\right),\end{equation}
 which is the $N$-body GHZ state, or the NOON state in quantum optics.
We emphasize that, for this state the phase cannot be estimated by
measuring the population difference $J_{z}$, since\begin{equation}
\left\langle J_{z}\right\rangle _{t=0}=\left\langle J_{x}\right\rangle _{t=0}=0,\end{equation}
 that means we cannot get any information about $\phi$, and the spin
squeezing parameter $\xi_{R}^{2}$ is divergent. On the other hand,
after the Ramsey process\begin{equation}
|\psi\left(\phi\right)\rangle_{\text{GHZ}}=U|\psi\left(0\right)\rangle_{\text{GHZ}}=\frac{1}{\sqrt{2}}\left(|j,j\rangle_{y}+e^{-iN\phi}|j,-j\rangle_{y}\right).\end{equation}
 By using Eq.~(\ref{FI1}) we find its QFI $F_{\phi}=N^{2}$, which
implies that the precision of $\phi$ could attain the Heisenberg
limit. To achieve this precision, we should measure the parity
operator~\cite{Bollinger1996,Meyer2001,Leibfried2004}\begin{equation}
P\equiv\prod_{i=1}^{N}\sigma_{iz}.\label{parity}\end{equation}
 The expectation values and variances of $P$ under the state $|\psi\left(\phi\right)\rangle_{\text{GHZ}}$
are \begin{align}
\left\langle P\right\rangle _{t} & =\cos N\phi,\quad\left\langle P^{2}\right\rangle _{t}=1,\end{align}
 thus the optimal phase uncertainty\begin{equation}
\left(\Delta\phi\right)_{\text{GHZ}}^{2}=\frac{1}{N^{2}}\end{equation}
 is attained when $N\phi\simeq\pi/2$, which is the Heisenberg-limit
uncertainty. Although the Heisenberg-limit precision is attained,
we should note that, the parity operator $P$ shown in Eq.~(\ref{parity})
is not easy to measure in experiment as compared with the angular
momentum operators, especially when $N$ is large. Additionally, it
is also difficult to prepare an $N$-body GHZ state with nowadays
techniques. In summary, both spin squeezing and QFI are related to
the precision in phase estimation. Metrology based on spin squeezing
is comparatively easy to implement in experiments, while the ultimate
precision is determined by the QFI.

In the end we give a brief review about the Mach-Zehnder interferometer,
which can be regarded as an optical version Ramsey interferometer.
It consists of two beam spliters (BS) and two mirrors, and the schematic
diagram is shown in Fig.~\ref{fig_Mach_Zehnder}. Light beams passing
through the Mach-Zehnder interferometer undergo transformations as
\begin{equation}
\left(\begin{array}{c}
a_{1\text{out}}\\
a_{2\text{out}}\end{array}\right)=V_{\text{BS}}V_{\text{\ensuremath{\phi}}}V_{\text{BS}}\left(\begin{array}{c}
a_{1\text{in}}\\
a_{2\text{in}}\end{array}\right),\end{equation}
 where $a_{i}$ ($i=1,2$) are the annihilation operators of the $i$th
path, and \begin{equation}
V_{\text{BS}}=\exp\left(i\frac{\alpha}{2}\sigma_{x}\right)=\left(\begin{array}{cc}
\cos\frac{\alpha}{2} & i\sin\frac{\alpha}{2}\\
i\sin\frac{\alpha}{2} & \cos\frac{\alpha}{2}\end{array}\right),\end{equation}
 denotes the transform of the BS, $R=\sin\left(\alpha/2\right)^{2}$
and $T=\cos\left(\alpha/2\right)^{2}$ are the transmission and reflection
rates, respectively. The image number $i$ arises from the half-wave
loss. Below, we consider the 50-50 BS, that is $\alpha=\pi/2$, which
acts as the two $\pi/2$ pulses in the Ramsey process. The difference
between the two path lengths gives rise to a phase difference $\phi$,
and can be represented by acting the transform\begin{equation}
V_{\phi}=\exp\left(-i\frac{\phi}{2}\sigma_{z}\right)=\left(\begin{array}{cc}
e^{-i\phi/2} & 0\\
0 & e^{i\phi/2}\end{array}\right)\end{equation}
 on modes $a_{1}$ and $a_{2}$. It is convenient to manipulate these
transformations in the Schwinger representation (\ref{Schwinger_Rep})
as \begin{align}
J_{\alpha,\text{out}} & =\left(a_{1\text{out}}^{\dagger},a_{2\text{out}}^{\dagger}\right)\frac{\sigma_{\alpha}}{2}\left(\begin{array}{c}
a_{1\text{out}}\\
a_{2\text{out}}\end{array}\right)=UJ_{\alpha,\text{in}}U^{\dagger},\quad\left(\alpha=x,\, y,\, z\right),\end{align}
 where the unitary transform $U$ is just as Eq.~(\ref{eq:ramsey_u}).
At the output-port, the information of the phase $\phi$ is obtained
via detecting the photon number difference. Since the Mach-Zehnder
interferometry is an optical instrument, it is not easy to prepare
well-defined number of photons, unlike the Ramsey interferometer,
where the atom number is conserved.

\section{Spin squeezing, quantum phase transitions, and quantum chaos\label{qpt}}

In this section, we discuss the applications of spin squeezing in
identifying QPTs~\cite{Sachdev1999} and quantum
chaos~\cite{Haake1991}. For this task, spin squeezing mainly has
three advantages: (i) It is comparatively easy to be measured. (ii)
It is an entanglement witness. (iii) It characterizes the
sensitivity of a state with respect to SU(2) rotations, thus is
promising to detect quantum chaos.

The QPTs occur at absolute zero temperature, and is driven purely by
quantum fluctuations. Conventionally, QPTs were first studied by
Landau's order parameter theory in the framework of statistics and
condensed matter physics. When a QPT occurs, the ground state can
change drastically, and the correlation length diverges. Therefore,
considering these intrinsic properties of QPT, researchers
investigated it by using concepts borrowed from quantum
information~\cite{Nielsen2000,Stolze2008}, such as quantum
entanglement~\cite{Amico2008,Guehne2009} and
fidelity~\cite{Gu2010}.. As discussed previously, spin squeezing is
closely related to entanglement, and it is easier to measure
experimentally. Therefore, it is desirable to study spin squeezing
in QPTs.

Then we discuss using spin squeezing as a signature of quantum
chaos. In classical regime, one of the most distinct feature of
chaos is the extreme sensitivity of the trajectories with respects
to perturbation, however, in the quantum world due to the unitarity
of quantum evolutions, the overlap (or fidelity) between two
initially separated states is invariant during the evolution, and
thus there is no well-accepted definition of quantum chaos. To solve
this problem, various signatures of quantum chaos have been
identified{~}\cite{Haake1991,Heller1984,Schack1994}. Entanglement
and spin squeezing~\cite{KITAGAWA1993,WINELAND1994}, which are pure
quantum effects, have also been identified as signatures of quantum
chaos. Recently, entanglement, measured by the linear entropy, as a
signature of quantum chaos has been demonstrated experimentally in
an atomic ensemble~\cite{Chaudhury2009}. The experimental results
and the theoretical predictions coincide very
well~\cite{Chaudhury2009}.

\subsection{Spin squeezing and quantum phase transitions in the Lipkin-Meshkov-Glick model\label{sub:LMG}}

Below, we discuss the spin squeezing for the ground state of the
Lipkin-Meshkov-Glick model~\cite{Lipkin1965}, which occurs a typical
second-order QPT~\cite{Vidal2006,Vidal2004a}. It has been widely
studied statistical mechanics of quantum spin
system~\cite{Botet1982}, Bose-Einstein condensations
\cite{Cirac1998}, and superconducting circuits~\cite{Tsomokos2008}.
It is an exactly solvable~\cite{Draayer1999,JonLinks2003} many-body
interacting quantum system, as well as one of the simplest to show a
quantum transition in the strong coupling regime. It is convenient
to discuss spin squeezing in this model, since it consists of
collective spin operators.

The Hamiltonian of the Lipkin-Meshkov-Glick model reads\begin{equation}
H=-\frac{\lambda}{N}\left(J_{x}^{2}+\gamma J_{y}^{2}\right)-hJ_{z},\label{lmg1}\end{equation}
 where $J_{\alpha}=\sum_{i=1}^{N}\sigma_{\alpha}^{i}/2$ ($\alpha=x,y,z$)
are the collective spin operators; $\sigma_{\alpha}^{i}$ are the
Pauli matrices; $N$ is the total spin number;
$\gamma\in\left[0,1\right]$ is the anisotropic parameter; $\lambda$
and $h$ are the spin-spin interaction strengths and the effective
external field, respectively. Here we set $\lambda=1$, where the
Lipkin-Meshkov-Glick model describes ferromagnetism. The QPT of this
model originates from the competition between the spin-spin
interaction and the external field. With mean-field
approach~\cite{Dusuel2005}; we can see that, when $h>1$, all spins
tends to be polarized in the field direction $\left(\left\langle
\sigma_{z}^{i}\right\rangle =1\right)$. However, when $h<1$, it is
two-fold degenerate with $\left\langle \sigma_{z}^{i}\right\rangle
=h$. Therefore, a spontaneous symmetry breaking occurs at $h=1$,
which is a second-order QPT point between the so-called symmetric
($h\geq1$) phase and symmetry broken ($h<1$) phase. However, by
considering the quantum effects, the exact ground state is not
degenerate in the symmetry broken phase ($\gamma\neq1$) . Since the
Hamiltonian is of spin-flip symmetry, i.e.,
$\left[H,\;\prod_{i=1}^{N}\sigma_{z}^{i}\right]=0$, we have
$\left\langle J_{x}\right\rangle =\left\langle J_{y}\right\rangle
=0$, $\left\langle J_{x}J_{z}\right\rangle =\left\langle
J_{y}J_{z}\right\rangle =0$, and the MSD is along the $z$ direction.
In addition, $\left[H,\mathbf{J}^{2}\right]=0$, and the ground state
lies in the $j=N/2$ symmetric sector.

\begin{figure}[ptb]
\begin{centering}
\includegraphics[width=12cm]{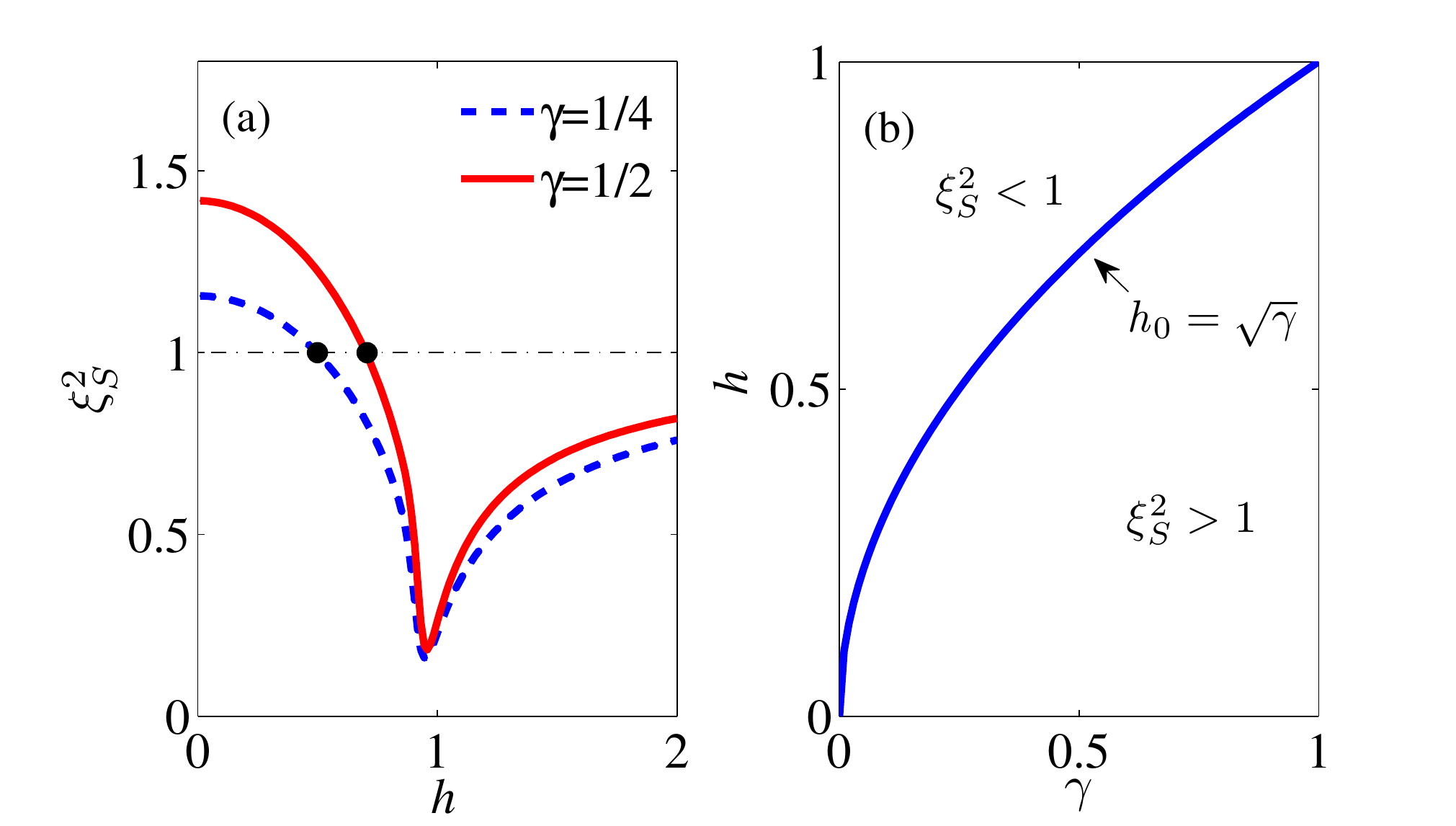}
\par\end{centering}

\caption{(Color online) Squeezing parameter $\xi_{S}^{2}$ versus the applied
magnetic field $h$ (a), and the `phase diagram' of spin squeezing
in the $h-\gamma$ plane (b), for the ground state of the Lipkin-Meshkov-Glick
model. The system size used here is $N=2^{7}$. The maximal squeezing
occurs at the critical point. The two black dots denote $\xi_{S}^{2}=1$,
at these points the ground states are coherent spin states. In (b),
we show the squeezed and non-squeezed regions that are separated by
$h_{0}=\sqrt{\gamma}$, $\xi_{S}^{2}=1$ in the thermodynamic limit.}

\label{fig_lmg_1}
\end{figure}

In the isotropic case, $\gamma=1$, the ground state is simply the
Dicke state, as discussed in Sec.~\ref{definition}, and $\xi_{S,R}^{2}\geq1$
for all Dicke state, indicating that there is no spin squeezing with
respect to $\xi_{S}^{2}$ and $\xi_{R}^{2}$. For the Dicke state
$|j,m\rangle$, we have\begin{equation}
\tilde{\xi}_{E}^{2}=\frac{m^{2}}{j^{2}}\leq1,\end{equation}
 thus the squeezing occurs along the $z$-axis direction with respect
to $\tilde{\xi}_{E}^{2}$.

In the anisotropic case, $\gamma\neq1$, there is a second-order QPT.
Since the system is of exchange symmetry, and the ground state has a
fixed parity, $\xi_{S}^{2}$ and $\tilde{\xi}_{E}^{2}$ are closely
related to the concurrence, as demonstrated in
Refs.~\cite{Vidal2006,Vidal2004a,Yin2009}. In the thermodynamic
limit, we could calculate the squeezing parameter by using the
Holstein-Primakoff transformation. The results
are~\cite{Ma2009}\begin{equation} \xi_{S}^{2}=\left\{
\begin{array}{c}
\sqrt{\left(h-1\right)/\left(h-\gamma\right)},\\
\sqrt{\left(1-h^{2}\right)/\left(1-\gamma\right)},\end{array}\begin{array}{c}
\text{for }h\geq1,\\
\text{for }h<1.\end{array}\right.\end{equation}
 As shown in Fig.~\ref{fig_lmg_1}, there is no spin squeezing when
$h$ is smaller than $h_{0}=\frac{N-1}{N}\sqrt{\gamma}.$ At the point
$h=h_{0},$ the ground state is a CSS, for which the squeezing parameters
are equal to one.

To demonstrate the QPT, we calculate the values of $\xi_{S}^{2}$ and
$\partial_{h}\xi_{S}^{2}$ in the vicinity of the critical point.
Reference \cite{Vidal2004a} found that at the critical
point\begin{align}
\xi_{S}^{2} & \sim1/N^{-0.33\pm0.01},\notag\\
\partial_{\lambda}\xi_{S}^{2} & \sim-1/N^{-0.33\pm0.03}.\end{align}
 The above results are immediate consequences of the concurrence $C$,
as shown in Ref.~\cite{Vidal2004a}. If $\xi_{S}^{2}\le1$, we find
$\xi_{S}^{2}=1-(N-1)C$ due to the exchange symmetry. Generalized
spin squeezing inequalities can be used to detect the ground state
entanglement for lattice models~\cite{Guehne2009}. However, as these
inequalities are not given in terms of parameters like
$\xi_{S}^{2}$, there are no such scaling laws, and we do not focus
on them here.

Besides spin squeezing and concurrence, other measures such as
negativity~\cite{Wichterich2010}, geometric
entanglement~\cite{Orus2008}, and entropy~\cite{Vidal2007} have also
been studied. Due to its symmetry, the spin-squeezing parameter
$\xi_{S}^{2}$ is closely related to the concurrence in this model.
At the critical point, the derivative of the parameter $\xi_{S}^{2}$
with respect to the external driving field strength tends to
diverge.

\subsection{Spin squeezing and quantum chaos in the Quantum kicked-top model\label{sub:QKT}}

In this section and the following, we discuss the behaviors of spin
squeezing in two typical quantum chaotic systems: the quantum kicked-top
model and the Dicke model. To facilitate the correspondence between
the classical and quantum regimes, the initial states are chosen as
CSSs. To connect the quantum dynamics to classical chaos, we choose
the initial states located in both chaotic and regular regions. The
dynamics of the squeezing parameters distinguish well between regular
and chaotic regions, and we conclude that spin squeezing could be
a good signature for quantum chaos.

\begin{figure}[ptb]
\begin{centering}
\includegraphics[width=8cm]{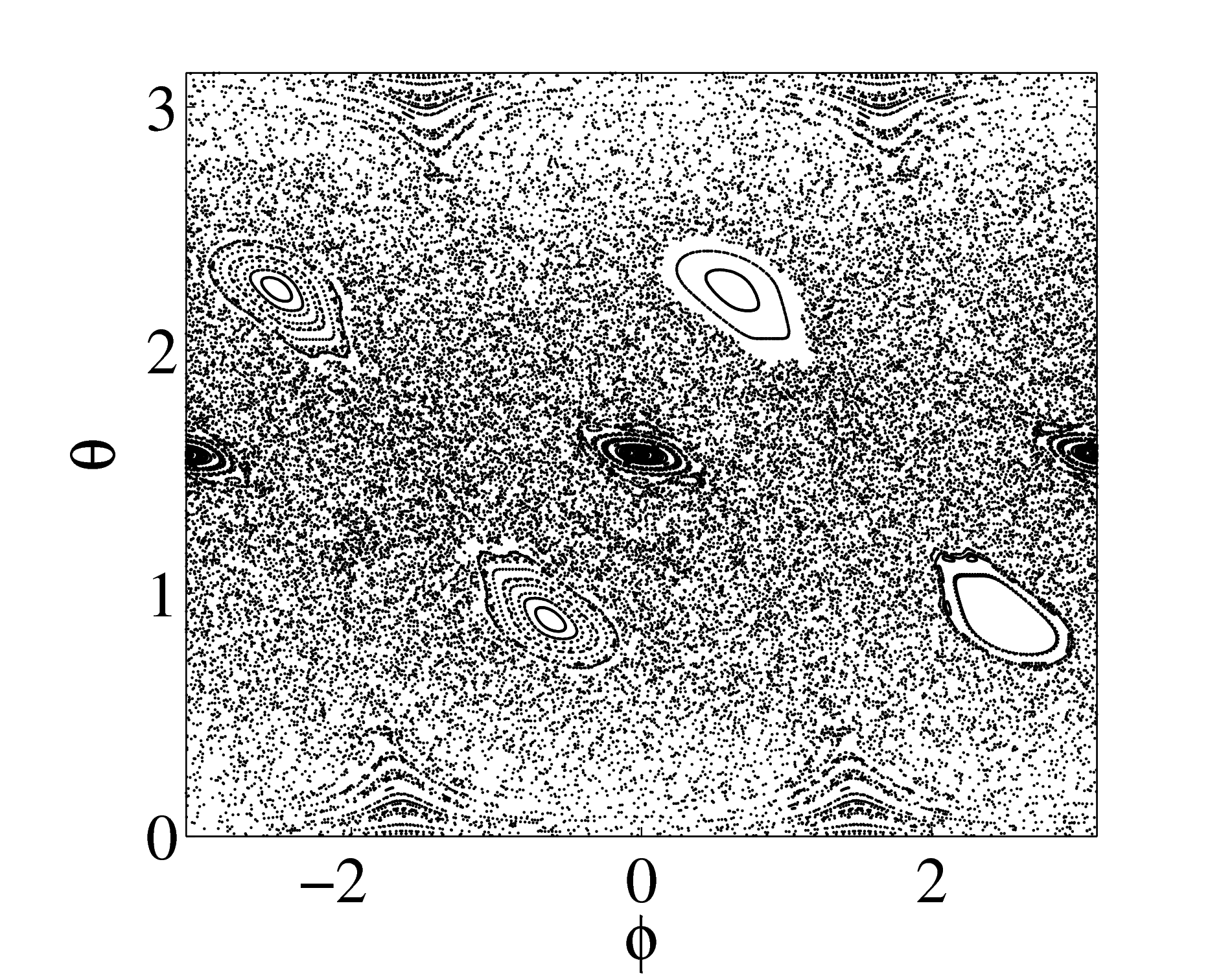}
\par\end{centering}

\caption{ Stroboscopic phase-space dynamics of the classical kicked
top for $\kappa=3$. Two hundred stroboscopic trajectories are
plotted, each for a duration of 233 kicks. This figure is from
Ref.~\protect\cite{Song2006}.}

\label{fig_qkt_poinsec}
\end{figure}

The quantum kicked-top model can be realized in cold atomic
ensembles~\cite{Smith2004,Chaudhury2007}, and it is familiar to us
since it has the form of a one-axis twisting Hamiltonian with a
periodic pulse,\begin{equation}
H=\frac{\kappa}{2j\tau}J_{z}^{2}+pJ_{y}\sum_{n=-\infty}^{\infty}\delta\left(t-n\tau\right),\end{equation}
 where the second term is the periodic driven pulse (with discrete
kicks). The dynamics of this model is described by the Floquent operator\begin{equation}
F=\exp\left(-i\frac{\kappa}{2j}J_{z}^{2}\right)\exp\left(-ipJ_{y}\right),\end{equation}
 where we choose $p=\pi/2$ and $\tau=1$ for convenience. The evolution
of an initial state $|\psi\left(0\right)\rangle$ is given by
$|\psi\left(n\right)\rangle=F^{n}|\psi\left(0\right)\rangle$. The
spin-squeezing dynamics can be obtained from this state. The chaotic
behavior of this model has been recently demonstrated experimentally
via linear entropy~\cite{Chaudhury2009}, by using the tensor part of
an effective Hamiltonian for the dispersive atom-field interaction,
and just in the same system, spin-squeezed states have been created
\cite{Smith2004,Chaudhury2007}. Therefore, the theoretical proposal
of using spin squeezing as a signature of quantum chaos could be
realized using current techniques. Below, we present numerical
results for the dynamics of the spin-squeezing parameters.

\begin{figure}[H]
\begin{centering}
\includegraphics[width=12cm]{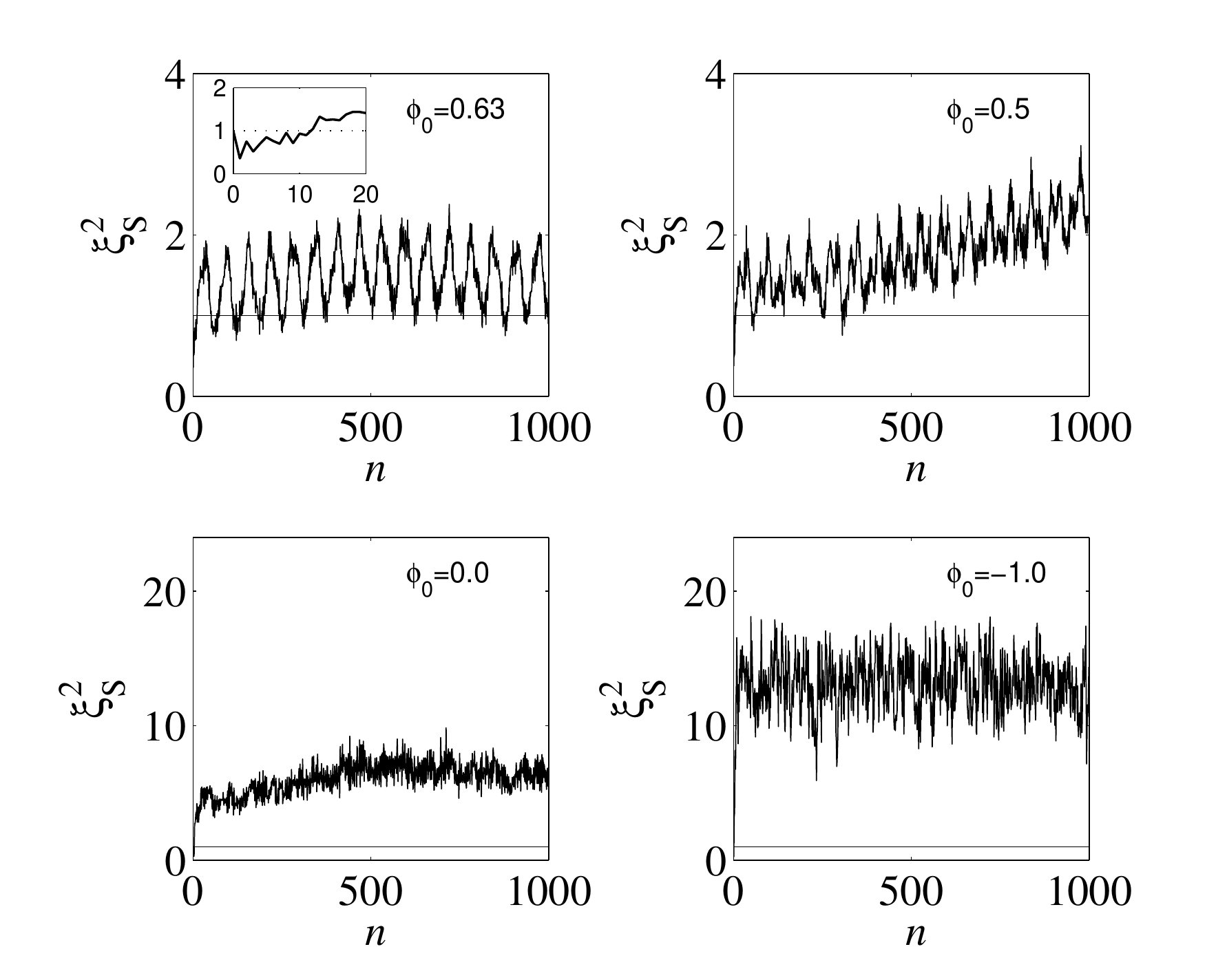}
\par\end{centering}

\caption{ Dynamical evolution of the spin squeezing parameter
$\xi_{S}^{2}$ for the initial CSS with $\theta_{0}=2.25$ for
different values of $\phi_{0}$. The other parameters are fixed here
at $\kappa=3$ and $j=25$. The inset displays the initial dynamics
when $\phi_{0}=0.63$. This figure is from
Ref.~\protect\cite{Song2006}.}

\label{fig_qkt_xi}
\end{figure}

To demonstrate the chaotic properties for this model, one could represent
the model Hamiltonian under the CSS in large-$j$ limit. The classical
limit of the quantum kicked top is obtained by expressing $\left(X,\, Y,\, Z\right)=\left(\left\langle J_{x}\right\rangle ,\,\left\langle J_{y}\right\rangle ,\,\left\langle J_{z}\right\rangle \right)/j,$
and factorizing all the second moments to products of first moments,
such as $\left\langle J_{x}J_{y}\right\rangle /j^{2}=XY.$ Then the
classical equations of motion can be obtained from the Heisenberg
operator equations of motion with the factorization rule. The classical
normalized angular momentum variables ($X,\, Y,\, Z$) can be parameterized
in spherical coordinates as $\left(X,\, Y,\, Z\right)=\left(\sin\theta\cos\phi,\,\sin\theta\sin\phi,\,\cos\theta\right)$,
where $\theta$ and $\phi$ are the polar and azimuthal angles, respectively.
Thus, the map is essentially two dimensional.

The stroboscopic plot with parameter $\kappa=3$ is displayed in Fig.~\ref{fig_qkt_poinsec}.
The parameter $\kappa$ is chosen such as to yield a mixture of regular
and chaotic areas. Elliptic fixed points surrounded by the chaotic
sea are evident. Two such elliptic fixed points have coordinates $\left(\theta,\,\phi\right)=\left(2.25,\,-2.5\right)$
and $\left(\theta,\,\phi\right)=\left(2.25,\,0.63\right)$. As we
will see, this phase-space structure of the classical kicked top determines
spin squeezing in the quantum kicked top. If the initial state $|\theta_{0},\phi_{0}\rangle$
is chosen at one of these two points, the classical dynamics is regular.

Now, we return to the quantum Hamiltonian. The initial state is also
a CSS, while the evolution is via the Floquent operator. The numerical
results for the squeezing parameter $\xi_{S}^{2}$ are shown in Fig.~\ref{fig_qkt_xi}.
The polar angle $\theta=2.25$ is chosen for all the initial states.
The four plots show very clear variations of the dynamics of $\xi_{S}^{2}$
for the fixed point ($\phi_{0}=0.63$), regular region ($\phi_{0}=0.5$),
and the chaotic region ($\phi_{0}=0.0$ and $\phi_{0}=-1.0$).

Spin squeezing is drastically suppressed in the chaotic region. As
shown in Fig.~\ref{fig_qkt_xi}, spin squeezing occurs frequently
for the initial state located at the elliptic fixed point $\phi_{0}=0.63$,
whereas spin squeezing vanishes after a very short time $\left(n=2\right)$
for the initial state is in the `deep' chaotic region ($\phi_{0}=-1.0$).
When the initial state is in the regular region ($\phi_{0}=0.5$),
spin squeezing exists over a relatively long time, and only when steps
$n>318$, the spin squeezing vanishes. When the initial state is centered
in the `shallow' chaotic sea $\phi_{0}=0$, spin squeezing disappears
after $n=4$ and never revives. The strong spin-squeezing oscillations
mainly originate from the periodic kicks. Without kicks, the spin
squeezing exhibits only periodic regular oscillations. Thus, we see
that spin squeezing is very sensitive to classical chaos, and classical
chaos strongly suppresses spin squeezing.

\subsection{Spin squeezing and quantum chaos in the Dicke model\label{sub:Dicke}}

In this section, we consider the Dicke model, which characterizes
the interaction between optical field and atoms. The Dicke model describes
an ensemble of two-level particles interacting with a single-mode
cavity bosonic field, and the corresponding Hamiltonian reads $\left(\hbar=1\right)$\begin{equation}
H=\omega_{0}J_{z}+\omega a^{\dagger}a+\frac{R}{\sqrt{2j}}\left(J_{+}a+J_{-}a^{\dagger}\right)+\frac{R^{\prime}}{\sqrt{2j}}\left(J_{+}a^{\dagger}+J_{-}a\right),\label{Dicke_Ham}\end{equation}
 where the collective pseudo-spin operators are $J_{\alpha}=\sum_{i=1}^{N}\sigma_{i\alpha}/2$,
($\alpha=x,y,z$) with $\sigma_{i\alpha}$ being the Pauli operator
for the $i$-th particle, and $R$, $R^{\prime}$ are the coupling
strengths. The denominator $\sqrt{2j}$ arises from that the dipole
coupling strength is proportional to $1/\sqrt{V}$, where $V$ is the
volume of the cavity. When we consider the ground state, the system
undergoes a second-order QPT between a normal phase and a
super-radiant phase \cite{Lambert2004,Lambert2009}. This Hamiltonian
is integrable under the rotating-wave approximation, i.e.
$R^{\prime}=0$, which is used to demonstrate the transfer of
squeezing between the field and atoms, while in this case no chaos
exists. In the non-rotating-wave-approximation regime,
$R^{\prime}\neq0$, its classical dynamics exhibits chaos. Therefore,
this system is a desirable model for studying the interface between
quantum and classical chaos.

\begin{figure}[ptb]
\begin{centering}
\includegraphics[width=10cm]{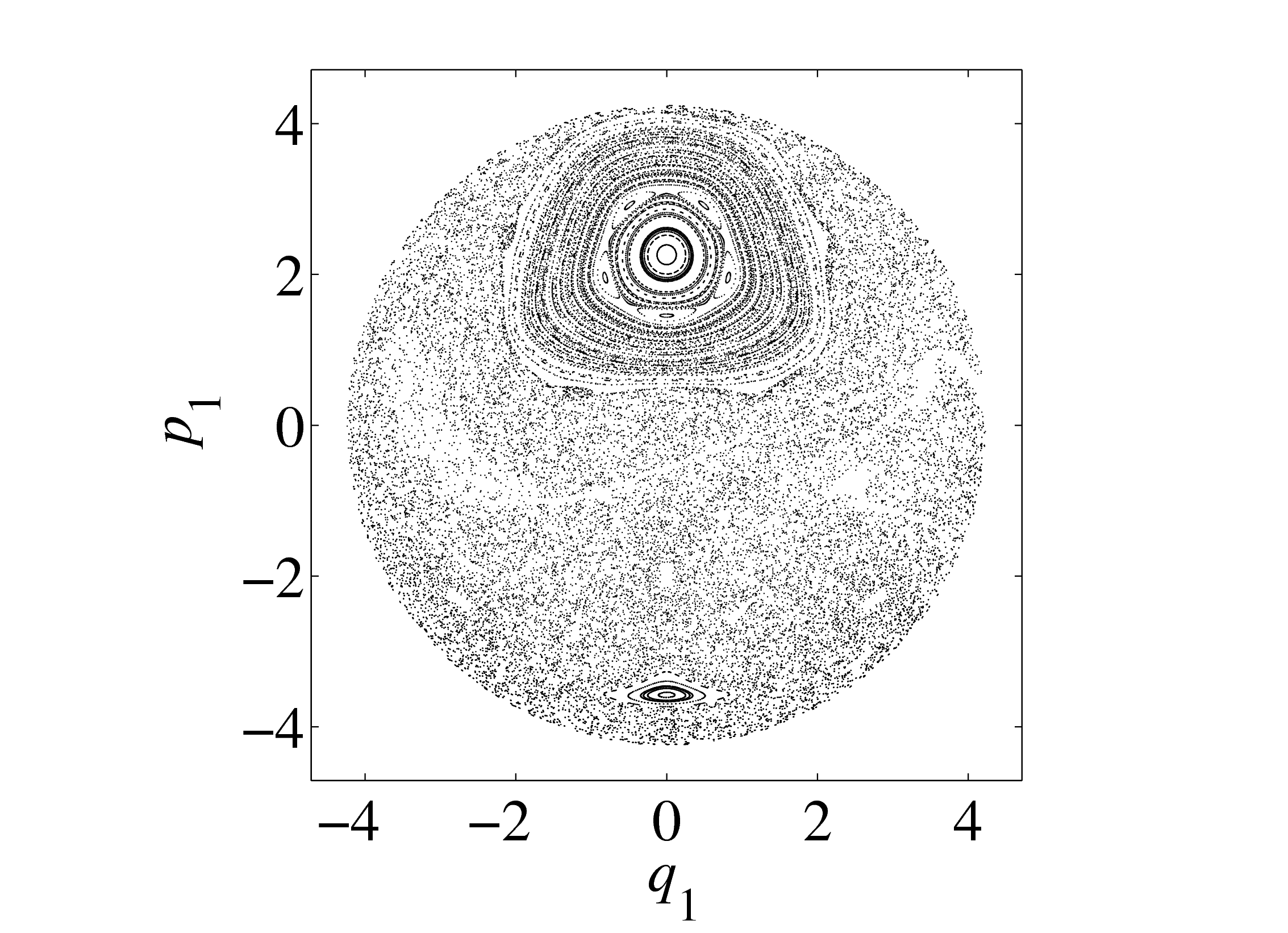}
\par\end{centering}

\caption{Poincaré section for the spin degrees of freedom (section
with $q_{2}=0.0$ and $p_{2}>0.0$) in the resonant case
$(\omega=\omega_{0}=1)$, energy $E=8.5$. This figure is from
Ref.~\protect\cite{Furuya1998}.}

\label{Dicke_poin}
\end{figure}

As discussed previously, to study the quantum chaos, we write the
Hamiltonian in the classical limit (in the coherent-state representation).
As this system consists of the atomic and optical parts, the initial
quantum states are chosen as follows: \begin{equation}
|{\psi}(0)\rangle=|{\eta}\rangle\otimes|{\alpha}\rangle,\end{equation}
 where $|\eta\rangle$ and $|\alpha\rangle$ are the coherent spin
and bosonic states, respectively. Here the variables $\eta$ and $\alpha$
can be written as functions of the classical variables in the corresponding
phase spaces, $(q_{1},p_{1})$ for the atomic degree of freedom, and
$(q_{2},p_{2})$ for the bosonic field $\eta=\left(p_{1}+iq_{1}\right)/\sqrt{4j-(p_{1}^{2}+q_{1}^{2})}$,
$\alpha=\frac{1}{\sqrt{2}}(p_{2}+iq_{2}),$ where $q_{1},p_{1},q_{2},p_{2}$
describe the phase space of the system under consideration, and the
indices 1 and 2 denote the atomic and field subsystem, respectively.
The classical Hamiltonian corresponding to Eq.~(\ref{Dicke_Ham})
can be obtained under the CSS representation.

\begin{figure}[H]

\begin{centering}
\includegraphics[width=8cm]{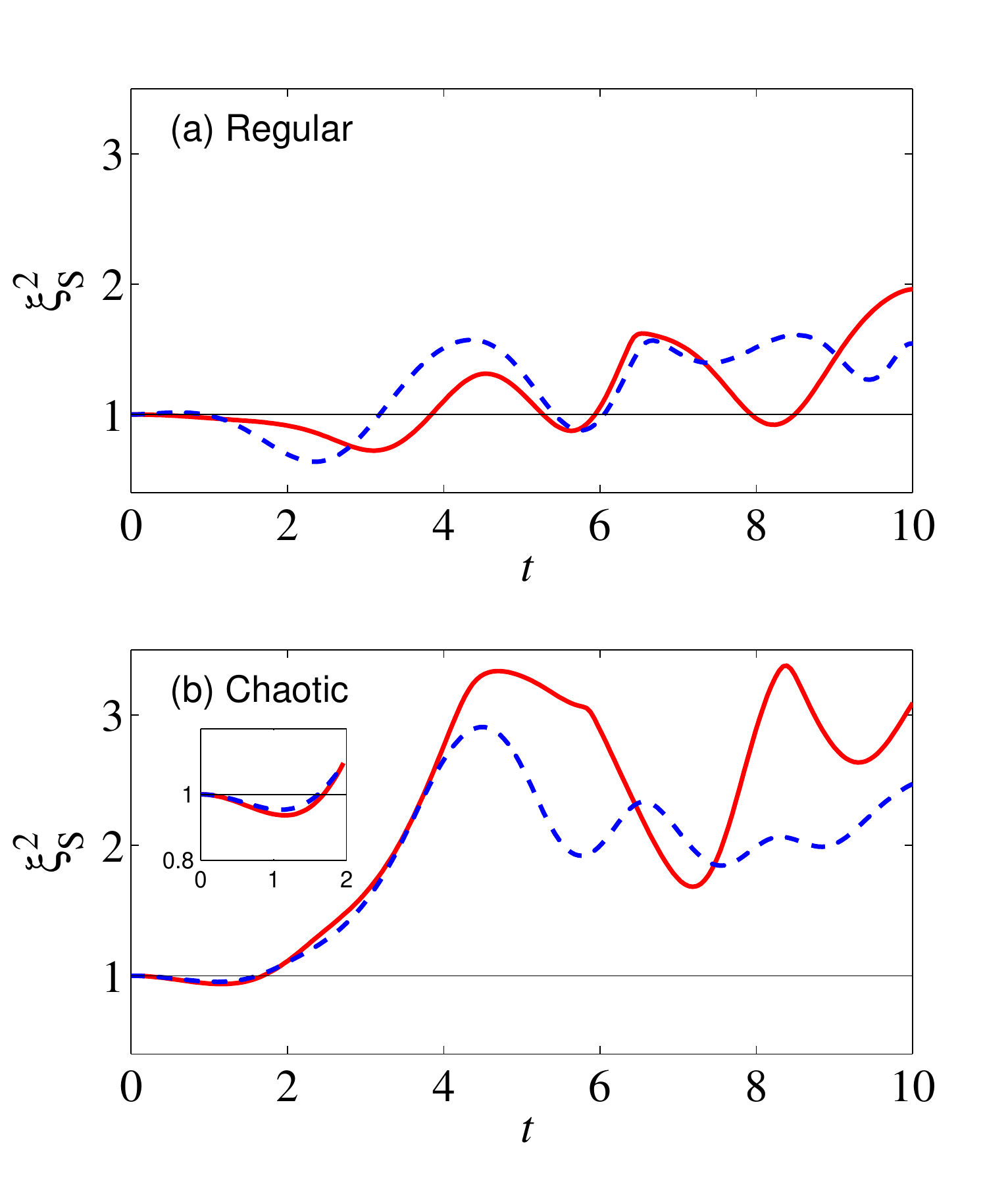}
\par\end{centering}

\caption{(Color online) Dynamical evolution of the spin-squeezing
parameter $\xi_{S}^{2}$ for the initial states with $q_{1}=0$ and
different $p_{1}$. (a) The regular region corresponding to
Fig.~\protect\ref{Dicke_poin}: $p_{1}=2.0$ (solid line) and
$p_{1}=-3.5$ (dashed line); (b) Chaotic region corresponding to
Fig.~\protect\ref{Dicke_poin}: $p_{1}=0$ (solid line) and
$p_{1}=-1.0$ (dashed line). This figure is from
Ref.~\protect\cite{Song2009}.}

\label{Dicke_xi_dyn}
\end{figure}

The classical dynamics associated with the Dicke Hamiltonian was
explored in Ref.~\cite{Deaguiar1992}. It was shown that the
integrable cases are recovered when either $R$ or $R^{\prime}$ is
zero, and the most chaotic dynamics is associated to the condition
$R=R^{\prime}$. In Fig.~\ref{Dicke_poin}, we show the Poincaré
section for the classical counterpart of the spin degrees of
freedom, defined by the section $q_{2}=0$ in the four-dimensional
phase space so that every time a trajectory pierces this section
with $p_{2}>0$, the corresponding point ($q_{1}$, $p_{1}$) is
plotted. Here, the total energy is fixed at $E=8.5$, $j=4.5$,
$\omega=\omega_{0}=1.$ The limit of atomic phase space is indicated
by a border of radius equal to $\sqrt{4j}$. In
Fig.~\ref{Dicke_poin}, we choose the coupling parameters $R=0.5$ and
$R^{\prime}=0.2,$ which yield a mixture of regular and chaotic areas
of significant sizes. Many fixed points and near-periodic orbits
surrounded by the chaotic sea are evident.

The dynamic behavior of the spin-squeezing parameter $\xi_{S}^{2}$
is computed by choosing the initial states in the chaotic and regular
regions with respect to the Poincaré section. The results are shown
in Fig.~\ref{Dicke_xi_dyn}. We choose two fixed points at $q_{1}=0$,
$p_{1}=2.0$ and $q_{1}=0$, $p_{1}=-3.5,$ which are in the regular
regions, and two additional points $q_{1}=0$, $p_{1}=0$ and $q_{1}=0$,
$p_{1}=-1.0$, well inside the chaotic sea.

For the regular regions {[}see Fig.~\ref{Dicke_xi_dyn} (a){]}, spin
squeezing vanishes after a relatively longer time ($t\approx9.0$),
whereas for the chaotic region {[}see Fig.~\ref{Dicke_xi_dyn}
(b){]}, the spin squeezing vanishes after a short time
($t\approx1.8$). In summary, spin squeezing is sensitive to and
suppressed by chaos. In contrast to this, chaos enhances the
bipartite entanglement quantified by the linear entropy in the Dicke
model~\cite{Furuya1998}. So, the underlying chaos affects
differently the two typical pure quantum mechanical phenomena, spin
squeezing and bipartite entanglement. These results are very similar
to the case of quantum kicked top model. Numerical results for the
quantum kicked top and Dicke models suggest that spin squeezing is a
good signature of chaos. Furthermore, by considering the advantage
of the spin-squeezing parameter in measurement, it may have good
applications in experiments.

\section{Spin squeezing under decoherence\label{decoherence}}

In this section, we take into account decoherence
\cite{Weiss1999,Gardiner2000,Breuer2002,Schlosshauer2007}, which is
induced by unavoidable technical noise, and interactions with the
environment. We shall discuss the effects of decoherence in three
aspects.

(i) Decoherence in the generation of spin squeezing. In Sec.~7.1 we
shall show that, if the decoherence effect is sufficiently weak,
although the ability of squeezing generation is weakened,
considerable large amount of squeezing can still be obtained.

(ii) The robustness and lifetime of spin squeezing in the presence
of decoherence. In Sec.~7.2, we show that spin-squeezed states are
more robust to particle-loss~\cite{Stockton2003} as compared to the
GHZ state. In Sec.~7.3, the lifetime of spin squeezing is studied in
three types of decoherence channels, which are quite general to
describe decoherence processes, and spin squeezing sudden death is
observed.

(iii) Decoherence in the quantum metrology process. In Sec.~7.4, we
discuss how decoherence affects the ability of spin-squeezed states
to the improvement of phase resolution in Ramsey processes. Although
the degree of spin squeezing is weakened by decoherence, sub-shot
noise phase resolution can still be achieved~\cite{Ulam-Orgikh2001},
in contrast with the maximal entangled state, with which we could
only attain the shot-noise level precision in the presence of
decoherence~\cite{Huelga1997}.

\subsection{Decoherence and spin squeezing generation}

In Sec.~3, we have discussed generating spin-squeezed states via
ideal one-axis twisting Hamiltonian, where decoherence effects are
not taken into account, and the best squeezing scales as $\xi_{R}^{2}\propto1/N^{2/3}$
for the one-axis twisting and $\xi_{R}^{2}\propto1/N^{2}$ for the
two-axis twisting. It is natural to expect that, decoherence is detrimental
to spin squeezing generation, while if the decoherence effect is sufficiently
weak, considerable large amount of squeezing can still be obtained.
Below, we mainly consider the particle loss induced decoherence, which
is an unavoidable source of decoherence in cold atom systems, such
as BEC, due to collisions of condensed atoms with the background gas.

We first consider the one-axis twisting Hamiltonian. In
Ref.~\cite{Sorensen2001}, to estimate the effect of particle losses
in a two-mode BEC, they have performed a Monte Carlo simulation of
the evolution of squeezing from the one-axis twisting Hamiltonian
$H=\chi J_{z}^{2}$. The particle loss is phenomenologically taken
into account by introducing a loss rate $\Gamma$. Their simulations
indicated that even under the conditions that approximately 10\%
atoms are lost, squeezing of $\xi_{R}^{2}\sim0.01$ may be obtained.
In Refs.~\cite{Li2008,Li2009}, the authors presented a more detailed
analysis of spin squeezing generation in the presence of particle
loss in a two-mode BEC. The time evolution of the squeezing
parameter is derived as\begin{equation}
\xi_{R}^{2}\left(t\right)=\xi_{R0}^{2}\left(t\right)\left[1+\frac{1}{3}\frac{\Gamma_{\text{sq}}t}{\xi_{R0}^{2}\left(t\right)}\right],\end{equation}
 where $\xi_{R}^{2}\left(t\right)$ is the squeezing without particle
loss, and\begin{equation}
\Gamma_{\text{sq}}=\sum_{m}\Gamma_{\text{sq}}^{\left(m\right)},\quad\text{and}\quad\Gamma_{\text{sq}}^{\left(m\right)}=m\Gamma^{\left(m\right)},\end{equation}
 where $\Gamma^{\left(m\right)}$ is the $m$-body loss rate. They
then found that, for one-body losses, the best obtainable spin squeezing
scales as\begin{align}
\xi_{R}^{2} & \propto N^{-4/15}\;\text{for}\; N\rightarrow\infty,\end{align}
 for two-body loss, in the limit $N\rightarrow\infty$ the best squeezing
is independent of $N$, and for three-body losses, the best squeezing
\begin{equation}
\xi_{R}^{2}\propto N^{4/15}\;\text{for}\; N\rightarrow\infty.\end{equation}
 Thus for one-body losses, we can obtain arbitrary small squeezing
parameter by increasing the particle number, while for three-body
losses case, there is a finite optimum number of particles for
squeezing. For two-axis twisting model, the decoherence effects
induced by particle loss was studied in Ref.~\cite{Andre2002a},
where they established an Raman-scattering based approach to
implement the two-axis twisting Hamiltonian as Eq.~(\ref{TAT}). They
found that, to achieve the Heisenberg limit squeezing, it is
required that $\chi\ge\Gamma$, where $\chi$ is the twisting strength
in Eq.~(\ref{TAT}), and $\Gamma$ is the one-body loss rate. To
achieve any squeezing, it requires $N\chi\ge\Gamma$, where $N$ is
the particle number. Quantum-limited metrology in the presence of
collisional dephasing was studied in \cite{Rey2007,Jin2010},
although these works are associated with nonlinear estimation
scheme, their analysis can also be extended to the discussion of
spin squeezing. Otherwise, a detailed analysis of decoherence in
quantum light-atom interfaces, such as QND-type interactions, was
studied in \cite{KoschorreckJP2009}.

\subsection{Spin-squeezed states under particle loss}

In the above subsection, we discussed how decoherence affects the
production of spin squeezing. Below, we consider the effects of
particle loss on the generated spin-squeezed states, which was
studied in Refs.~\cite{Stockton2003,Micheli2003,Li2008}. For
convenience, the system is now assumed to be symmetric under
particle exchange. Based on this assumption, the spin squeezing
after particle loss is equivalent to the spin squeezing of the
subsystem composed of the remaining particles. The state after
particle loss is obtained by tracing over the degrees of freedom of
the lost particles, and this procedure is what we used when deriving
the reduced density matrix of the state for the subsystem.
Therefore, we immediately arrive at the conclusion that concurrence
is not affected by particle loss.

Consider an $N$-body system with exchange symmetry, represented by
a density matrix $\rho_{N}$. The reduced states also have exchange
symmetry. After particle loss, the reduced state for the remaining
$N_{r}$ particles becomes\begin{equation}
\rho_{_{N_{r}}}=\text{Tr}_{(N-N_{r})}\{\rho_{_{N}}\},\end{equation}
 where we trace over the $(N-N_{r})$ lost particles. As the state
is of exchange symmetry, we have\begin{align}
\left\langle \sigma_{1\alpha}\right\rangle  & \,\equiv\,\left\langle \sigma_{i\alpha}\right\rangle _{N}=\left\langle \sigma_{i\alpha}\right\rangle _{N_{r}},\notag\\
\left\langle \sigma_{1\alpha}\sigma_{2\alpha}\right\rangle  & \,\equiv\,\left\langle \sigma_{i\alpha}\sigma_{j\alpha}\right\rangle _{N}=\left\langle \sigma_{i\alpha}\sigma_{j\alpha}\right\rangle _{N_{r}},\label{partloss0}\end{align}
 where the subscripts $N$ and $N_{r}$ denote the state before and
after particle loss, respectively. The indices $i,$ $j$ denote any
two remaining particles. Note that, according to Eq.~(\ref{partloss0}),
the MSD is invariant under particle loss.

From the expression $J_{\vec{n}}=\frac{1}{2}\sum_{i=1}^{N}\sigma_{i\vec{n}},$
one can verify that \begin{align}
\langle J_{\vec{n}}^{2}\rangle_{N}\;=\;\frac{N}{4}+\frac{N^{2}-N}{4}\langle\sigma_{1\vec{n}}\sigma_{2\vec{n}}\rangle,\end{align}
 where we have used the symmetry of the state. Conversely, one finds\begin{equation}
\langle\sigma_{1\vec{n}}\sigma_{2\vec{n}}\rangle=\frac{1}{N-1}\left({\frac{4\langle J_{\vec{n}}^{2}\rangle_{N}}{N}-1}\right).\label{n1}\end{equation}
 For the state after loss, similarly, we obtain\begin{equation}
\langle\sigma_{1\vec{n}}\sigma_{2\vec{n}}\rangle=\frac{1}{N_{r}-1}\left({\frac{4\langle J_{\vec{n}}^{2}\rangle_{N_{r}}}{N_{r}}-1}\right).\label{nr1}\end{equation}
 Thus, from Eqs.~(\ref{n1}) and (\ref{nr1}), we obtain the following
relation\begin{equation}
\frac{4\langle J_{\vec{n}}^{2}\rangle_{N_{r}}}{N_{r}}\;=\;\frac{N_{r}-1}{N-1}\frac{4\langle J_{\vec{n}}^{2}\rangle_{N}}{N}+\frac{N-N_{r}}{N-1},\end{equation}
 which is applicable to an arbitrary direction $\vec{n}.$ Then, we
have a similar expression for the direction $\vec{n}_{\perp}$ (perpendicular
to the invariant mean-spin direction),\begin{equation}
\frac{4\langle J_{\vec{n}_{\perp}}^{2}\rangle_{N_{r}}}{N_{r}}=\frac{N_{r}-1}{N-1}\frac{4\langle J_{\vec{n}_{\perp}}^{2}\rangle_{N}}{N}+\frac{N-N_{r}}{N-1}.\end{equation}
 The two quantities ${4\langle J_{\vec{n}_{\perp}}^{2}\rangle_{N_{r}}}/{N_{r}}$
and ${4\langle J_{\vec{n}_{\perp}}^{2}\rangle_{N}}/{N}$ are just
linearly related to each other, and $N_{r}-1>0$. So, if we choose
$\vec{n}_{\perp}$ be the direction where the minimal variance is
obtained for the state before loss, the direction $\vec{n}_{\perp}$
also corresponds to the minimal variance after loss. Finally, from
the definition of the squeezing parameter $\xi_{S}^{2}$ (Table
\ref{tab_definition})$,$ we obtain \cite{Stockton2003}
\begin{equation}
\xi_{S,N_{r}}^{2}\;=\;\frac{N_{r}-1}{N-1}\xi_{S,N}^{2}+\frac{N-N_{r}}{N-1},\label{aaa123}\end{equation}
 which gives the relation of the spin-squeezing parameters before
and after loss.

Now, we discuss the spin-squeezing parameter $\xi_{R}^{2}$, which
is simply related to the parameter $\xi_{S}^{2}$ via\begin{align}
\xi_{R,N}^{2}\;=\;\frac{N^{2}}{4\langle\vec{J}\rangle_{N}^{2}}\xi_{S,N}^{2}\;=\;\frac{1}{\langle\vec{\sigma}_{1}\rangle^{2}}\xi_{S,N}^{2}.\end{align}
 A similar expression holds for the squeezing parameters after loss,
and it is given by\begin{align}
\xi_{R,N_{r}}^{2}\;=\;\frac{1}{\langle\vec{\sigma}_{1}\rangle^{2}}\xi_{S,N_{r}}^{2}.\end{align}
 Multiplying Eq.~(\ref{aaa123}) by ${\langle\vec{\sigma}_{1}\rangle^{-2}},$
one immediately obtains \cite{Stockton2003} \begin{equation}
\xi_{R,N_{r}}^{2}\;=\;\frac{N_{r}-1}{N-1}\,\xi_{R,N}^{2}+\frac{N-N_{r}}{N-1}\frac{1}{\langle\vec{\sigma}_{1}\rangle^{2}}.\end{equation}
 The squeezing parameter $\xi_{S,N_{r}}^{2}$ only requires the knowledge
of $\xi_{S,N}^{2}$, while the parameter $\xi_{R,N_{r}}^{2}$ is related
to both $\xi_{R,N}^{2}$ and $\langle\vec{\sigma}_{1}\rangle^{2}$.

Note that the GHZ state, which is maximally entangled, is extremely
fragile under particle loss. Indeed, the entanglement is totally
destroyed when only one particle is lost. However, the spin-squeezed
states, which are only partially correlated (e.g., pairwise
correlated) are more robust. We emphasize that the robustness is
different in the depolarizing channel. It has been found~that in the
limit $N\rightarrow\infty$, GHZ states can stand more than 55\%
local depolarization, while for spin-squeezed states it can stand
more than 29\% local depolarization \cite{Simon2002}.

\subsection{Spin squeezing under decoherence channels}

In the previous subsection, the decoherence is induced by particle
loss. Here, the decoherence effects are described by three types of
decoherence channels: the amplitude damping channel (ADC), the phase
damping channel (PDC), and the depolarizing channel (DPC). In
general, decoherence processes can be described by these three
typical channels. They are prototype models of decoherence relevant
to various experimental systems. They provide {}``a revealing
caricature of decoherence in realistic physical situations, with all
inessential mathematical details stripped away\textquotedblright{}\
\cite{Preskill1999}. But yet this {}``caricature of
decoherence\textquotedblright{}\ leads to theoretical predictions
being often in good agreement with experimental data. Examples
include multiphoton systems, ion traps, atomic ensembles, or a
solid-state spin systems such as quantum dots or nitrogen-vacancy
centers, where qubits are encoded in electron or nuclear spins. Let
us first introduce the three decoherence channels.

\subsubsection{Decoherence channels}

\emph{Amplitude-damping channel.}--- The ADC is defined by \begin{equation}
\mathcal{E}_{\text{ADC}}(\rho)=E_{0}\rho E_{0}^{\dagger}+E_{1}\rho E_{1}^{\dagger},\end{equation}
 where $E_{0}=\sqrt{s}|0\rangle\langle0|+|1\rangle\langle1|$ and
$E_{1}=\sqrt{p}|1\rangle\langle0|$ are the Kraus operators, $p=1-s$,
$s=\exp(-\gamma t/2)$, and $\gamma$ is the damping rate. In the
Bloch representation, the ADC squeezes the Bloch sphere into an ellipsoid
and shifts it toward the north pole. The radius in the $x$-$y$ plane
is reduced by a factor $\sqrt{s}$, while in the $z$-direction it
is rescaled by a factor $s$. The ADC is a prototype model of a dissipative
interaction between a qubit and its environment. For example, the
ADC model can be applied to describe the spontaneous emission of a
photon by a two-level system into an environment of photon or phonon
modes at zero (or very low) temperature in (usually) the weak Born-Markov
approximation.

\emph{Phase-damping channel.}---The PDC is a prototype model of dephasing
or pure decoherence, i.e., loss of coherence of a two-level state
without any loss of the system's energy. The PDC is described by the
map \begin{equation}
\mathcal{E}_{\text{PDC}}(\rho)=s\rho+p\left(\rho_{00}|0\rangle\langle0|+\rho_{11}|1\rangle\langle1|\right),\end{equation}
 and obviously the three Kraus operators are given by $E_{0}=\sqrt{s}\mathbb{I}$,
$E_{1}=\sqrt{p}|0\rangle\langle0|$ and $E_{2}=\sqrt{p}|1\rangle\langle1|$,
where $\mathbb{I}$ is the identity operator. For the PDC, there is
no energy change and a loss of decoherence occurs with probability
$p.$ As a result of the action of the PDC, the Bloch sphere is compressed
by a factor ($1-2p$) in the $x$-$y$ plane. It is evident that the
action of the PDC is nondissipative. This means that, in the standard
computational basis $|0\rangle$ and $|1\rangle$, the diagonal elements
of the density matrix $\rho$ remain unchanged, while the off-diagonal
elements are suppressed. Moreover, the qubit states $|0\rangle$ and
$|1\rangle$ are also unchanged under the action of the PDC, although
any superposition of them (i.e., any point in the Bloch sphere, except
the poles) becomes entangled with the environment. The PDC can be
interpreted as elastic scattering between a (two-level) system and
a reservoir. The PDC is also a suitable model to describe $T_{2}$
relaxation in spin resonance.

\emph{Depolarizing channel.}---The definition of the DPC is given
by the map \begin{equation}
\mathcal{E}_{\text{DPC}}(\rho)=s\rho+p\frac{\mathbb{I}}{2}.\end{equation}
 We see that for the DPC, the spin is unchanged with probability $s=1-p$,
or is depolarized to the maximally-mixed state $\mathbb{I}/2\,\ $with
probability $p.$ It is seen that due to the action of the DPC, the
radius of the Bloch sphere is reduced by a factor $s$, but its shape
remains unchanged. 

\begin{table}[H]
 \caption{Analytical results for the time-evolutions of all relevant expectations,
correlations, spin-squeezing parameters, and concurrence, as well as
the critical values $p_{c}$ of the decoherence strength
$p$~\protect\cite{Wang2010}. This is done for the three decoherence
channels considered here. For the concurrence $C$, we give the
expression for $C_{r}^{\prime}$, which is related to the rescaled
concurrence $C_{r}$ via $C_{r}=\max(0,C_{r}^{\prime})$. The
parameters $x_{0}=1+2\left\langle \sigma_{z}\right\rangle
_{0}+\left\langle \sigma_{1z}\sigma_{2z}\right\rangle _{0}$,
$a_{0}=\left(N-1\right)\left(1-\left\langle
\sigma_{1z}\sigma_{2z}\right\rangle _{0}\right)$. The critical
values $p_{c}^{(1)}$ , $p_{c}^{(2)}$, and $p_{c}^{(3)}$ correspond
to the concurrence, squeezing parameter $\zeta_{R}^{2}$, and
$\zeta_{E}^{2}$, respectively. See Ref.~\protect\cite{Wang2010}}

\label{tab_decoherence} 

\centering{}\begin{tabular}{c||c|c|c}
\hline
 & Amplitude-damping channel  & Phase-damping channel  & Depolarizing channel \tabularnewline
 & (ADC)  & (PDC)  & (DPC) \tabularnewline
\hline
\hline
\parbox[c]{1.5cm}{%
\vspace{0.3cm}
 $\langle\sigma_{1z}\rangle$\vspace{0.3cm}
}  & %
\parbox[c]{4cm}{%
$s\langle\sigma_{1z}\rangle_{0}-p$%
}  & %
\parbox[c]{4cm}{%
$\langle\sigma_{1z}\rangle_{0}$%
}  & %
\parbox[c]{4cm}{%
$s\langle\sigma_{1z}\rangle_{0}$%
} \tabularnewline
\hline
\parbox[c]{1.5cm}{%
\vspace{0.3cm}
 $\langle\sigma_{1z}\sigma_{2z}\rangle$\vspace{0.3cm}
}  & %
\parbox[c]{4.3cm}{%
$s^{2}\langle\sigma_{1z}\sigma_{2z}\rangle_{0}-2sp\langle\sigma_{1z}\rangle_{0}+p^{2}$%
}  & %
\parbox[c]{4cm}{%
$\langle\sigma_{1z}\sigma_{2z}\rangle_{0}$%
}  & %
\parbox[c]{4cm}{%
$s^{2}\langle\sigma_{1z}\sigma_{2z}\rangle_{0}$%
} \tabularnewline
\hline
\parbox[c]{1.5cm}{%
\vspace{0.3cm}
 $\langle\sigma_{1+}\sigma_{2-}\rangle$\vspace{0.3cm}
}  & %
\parbox[c]{4cm}{%
$s\langle\sigma_{1+}\sigma_{2-}\rangle_{0}$%
}  & %
\parbox[c]{4cm}{%
$s^{2}\langle\sigma_{1+}\sigma_{2-}\rangle_{0}$%
}  & %
\parbox[c]{4cm}{%
$s^{2}\langle\sigma_{1+}\sigma_{2-}\rangle_{0}$%
} \tabularnewline
\hline
\parbox[c]{1.5cm}{%
\vspace{0.3cm}
 $\langle\sigma_{1-}\sigma_{2-}\rangle$\vspace{0.3cm}
}  & %
\parbox[c]{4cm}{%
$s\langle\sigma_{1-}\sigma_{2-}\rangle_{0}$%
}  & %
\parbox[c]{4cm}{%
$s^{2}\langle\sigma_{1-}\sigma_{2-}\rangle_{0}$%
}  & %
\parbox[c]{4cm}{%
$s^{2}\langle\sigma_{1-}\sigma_{2-}\rangle_{0}$%
} \tabularnewline
\hline
\parbox[c]{1.5cm}{%
\vspace{0.3cm}
 $\langle\vec{\sigma}_{1}\cdot\vec{\sigma}_{2}\rangle$\vspace{0.3cm}
}  & %
\parbox[c]{4cm}{%
{$1-s\, p\, x_{0}$} %
}  & %
\parbox[c]{5cm}{%
$s^{2}(1-\langle\sigma_{1z}\sigma_{2z}\rangle_{0})+\langle\sigma_{1z}\sigma_{2z}\rangle_{0}$%
}  & %
\parbox[c]{4cm}{%
$s^{2}$%
} \tabularnewline
\hline
\parbox[c]{1.5cm}{%
\vspace{0.3cm}
 ${\cal C}_{zz}$\vspace{0.3cm}
}  & %
\parbox[c]{4cm}{%
$s^{2}{\cal C}_{zz}(0)$%
}  & %
\parbox[c]{4cm}{%
${\cal C}_{zz}(0)$%
}  & %
\parbox[c]{4cm}{%
$s^{2}{\cal C}_{zz}(0)$%
} \tabularnewline
\hline
\parbox[c]{1.5cm}{%
\vspace{0.3cm}
 $\xi_{S}^{2}$\vspace{0.3cm}
}  & %
\parbox[c]{4cm}{%
{$1-sC_{r}(0)$} %
}  & %
\parbox[c]{4cm}{%
$1-s^{2}C_{r}(0)$%
}  & %
\parbox[c]{4cm}{%
$1-s^{2}C_{r}(0)$%
} \tabularnewline
\hline
\parbox[c]{1.5cm}{%
\vspace{0.3cm}
 $\xi_{R}^{2}$\vspace{0.3cm}
}  & %
\parbox[c]{4cm}{%
{${\displaystyle \frac{1-sC_{r}(0)}{(s\langle\sigma_{1z}\rangle_{0}-p)^{2}}}$} %
}  & %
\parbox[c]{4cm}{%
\vspace{0.15cm}
 ${\displaystyle \frac{1-s^{2}C_{r}(0)}{\langle\sigma_{1z}\rangle_{0}^{2}}}$\vspace{0.15cm}
}  & %
\parbox[c]{4cm}{%
${\displaystyle \frac{1-s^{2}C_{r}(0)}{s^{2}\langle\sigma_{1z}\rangle_{0}^{2}}}$%
} \tabularnewline
\hline
\parbox[c]{1.5cm}{%
\vspace{0.3cm}
 $\xi_{E}^{2}$\vspace{0.3cm}
}  & %
\parbox[c]{4cm}{%
{${\displaystyle \frac{1-sC_{r}(0)}{1+(1+N^{-1})s\, p\, x_{0}}}$} %
}  & %
\parbox[c]{5.5cm}{%
{\vspace{0.15cm}
 ${\displaystyle \frac{1-s^{2}C_{r}(0)}{(1-\frac{1}{N})[s^{2}+(1-s^{2})\langle\sigma_{1z}\sigma_{2z}\rangle_{0}]+\frac{1}{N}}}$\vspace{0.2cm}
 } %
}  & %
\parbox[c]{4cm}{%
${\displaystyle \frac{1-s^{2}C_{r}(0)}{(1-N^{-1})s^{2}+N^{-1}}}$%
} \tabularnewline
\hline
\parbox[c]{1.5cm}{%
\vspace{0.3cm}
 $C_{r}'$\vspace{0.3cm}
}  & %
\parbox[c]{4cm}{%
{$sC_{r}(0)-(N-1)s\, p\, x_{0}/2$} %
}  & %
\parbox[c]{4cm}{%
{$s^{2}C_{r}(0)+{a_{0}(s^{2}-1)}/{2}$}%
}  & %
\parbox[c]{4cm}{%
$s^{2}C_{r}(0)+\frac{N-1}{2}(s^{2}-1)$%
} \tabularnewline
\hline
\parbox[c]{1.5cm}{%
\vspace{0.3cm}
 $p_{c}^{(1)}$\vspace{0.3cm}
}  & %
\parbox[c]{4cm}{%
{${\displaystyle \frac{2C_{r}(0)}{\left(N-1\right)x_{0}}}$} %
}  & %
\parbox[c]{4cm}{%
${\displaystyle 1-\left(\frac{a_{0}}{2C_{r}(0)+a_{0}}\right)^{\frac{1}{2}}}$%
}  & %
\parbox[c]{4cm}{%
${\displaystyle 1-\left(\frac{N-1}{2C_{r}(0)+N-1}\right)^{\frac{1}{2}}}$%
} \tabularnewline
\hline
\parbox[c]{1.5cm}{%
\vspace{0.3cm}
 $p_{c}^{(2)}$\vspace{0.3cm}
}  & %
\parbox[c]{4cm}{%
${\displaystyle \frac{\langle\sigma_{1z}\rangle_{0}^{2}+C_{r}(0)-1}{1+2\langle\sigma_{1z}\rangle_{0}+\langle\sigma_{z}\rangle_{0}^{2}}}$%
}  & %
\parbox[c]{4cm}{%
${\displaystyle 1-\left(\frac{1-\langle\sigma_{1z}\rangle_{0}^{2}}{C_{r}(0)}\right)^{\frac{1}{2}}}$%
}  & %
\parbox[c]{4cm}{%
${\displaystyle 1-\left(\frac{1}{C_{r}(0)+\langle\sigma_{1z}\rangle_{0}^{2}}\right)^{\frac{1}{2}}}$%
} \tabularnewline
\hline
\parbox[c]{1.5cm}{%
\vspace{0.3cm}
 $p_{c}^{(3)}$ $\vspace{0.3cm}$%
}  & %
\parbox[c]{4cm}{%
${\displaystyle \frac{NC_{r}(0)}{\left(N-1\right)x_{0}}}$%
}  & %
\parbox[c]{4cm}{%
${\displaystyle 1-\left(\frac{a_{0}}{NC_{r}(0)+a_{0}}\right)^{\frac{1}{2}}}$%
}  & %
\parbox[c]{4cm}{%
${\displaystyle 1-\left(\frac{N-1}{NC_{r}(0)+N-1}\right)^{\frac{1}{2}}}$%
} \tabularnewline
\hline
\end{tabular}
\end{table}

\subsubsection{Spin squeezing parameters under decoherence}

Now we begin to study spin squeezing under the above three different
decoherence channels. The spin-squeezed states are prepared by a one-axis
twisting Hamiltonian, and the spin-squeezing parameters $\xi_{S}^{2}$,
$\xi_{R}^{2}$ and $\tilde{\xi}_{E}^{2}$ can be evaluated by using
Eqs.~(\ref{spsqp}) and (\ref{xi_t_local}), since the system has
exchange symmetry, and the decoherence channels act independently
on each particle. Therefore, all the spin-squeezing parameters and
the concurrence are determined by some correlation functions and expectations,
and they can be calculated via the Heisenberg approach shown below.

The spin-squeezing parameters and concurrence for the initial state
are shown in Table~\ref{tab_decoherence}. The state under decoherence
is given by \begin{equation}
\mathcal{E}\left(\rho\right)=\sum_{\mu_{1},...,\mu_{N}}\left(\otimes_{i=1}^{N}K_{\mu_{i}}\right)\rho\left(\otimes_{i=1}^{N}K_{\mu_{i}}^{\dagger}\right),\end{equation}
 where $K_{\mu_{i}}$ denotes the Kraus operator for the $i$-th particle.
The spin-squeezing parameters and concurrence consist of some correlation
functions and expectations, and we need just to calculate the evolution
of these quantities by employing Heisenberg picture without writing
the state explicitly. The evolution of operator $A$ can be put forward
by employing the Heisenberg picture with \begin{equation}
\left\langle A\right\rangle =\text{Tr}\left[A\mathcal{E}(\rho)\right]=\text{Tr}\left[\mathcal{E}^{\dagger}\left(A\right)\rho\right],\end{equation}
 where \begin{equation}
\mathcal{E}^{\dagger}\left(A\right)=\sum_{\mu_{1},...,\mu_{N}}\left(\otimes_{i=1}^{N}K_{\mu_{i}}^{\dagger}\right)A\left(\otimes_{i=1}^{N}K_{\mu_{i}}\right)\end{equation}
 is relatively easy to calculate for our cases.

\begin{figure}[H]
\begin{centering}
\includegraphics[width=10cm]{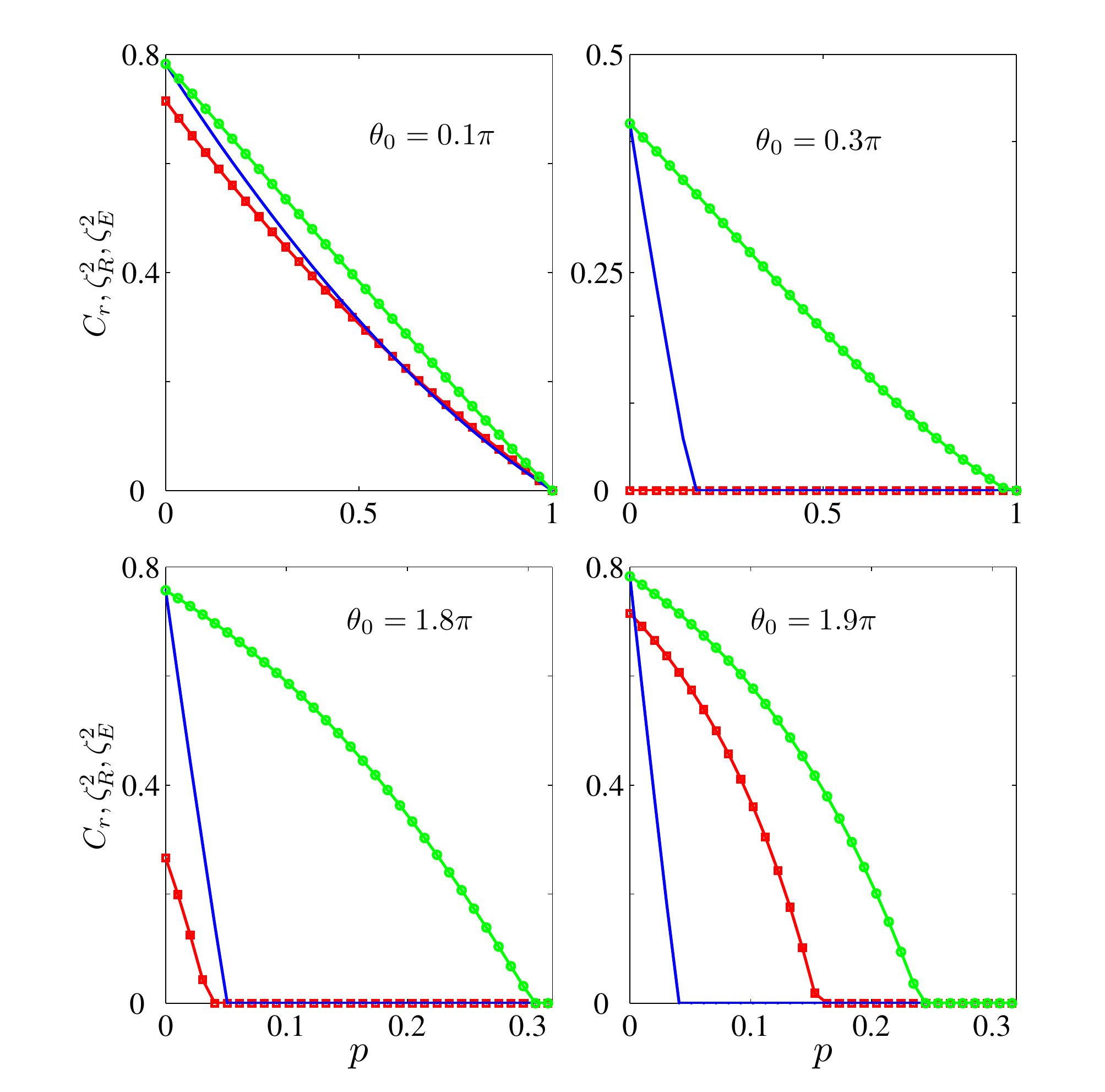}
\par\end{centering}

\caption{(Color online) Spin-squeezing parameters $\zeta_{R}^{2}$
(red curve with squares), $\zeta_{E}^{2}$ (top green curve with
circles), and the concurrence $C_{r}$ (solid curve) versus the
decoherence strength $p=1-\exp(-\gamma t)$ for the amplitude damping
channel, where $\gamma$ is the damping rate. Here, $\theta_{0}$ is
the initial twist angle given by Eq.~(\protect\ref{angle-1}). In all
figures, we consider an ensemble of $N=12$ spins. Note that for
small initial twist angle $\theta_{0}$ (e.g., $\theta_{0}=0.1\pi$),
the two squeezing parameters and the concurrence all concur. For
larger values of $\theta_{0}$, then $\zeta_{R}^{2}$,
$\zeta_{E}^{2}$, and $C$ become quite different, and all vanish for
sufficiently large values of the decoherence strength. This figure
is from Ref.~\protect\cite{Wang2010}.}

\label{ssssv1}
\end{figure}

In order to characterize spin squeezing more conveniently, we now
define the following squeezing parameters: \begin{equation}
\zeta_{k}^{2}=\max(0,1-\xi_{k}^{2}),\;\; k\in\{S,R,E\}.\end{equation}
 Spin squeezing appears when $\zeta_{k}^{2}>0$, and there is no squeezing
when $\zeta_{k}^{2}$ vanishes. Thus, the definition of the first
parameter $\zeta_{S}^{2}$ has a clear meaning: namely, it is the
\emph{strength} of the negative correlations as seen from Eq.~(\ref{kita_pc}).
More explicitly, for the initial state, we have $\xi_{S}^{2}=1-(N-1)C_{0}$
as shown in Eq.~(\ref{xis_conc}), so $\zeta_{S}^{2}$ is just the
\emph{rescaled concurrence} \begin{equation}
C_{r}(0)=(N-1)C_{0}.\end{equation}

To study the decoherence of spin squeezing, we choose a state which
is initially squeezed. One typical class of such spin-squeezed
states is the one-axis twisting collective spin
state~\cite{KITAGAWA1993},
\begin{equation}
|\Psi(\theta_{0})\rangle=e^{-i\theta_{0}J_{x}^{2}/2}|1\rangle^{\otimes N},\label{initial-1}\end{equation}
 which could be prepared by the one-axis twisting Hamiltonian $H=\chi J_{x}^{2},$
where \begin{equation}
\theta_{0}=2\chi t\label{angle-1}\end{equation}
 is the \textit{one-axis twist angle} and $\chi$ is the coupling
constant.

The numerical results for the squeezing parameters and concurrence
are illustrated in Fig.~\ref{ssssv1} for different initial values
of $\theta$. For the smaller value of $\theta_{0}$, e.g., $\theta_{0}=\pi/10$,
we see that there is no entanglement sudden death (ESD) and spin squeezing
sudden death (SSSD). The spin squeezing and the pairwise entanglement
are completely robust against decoherence. Intuitively, the larger
is the squeezing, the larger is the vanishing time for the squeezing.
However, here, in contrast to this, no matter how small the squeezing
parameters and concurrence are, they vanish only in the asymptotic
limit. This results from the complex correlations in the initial state
and the special characteristics of the ADC. For larger values of $\theta_{0},$
as the decoherence strength $p$ increases, the spin squeezing decreases
until it suddenly vanishes, so the phenomenon of SSSD occurs. There
exists a critical value $p_{c},$ after which there is no spin squeezing.
The vanishing time of $\tilde{\xi}_{E}^{2}$ is always larger than
those of $\xi_{R}^{2}$ and the concurrence. We note that depending
on the initial state, the concurrence can vanish before or after $\xi_{R}^{2}$.
This means that in our model, the parameter $\tilde{\xi}_{E}^{2}<1$
implies the existence of pairwise entanglement, while $\xi_{R}^{2}$
does not.

We can calculate all the relevant correlation functions, squeezing
parameters, concurrence and the critical values of the decoherence
strength $p,$ which are given in Table~\ref{tab_decoherence}. Initially,
the state is spin-squeezed, i.e., $\xi_{S}^{2}(0)<1$ or $C_{r}(0)>0$.
As seen from the Table, one can find that $\xi_{S}^{2}<1$, except
in the asymptotic limit of $p=1.$ Thus, we conclude that according
to $\xi_{S}^{2}$, the initially spin-squeezed state is always squeezed
for $p\neq1$, irrespective of both the decoherence strength and decoherence
models. In other words, there exists no SSSD if we quantify spin squeezing
by the first parameter $\xi_{S}^{2}$.

\begin{figure}[H]

\begin{centering}
\includegraphics[width=8cm]{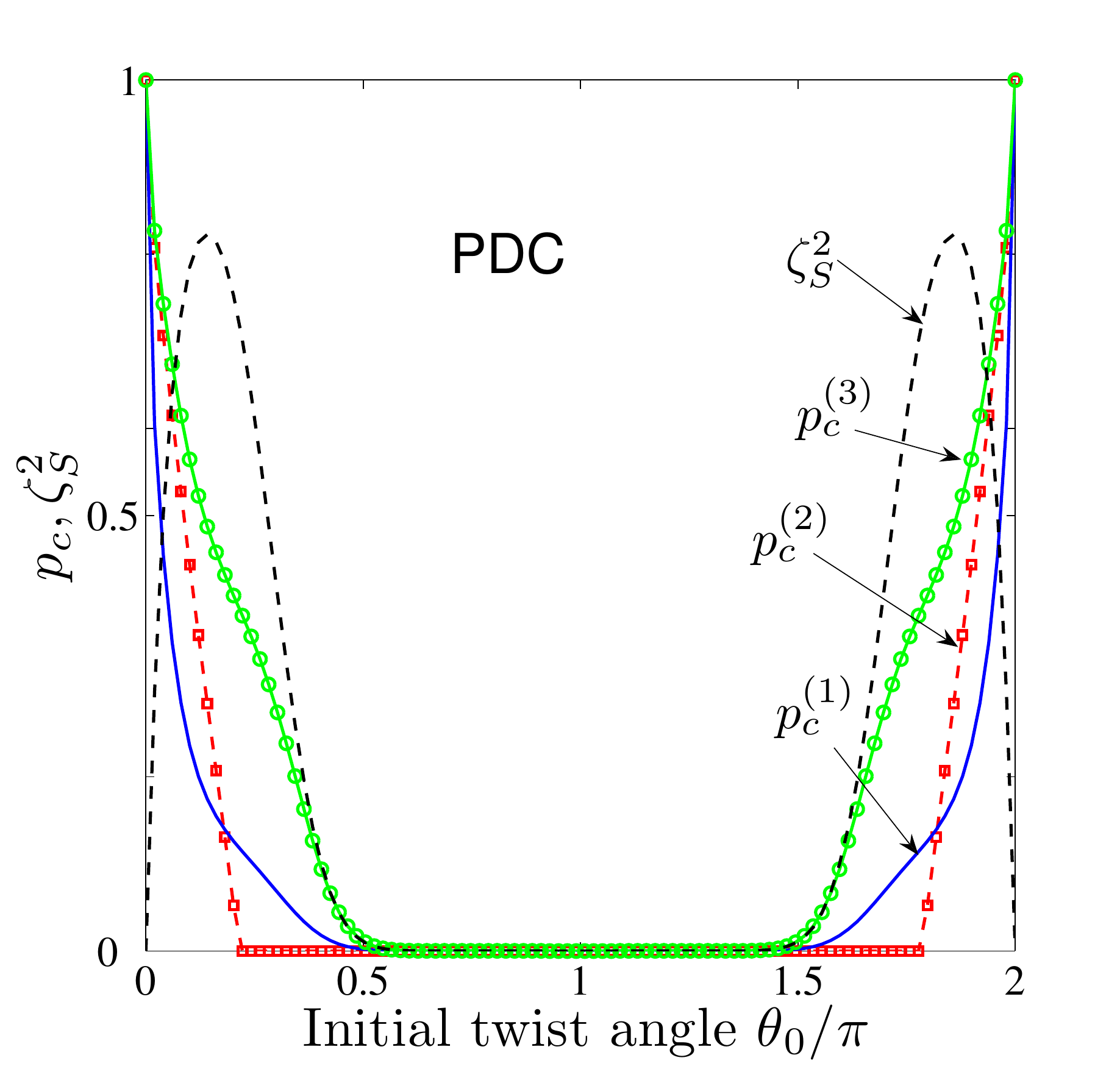}
\par\end{centering}

\caption{(Color online) Critical values of the decoherence strength
$p_{c}^{(1)}$ (blue solid curve), $p_{c}^{(2)}$ (red curve with
squares), $p_{c}^{(3)}$ (top green curve with circles), and the
squeezing parameter $\zeta_{S}^{2}$ (black dashed curve) versus the
initial twist angle $\theta_{0}$ given by
Eq.~(\protect\ref{angle-1}) for the amplitude-damping channel, PDC.
Here, $p_{c}$ is related to the vanishing time $t_{v}$ via
$p_{c}=1-\exp(-\gamma t_{v})$. At vanishing times, SSSD occurs. This
figure is from Ref.~\protect\cite{Wang2010}.}

\label{fig_pc_pdc}
\end{figure}

The SSSD for the three decoherence channels with respect to
different initial twisting angles was also studied \cite{Wang2010},
here, we only take the PDC for example. In Fig.~\ref{fig_pc_pdc}, we
plot the decoherence strength $p_{c}$ versus the twist angle
$\theta_{0}$ of the initial state for the PDC. For this decoherence
channel, the critical values $p_{c}$'s first decrease, until they
reach zero. Also, it is symmetric with respect to $\theta_{0}=\pi,$
which is in contrast to the ADC. There are also intersections
between the concurrence and parameter $\xi_{R}^{2},$ and the
critical value $p_{c}^{(3)}$ is always larger than $p_{c}^{(1)}$ and
$p_{c}^{(2)}.$

As summarized in Ref.~\cite{Wang2010}, the common features of the
three decoherence channels are: (i) The critical value
$p_{c}^{\left(3\right)}$ is always larger than or equal to the other
two, namely, the spin-squeezing correlations according to
$\tilde{\xi}_{E}^{2}$ are more robust; (ii) There always exist two
intersections between the concurrence and the parameter
$\xi_{R}^{2},$ for $\theta_{0}$ from 0 to $2\pi$, irrespective of
the decoherence channels; (iii) When there is no squeezing (central
area of Fig.~\ref{fig_pc_pdc}), all vanishing times are zero.
Table~\ref{tab_decoherence} conveniently lists all the analytical
results obtained in this section.

\subsection{Effects of decoherence on spin squeezing in Ramsey processes}

In this section, we present the effects of decoherence on spin-squeezed
states in Ramsey interferometry, which was discussed in Sec.~\ref{sub:metrology}.
Now we take into account decoherence, and each spin-1/2 particle is
assumed to evolve independently via the master equation\begin{equation}
\frac{\partial\rho}{\partial t}=-\frac{i}{2}\left[\left(\omega_{0}-\omega\right)\sigma_{z},\rho\right]+\frac{\gamma}{2}\left(\sigma_{z}\rho\sigma_{z}-\rho\right),\end{equation}
 where $\gamma$ is the decoherence rate. Under the above master equation,
which describes a dephasing process, nondiagonal elements of $\rho$
decay as $\exp\left(-\gamma t\right)$, while the diagonal terms are
invariant. Spin squeezing weakened by dissipation effects was
studied in Refs.~\cite{Civitarese2009,Kominis2008}. By using the
results obtain in Table~\ref{tab_decoherence}, we now
derive\begin{align}
\left\langle J_{z}\right\rangle _{t}= & -\cos\phi\left\langle J_{z}\right\rangle _{t=0}\exp({-\gamma t}),\notag\\
\left(\Delta J_{z}^{2}\right)_{t}= & \left[\sin^{2}\phi\left\langle J_{x}^{2}\right\rangle _{t=0}+\cos^{2}\phi\left(\Delta J_{z}\right)_{t=0}^{2}-\sin(2\phi)\text{Cov}(J_{x},J_{z})_{t=0}\right]\exp({-2\gamma t})\notag\\
 & +\frac{N}{4}\left[1-\exp(-2\gamma t)\right],\label{ramsey_Jz}\end{align}
 where the exponential terms come from the free evolution, when the
system is rotating around the $z$-axis. Thus, from the above equations,
the decoherence effects on the spin-squeezing parameter $\xi_{R}^{2}$
are clear: the length of the spin becomes shorter, while the fluctuation,
which maybe squeezed at the beginning, becomes larger. Therefore,
the degree of squeezing becomes smaller.

Since $\phi$ is time dependent, it is more convenient to estimate
$\omega_{0}$. By using Eqs.~(\ref{eq:dphi}) and (\ref{ramsey_Jz}),
we obtain\begin{equation}
\Delta\omega_{0}=\left[\frac{\sin^{2}\phi\left\langle J_{x}^{2}\right\rangle +\cos^{2}\phi\left(\Delta J_{z}\right)^{2}+\frac{N}{4}\left[\exp({2\gamma t})-1\right]}{tT\sin^{2}\phi\left\langle J_{z}\right\rangle ^{2}}\right]^{1/2},\end{equation}
 where we omit the subscript $t=0$, and assume $\text{Cov}\left(J_{x},J_{z}\right)_{t=0}=0$,
which is usually satisfied when the initial states have parity, e.g.
the spin-squeezed states generated via the twisting Hamiltonian.

Now, we search for the point $\left(\phi_{\text{opt}},t_{\text{opt}}\right)$
where the minimum of $\Delta\omega_{0}$ is attained. The minimization
of $\Delta\omega_{0}$ is equivalent to the minimization of ($\Delta\omega_{0})^{2},$
which is given by\begin{equation}
\left(\Delta\omega_{0}\right)^{2}=\frac{\left\langle J_{x}^{2}\right\rangle -\left(\Delta J_{z}\right)^{2}}{tT\left\langle J_{z}\right\rangle ^{2}}+\frac{\left(\Delta J_{z}\right)^{2}+\frac{N}{4}\left[\exp({2\gamma t})-1\right]}{tT\sin^{2}\phi\left\langle J_{z}\right\rangle ^{2}}.\end{equation}
 It is evident that ($\Delta\omega_{0})^{2}$ decreases monotonically
with the increase of $\sin\phi$, thus\begin{equation}
\phi_{\text{opt}}=\frac{k\pi}{2}~~\left(k\text{ odd}\right).\end{equation}
 And at this point, $\left(\Delta\omega_{0}\right)^{2}$ becomes\begin{equation}
\left(\Delta\omega_{0}\right)^{2}\left(\phi=\phi_{\text{opt}}\right)=\frac{\left\langle J_{x}^{2}\right\rangle +\frac{N}{4}\left[\exp(2\gamma t)-1)\right]}{tT\left\langle J_{z}\right\rangle ^{2}}.\label{deltaomega}\end{equation}
 It is interesting to see that this minimal value is independent of
the variance $\left(\Delta J_{z}\right)^{2},$ irrespective of decoherence.
Now, we minimize $\left(\Delta\omega_{0}\right)^{2}$ with respect
to $t.$ The value of $t_{\text{opt}}$ is determined by the conditions\begin{align}
\frac{\partial}{\partial t}\left[\left(\Delta\omega_{0}\right)^{2}\left(\phi=\phi_{\text{opt}}\right)\right] & =0,\notag\\
\frac{\partial^{2}}{\partial
t^{2}}\left[\left(\Delta\omega_{0}\right)^{2}\left(\phi=\phi_{\text{opt}}\right)\right]
& >0,\end{align}
 and thus $t_{\text{opt}}$ is the solution of\begin{equation}
\left(2\gamma t-1\right)\exp(2\gamma t)+1=\xi_{x}^{2},\label{tau}\end{equation}
 where\begin{equation}
\xi_{x}^{2}=\frac{4\left\langle J_{x}^{2}\right\rangle }{N},\end{equation}
 which is equal to $\xi_{S}^{2}$ if $\left\langle J_{x}^{2}\right\rangle $
is the minimal spin variance in the $\vec{n}_{\perp}$ direction.
When $t=t_{\text{opt}},$ one finds\begin{equation}
\frac{\partial^{2}}{\partial{t^{2}}}\left[\left(\Delta\omega_{0}\right)^{2}\left(\phi=\phi_{\text{opt}}\right)\right]=\frac{N\gamma^{2}\exp({2\gamma t})}{Tt\left\langle J_{z}\right\rangle ^{2}},\end{equation}
 which is obviously larger than zero. Thus, at the optimal point,
we obtain the minimal value.

From Eqs.~(\ref{deltaomega}) and (\ref{tau}), we obtain the minimal
$\Delta\omega_{0}$ as\begin{equation}
\min\left(\Delta\omega_{0}\right)=\sqrt{\frac{2\gamma\exp\left(2\gamma t_{\text{opt}}\right)}{TN\eta_{z}}},\label{min_omega}\end{equation}
 where $\eta_{z}=4\left\langle J_{z}\right\rangle ^{2}/N^{2}$, which
is always less than or equal to one. For the CSS $|j,-j\rangle$,
we have $\eta_{z}=1$, $\xi_{x}^{2}=1$ and $t_{\text{opt}}=1/\left(2\gamma\right)$.
Thus according to Eq. (\ref{min_omega}), we have \begin{equation}
\min\left(\Delta\omega_{0}\right)_{\text{CSS}}=\sqrt{\frac{2\gamma\exp\left(1\right)}{TN}}.\end{equation}
 A relative improvement in the precision over the above minimal uncertainty
may be given by\begin{equation}
P\left(t_{\text{opt}},\eta_{z}\right)=1-\frac{\min\left(\Delta\omega_{0}\right)}{\min\left(\Delta\omega_{0}\right)_{\text{CSS}}}=1-\sqrt{\frac{\exp\left(2\gamma t_{\text{opt}}\right)}{\exp\left(1\right)\eta_{z}}}.\end{equation}
 Then in the idealized case, $t_{\text{opt}}\rightarrow0$ and $\eta_{z}\rightarrow1$,
the theoretical absolute decoherence limit in the precision improvement
is\begin{equation}
P_{\text{abs}}=1-\exp\left(-\frac{1}{2}\right)\simeq0.39.\end{equation}
which is independent of $N$. Here, we emphasize that, the limit $t\rightarrow0$
implies $\xi_{x}^{2}\rightarrow0$ according to Eq.~(\ref{tau}).

In Ref.~\cite{Huelga1997}, a similar issue was studied with the
maximally entangled state, the GHZ state, which is not spin-squeezed
but can be used to achieve the Heisenberg
limit~\cite{Bollinger1996,Meyer2001,Leibfried2004}. Compared with
spin-squeezed states, for Markovian dephasing processes as discussed
in this section, the best precision given by the GHZ state is the
shot-noise limit, the same as a CSS. Recently, the lifetime of
spin-squeezed states under the influence of dephasing noise was
measured in Ref.~\cite{Leroux2010a}, where the spin-squeezed state
was generated in a collection of atoms in a cavity via feedback
mechanism. For experimental conditions in \cite{Leroux2010a}, the
lifetime is $\tau\simeq600\,\mu\text{s}$ for an initial
spin-squeezed state with
$\left(\xi_{R}^{2}\right)^{-1}\simeq4\,\text{dB}$. Since the CSS is
also affected by the same dephasing noise, at the time,
$\tau=600\,\mu\text{s}$, even though the initially squeezed states
becomes unsqueezed, $\xi_{R}^{2}=1$, it still improves the
signal-to-noise ratio by $\sim3\,\text{dB}$ over that of an
initially CSS.

\section{Theoretical proposals and experimental realizations of squeezing generation\label{experiment}}

Above we have discussed many theoretical works about spin squeezing
and its applications. Below, we review the generation of squeezing
in several physical systems, with both theoretical proposals and experimental
realizations. In Sec.~8.1, we review experimental progresses of generating
spin squeezing in the BEC, for which the theoretical proposals have
been discussed in Sec.~\ref{sub:Mapping-a-two-component}. Then in
Sec.~8.2, we discuss the transfer of squeezing from squeezed lights
to atoms. Afterwards, spin squeezing produced by QND measurement is
reviewed in Sec.~8.3.

\subsection{Generating spin squeezing in Bose-Einstein condensations\label{experiment-BEC}}

In this section, we discuss how to generate spin-squeezed states in
BEC. There are two main advantages for producing spin-squeezed
states in this system: the considerable long coherence times, and
the strong atom-atom interactions, which induce nonlinearity and
squeezing. The simplest and most widely studied scheme to generate
spin-squeezed states in a two-component BEC is to utilize particle
collisions. Under the single-mode approximation, the BEC Hamiltonian
can be mapped to the one-axis twisting Hamiltonian in the Schwinger
representation~\cite{Sorensen2001,Orzel2001,Law2001,Poulsen2001a,Raghavan2001,Jenkins2002,Jaksch2002,Jing2002,Sorensen2002,Micheli2003,Molmer2003,Choi2005,Jaaskelainen2005,Jin2007,Jin2007a,Thanvanthri2007,Esteve2008,Jin2008,Grond2009,Grond2009a},
which was realized in
experiments~\cite{Orzel2001,Esteve2008,Gross2010,Riedel2010}. The
validity of the two-mode approximation has been confirmed by using
perturbation-based Bogoliubov theory~\cite{Sorensen2002}, and also
by employing the positive-$P$ method~\cite{Poulsen2001a}, both gave
a more detailed analysis than using a single-mode approximation.
References \cite{Helmerson2001,Zhang2003a} presented an effective
two-axis twisting Hamiltonian resulting from Raman scattering. Spin
squeezing in a spinor-1 BEC was studied in
Refs.~\cite{Pu2000,Duan2000,Duan2002,Mustecaplioglu2002,Yi2002,Yi2006,Mutecaplioglu2007}.
There, spin-squeezed states could be generated via spin-exchange
interactions~\cite{Pu2000,Duan2000}. For a spinor-1 BEC, which
should be described by the SU(3) algebra, spin-squeezed states can
be created in an SU(2) subspace
\cite{Duan2002,Mustecaplioglu2002,Mutecaplioglu2007}, and even
simultaneously in all the three SU(2) subspaces
\cite{Mustecaplioglu2002}. Decoherence in the generation of spin
squeezing was studied in the cases of particle collisions
\cite{Gasenzer2002,Gasenzer2002a}, and particle losses
\cite{Micheli2003,Li2008,Li2009}. Below, we present some recent
experiments about producing spin-squeezed states.

\subsubsection{Experimental realizations}

In the BEC system, we use the concept of spin squeezing since
physical observables associated with the $N$ two-level atoms can be
described by a fictitious spin $J=N/2$ in the Schwinger
representation, shown in Eq.~(\ref{Schwinger_Rep}). Below, we show
some very recent experimental results~\cite{Gross2010,Riedel2010}.
To illustrate these results, we explain the so-called
number-squeezing parameter, which is defined as\begin{equation}
\xi_{N}^{2}=\frac{4\left(\Delta J_{z}\right)^{2}}{N}.\end{equation}
 It is called number-squeezing parameter because $J_{z}$ measures
the number difference between the two components, it is related to
the metrological squeezing parameter as\begin{equation}
\xi_{R}^{2}=\frac{\xi_{N}^{2}}{C^{2}},\end{equation}
 where \begin{equation}
C=\frac{2|\langle\vec{J}\rangle|}{N},\end{equation}
 is the contrast of the Ramsey interferometer.

The intrinsic physics of the generation of squeezed states in BEC is
the particle collisions, as illustrated in Sec.~3, where we employed
a simple two-component BEC, for which an effective one-axis twisting
Hamiltonian is derived. In experiments, the BEC can be loaded in a
optical lattice, and is described by a Bose-Hubbard model, of which
the atom number statistics undergoes a drastic change from a
superfluid phase to a Mott insulator phase \cite{Gerbier2006}, where
the atom number fluctuation is strongly suppressed. The atomic
number squeezing has been implemented in Ref.~\cite{Orzel2001}. In
this experiment, the condensates were loaded in an optical lattice,
and the system can be described by a Bose-Hubbard Hamiltonian, of
which the atom number statistics undergoes a drastic change from a
superfluid phase to a Mott insulator phase \cite{Gerbier2006}. For
$N$ atoms loaded in an $M$-site lattice, The Bose-Hubbard
Hamiltonian reads \begin{equation} H=\gamma\sum_{\langle
i,j\rangle}\hat{a}_{i}^{\dagger}\hat{a}_{j}+\sum_{i}\epsilon_{i}\hat{n}_{i}+\frac{g\beta}{2}\sum_{i}\hat{n}_{i}\left(\hat{n}_{i}-1\right),\label{eq:Bose-Hubbard}\end{equation}
 where $\hat{a}_{i}$ and $\hat{a}_{i}^{\dagger}$ are bosonic operators
of atoms on the $i$th site, $\hat{n}_{i}=\hat{a}_{i}^{\dagger}\hat{a}_{i}$
is the corresponding number operator, and $\epsilon_{i}$ denotes
the external potential term. The interaction strength $g$ is associated
with the $s$-wave scattering length, and is shown in Eq.~(\ref{eq:BEC_g}),
and \begin{equation}
\beta\equiv\int d^{3}r\,|\phi\left(r\right)|^{4}.\end{equation}
 The tunneling amplitude $\gamma$ between the $i$ and $j$ sites
is\begin{equation}
\gamma=\int d^{3}r\,\phi\left(r-r_{i}\right)\left[-\frac{\hbar^{2}}{2m}\nabla^{2}+U\left(r\right)\right]\phi\left(r-r_{j}\right).\end{equation}
 In the strong tunneling regime $g\beta/\gamma\ll1$, i.e., the superfluid
phase, the interaction term is negligible and the single atomic
state will spread the lattice. The many-body ground state is given
as \cite{Greiner2002}\begin{equation}
|\psi_{\text{SF}}\rangle_{g\beta=0}\propto\left(\sum_{i=1}^{M}\hat{a}_{i}^{\dagger}\right)^{N}|0\rangle,\end{equation}
 of which the on-site number fluctuation obeys Poissonian distribution,
i.e., \begin{equation}
\Delta\hat{n}_{i}=\langle\hat{n}_{i}\rangle.\end{equation}
 Conversely, if the ratio $g\beta/\gamma$ overcomes a critical point,
the ultracold gas enters the Mott insulator phase, where the atomic
states are localized and the many-body ground state wavefunction becomes
\begin{equation}
|\psi_{\text{MI}}\rangle_{\gamma=0}\propto\prod_{i=1}^{M}\left(\hat{a}_{i}^{\dagger}\right)^{n_{i}}|0\rangle,\end{equation}
 which is a product of local Fock states for each lattice site, with
$n_{i}$ the on-site atom number, and the corresponding number fluctuations
vanish.

In Ref.~\cite{Orzel2001}, atom number squeezing was observed in the
Mott insulator phase. They loaded about $10^{8}$ $^{87}$Rb atoms
into a time-orbiting potential (TOP) trap and produced about
$10^{4}$ condensates of $F=2$, $m_{f}=2$ atoms. After an adiabatic
relaxation, the condensate is then loaded in a one-dimensional
vertically oriented optical lattice, which consists of $\sim$12
weakly linked mesoscopic wells. The squeezing of the particle number
was indirectly observed through detecting the phase-sensitivity
interference of atoms released from the lattice. The phase variance
$\sigma_{\phi}$ is related to the number variance $\sigma_{n}$ at
each lattice as \begin{equation}
\sigma_{n}\propto\frac{1}{\sigma_{\phi}},\end{equation}
 due to the phase number uncertainty relation. The increase of the
phase variance indicates the reduction of the number variance. As
discussed in Sec.~3.2.3, the spin squeezing can be produced in BEC
due to nonlinear particle collisions, while in this experiment, besides
particle collisions in the same well, there exists a tunneling between
neighboring wells. The physics of the formation of number squeezing
can be illustrated by considering a simple two-well version of the
Bose-Hubbard Hamiltonian (\ref{eq:Bose-Hubbard}), written as \begin{equation}
H=\gamma\left(a_{\text{L}}^{\dagger}a_{\text{R}}+a_{\text{R}}^{\dagger}a_{\text{L}}\right)+\frac{g\beta}{2}\left[\left(a_{\text{L}}^{\dagger}a_{\text{L}}\right)^{2}+\left(a_{\text{R}}^{\dagger}a_{\text{R}}\right)^{2}\right],\end{equation}
 where the subscripts L and R represent the left and right well, respectively.
In the Schwinger representation (\ref{Schwinger_Rep}) the above Hamiltonian
becomes \begin{equation}
H=\gamma J_{x}+\chi J_{z}^{2}+\frac{\chi N^{2}}{2},\end{equation}
 where $N$ is the total particle number, and the squeezing strength
$\chi=g\beta$. To create number-squeezed states, they increase the
ratio $N\chi/\gamma$ by varying the intensity of the lattice laser
and the initial condensate density. The atomic interference is realized
by releasing the atoms to fall under the gravity. The contrast of
the resulting interference pattern is used as a phase probe to distinguish
between the limits of coherent and Fock states. With the adiabatic
increase of the lattice potential depth, they found the increase of
the phase variance, characterized by the squeezing factor $S$, defined
by \begin{equation}
\sigma_{\phi}^{2}=S\sigma_{\phi0}^{2},\end{equation}
 where $\sigma_{\phi0}^{2}$ corresponds to non-squeezed case. In
the case \begin{equation}
Ng\beta/\gamma=10^{5},\end{equation}
 where $N$ refers to the particle number in the central well, they
found that the squeezing factor\begin{equation}
S_{\text{dB}}=10\log_{10}\frac{\sigma_{\phi}^{2}}{\sigma_{\phi0}^{2}}\simeq25\,\text{dB},\end{equation}
 corresponding to a number variance $\sigma_{n}\simeq1$ in the central
well.

In Ref.~\cite{Gerbier2006}, by using a probe that is sensitive only
to the presence of atom pairs at a given lattice, they observed a
continuous suppression of number fluctuations when the ultracold
sample evolves from the superfluid regime to the Mott insulator
regime. In this experiment, the specific two-particle probe are
spin-changing collisions, which convert spin $f=1$ atoms in the
$m=0$ Zeeman sublevel to pairs with one atom in $m=+1$ and the other
in the $m=-1$. The advantage of the spin-collisional based probes is
that they are nondestructive, and they can be resonantly controlled
using the differential shift between Zeeman sublevels induced by an
off-resonant microwave field. The experiments confirmed that the
atom number exhibits near-Poissonian fluctuations for shallow
lattices, and strongly suppressed fluctuations for deep lattice.
Besides, their results indicate that number squeezing is robust with
respect to experimental manipulations.

Experiments of Ref.~\cite{Li2007} presented extended coherence times
by a factor of 2 over those expected with coherent state BEC
interferometry. The coherent time, which is limited due to the
mean-field interaction induced decoherence, is probed by using the
decay of Bloch oscillations. The theoretical treatment for a BEC in
an optical lattice begins with the Bose-Hubbard Hamiltonian
(\ref{eq:Bose-Hubbard}). The tunneling is weakened by applying a
large energy gradient across the array, which benefits from the
advantage that the on-site mean-field energy is unaffected. They
have demonstrated that, for number squeezed states prepared at
potential depth $U=22.5E_{R}$, where $E_{R}=\hbar^{2}k^{2}/2m$ and
$k=2\pi/\lambda$, with $\lambda=852$ nm, the coherence time
$\tau_{c}=19.3\pm3.5$ ms, which represents an increase of a factor
of 2.1 over the expected decay time of an array of coherent states
in the same lattice potential. It is interesting to note that here
squeezing extends the coherence time: typically, the enhanced
fragility of squeezed states to loss mechanisms results in reduced
coherence times.\textbf{\emph{ }}Then in Ref.~\cite{Jo2007}, phase
coherence of two BEC confined in a double-well potential on an
atomic chip was observed. It holds for times up to $\sim$200 ms
after splitting, a factor of 10 longer than the phase diffusion time
expected for a coherent state for the experimental
conditions.\textbf{ }

In the previous experiments
\cite{Orzel2001,Greiner2002,Gerbier2006,Jo2007,Li2007}, the
suppression of atom number fluctuations in a BEC was observed
indirectly. The direct determination of squeezing parameter requires
access to the on-site atom occupations. This was demonstrated in
Ref.~\cite{Esteve2008} by imaging the condensate with a resolution
of 1~$\mu$m (full-width at half-maximum), which is well below the
lattice spacing of 5.7~$\mu$m, and fulfils the criterion of local
measurement. The condensate was loaded in an optical lattice of two
to six wells, and the occupation number per site ranges from 100 to
1,100 atoms. The wells of the lattice are fully resolved, and thus
the atom number in each lattice site can be determined by direct
integration of the atomic density obtained by absorption imaging.
The observed spin-squeezed states allow a precision gain of 2.3~dB
(two wells) and 3.8~dB (six wells) over the standard quantum limit
for interferometric measurements.

\begin{figure}[ptb]

\begin{centering}
\includegraphics[width=14cm]{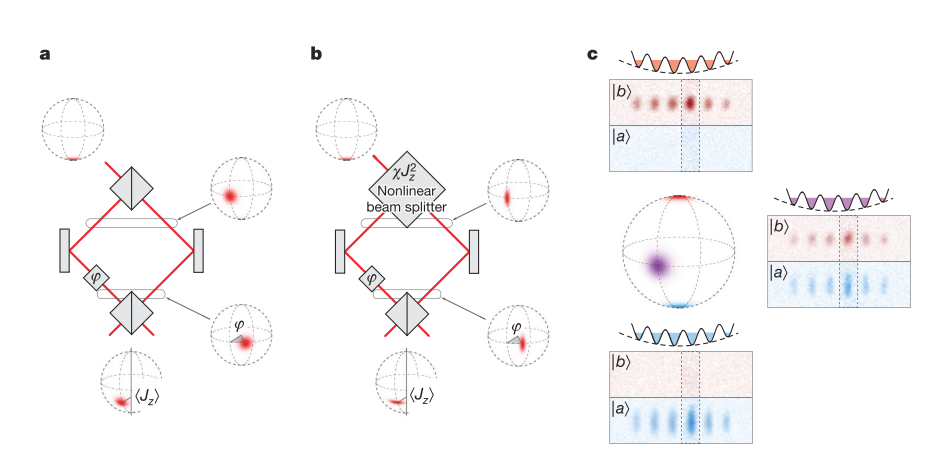}
\par\end{centering}

\caption{(Color online) Comparison of linear and nonlinear
interferometry~\protect\cite{Gross2010}. \textbf{a, }Classical
Mach-Zehnder interferometer. The two-mode field is mapped to a
Bloch-sphere under the Schwinger representation
(\protect\ref{Schwinger_Rep}). Here the input state is equivalent to
a CSS. \textbf{b,} By using a nonlinear beam splitter, which acts as
one-axis twisting Hamiltonian and generates spin-squeezed state, the
precision of interferometry was improved. \textbf{c, }Six
independent BECs of $^{87}$Rb were prepared in a one-dimensional
optical lattice~\protect\cite{Gross2010}. State-selective
time-delayed imaging causes the atomic clouds to have different
shapes. For this trap geometry, the dynamics is described by a
one-axis twisting Hamiltonian. This figure is from
Ref.~\protect\cite{Gross2010}.}

\label{fig_BEC_exp1a}
\end{figure}

Recently, some notable improvements have been
reported~\cite{Gross2010,Riedel2010}. Although squeezed states have
been generated previously, Ref.~\cite{Gross2010} first performed a
direct experimental demonstration of interferometric phase precision
beyond the standard quantum limit in a novel nonlinear Ramsey
interferometer. The experimental scheme, shown in
Fig.~\ref{fig_BEC_exp1a}, involved six independent BECs of $^{87}$Rb
(total number 2,300) prepared in a one-dimensional lattice, enables
parallel experiments performed on six wells independently, which
results in increased statistics for a given measurement time. The
hyperfine states $|a\rangle=|F=1,m_{F}=1\rangle$ and
$|b\rangle=|F=2,m_{F}=-1\rangle$ form a two-level system. The single
traps are almost spherical and have dipole frequencies of
$\omega_{\text{trap}}=2\pi\times425\,\text{Hz}$. For this trap
geometry, the single mode approximation is well justified, and the
dynamics is governed by a one-axis twisting
Hamiltonian\begin{equation} H/\hbar=\Delta\omega_{0}J_{z}+\chi
J_{z}^{2}+\Omega J_{\phi},\label{BEC_Ham_Exp1}\end{equation}
 where the third term describes spin rotations around an axis \begin{equation}
J_{\phi}=J_{x}\cos\phi+J_{y}\sin\phi,\end{equation}
 due to the coupling of $|a\rangle$ and $|b\rangle$ by using two-photon
combined microwave and radio-frequency pulses with a Rabi frequency
of $\Omega$ and phase $\phi$. In the experiment, the Rabi frequency
$\Omega$ can be switched rapidly between $0$ and $2\pi\times600$
Hz, allowing for fast diabatic coupling of the states.

Then we consider the nonlinear interaction strength $\chi$. When
the two states $|a\rangle$ and $|b\rangle$ are complete overlap
spatially, according to the discussion in Sec.~3.2.3 we have\begin{equation}
\chi\propto a_{aa}+a_{bb}-2a_{ab},\end{equation}
 where the $s$-wave scattering lengths in this experiment \cite{Gross2010}
satisfy\begin{equation}
a_{aa}\,:\, a_{bb}\,:\, a_{ab}=100\,:\,97.7\,:\,95,\end{equation}
 which results in very small nonlinear interaction strength, $\chi\simeq0$.
To overcome this problem, they used the Feshbach resonance to control
scattering lengths. By choosing the magnetic field $B=9.10$ G, $a_{ab}$
can be decreased by a narrow Feshbach resonance and they achieved
$\chi=2\pi\times0.063$ Hz. The magnetic field is preserved constant
throughout the whole interferometric sequence, while during the coupling
pulses the interaction strength $\chi$ is negligible since the system
is in the Rabi regime, that is, $\chi N/\Omega\ll1$.

\begin{figure}[ptb]

\begin{centering}
\includegraphics[width=14cm]{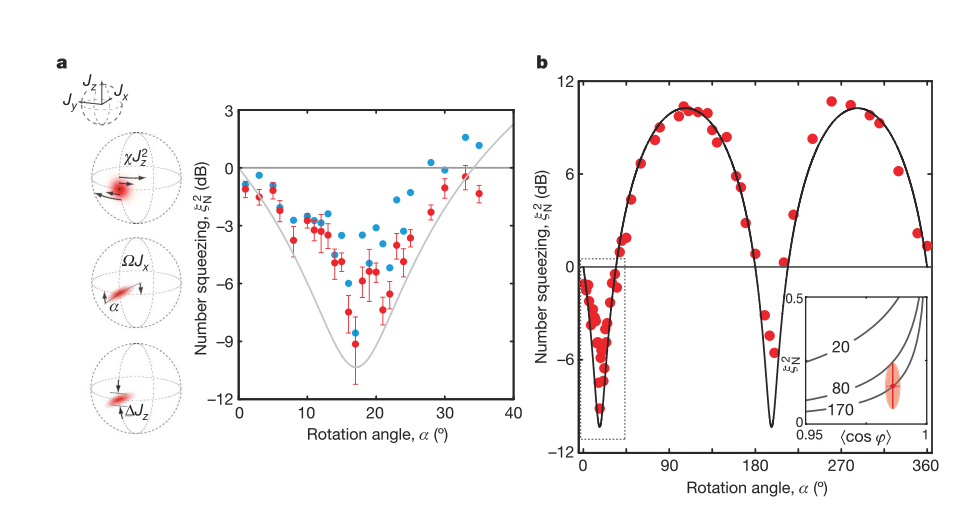}
\par\end{centering}

\caption{(Color online) Characterization of the quantum state within
a nonlinear interferometer~\protect\cite{Gross2010}. \textbf{a,} In
the left, the twisting effects of the one-axis twisting Hamiltonian.
In the right, the blue data have been corrected for photon shot
noise and the red data additionally take the technical noise into
account~\protect\cite{Gross2010}. The best number-squeezing factor
was $\xi_{N}^{2}=-8.2_{-1.2}^{+0.9}$~dB. \textbf{b,} Noise
tomography of the output state of the nonlinear beam splitter. The
dotted box indicates the region detailed in \textbf{a}. The largest
fluctuations measured have a number-squeezing factor of
$\xi_{N,\max}^{2}=+10.3_{-0.4}^{+0.3}$~dB. This figure is from
Ref.~\protect\cite{Gross2010}.}

\label{fig_BEC_exp1b}
\end{figure}

The experiment begins with a CSS that polarized to the $-z$
direction according to the chosen coordinate system. After a $\pi/2$
pulse the state turns to be polarized in the $x$ direction, thus
$\langle J_{z}\rangle=\langle J_{y}\rangle=0$. Next, by means of
Feshbach resonance, the state evolves under the one-axis twisting
Hamiltonian, which causes a shearing effect, shown in
Fig.~\ref{fig_qpd_oat_a}, and finally the state has a squeezing
angle $\alpha_{0}$ with respect to the $z$ direction. As shown in
Fig.~\ref{fig_BEC_exp1a}, to make the state useful for squeezing
metrology, it is first rotated around the center of the uncertainty
ellipse by $\alpha=\alpha_{0}+\pi/2$, and then evolves freely for
$\tau=2\,\mu\text{s}$, during which a phase difference $\varphi$ is
gained between the two states $|a\rangle$ and $|b\rangle$. Followed
by a $\pi/2$ pulse, this phase difference is estimated by a
population measurement. This standard Ramsey process is illustrated
in Fig.~\ref{fig_BEC_exp1a}. The experimental results for number
squeezing are shown in Fig.~\ref{fig_BEC_exp1b}. The minimal
fluctuations are found for rotation angle $\alpha=16\,^{\circ}$,
where the corresponding number squeezing factor is\begin{equation}
\xi_{N}^{2}=\frac{4\left(\Delta
J_{z}\right)^{2}}{N}=-6.9_{-0.9}^{+0.8}\,\text{dB},\end{equation}
 for which the photon shot noise due to the imaging process is removed.
By additionally taking into account the technical noise due to
coupling-pulse imperfections, the squeezing factor can be improved
to \begin{equation}
\xi_{N}^{2}=-8.2_{-1.2}^{+0.8}\,\text{dB},\end{equation} which is
close to the atom-loss-limited theoretical optimum for this system
\cite{Li2009}. Furthermore, they pointed out that the best spin
squeezing factor $\xi_{R}^{2}=\xi_{N}^{2}/\cos\varphi=-8.2$~dB is
obtainable, which implies the entanglement of $170$
atoms~\cite{Sorensen2001a}.

\begin{figure}[ptb]

\begin{centering}
\includegraphics[width=14cm]{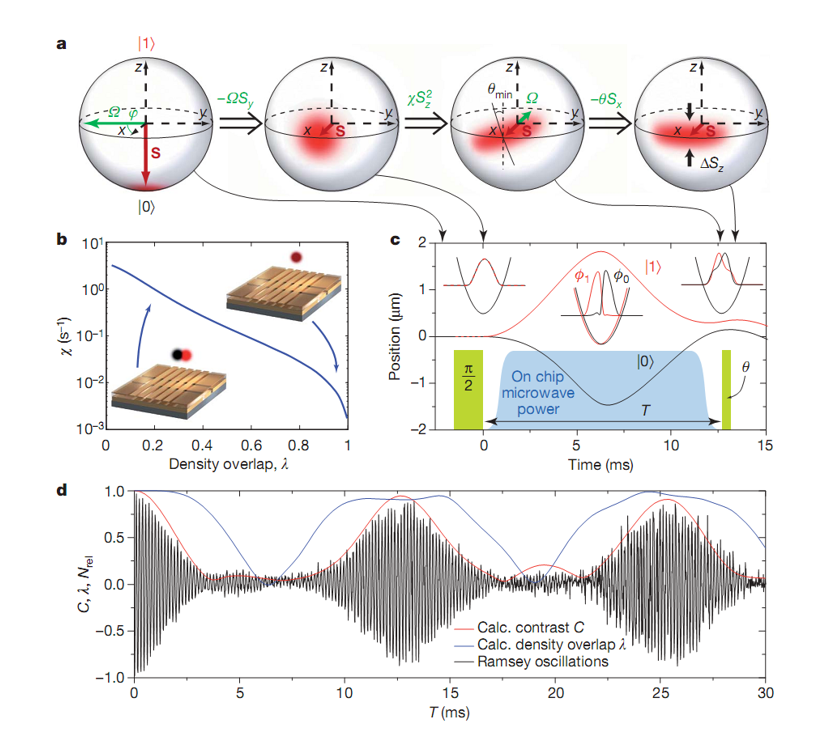}
\par\end{centering}

\caption{(Color online) Spin squeezing and entanglement through
controlled interactions on an atom chip~\protect\cite{Riedel2010}.
\textbf{a,} Bloch sphere representations of the dynamic evolution of
the BEC internal states. The red disks represent spin noise. The
one-axis twisting Hamiltonian $\chi S_{z}^{2}$ reduces the spin
noise at an angle $\theta_{\min}$, and the subsequent pulse rotates
the state around $-x$ by an angle
$\theta$~\protect\cite{Riedel2010}. \textbf{b}, Control of the
nonlinearity $\chi$ on the atomic chip. $\chi$ depends on the
difference of intra- and inter-state atomic interactions.
\textbf{c,} Experimental sequence and motion of the two BEC
components~\protect\cite{Riedel2010}. \textbf{d,} Measured Ramsey
fringes in the normalized population difference. This figure is from
Ref.~\protect\cite{Riedel2010}.}

\label{fig_BEC_exp2a}
\end{figure}

\begin{figure}[ptb]

\begin{centering}
\includegraphics[width=14cm]{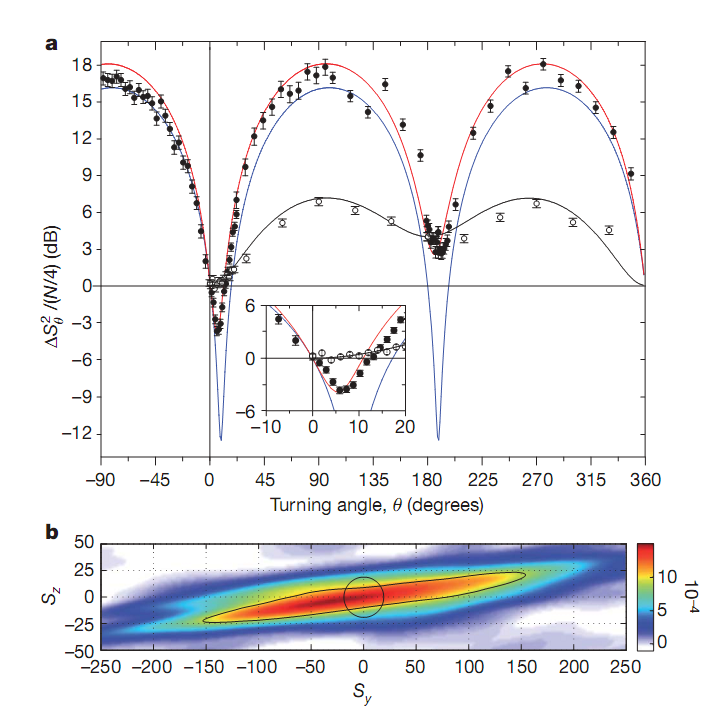}
\par\end{centering}

\caption{(Color online) Spin noise tomography and reconstructed
Wigner function of the spin-squeezed BEC~\protect\cite{Riedel2010}.
\textbf{a}, Observed spin noise for the spin-squeezed state (filled
circles) and for a CSS (open circles). In the squeezed state, a spin
noise reduction of $-3.7\pm0.4$~dB is observed for
$\theta_{\min}=6{^{\circ}}$, corresponding to
$\xi_{R}^{2}=-2.5\pm0.6$~dB of metrologically useful squeezing for
Ramsey spectroscopy~\protect\cite{Riedel2010}. The blue line
describes the simulation results which take account of particle loss
but not technical noise, give rise to a minimum variance of
-12.8~dB. The technical noise is included in the red line, which is
in good agreement with the experimental data.\textbf{ b},
Reconstructed Wigner function. The black contour line indicates
where the Wigner function has fallen to $1/\sqrt{e}$ of its maximum.
For comparison, the circular $1/\sqrt{e}$ contour of an ideal
coherent spin state is shown. This figure is from
Ref.~\protect\cite{Riedel2010}.}

\label{fig_BEC_exp2b}
\end{figure}

In another recent work~\cite{Riedel2010}, spin squeezing was
generated in an atom chip by controlling the elastic collision
interactions through the wavefunction overlap of the two states with
a state-dependent microwave potential. The hyperfine states
$|0\rangle\equiv|F=1,m_{F}=-1\rangle$ and
$|1\rangle\equiv|F=2,m_{F}=1\rangle$ of $^{87}$Rb forms a two-level
system. The system Hamiltonian is the same as in
Eq.~(\ref{BEC_Ham_Exp1}), while here the notation of the spin
operator $J$ is replaced with $S$. A key problem in this experiment
is also to produce nonvanishing nonlinear interaction strength,
since the $s$-wave scattering lengths here satisfy\begin{equation}
a_{00}\,:\, a_{01}\,:\,
a_{11}=100.4\,:\,97.7\,:\,95.0.\end{equation}
 They realized $\chi=0.49$ Hz though spatially separating the two
components of the BEC with a state-dependent potential, during which
the $U_{AB}$ term in Eq.~(\ref{eq:BEC_chi}) is zero. The Ramsey
process here is similar to that in Ref.~\cite{Gross2010}, and is
displayed in Fig.~\ref{fig_BEC_exp2a}. At first, a CSS is prepared
by a $\pi/2$ pulse of duration 120
$\text{\ensuremath{\mu}}\text{s}$. After that, the two components
are separated, and the nonlinear interaction is active for a best
squeezing time to avoid `oversqueezing'. The two states overlap
again after 12.7 ms, and the nonlinear interaction stops. As shown
in Fig.~\ref{fig_BEC_exp2a}, the transverse spin
component\begin{equation}
S_{\theta}=S_{z}\cos\theta-S_{y}\sin\theta,\end{equation}
 is measured by rotating the state in the $y$-$z$ plane by $\theta$
before detecting $S_{z}$. The experiment results are shown in Fig.~\ref{fig_BEC_exp2b},
where comparisons between squeezed and coherent states are presented.
The normalized variance $\Delta_{n}S_{\theta}^{2}=4\Delta S_{\theta}^{2}/\langle N\rangle$
is plotted so that $\Delta_{n}S_{\theta}^{2}=0\,\text{dB}$ corresponds
to the shot noise limit. For squeezed state, a minimum of \begin{equation}
\Delta_{n}S_{\theta}^{2}=-3.7\pm0.4\,\text{dB}\end{equation}
 is found at $\theta_{\min}=6\,^{\circ}$, with interference contrast
$C=2|\langle S_{x}\rangle|/N=(88\pm3)\%$, indicating a squeezing
parameter of \begin{equation}
\xi_{R}^{2}=-2.5\pm0.6\,\text{dB},\end{equation}
 and reduction of spin noise by $-3.7\pm0.4$~dB was the observed.
This could be used to improve interferometric measurements by
$-2.5\pm0.6$~dB over the standard quantum limit. The experimental
scheme and results~\cite{Riedel2010} are summarized in
Figs.~\ref{fig_BEC_exp2a} and \ref{fig_BEC_exp2b}. It is pointed out
that, the technique for tuning of interactions in a BEC through
wavefunction overlap also works in magnetic traps and for atomic
state pairs where no convenient Feshbach resonance exists.

\subsection{Squeezing transferred from light to atoms}

Squeezing of photons is not easy to preserve, hence researchers
considered how to transfer squeezing from photons to atomic
ensembles~\cite{Banerjee1996,Ueda1996,Kuzmich1997,Hald1999,Hald2000},
which are more convenient for storing quantum information. Many
efforts have been devoted to use atomic ensembles to realize quantum
memories~\cite{Hammerer2010,Sunprl2003,Wanghf2001}, and using
photons as information carriers. A practical analysis of generating
spin squeezing in an ensemble of atoms, by absorption of squeezed
light, was studied theoretically in~\cite{Kuzmich1997}, and soon
after it was realized in experiments~\cite{Hald1999}. Spin-squeezed
states can also be generated via electromagnetically induced
transparency
\cite{Fleischhauer2000,Dantan2003,Akamatsu2004,Dantan2004,Dantan2005}.
In this section, we present the experimental~\cite{Hald1999} results
for generating spin-squeezed states when the atomic ensemble absorbs
the squeezed vacuum field of the photons.

\subsubsection{Theoretical proposals}

\begin{figure}[ptb]

\begin{centering}
\includegraphics[width=16cm]{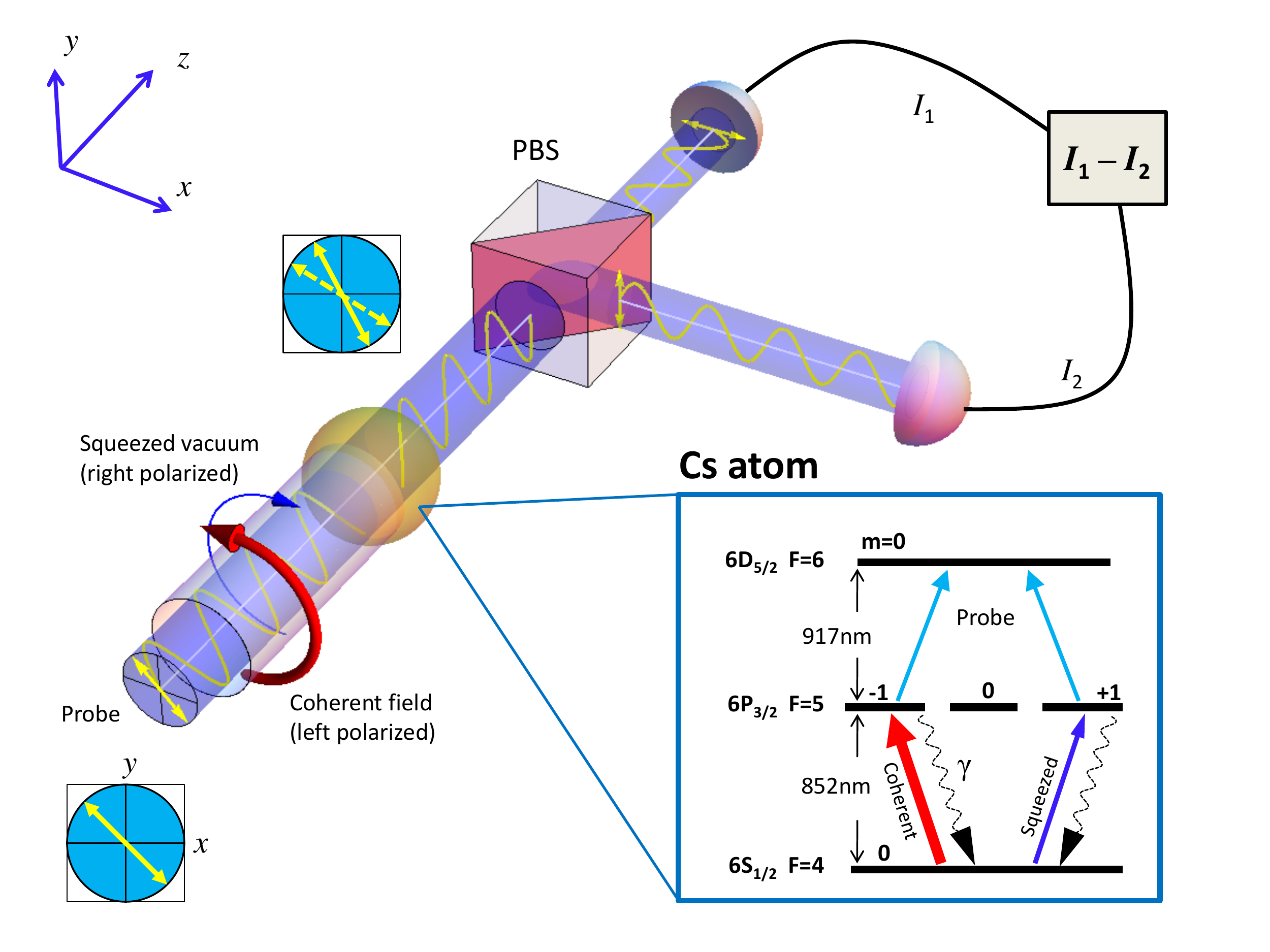}
\par\end{centering}

\caption{(Color online) Schematic diagram for a polarization
measurement~\protect\cite{Kuzmich1997}. Three input beams propagate
through a Cs atom cloud along the $z$-direction, a $\pi/4$-polarized
probe beam which is nearly resonant with $6P_{3/2}$ to $6D_{5/2}$
transition, a left-polarized coherent beam that induces
transition~\protect\cite{Kuzmich1997} between levels $|0\rangle$
($6S_{1/2},$ $m=0$) and $|1\rangle$ ($6P_{3/2},$ $m=-1$), and a
right-polarized squeezed vacuum which coupled with levels
$|0\rangle$ and $|2\rangle$ ($6P_{3/2},m=1$). The coherent field is
much stronger than the squeezed vacuum. The atom cloud is optically
thick for the squeezed vacuum and coherent field, and we assume that
these two fields are completely absorbed, while the probe beam is
only partially absorbed. Due to the population differences between
levels $|1\rangle$ and $|2\rangle$, the refraction index for left-
and right-polarization components of the probe beam are different,
thus the polarization plane of the probe beam is rotated along the
$z$-direction. After a polarization beam splitter (PBS), the probe
beam is split into $x$- and $y$-polarized beams and is detected by
two photon detectors. As the probe beam is $\pi/4$ polarized, for
the case of no atom clouds, the photocurrent difference is zero.
Therefore, when an atom could exist, the photocurrent different is
measured to gain information about the population fluctuations of
levels $|1\rangle$ and $|2\rangle$. As the absorption is complete, a
$50\%$ degree of spin squeezing can be obtained.}

\label{fig_light-matter}
\end{figure}

Here discuss the theoretical proposal~\cite{Kuzmich1997} of creating
spin-squeezed states in atomic ensembles via the absorption of
squeezed vacuum, which is illustrated in
Fig.~\ref{fig_light-matter}. The experiment is realized in
Ref.~\cite{Hald1999}. We consider a cloud of Cs
atoms~\cite{Kuzmich1997}. The energy level structure is shown in
Fig.~\ref{fig_light-matter}, where we only consider levels
$|6S_{1/2},\, F=4\rangle$, $|6P_{3/2},\, F=5\rangle$ and
$|6D_{5/2},\, F=6\rangle$. There are two transition configurations:
A $V$-type configuration consisting of levels $|0\rangle$
($|6S_{1/2},\, m=0\rangle$), $|1\rangle$ ($|6P_{3/2},\,
m=-1\rangle$) and $|2\rangle$ ($|6P_{3/2},\, m=1\rangle$), which is
used to generate spin-squeezed states; a $\Lambda$-type
configuration consisting of levels $|1\rangle$, $|2\rangle$ and
$|6D_{5/2},\, m=0\rangle$, which is used for polarization
measurement.

Below we mainly consider the $V$-type structure and only investigate
the subspace spanned by levels $|1\rangle$ and $|2\rangle$. The
transition between these two levels is forbidden due to symmetry.
However, effective transition processes between these two levels are
realized via the level $|0\rangle$, and the processes are described
by the operators $F_{12}$ and $F_{21}$. Therefore, we should first
derive the dynamical evolution between levels $|1\rangle$ (or
$|2\rangle$) and $|0\rangle$, which involves some assumptions and
the Heisenberg-Langevin theory~\cite{Kuzmich1997}.

The pseudospin operators are given by\begin{align}
F_{x} & =\frac{1}{2}\left(F_{12}+F_{21}\right),\notag\\
F_{y} & =\frac{1}{2i}\left(F_{12}-F_{21}\right),\notag\\
F_{z} & =\frac{1}{2}\left(F_{11}-F_{22}\right),\label{pesudospin}\end{align}
where the collective operators $F_{ij}=\sum_{\mu}\sigma_{ij}^{(\mu)}$,
$(i,\, j=1,2)$, and $\mu$ is the particle index and $\sigma_{ij}=|i\rangle\langle j|$.
Note that, since we use the hyperfine states, the angular momentum
operators are denoted by $F$ instead of $J$.

As shown in Fig.~\ref{fig_light-matter}, a left-circularly-polarized
coherent field and a right-circularly-polarized squeezed vacuum
field propagate along the $z$-direction through a cloud of Cs atoms.
Due to the selection rule, the coherent field is coupled to levels
$|0\rangle$ and $|1\rangle$, and the squeezed vacuum is coupled to
$|0\rangle$ and $|2\rangle$. The atomic decay rate $\gamma$ is
assumed to be equal for the two upper states $|1\rangle$ and
$|2\rangle$. To describe the interaction between field and atoms,
Ref.~\cite{Kuzmich1997} employs the continuous field operators
$a\left(\vec{r},t\right)$ and atomic operators
$\sigma_{ij}\left(\vec{r},t\right)$. The continuous atomic operators
are given by~\cite{Kuzmich1997,Fleis1995}\begin{equation}
\sigma_{ij}\left(\vec{r},t\right)=\frac{1}{\rho\Delta
V}\sum_{\mu}\exp\left[i\frac{\omega_{ij}}{c}\left(z-z^{(\mu)}\right)\right]\sigma_{ij}^{(\mu)},\label{def_dipole}\end{equation}
 where $\rho$ is the atomic density, and the sum is performed over
atoms enclosed in the volume $\Delta V$ around the position $\vec{r}$.
The volume $\Delta V$ is only a small fraction of the entire atom
cloud interacting with the field. The frequency between levels $|i\rangle$
and $|j\rangle$ is $\omega_{ij}=(E_{i}-E_{j})/\hbar$.

Now we consider an important assumption
$\lambda~\ll~L~\sim~\left(\Delta
V\right)^{1/3}~\ll~\lambda^{\prime}~\lesssim~l$, where $\lambda$ is
the wavelength of the field, $\lambda^{\prime}$ is the length-scale
of the field changes due to absorption and dispersion, and $l$ is
the length of the atomic cloud along the field propagation
direction. This assumption ensures a slow-varying field
equation~\cite{Kuzmich1997,Fleis1995}\begin{equation}
\left(\frac{1}{c}\frac{\partial}{\partial
t}+\frac{\partial}{\partial z}\right)a\left(z,t\right)=g\rho\int\!
dx\int\!
dy\text{~}u\!\left(x,y\right)\sigma_{01}\!\left(\vec{r},t\right),\label{slow_field_eq}\end{equation}
 which separates the transverse component of the field $a\left(\vec{r},t\right)$.
Here, $u\left(x,y\right)$ is the transverse mode function satisfying
the normalization condition $\int dx\!\int\! dy\text{~}u^{2}\!\left(x,y\right)=1$.
The interaction strength $g$ is determined by the atomic dipole moment.

The Heisenberg equation of motion for the continuous atomic dipole
operator is given by~\cite{Fleis1995}\begin{align}
\frac{\partial}{\partial t}\sigma_{01}\left(\vec{r},t\right)= & -\left(\frac{1}{2}\gamma+i\omega_{01}\right)\sigma_{01}\left(\vec{r},t\right)+g~u\left(x,y\right)a\left(z,t\right)\left[\sigma_{11}\left(\vec{r},t\right)-\sigma_{00}\left(\vec{r},t\right)\right]\notag\\
 & +\sqrt{\gamma}\left[\sigma_{11}\left(\vec{r},t\right)-\sigma_{00}\left(\vec{r},t\right)\right]b^{\text{in}}\left(\vec{r},t\right),\label{dipole_eq}\end{align}
 where $\gamma$ is the atom decay rate, $b^{\text{in}}\left(\vec{r},t\right)$
is the continuous operator of the atomic noise. For different atoms,
the noise operators are independent, and thus obey the commutation
relations\begin{equation}
\rho\left[b^{\text{in}}\left(\vec{r},t\right),b^{\text{in}\dagger}\left(\vec{r}_{1},t_{1}\right)\right]=\delta\left(\vec{r}-\vec{r}_{1}\right)\delta\left(t-t_{1}\right).\end{equation}
 Now we assume the field to be so weak that the number of excited
atoms is negligibly small, and replace $\sigma_{00}\left(\vec{r},t\right)-\sigma_{11}\left(\vec{r},t\right)=1$
in Eq.~(\ref{dipole_eq}). Then we obtain\begin{equation}
\frac{\partial}{\partial t}\sigma_{01}\left(\vec{r},t\right)=-\left(\frac{1}{2}\gamma+i\omega_{01}\right)\sigma_{01}\left(\vec{r},t\right)-g~u\!\left(x,y\right)a\!\left(z,t\right)-\sqrt{\gamma}~b^{\text{in}}\left(\vec{r},t\right),\label{dipole}\end{equation}

The solutions of Eqs.~(\ref{dipole}) and (\ref{slow_field_eq}) can
be obtained~\cite{Kuzmich1997} by using the Fourier transform of
$\sigma_{01}$, $b^{\text{in}}$ and $a$,\begin{align}
\sigma_{01}\left(\vec{r},\omega\right) & =-\frac{g~u\left(x,y\right)a\left(z,\omega\right)+\sqrt{\gamma}~b^{\text{in}}\left(\vec{r},\omega\right)}{\gamma/2-i\left(\omega-\omega_{01}\right)},\label{sol_dipole}\\
a\left(z,\omega\right) & =a\left(0,\omega\right)e^{ik\left(\omega\right)z}-\int_{0}^{z}dz_{1}\frac{\sqrt{\gamma}~g~\rho~e^{ik\left(\omega\right)\left(z-z_{1}\right)}}{\gamma/2-i\left(\omega-\omega_{01}\right)}b^{\text{in}}\left(z_{1},\omega\right),\label{sol_field}\end{align}
 with the new dispersion relation \begin{equation}
k\left(\omega\right)=\frac{\omega}{c}+\frac{ig^{2}\rho}{\gamma/2-i\left(\omega-\omega_{01}\right)}.\end{equation}
 The redefined continuous noise operators become\begin{equation}
b^{\text{in}}\left(z,\omega\right)=\int dx\int dy\text{~}u\left(x,y\right)b^{\text{in}}\left(\vec{r},\omega\right),\end{equation}
 with commutation relation $\rho\left[b^{\text{in}}\left(z,\omega\right),b^{\text{in}\dagger}\left(z_{1},\omega_{1}\right)\right]=\delta\left(z-z_{1}\right)\delta\left(\omega-\omega_{1}\right)$.
For a specific field, the expectation values of
Eq.~(\ref{sol_field}) can be calculated, and inserted into
Eq.~(\ref{sol_dipole}) to obtain the dynamics of the local dipole
operators. Finally, by using Eq.~(\ref{def_dipole}) the collective
atomic dipole of the sample can be obtained~\cite{Kuzmich1997}.

The above results can be used to compute the collective operators\begin{equation}
F_{ij}=\sum_{\mu}\sigma_{ij}^{(\mu)}=\rho\int d\vec{r}~\sigma_{ij}\left(\vec{r}\right),\text{ \ }\left(i,j=1,2\right).\end{equation}
 Assume now that the coherent field $a_{1}$ and squeezed vacuum $a_{2}$
have an identical transverse mode function $u\left(x,y\right)$, and
also that the coherent amplitude is much stronger than the squeezed
vacuum fluctuations. Thus the operator products can be linearized
as\begin{align}
\sigma_{12}\left(\vec{r},t\right) & =\left\langle \sigma_{10}\left(\vec{r},t\right)\right\rangle ~\sigma_{02}\left(\vec{r},t\right),\notag\\
\sigma_{12}\left(\vec{r},\Delta=\omega-\omega_{1}\right) & =\left\langle \sigma_{10}\left(\vec{r},t=0\right)\right\rangle ~\sigma_{02}\left(\vec{r},\omega\right),\end{align}
 where $\left\langle \sigma_{10}\left(\vec{r},t\right)\right\rangle =\left\langle \sigma_{10}\left(\vec{r},t=0\right)\right\rangle \exp({-i\omega_{1}t})$.
Now we assume $\omega_{1}=\omega_{01}=\omega_{02}$, and consider
the practical condition\begin{equation}
g^{2}\rho\gg\frac{\gamma\Delta}{c},\end{equation}
 which implies complete absorption, and that the light traveling time
is shorter than the atomic decay time. Finally, the collective
operator $F_{12}$ can be obtained~\cite{Kuzmich1997}\begin{align}
F_{12}\left(\Delta\right) & \simeq\frac{\alpha^{\ast}}{\gamma-i\Delta}\left[a_{2}^{\text{in}}\left(\omega_{1}+\Delta\right)+d_{2}^{\text{in}}\left(\omega_{1}+\Delta\right)\right]\notag\\
 & =\kappa e^{-i\phi}\left[a_{2}^{\text{in}}\left(\omega_{1}+\Delta\right)+d_{2}^{\text{in}}\left(\omega_{1}+\Delta\right)\right],\end{align}
 where $a_{2}^{\text{in}}\left(t\right)\equiv a_{2}^{\text{in}}\left(\vec{r}=0,t\right)$,
and $\kappa$, $\phi$ are the argument and phase of $\alpha^{\ast}/\left(\gamma-i\Delta\right)$,
respectively. The spin squeezing is obtained from\begin{equation}
\left\langle F_{x,y}^{2}\right\rangle =\frac{1}{4}\left\langle F_{z}\right\rangle \left(4X_{+,-}^{2}+1\right),\end{equation}
 where $\left\langle F_{z}\right\rangle =\kappa^{2}=\frac{\left\vert \alpha\right\vert ^{2}}{\gamma^{2}+\Delta^{2}}$,
and $X_{+,-}$ are the quadrature phase amplitudes defined as\begin{align}
X_{+} & =\frac{1}{2}\left[e^{-i\phi}a_{2}^{\text{in}}+e^{i\phi}a_{2}^{\text{in}\dagger}\right],\notag\\
X_{-} & =\frac{1}{2i}\left[e^{-i\phi}a_{2}^{\text{in}}-e^{i\phi}a_{2}^{\text{in}\dagger}\right].\end{align}
 For the vacuum input field $a_{2}$, the variance is $X_{+,-}^{2}=1/4$,
and thus $\left\langle F_{x,y}^{2}\right\rangle
=\frac{1}{2}\left\langle F_{z}\right\rangle =\frac{1}{2}\left\langle
F\right\rangle $ for coherent spin state~\cite{Kuzmich1997}. For
squeezed vacuum, as discussed in Sec.~\ref{definition},
\begin{equation} 4X_{+}^{2}=\exp(-2\left\vert \eta\right\vert
)\text{, }\quad4X_{-}^{2}=\exp({2\left\vert \eta\right\vert
}),\end{equation}
 where $\eta=\left\vert \eta\right\vert \exp({i2\phi})$ is the bosonic
squeezing parameter. Therefore, if the input field is a broadband
vacuum, in the limit $X_{+}^{2}=0$, then $\left\langle F_{x}^{2}\right\rangle =\left\langle F_{z}\right\rangle /4$
and the spin-squeezing parameter becomes (see Table \ref{tab_definition})
\begin{equation}
\xi_{H^{\prime\prime}}^{2}=\frac{1}{2}.\end{equation}
 Thus for an optically thick atomic ensemble, it is possible to achieve
$50\%$ spin squeezing when the squeezed vacuum is completely absorbed.

\subsubsection{Experimental realizations}

\begin{figure}[ptb]

\begin{centering}
\includegraphics[width=16cm]{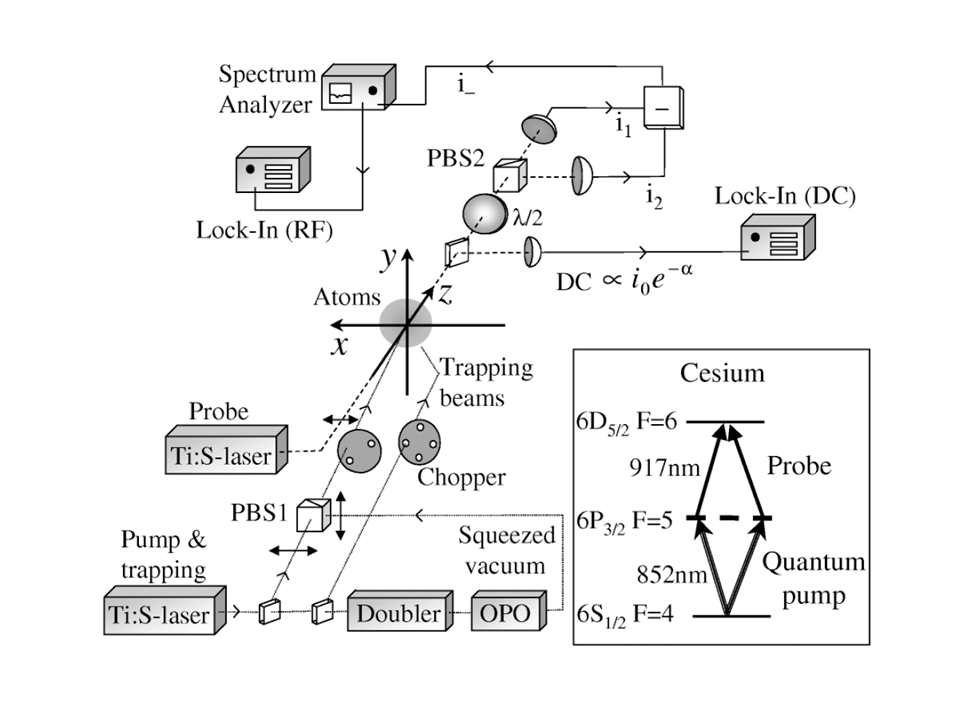}
\par\end{centering}

\caption{(Color online) Schematic diagram of the experiment in
Ref.~\protect\cite{Hald1999}. The squeezed vacuum is $y$-polarized,
while the pumping beam is $x$-polarized. The probe is also
$x$-polarized, and after the $\lambda/2$ wave plate, it turns to
$\pi/4$ polarized. This is from Ref.~\protect\cite{Hald1999}.}

\label{fig_PBS_exp}
\end{figure}

Soon after the above theoretical proposal~\cite{Kuzmich1997}, an
experiment was carried out in Ref.~\cite{Hald1999}, with the
experimental setup shown in Fig.~\ref{fig_PBS_exp}. They observed a
spin-squeezed ensemble of $10^{7}$ cold Cs atoms, which is produced
via transferring a state of free propagating squeezed light to the
atomic ensemble. This experiment also demonstrated storage of
quantum information of light in atoms. In this experiment, they
prepared about $10^{9}$ Cs atoms in a magneto-optical trap, and then
excited the $6P$ collective spin with the coherent pump polarized
along the $x$-direction to establish the quantum limit of the spin
noise. The coherent pump is of 50~mW with a beam diameter of 4.0~mm,
resonant with the 852 nm transition between $|6S_{1/2},\,
F=4\rangle$ and $|6D_{5/2},\, F=6\rangle$, provides weak excitation
for the $6P$ collective spin of about $10^{7}$ atoms. Then the atom
states are detected by a probe beam, which is linearly polarized
parallel to the coherent component of the pump and near resonant
with the $|6P_{3/2},\, F=5\rangle$ $\rightarrow$ $|6D_{5/2},\,
F=6\rangle$ (917 nm) transition. The properties of atoms are related
to the change in the probe differential photocurrent $\delta i$. In
the absence of atoms, the differential photocurrent is zero on
average, and fluctuates due to shot noise. In the experiment, the
fluctuation of the change of the differential photocurrent
\cite{Sorensen1998}\begin{equation} (\delta
i)^{2}\left(\Delta\right)=-\left[1-\exp\left(-\alpha_{\Delta}\right)\right]+s~\alpha_{0}~\exp\left(-2\alpha_{\Delta}\right)\left(\delta\tilde{J}\right)^{2}\end{equation}
 is measured, where $(\delta\tilde{J})^{2}$ is the atomic noise contribution
per atom depending on the geometry of the experiment, $s$ is the
probe saturation parameter, and \begin{equation}
\alpha_{\Delta}=\alpha_{0}\frac{\left(\gamma/2\right)^{2}}{\Delta^{2}+\left(\gamma/2\right)^{2}}=\frac{\sigma}{A}N\frac{\left(\gamma/2\right)^{2}}{\Delta^{2}+\left(\gamma/2\right)^{2}}\end{equation}
 is the probe optical depth at the detuning $\Delta$~\cite{Hald1999,Kuzmich1999},
with $N$ the number of atoms interacting with the probe, $\sigma$
the atomic cross section for unpolarized light, $A$ the cross section
of the probe beam, and $\gamma$ the atomic decay rate.\emph{ }

\begin{figure}[ptb]
\begin{centering}
\includegraphics[width=10cm]{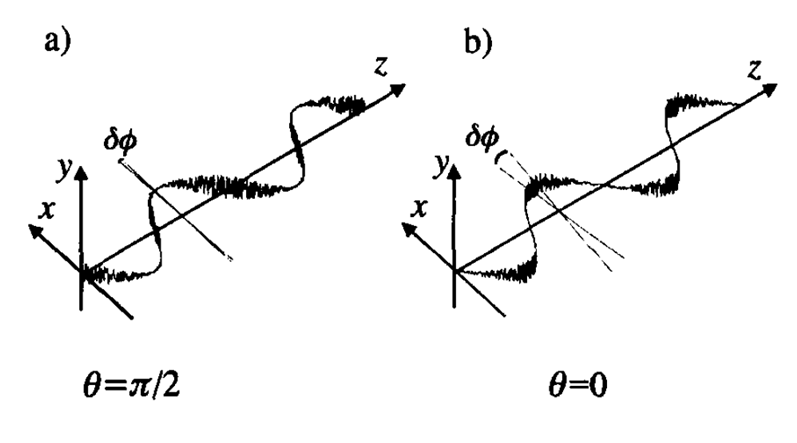}
\par\end{centering}

\caption{(Color online) Schematic diagram of the electric field for
the mixed output beam after PBS1 in
Fig.~\protect\ref{fig_PBS_exp}~\protect\cite{Hald2000}. The
fluctuations in the composite field of the coherent pumping beam and
the squeezed vacuum depend on their phase difference. If the phase
difference $\theta=0$, the so-called out-of-phase case, the field
will have reduced fluctuations in the polarization axis direction,
while if $\theta=\pi$, the so-called in-phase case, the field will
have reduced fluctuations. This figure is from
Ref.~\protect\cite{Hald2000}.}

\label{coh_squ_mix}
\end{figure}

The probe beam diameter is 3.7~mm and the power is about 250~mW for
the off-resonant spin squeezing measurements $s=0.4$. The squeezing
transfer is performed via injecting squeezed vacuum which is
linearly polarized along the $y$-direction. After the first
polarized beam splitter (PBS1), the coherent pumping beam and the
squeezed vacuum are mixed, and thus the output field will have
fluctuations in either the polarization axis direction or in the
ellipticity (depending on the phase as shown in
Fig.~\ref{coh_squ_mix}) reduced below the standard quantum
limit~\cite{Hald1999}.

In the case of squeezed light, the fluctuations in the intensity difference
between the right- and left-polarized components of the pump field
is\begin{equation}
\Delta S_{z}^{2}=\frac{n}{4}\left(R_{-}\cos^{2}\theta+R_{+}\sin^{2}\theta\right),\end{equation}
 where $S_{z}$ is the Stokes operator (\ref{Stokes_Op}), $\theta$
is the phase difference between the squeezed vacuum and the coherent
pumping field, and $R_{\pm}$ are determined by the efficiency of the
squeezing source, for coherent states $R_{\pm}=1$. Therefore, if
$R_{-}<1<R_{+}$ and in the case of $\theta=0$, the output field has
ellipticity fluctuations, and the fluctuations of the intensity
difference between the right- and left-polarized components of the
output field are reduced below the shot-noise limit \cite{Hald1999}.
Intuitively, this reduces the fluctuations in the population
difference between the $+m$ and $-m$ Zeeman levels.

\begin{figure}[ptb]

\begin{centering}
\includegraphics[width=10cm]{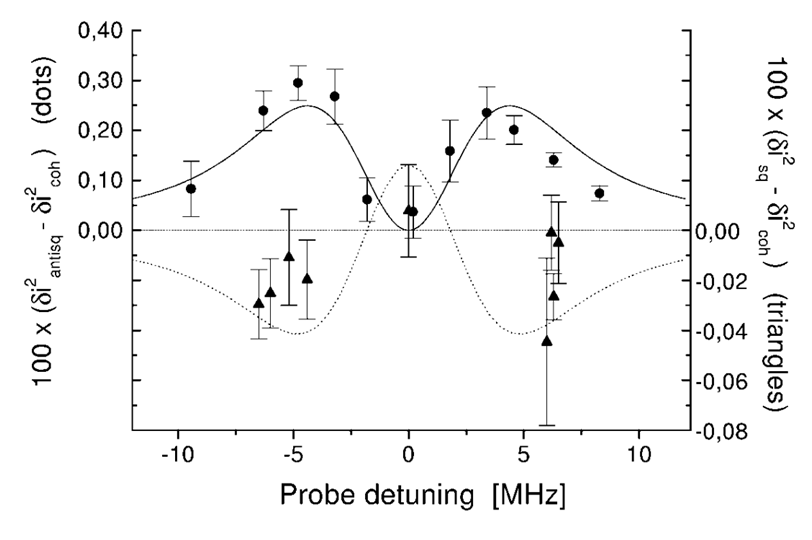}
\par\end{centering}

\caption{Squeezed spin noise of atoms~\protect\cite{Hald1999}.
Dashed line at zero: spin noise for uncorrelated atoms. Dots (left
axis): antisqueezed spin noise. Triangles (right axis): squeezed
spin noise. Solid line: square Doppler broadened dispersion
function. Dotted line: expected spin squeezing spectrum. The
detuning uncertainty is 0.5~MHz. Note that the scales for spin
squeezing and spin antisqueezing are different. This figure is from
Ref.~\protect\cite{Hald1999}.}

\label{fig_squ_light_matter}
\end{figure}

With the squeezed vacuum in phase, the average squeezing of the pump
light available at the trap site is $\left(-1.8\pm0.2\right)$~dB.
The squeezing parameter is defined
as~\cite{Hald1999}\begin{equation} \xi=\frac{(\delta
i)_{\text{sq}}^{2}-(\delta i)_{\text{coh}}^{2}}{(\delta
i)_{\text{coh}}^{2}},\end{equation}
 where $\delta i^{2}$ is the change in the probe differential photocurrent
noise caused by the atoms. The subscripts
{}``sq\textquotedblright{}\ and {}``coh\textquotedblright{}\ denote
the spin-squeezed state and CSS, respectively. The best squeezing
shown in Fig.~\ref{fig_squ_light_matter} is $-3.0\%$ at $\pm6$~MHz.
The average value for all of the $183$ individual
measurements~\cite{Hald1999} at $\pm6$~MHz is
$\xi=-\left(1.4\pm0.4\right)\%$. The actual degree of squeezing
should be greater, since the relative contribution of the readout
efficiency is not known from the overall efficiency $\eta$.

\subsection{Generating spin squeezing via quantum nondemolition measurements}

Quantum nondemolition measurements have been studied to generate
non-classical photon states~\cite{Scully1997}, and also
spin-squeezed
states~\cite{Kuzmich1998,Kuzmich1999,Molmer1999,Takahashi1999,Kuzmich2000,Thomsen2002,Thomsen2002a,Zhang2003,Auzinsh2004,Oblak2005,Meiser2008,Teper2008,Appel2009,Saffman2009,Takano2009,Windpassinger2009,Shah2010}.
At first, we briefly introduce the criteria for QND measurements.
Then we present a typical QND-type Hamiltonian, implemented via
dispersive interactions between light field and atoms. Utilizing
this Hamiltonian, spin squeezing was generated in
experiments~\cite{Kuzmich1999,Appel2009,Takano2009}. In this
QND-type Hamiltonian, the light field plays the role of a probe, by
readout of the light field, the atomic spin state could be squeezed,
but is conditioned to the specific readout of the light field, and
thus is called conditional spin-squeezed state. By using feedback
techniques~\cite{Molmer1999,Thomsen2002,Berry2002a,Thomsen2002a,Smith2003,Kurucz2010},
unconditional spin-squeezed states can be generated.

\subsubsection{Theoretical background}

We first briefly discuss the basic requirements for a QND
measurement (see, e.g. Ref.~\cite{Scully1997}). Consider two systems
$A$ and $P$, where $A$ is the signal, and $P$ acts as a probe
coupled to the signal. By detecting the probe (e.g., measuring a
physical quantity $P_{\alpha}\in\left\{ P_{i}\right\} $ that belongs
to $P$), we could gain information about a signal operator
$A_{\alpha}\in\left\{ A_{i}\right\} $ without perturbing its
subsequent evolution. The two sets $\left\{ A_{i}\right\} $ and
$\left\{ P_{i}\right\} $ contain conjugate operators of the probe
and signal systems, respectively. For example, for a free particle,
the two conjugate operators could be the position operator $x$ and
momentum operator $p$. The total Hamiltonian for the signal-probe
system reads\begin{equation} H=H_{A}+H_{P}+H_{I}.\end{equation}
 The conditions for a QND-type measurement of a signal operator $A_{\alpha}$
via detecting a probe operator $P_{\beta}$ are listed below:

(i) For a QND-type measurement of $A_{\alpha}$, the interaction term
$H_{I}$ must contain the operator $A_{\alpha}$.

(ii) Backaction-evasion in the measurement of the operator $A_{\alpha}$
requires\begin{equation}
\left[H_{I},\, A_{\alpha}\right]=0.\end{equation}

(iii) The dynamics of $P_{\beta}$ should be associated with $A_{\alpha}$,
so that the readout of $P_{\beta}$ contains information about $A_{\alpha}$;
therefore\begin{equation}
\left[H_{I},\, P_{\beta}\right]\neq0.\end{equation}

(iv) Since the successive measurements of $A_{\alpha}$ should be
predictable, then\begin{equation}
\left[H_{A},\, A_{\alpha}\right]=0.\end{equation}

\begin{figure}[ptb]

\begin{centering}
\includegraphics[width=16.5cm]{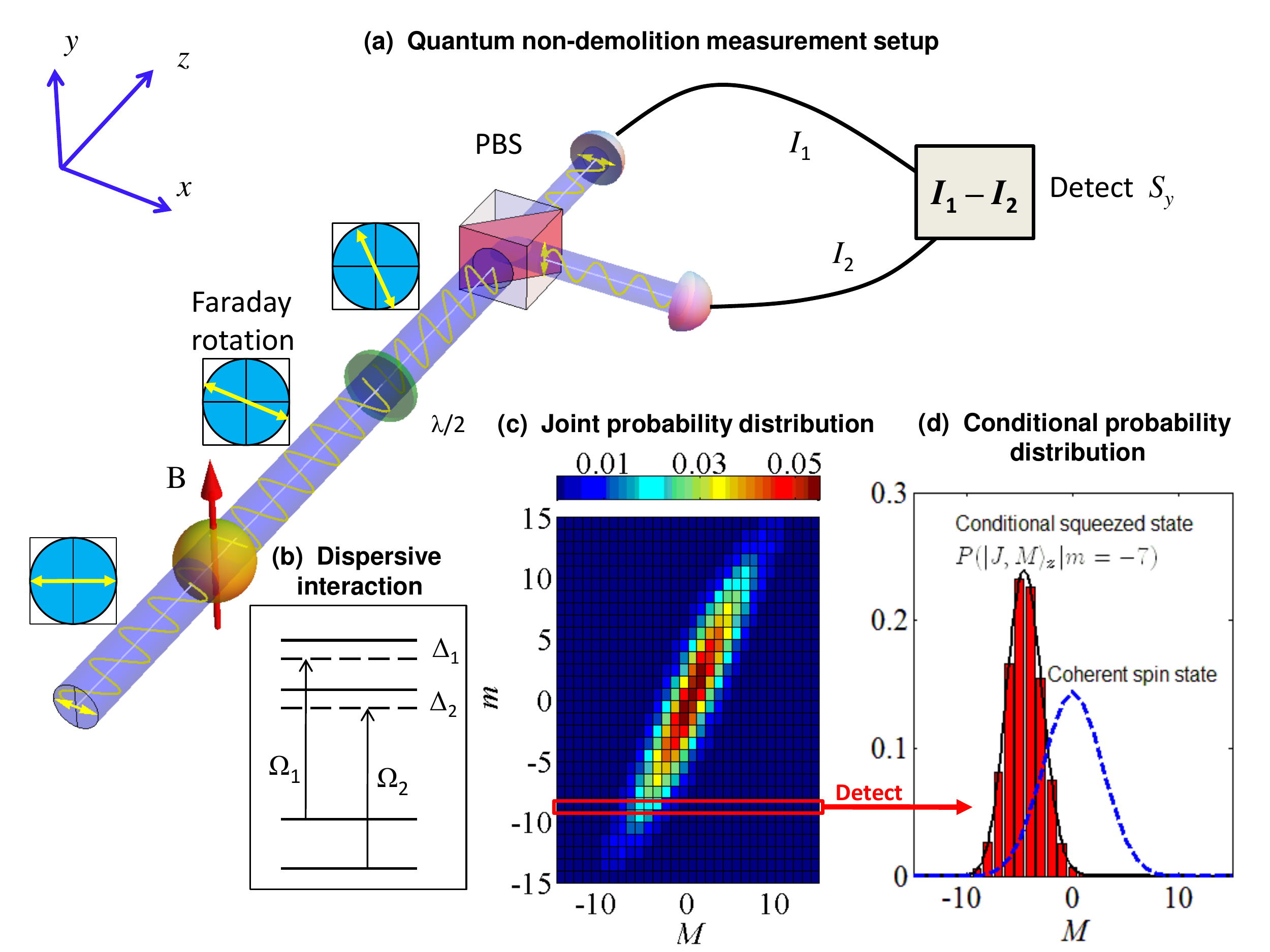}
\par\end{centering}

\caption{(Color online) (a) Schematic diagram for the QND measurement. The
input probe beam is $x$-polarized, and the atoms are initially polarized
along the $x$-axis. The two circular polarized components of the
probe beam interact dispersively with atoms, with detuning $\Delta_{1}$
and $\Delta_{2}$, as shown in (b), $\Omega_{1}$ and $\Omega_{2}$
are atom-field couplings. The dispersive interactions induce a rotation
of the polarization plane, the Faraday rotation. The $\lambda/2$
wave plate together with the following PBS and two detectors play
the role of QND measurement of $J_{z}$ by detecting $S_{y}$, which
is readout by the photocurrent difference. (c) Joint probability distributions
for a system with $N=n=30$ and $\chi=0.1$. (d) When the readout
of $S_{y}$ gives $m=-7$, the conditional squeezed state is plotted
and compared with the coherent spin state.}

\label{fig_QND_sketch}
\end{figure}

From the criteria given above, one of the simplest QND-type Hamiltonians
is\begin{equation}
H_{I}=A_{\alpha}P_{\alpha},\end{equation}
 and the probe operator to be detected is $P_{\beta}$ with $\beta\neq\alpha$.
This QND-type Hamiltonian can be implemented through far-off
resonant dispersive interactions between light and atomic ensembles,
as shown in Fig.~\ref{fig_QND_sketch}(b). Two probe light beams,
$a_{1}$ and $a_{2}$, \ are coupled to different hyperfine levels,
with detuning $\Delta_{1}$ and $\Delta_{2}$, respectively. In the
large detuning case, the effective Hamiltonian
is~\cite{Appel2009}\begin{equation}
H=\hbar\sum_{i=1}^{2}\frac{g_{i}^{2}\Delta_{i}}{\gamma_{i}^{2}/4+\Delta_{i}^{2}}~a_{i}^{\dagger}~a_{i}~J_{ii}=\hbar\Omega\left(S_{z}J_{z}+\frac{1}{4}nN\right),\label{QND_Ham}\end{equation}
 where the $\gamma_{i}$'s are the spontaneous decay rates for the
excited levels, and \begin{equation}
J_{ii}=\sum_{\mu}|i\rangle^{\mu~\mu}\langle i|\end{equation}
 measures the population of atoms in the level $|i\rangle^{\mu}$,
and $n$, $N$ are the photon and atom numbers, respectively. The
Stokes operator $S_{z}$ (\ref{Stokes_Op}) measures the population
difference between the left- and right-polarized states. In
experiment~\cite{Appel2009}, the detuning $\Delta_{1}$ and
$\Delta_{2}$ are chosen in such a way that\begin{equation}
\frac{g_{1}^{2}\Delta_{1}}{\gamma_{1}^{2}/4+\Delta_{1}^{2}}~=~\frac{g_{2}^{2}\Delta_{2}}{\gamma_{2}^{2}/4+\Delta_{2}^{2}}~=~\frac{\Omega}{2}.\end{equation}
 This Hamiltonian acts as a Faraday rotation to the photon polarization,
and as a fictitious magnetic field to atoms. The Heisenberg equations
for the operators $\vec{S}$ and $\vec{J}$ are\begin{align}
\begin{pmatrix}S_{x}\\
S_{y}\\
S_{z}\end{pmatrix}_{t} & =\begin{pmatrix}\cos\left(\chi J_{z}\right) & -\sin\left(\chi J_{z}\right) & 0\\
\sin\left(\chi J_{z}\right) & \cos\left(\chi J_{z}\right) & 0\\
0 & 0 & 1\end{pmatrix}\begin{pmatrix}S_{x}\\
S_{y}\\
S_{z}\end{pmatrix}_{\text{in}},\notag\\
\begin{pmatrix}J_{x}\\
J_{y}\\
J_{z}\end{pmatrix}_{t} & =\begin{pmatrix}\cos\left(\chi S_{z}\right) & -\sin\left(\chi S_{z}\right) & 0\\
\sin\left(\chi S_{z}\right) & \cos\left(\chi S_{z}\right) & 0\\
0 & 0 & 1\end{pmatrix}\begin{pmatrix}J_{x}\\
J_{y}\\
J_{z}\end{pmatrix}_{\text{in}},\end{align}
 where $\chi=\Omega~t$. Therefore, a QND measurement on $J_{z}$
is performed through the readout of $S_{x}\left(t\right)$ {[}or $S_{y}\left(t\right)${]}.

Now, we consider a specific case, where the initial states for the
light field and atoms are both CSSs along the $x$-direction,\begin{equation}
|\varphi\left(0\right)\rangle=|S,S\rangle_{x}|J,J\rangle_{x},\end{equation}
 where $S=n/2$, $J=N/2$, and the subscript $x$ means that the states
are represented in the $S_{x}$ and $J_{x}$ basis. A QND measurement
on $J_{z}$ is performed by detecting $S_{y}$ (or $S_{x}$), the
readout of $S_{y}\left(t\right)$ becomes\begin{equation}
\left\langle S_{y}\left(t\right)\right\rangle =\frac{n}{2}{\large\langle\sin\left(\chi J_{z}\right)\rangle\simeq\left\langle S_{y}\right\rangle +\frac{n\chi}{2}\left\langle J_{z}\right\rangle ,}\label{qnd_in_out}\end{equation}
 where the average $\left\langle \cdot\right\rangle $ is carried
out under the initial states since we are using in the Heisenberg
picture. Thus we see that $\left\langle J_{z}\right\rangle $ and
$\left\langle S_{y}\right\rangle $ are related, the mean value of
$\left\langle S_{y}\left(t\right)\right\rangle $ is approximately
zero.\ From (\ref{qnd_in_out}), the fluctuation of
$S_{y}\left(t\right)$ is also related to $J_{z}$~\cite{Kuzmich1998}.

Below, we demonstrate the QND-measurement-induced spin squeezing from
the point of view of state evolution, shown in Fig.~\ref{fig_QND_sketch}(c)-(d).
As presented in Fig.~\ref{fig_spin_bloch}, for a coherent spin state
$|J,J\rangle_{x}$, if we measure $J_{z}$, the probability of the
readout of states $|J,M\rangle_{z}$ obeys a binomial distribution\begin{equation}
P_{J}\left(M\right)\equiv P\Big(|J,M\rangle_{z}\Big{|}|J,J\rangle_{x}\Big)=\Big\vert~_{z}\!\!\left\langle J,M|J,J\right\rangle _{x}\Big\vert^{2}=\frac{1}{2^{N}}\left(\begin{array}{c}
N\\
N/2+M\end{array}\right),\end{equation}
 where $P\left(|J,M\rangle_{z}\big{|}|J,J\rangle_{x}\right)$ is a
conditional distribution. Similarly, in the basis of $|S,m\rangle_{y}$,
the measurement basis, we have\begin{equation}
P_{S}\left(m\right)\equiv P\Big(|S,m\rangle_{y}\Big{|}|S,S\rangle_{x}\Big)=\Big\vert~_{y}\!\!\left\langle S,m|S,S\right\rangle _{x}\Big\vert^{2}=\frac{1}{2^{n}}\left(\begin{array}{c}
n\\
n/2+m\end{array}\right).\end{equation}
 The evolution of the state becomes\begin{align}
|\varphi\left(t\right)\rangle & =\exp\left(-i\chi S_{z}J_{z}\right)|S,S\rangle_{x}|J,J\rangle_{x}\notag\\
 & =\sum_{M=-J}^{J}\exp\left(-i\chi MS_{z}\right)\sqrt{P_{J}\left(M\right)}|S,S\rangle_{x}|J,M\rangle_{z}.\label{QND_dyn}\end{align}

At time $t$, we measure $S_{y}$. For a single trial measurement,
the readout will fall in an eigenvalue $m$ of $S_{y}$, and the photon
state collapses into the corresponding eigenstate $|S,m\rangle$.
After this readout, the atomic state is also changed according to
the measurement result of $S_{y}$. Therefore, the focus is on the
conditional distribution $P\left(|J,M\rangle_{z}\big{|}|S,m\rangle_{y}\right)$,
which gives information of the atomic state after a single trial measurement
of $S_{y}$. To analyze this problem, we consider the system size
to be very large, in which case the binomial distribution becomes
a Gaussian normal distribution,\begin{align}
P_{J}\left(M\right) & =\frac{1}{\sqrt{\pi N/2}}\exp\left(-\frac{2M^{2}}{N}\right),\notag\\
P_{S}\left(m\right) & =\frac{1}{\sqrt{\pi n/2}}\exp\left(-\frac{2m^{2}}{n}\right).\end{align}
 With the Bayes theorem\begin{equation}
P\left(|J,M\rangle_{z}\big{|}m\right)P_{S}\left(m\right)=P\left(|S,m\rangle_{y}\big{|}M\right)P_{J}\left(M\right),\end{equation}
 the conditional probability distribution for $P\left(|S,m\rangle_{y}\big{|}M\right)$
is easy to obtain. If the atomic system is in the state $|J,M\rangle_{z}$,
according to Eq.~(\ref{QND_dyn}) the photon state is rotated around
the $z$-axis with an angle $\phi=\chi M$. If the rotation angle
is very small, then the photon state after the rotation is still a
Gaussian distribution with mean value at $\sin\left(\phi\right)n/2\simeq\phi n/2$,
and \begin{equation}
P\left(|S,m\rangle_{y}\big{|}M\right)=\frac{1}{\sqrt{\pi n/2}}\exp\left[-\frac{\left(m-\phi n/2\right)^{2}}{n/2}\right].\end{equation}
 Therefore, the conditional probability distribution of the atomic
state becomes\begin{equation}
P\left(|J,M\rangle_{z}\big{|}m\right)=\frac{1}{\sqrt{\pi\xi_{R}^{2}N/2}}\exp\left[-\frac{(M-\chi m\xi_{R}^{2}N/2)^{2}}{\xi_{R}^{2}N/2}\right],\end{equation}
 where \begin{equation}
\xi_{R}^{2}=\frac{1}{1+\kappa^{2}}\text{,}\label{QND_spinsqu}\end{equation}
 with\begin{equation}
\kappa^{2}=nN\chi^{2}/4,\end{equation}
 and the state of the atomic system after detecting $S_{y}$ is squeezed,
while the expectation value of $J_{z}$ is shifted to $\chi m\xi_{R}^{2}N/2$.

Note that the spin length is generally shortened due to photon
scattering and other noises during the QND measurement, so the above
Eq.~(\ref{QND_spinsqu}) holds under the condition that the spin
length is not affected too much after the QND\ measurement. An
illustrative description of the above procedure is displayed in
Fig.~\ref{fig_QND_sketch}. The spin state derived above is the
so-called conditional squeezed state, which is conditioned to the
first measurement output of $S_{y}$. In a {}``single
trial\textquotedblright{}\ measurement of $S_{y}$, the output value
is random with respect to a specific probability distribution. The
conditional squeezed state is generated randomly, so it is not
repeatable. To obtain an unconditional squeezed state, we could
employ a feedback or filter scheme to shift $\left\langle
J_{z}\right\rangle $ back to zero, which is demonstrated in
Refs.~\cite{Molmer1999,Thomsen2002,Berry2002a,Thomsen2002a,Smith2003,Kurucz2010}.

\subsubsection{Experimental results}

Below, we first give a brief introduction about some experimental
works, then we present a detailed discussion about a recent notable
improvement. In Ref.~\cite{Kuzmich1999}, QND measurement was
performed on a collective spin of $\sim10^{6}$ Cs atoms, and
sub-shot-noise fluctuations of the collective spin were observed. In
this experiment, the probe beam was short pulses of laser light that
linearly polarized at $\pm45^{\circ}$. Then, in
Ref.~\cite{Kuzmich2000}, the authors performed continuous QND
measurement on a collection of $\sim10^{7}$ Cs atomic spins with an
off-resonant laser beam. The continuous QND measurement is an
important feature of this work, and the probe laser also serves for
measurement of the spin rotation signal.\emph{ }They observed
squeezed atomic spin states with spin noise reduction to $70\%$
below the shot-noise limit, and measured a small spin rotation with
an accuracy exceeding the shot-noise limit of the phase measurement.
Reference~\cite{Oblak2005} proposed a sequence of QND measurements
and spin rotations, which require interferometric QND measurement of
the atomic population with a sensitivity at the projection-noise
level. To achieve this they demonstrated experimentally that a
shot-noise-limited fiber optical interferometer at the white-light
setting can reach a sensitivity sufficient to detect the projection
noise under conditions not far from those of the QND measurement.
Increasing the optical density can significantly improve the QND
figure of merit $\kappa^{2}$, and thus will enhance the degree of
squeezing. In Ref.~\cite{Teper2008}, the authors demonstrated QND
measurement of atomic ensembles in a high-finesse optical cavity,
which can enhance the optical depth of the atomic sample. In this
experiment, a cloud of $^{87}$Rb atoms was cooled in a MOT, located
at the center of a hemispherical cavity. The maximum atom-cavity
coupling, $g/\left(2\pi\right)=53\,\text{kHz}$, is realized for the
$|F=2,\, m_{F}=2\rangle\rightarrow|F'=3,\, m_{F}=3\rangle$ cycling
transition. The atoms are released from the MOT, further cooled by
optical molasses, prepared in an equal superposition of the $|F=1,\,
m_{F}=0\rangle$ and $|F=2,\, m_{F}=0\rangle$ clock states, and
probed by a standing wave of intracavity light. They observed
$3.8\,\text{dB}$ in the variance of $J_{z}$ below the projection
noise with spontaneous emission of $\le30\%$.

Reference \cite{KoschorreckPRL2010a} demonstrated
sub-projection-noise sensitivity of QND spin measurements of
$1\times10^{6}$ cold $^{87}$Rb atoms in a broadband atomic
magnetometer. The high-bandwidth system avoids decoherence effects
through rapid probing. The spin readout noise is 2.8~dB below the
thermal spin noise and 1.6~dB below the CSS projection noise. Since
the Hamiltonian (\ref{QND_Ham}) only appropriate for spin-1/2 or
other two-level systems as pointed out by studies of
Ref.~\cite{KoschorreckPRL2010b}, where they considered larger spins
and proposed a decoupling mechanism based on a two-polarization
probing technique to recover the ideal QND behavior. The decoupled
QND measurement achieves a sensitivity 5.7(6)~dB better than the
spin projection noise.

\begin{figure}[H]

\begin{centering}
\includegraphics[width=14cm]{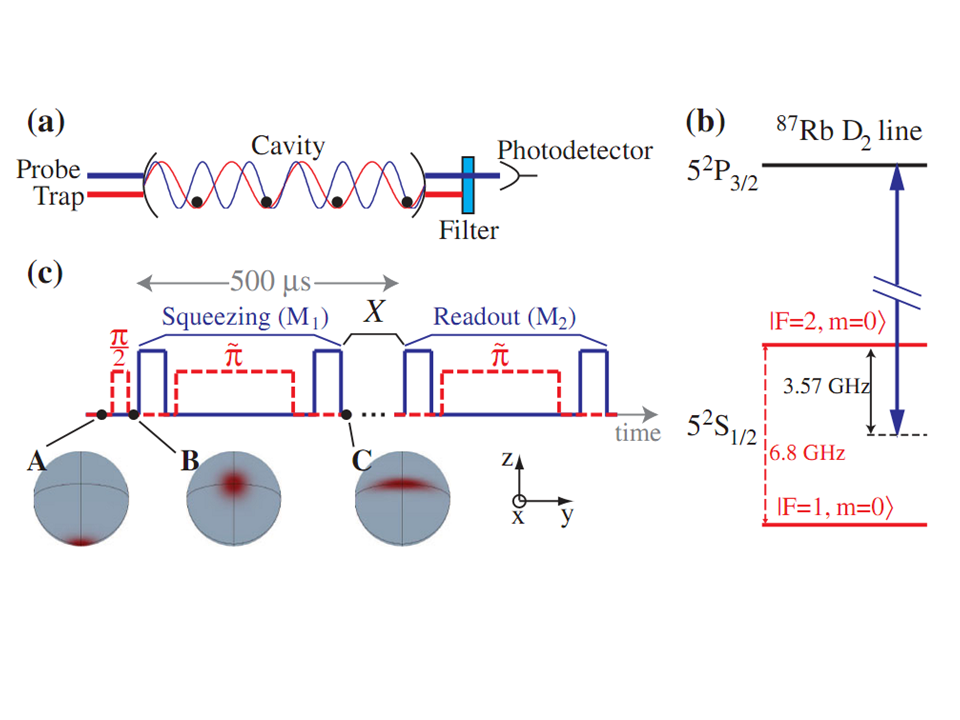}
\par\end{centering}

\caption{(Color online). Experimental schematic diagram of
Ref.~\cite{Schleier-Smith2010a}. (a) Experimental setup. (b) Level
structure of rubidium atom. (c) Experimental sequences. Dashed lines
denote the microwave pules while solid lines denote the probe
pulses. The $\tilde{\pi}$ pulse represents a composite $\pi$ pulses.
Various operations are performed between two readouts ($M_{1}$ and
$M_{2}$), denoted by $X$, such as the measurement of the variance of
the CSS, the operation of a clock. \textbf{A}-\textbf{C}:
Semiclassical probability distribution functions of the state. This
figure is from Ref.~\cite{Schleier-Smith2010a}. }

\label{fig_QND_3}
\end{figure}

Recently, Ref.~\cite{Schleier-Smith2010a} demonstrated the
generation of spin-squeezed state of $5\times10^{4}$ trapped
$^{87}$Rb atoms on an atom-clock transition by resonator-aided QND
measurement. For photon number $p=3\times10^{5}$, they observed
noise reduction of $3.0(8)\,\text{dB}$ below the shot-noise limit.
The hyperfine clock states used are $|1\rangle=|5^{2}S_{1/2},\,
F=1,\, m_{F}=0\rangle$ and $|2\rangle=|5^{2}S_{1/2},\, F=2,\,
m_{F}=0\rangle$. They used the metrological spin-squeezing parameter
defined as \begin{equation}
\zeta_{m}=\frac{2S_{\text{in}}\left(\Delta
S_{z}\right)^{2}}{\left|\left\langle \vec{S}\right\rangle
\right|^{2}},\end{equation}
 where $S_{\text{in}}$ is the initial spin of the uncorrelated ensemble.
Actually, this parameter is equal to $\xi_{R}^{2}$ when
$S_{\text{in}}=N/2$. The experimental setup is shown in
Fig.~\ref{fig_QND_3}. A strong ensemble-light coupling was achieved
by using a near-confocal optical resonator, and the experiments are
performed on an ensemble containing up to $N_{a}=5\times10^{4}$
laser-cooled $^{87}$Rb atoms optically trapped inside the resonator
in a standing wave of 851-nm light. The wavelength of the probe
light is $2\pi/k=780$~nm, and the cavity has a finesse
$\mathcal{F}=5.6(2)\times10^{3}$, a linewidth
$\kappa=2\pi\times1.01(3)$~MHz, and a mode waist $w=56.9(4)$~$\mu$m
at the atoms' position, corresponding to a maximal single-atom
cooperativity $\eta_{0}=24\mathcal{F}/\left(\pi
k^{2}w^{2}\right)=0.203(7)$\emph{.} One resonator mode is tuned
$3.57(1)$~GHz to the blue of the
$|5^{2}S_{1/2},F=2\rangle\rightarrow|5^{2}P_{3/2},F^{\prime}=3\rangle$
transition in $^{87}$Rb to result in a mode frequency shift $\omega$
proportional to the population difference $N=N_{2}-N_{1}$ between
the hyperfine clock states \begin{align}
|1\rangle & =|5^{2}S_{1/2},F=1,m_{F}=0\rangle,\nonumber \\
|2\rangle & =|5^{2}S_{1/2},F=1,m_{F}=0\rangle.\end{align}

\begin{figure}[ptb]

\begin{centering}
\includegraphics[width=10cm]{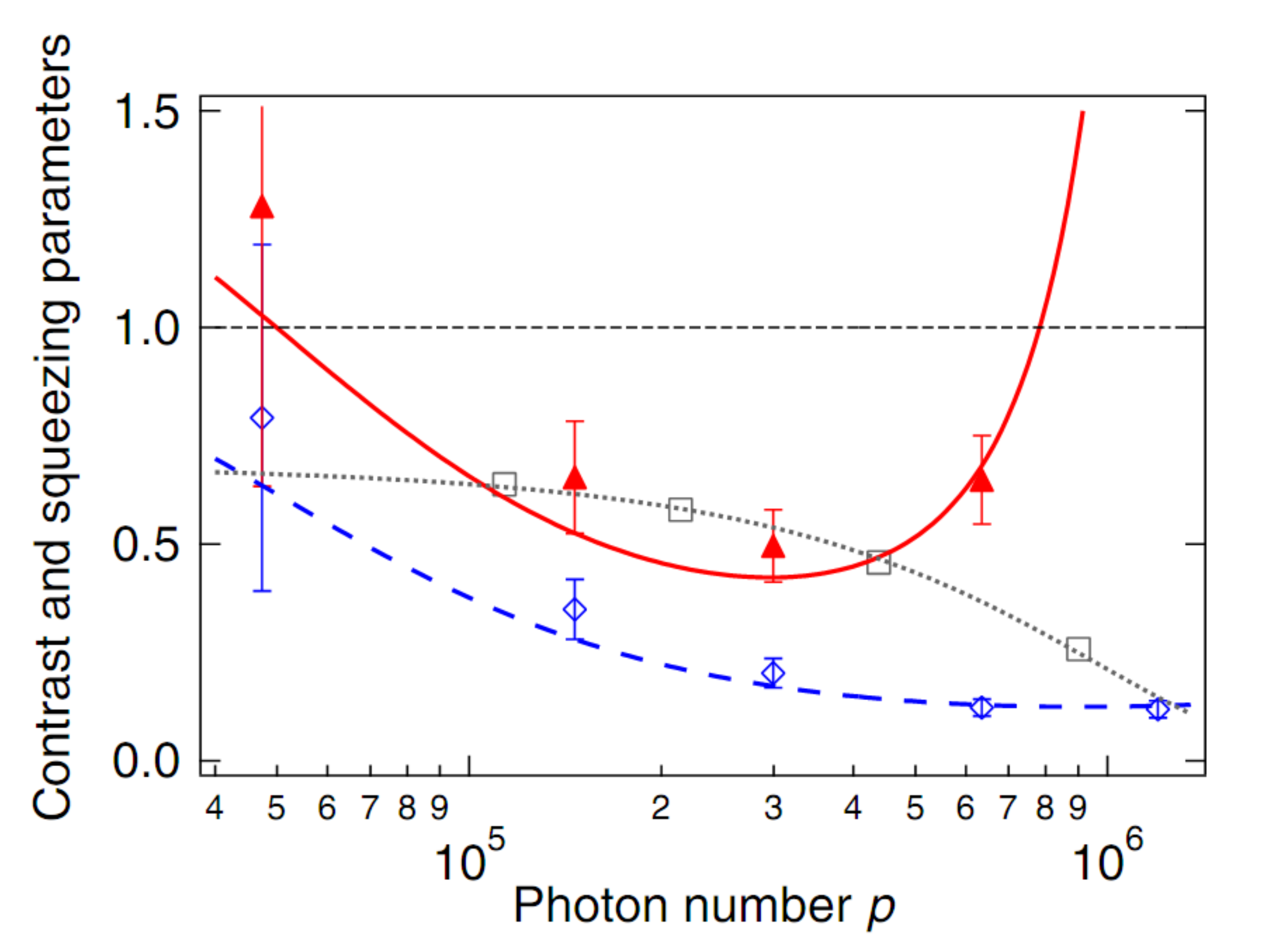}
\par\end{centering}

\caption{(Color online). Normalized spin noise
$\sigma^{2}=\left(\Delta S_{z}^{2}\right)_{M_{1}}/\left(\Delta
S_{z}^{2}\right)_{\text{CSS}}$ (open diamonds), contrast
$\mathcal{C}$ (open square), and metrological squeezing parameter
$\zeta_{m}$ (solid triangles). Solid line denotes the fits. This
figure is from Ref.~\cite{Schleier-Smith2010a}.}

\label{fig_QND_4}
\end{figure}

The experimental processes are shown in Fig.~\ref{fig_QND_3}(c).
At first, the atomic ensemble is prepared in state $|1\rangle$, after
a $\pi/2$ pulse, the atomic state polarizes in the $x$-axis, then
after measurements $M_{1}$, the atomic ensemble is conditional spin
squeezed as Fig.~\ref{fig_QND_3}\textbf{C}. To quantify the spin
squeezing, they measured $\Delta S_{z}^{2}$ and $\left|\left\langle \text{\ensuremath{\vec{S}}}\right\rangle \right|$,
the latter is obtained from the observed contrast $C$ of Rabi oscillations
since $\left|\left\langle \vec{S}\right\rangle \right|=CS_{0}$, as
shown in Fig.~\ref{fig_QND_4}, from which the metrological squeezing
parameter achieves\[
\zeta_{m}^{-1}=3.0(8)\,\text{dB}\]
 for photon number $p=3\times10^{5}$.

The QND measurements discussed above were all performed on alkali
atoms, which have complicated multilevel structures. In
Ref.~\cite{Takano2009}, the authors demonstrated a QND measurement
with a collective spin of cold ytterbium atoms, $^{171}$Yb, which
have the simplest ground state with a nuclear spin of one-half and
have no electron spin, and thus the system is robust against stray
magnetic fields since the magnetic moment of nuclear spin is a
thousandth of that of electron spin. By using short light pulses
with a width of 100~ns, more than a hundred time operations are
expected to be performed within the coherence time, and experimental
results showed $1.8_{-1.5}^{+2.4}\,\text{dB}$ spin squeezing.

\begin{figure}[ptb]

\begin{centering}
\includegraphics[width=17cm]{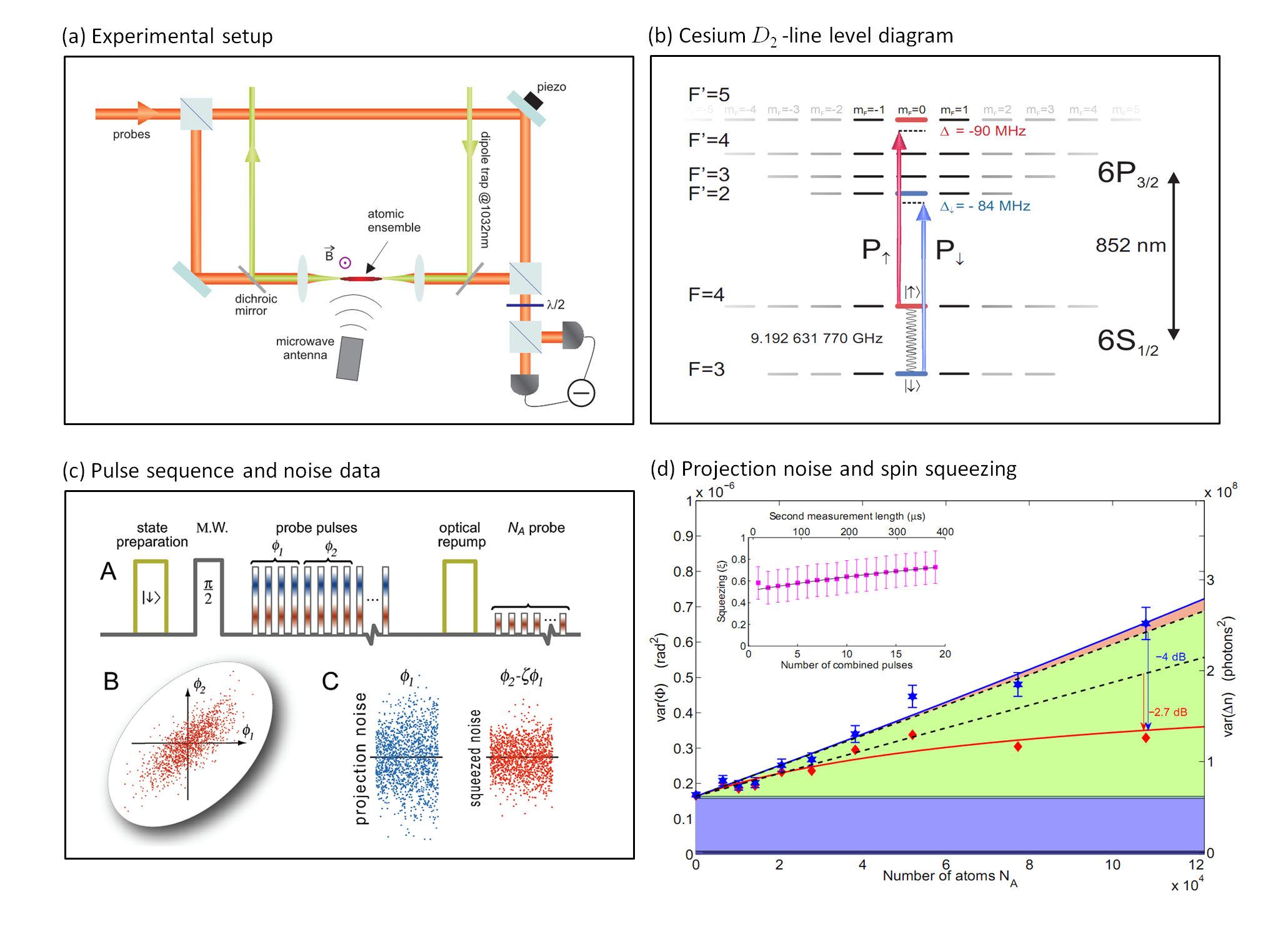}
\par\end{centering}

\caption{(Color online) (a) Schematic diagram of the QND measurement
experimental setup, a Mach-Zehnder
interferometer~\protect\cite{Louchet-Chauvet2010,Appel2009}. The
elongated Cesium atomic cloud is placed in the probe arm. The probe
beam consists of two identical linearly polarized light beams
$P_{\uparrow}$ and $P_{\downarrow}$, which acquire phase shifts
proportional to the number of atoms in the clock states
$N_{\uparrow}$ and $N_{\downarrow}$, respectively, as also shown in
(b). In (c) A, we show an experimental sequence for generating the
state $|J,J\rangle_{x}$ and QND measurement of $S_{y}$, and in B and
C, noise data of $\phi_{2}$ and $\phi_{1}$ show correlations between
$\phi_{2}$ and $\phi_{1}$, and reduced noise of
$(\phi_{2}-\zeta\phi_{1})$. In (d), we display the experimental
results of the projection noise and spin squeezing. Blue stars:
variance of the second measurement $\Delta^{2}\phi_{2}$ . Solid blue
line: quadratic fit to $\Delta^{2}\phi_{2}$ . Green area: atomic
projection noise of the CSS. Red diamonds: conditional variance
$\Delta^{2}\left(\phi_{2}-\zeta\phi_{1}\right)$. Red line: reduced
noise as predicted by the fits to the noise data. This data was
obtained by acquiring 4800 experimental runs. Inset: spin squeezing
$\xi_{R}^{2}$ as a function of the number of pulses combined to form
the second measurement. The data is fitted with an exponential decay
(solid line). Figures (a), (b), and (d) are from
Ref.~\protect\cite{Louchet-Chauvet2010}, and figure (c) is from
Ref.~\protect\cite{Appel2009}. }

\label{fig_qnd_exp}
\end{figure}

With the above experimental efforts, recently, two successive
notable improvements in QND experiment were
reported~\cite{Appel2009,Louchet-Chauvet2010}, where a complete
sub-shot-noise Ramsey process was demonstrated by means of QND
measurement. The experimental schemes for these two works
\cite{Appel2009,Louchet-Chauvet2010} are similar, thus here, we
mainly discuss the very recent one \cite{Louchet-Chauvet2010}. The
experimental setup is shown in Fig.~\ref{fig_qnd_exp}(a). An
ensemble of $\sim10^{5}$ cold Cs atoms are located in one of the
arms of the Mach-Zehnder interferometer. The atoms are first loaded
into a standard MOT, and are then transferred into a far-off
resonant optical dipole trap (FORT), generated by a Versadisk laser
with a wavelength of 1032 nm and a power of 2.3 W. After loading the
FORT, the MOT is turned off and a 1.22 Gauss magnetic field is
applied, which defines a quantization axis orthogonal to the
trapping beam. The ground levels \begin{align}
|\downarrow\rangle & \equiv|6S_{1/2},F=3,m_{F}=0\rangle,\nonumber \\
|\uparrow\rangle & \equiv|6S_{1/2},F=4,m_{F}=0\rangle,\end{align}
 are referred to the clock levels, and by optical pumping, the atoms
are prepared in the clock level $|\downarrow\rangle$. As described
in Ref.~\cite{Appel2009}, atoms remaining in states other than
$|\downarrow\rangle$ due to imperfect optical pumping are
subsequently pushed out of the trap.\emph{ }To prepare the CSS
$\otimes_{i=1}^{N_{A}}\left[\frac{1}{\sqrt{2}}\left(|\downarrow\rangle+|\uparrow\rangle\right)\right]_{i}$,
a resonant $\pi/2$ microwave pulse at the clock frequency is
applied. Then, successive QND measurements of the population
difference, $N_{\uparrow}-N_{\downarrow}$, will be performed by
measuring the phase shift of the probe light in a balanced homodyne
configuration. After the QND measurement, all atoms are pumped into
the $F=4$ level to determine the total atom number $N_{A}$. The
sequence, shown in Fig.~\ref{fig_qnd_exp}(c), is repeated several
thousand times with a cycle time of $\simeq5$ s.

As shown in Fig.~\ref{fig_qnd_exp}(a), two identical linear polarized
beams $P_{\uparrow}$ and $P_{\downarrow}$, which are generated by
two extended-cavity diode lasers, enters the interferometer from the
same port {[}Fig.~\ref{fig_qnd_exp}(a){]}, off-resonantly probe
transitions $|F=3\rangle\rightarrow|F^{\prime}=4\rangle$ and $|F=4\rangle\rightarrow|F^{\prime}=5\rangle$,
respectively. Each beam gains a phase shift \begin{equation}
\phi_{\uparrow\left(\downarrow\right)}=k_{\uparrow\left(\downarrow\right)}N_{\uparrow\left(\downarrow\right)},\end{equation}
 which is proportional to the number of atoms $N_{\uparrow}$ and
$N_{\downarrow}$ corresponding to the clock states
$|\uparrow\rangle$ and $|\downarrow\rangle$, respectively. The
detuning $\Delta_{\uparrow}$ and $\Delta_{\downarrow}$ are tuned to
ensure $k_{\uparrow}=k_{\downarrow}$, such that the Hamiltonian for
this system is Eq.~(\ref{QND_Ham}), and thus\begin{equation}
J_{z}=\left(N_{\uparrow}-N_{\downarrow}\right)/2,\end{equation}
 where $N_{\uparrow\left(\downarrow\right)}$ are the photon number
operators, $N_{A}=N_{\uparrow}+N_{\downarrow}$ is the total atom
number. The photocurrent difference is equivalent to measuring $S_{y,\text{out}}$,
which is derived from Eq.~(\ref{qnd_in_out}) as\begin{equation}
S_{y,\text{out}}=S_{y,\text{in}}+\frac{n}{2}\chi J_{z,\text{in}}.\end{equation}

The phase difference between the two arms of the Mach-Zehnder interferometer
is defined as\begin{equation}
\phi\equiv\frac{S_{y,\text{out}}}{n},\end{equation}
 where $n=n_{\uparrow}+n_{\downarrow}$ is the total photon number.
In experiments~\cite{Appel2009,Louchet-Chauvet2010}, the photon
numbers of the two probe beams $n_{\uparrow}$ and $n_{\downarrow}$
are set to be almost equal. If the probe beam is in a coherent
state, the variance of the phase difference is\begin{equation}
\Delta^{2}\phi=\frac{1}{n}+\chi^{2}\left(\Delta
J_{z,\text{in}}\right)^{2},\end{equation}
 which is related to both photon shot-noise $1/n$, and atomic spin
noise $(\Delta J_{z,\text{in}})^{2}$. As the atomic ensemble is prepared
in a CSS, $\Delta J_{z,\text{in}}=\sqrt{N_{A}}/2$. From Fig.~\ref{fig_qnd_exp}(c),
two successive measurements of $\phi_{1}$ and $\phi_{2}$ are performed,
using probe photons $n_{1}$ and $n_{2}$. Thus,\begin{equation}
\Delta^{2}\phi_{1(2)}=\frac{1}{n_{1,(2)}}+\chi^{2}\frac{N_{A}}{4},\quad\text{ Cov}\left(\phi_{1},\phi_{2}\right)=\chi^{2}\frac{N_{A}}{4}.\label{QND_var_cov}\end{equation}
 The covariance (\ref{QND_var_cov}) is calculated for the atomic
ensemble, which is the same for the two measurements, while the two
probe beams have no correlations. As $\phi_{2}$ is conditioned to
$\phi_{1}$ and can be predicted with a better precision, after the
measurement of $\phi_{1}$, an appropriate estimate of $\phi_{2}$
is chosen as $\zeta\phi_{1}$, and the variance of their difference
becomes\begin{align}
\Delta^{2}\left(\phi_{2}-\zeta\phi_{1}\right) & ~=~\left(\Delta^{2}\phi_{1}\right)\zeta^{2}-2~\text{Cov}\left(\phi_{1},\phi_{2}\right)\zeta+\Delta^{2}\phi_{2}\notag\\
 & ~=~\frac{1}{n_{2}}+\frac{1}{1+\kappa^{2}}\chi^{2}\frac{N_{A}}{4},\label{qnd_squeezing}\end{align}
 where the second line is the minimum of $\Delta^{2}\left(\phi_{2}-\zeta\phi_{1}\right)$
at\begin{equation}
\zeta=\frac{\text{Cov}\left(\phi_{1},\phi_{2}\right)}{\Delta^{2}\phi_{1}}=\frac{\kappa^{2}}{1+\kappa^{2}}.\end{equation}
 Here, \begin{equation}
\kappa^{2}=n_{1}N\chi^{2}/4,\end{equation}
 which is the same result derived in Eq.~(\ref{QND_spinsqu}). Now,
according to Eq.~(\ref{qnd_squeezing}), the spin-squeezing parameter\begin{equation}
\xi_{R}^{2}=\frac{1}{1+\kappa^{2}}.\end{equation}
 A measurement of $\kappa^{2}=1.6$ for $N_{A}=1.2\times10^{5}$ atoms
gives rise to a reduction of the projection noise by $-4\,\text{dB}$
compared to the CSS projection noise, displayed in Fig.~\ref{fig_qnd_exp}(d).

Now consider the decoherence during a QND measurement. In
experiments~\cite{Appel2009,Louchet-Chauvet2010}, the employ of
two-color QND can suppress the dephasing effects induced by an
inhomogeneous AC stark shift due to the transverse intensity profile
of the probe beam \cite{Saffman2009}. Otherwise, since the probe
interacts dispersively with the atoms, decoherence is inevitable
induced by spontaneous photon
scattering~\cite{Appel2009,Louchet-Chauvet2010}, which shortens the
mean-spin length as\begin{equation}
\langle\vec{J}\,\rangle~\rightarrow~\left(1-\eta\right)\langle\vec{J}\,\rangle,\end{equation}
 where $\eta=1-\exp\left(n\alpha\right),$ with $n$ the total number
of photons in the probe pulse and the parameter
$\alpha=-2.39\times10^{-8}$ \cite{Louchet-Chauvet2010}. The
coherence between the clock levels decreases while the phase shifts
are almost unaffected, so in the absence of the classical noise, we
have\begin{equation}
\xi_{R}^{2}=\frac{1}{\left(1-\eta\right)^{2}\left(1+\kappa^{2}\right)}.\label{eq:QND_xi_R}\end{equation}
 When using $n_{1}/2=3\times10^{6}$ photons per probe, the reduction
is $\eta=14\%$, and experimental data gives $1/\xi_{R}^{2}=2.7\,\text{dB}$
as shown in Fig.~\ref{fig_qnd_exp}(d). This is very close to the
theoretical prediction from Eq.~(\ref{eq:QND_xi_R}), $1/\xi_{R}^{2}=2.8\,\text{dB}$.
The above results indicate that experiments must carefully choose
photon numbers for optimal spin squeezing.

\section{Conclusion}

We have reviewed both basic notions and recent progresses on spin
squeezing. After a short introduction on bosonic squeezing, in
Sec.~\ref{definition}, we reviewed various definitions of
spin-squeezing parameters and their physical significance. In order
to understand these parameters better, we derived explicit
analytical expressions of them for states with parity. At the end of
Sec.~\ref{definition}, we showed that, in the limit of a large
number of atoms and small number of excitations, spin squeezing
reduces to bosonic squeezing.


In Sec.~\ref{states}, we reviewed theoretical approaches toward
producing spin-squeezed states via one-axis twisting and two-axis
twisting Hamiltonians. The one-axis twisting Hamiltonian has
attracted considerable attentions, and has been implemented in BEC,
as reviewed in Sec.~\ref{experiment}. For this model, we presented
analytical results for the spin-squeezing parameters $\xi_{S}^{2}$
and $\xi_{R}^{2}$, which scale with the particle number as
$1/N^{2/3}$.


The two-axis twisting Hamiltonian can produce more squeezing than
the one-axis case. Moreover, its optimal squeezing angle does not
vary with time. However, analytical results are only available for
particle number $N\le3$, and numerical results showed that $\xi_{S}^{2}$
and $\xi_{R}^{2}$ scale as $1/N$. This two-axis twisting Hamiltonian
can be realized in spinor BEC.


Section~\ref{entanglement} showed that spin squeezing has very close
relations to both entanglement and negative pairwise correlations. A
system consisting of spin-1/2 particles is entangled if
$\xi_{R}^{2}<1$. For states with parity and exchange symmetry, close
relations between the spin-squeezing parameters $\xi_{S}^{2}$,
$\tilde{\xi}_{E}^{2}$, and the concurrence $C$ have been
established. In this case, spin squeezing according to
$\tilde{\xi}_{E}^{2}$ is qualitatively equivalent to the existence
of pairwise entanglement. Moreover, recent results generalized the
proposal of using spin squeezing to detect entanglement, and also
developed a set of spin inequalities. In the end, the two-mode spin
squeezing is reviewed. It can detect the entanglement between spin
variables of two separated atomic samples, and was proposed to be a
valuable resource of quantum information processing. As explained in
this section, spin-squeezing-based entanglement witnesses should be
useful in measurements and are then expected to be of wide
applications in experiments.


In Sec.~\ref{Fisher}, we discussed the relation between spin
squeezing and QFI, as well as their applications in quantum
metrology. Applications of spin squeezing also deserve more
considerations, and in Sec.~\ref{qpt}, we explained how spin
squeezing can be used to detect quantum chaos and QPTs.


In Sec.~\ref{decoherence}, we first consider how decoherence affects
the generation of spin squeezing. If the decoherence effect is
sufficiently weak, considerable large amount of squeezing can still
be obtained. Then we showed that spin-squeezed states are more
robust than the GHZ state in the presence of particle loss. The
lifetime of squeezing is discussed for three typical decoherence
channels, and the spin-squeezing sudden death was derived and
analyzed. These channels are prototype models, however, are quite
general to describe decoherence processes. The results derived by
using these channels still make sense when using concrete physical
models. In addition, the one parameter in the dephasing,
dissipation, or the depolarizing channels may be a very complex
function depending on the realistic physical models of decoherence.
Afterwards, we show that in the Ramsey spectroscopy, decoherence
reduces the ability of spin-squeezed states to help in
precision-improvement, however, compared with the GHZ states,
spin-squeezed states are more robust to decoherence.


In Sec.~\ref{experiment} we reviewed several recent experiments, as
well as their theoretical backgrounds. Spin-squeezed states were
realized in BEC and atomic ensembles. In the past two decades, spin
squeezing has been studied extensively. It was first proposed in
order to overcome the classical shot-noise limit and make
high-precision atomic interferometers. This encouraged many
theoretical and experimental efforts in the control and manipulation
of atomic systems. Afterwards, it was realized that the notion of
spin squeezing can provide an insight on entanglement, and this has
attracted considerable recent interests. Experiments can now produce
spin-squeezed states, and even use them to perform sub-shot noise
interference. However, in practice, using spin-squeezed states in an
atomic clock or a gravity interferometer still faces many
challenges, including how to suppress the decoherence and enlarge
the particle number of the spin-squeezed states.


Finally, we summarize and list several main results, some of which
are of broad interest:

(a) There are several definitions of squeezing parameters (see Table
1), which one is preferable depends on various situations. For example,
when studying the accuracy of Ramsey spectroscopy, the parameter $\xi_{R}^{2}$
is more suitable. In this review, we mainly focused on two parameters,
$\xi_{S}^{2}$ and $\xi_{R}^{2}$, which are widely studied in the
literatures.

(b) In general, spin-squeezed states are entangled; however this depends
on specific squeezing parameters. From Sec.~\ref{entanglement},
we know that, for the parameter $\xi_{R}^{2}$, states are entangled
if $\xi_{R}^{2}<1$; while for the parameter $\tilde{\xi}_{E}^{2}$,
states with parity have pairwise entanglement if $\tilde{\xi}_{E}^{2}<1$.
Generalized spin inequalities have also been developed to detect entanglement.

(c) Since spin-squeezing parameters only involve expectations and
variances of collective spin operators, they can be measured relatively
easy in many physical systems, e.g., BEC and atomic ensembles, while
addressing individual atoms is an arduous task.

(d) Among proposals for generating spin-squeezed states, the twisting
Hamiltonian, which describes nonlinear two-body interactions, plays
an important role. This type of Hamiltonian has been studied and implemented
in physical systems like BEC.

(e) In parameter-estimation processes, compared with the GHZ state,
spin-squeezed states can also attain the Heisenberg limit, but are
more robust to decoherence.


In summary, spin squeezing has provided very useful insight into
quantum processes, and will continue to considerably impact quantum
optics and quantum information science in the future.

\textbf{Acknowledgements} We would like to thank Heng-Na Xiong, Xiaoqian
Wang, Xiao-Ming Lu, and Xiaolei Yin for useful discussions. FN acknowledges
partial support from the Laboratory of Physical Sciences (LPS), National
Security Agency (NSA), Army Research Office (ARO), Defense Advanced
Research Projects Agency (DARPA), Air Force Office of Scientific Research
(AFOSR), National Science Foundation (NSF) grant No. 0726909, JSPSRFBR
contract No.~09-02-92114, Grant-in-Aid for Scientific Research (S),
MEXT Kakenhi on Quantum Cybernetics, and the Funding Program for Innovative
R\&D on Science and Technology (FIRST). Xiaoguang Wang acknowledges
support from the NSFC with grant No.~11025527, 10874151, and 10935010.
Jian Ma acknowledges support from the Scholarship Award for Excellent
Doctoral Student granted by Ministry of Education.
\newpage
\begin{appendix}

\section{Principal squeezing parameter\label{principle}}

From Eqs.~(\ref{xtheta}) and (\ref{principle_squ}), we have\begin{align}
\zeta_{B}^{2} & =\min_{\theta\in\lbrack0,2\pi)}\Big[\cos^{2}\theta(\Delta X)^{2}+\sin^{2}\theta(\Delta P)^{2}+\sin(2\theta)~\text{Cov}(X,P)\Big]\nonumber \\
 & =\min_{\theta\in\lbrack0,2\pi)}\left\{ \frac{(\Delta X)^{2}+(\Delta P)^{2}}{2}+\frac{1}{2}\left\{ \cos(2\theta)\left[(\Delta X)^{2}-(\Delta P)^{2}\right]+2\sin(2\theta)~\text{Cov}(X,P)\right\} \right\} \nonumber \\
 & =\frac{1}{2}\left\{ \text{Var}(X)+\text{Var}(P)-\sqrt{\left[\text{Var}(X)-\text{Var}(P)\right]^{2}+4\text{Cov}^{2}(X,P)}\right\} ,\label{xib}\end{align}
 where the variance \begin{equation}
\text{Var}(X)=\langle X^{2}\rangle-\langle X\rangle^{2}\end{equation}
 and the covariance \begin{equation}
\text{Cov}(X,P)=\frac{1}{2}\langle XP+PX\rangle-\langle X\rangle\langle P\rangle.\end{equation}

From the expressions for $X$ and $P$ in Eq.~(\ref{eq:xp}), one
finds \begin{equation}
\frac{XP+PX}{2}=\frac{a^{2}-a^{\dagger2}}{i},\end{equation}
 which leads to the covariance\begin{equation}
\text{Cov}(X,P)=\frac{1}{i}\Big[\text{Var}(a)-\text{Var}(a^{\dagger})\Big].\label{covaa}\end{equation}
 From the identity \begin{equation}
\text{Var}\left(\sum_{i=1}^{N}X_{i}\right)\text{=}\sum_{i=1}^{N}\text{Var(}X_{i}\text{)}+2\sum_{i<j}\text{Cov}(X_{i},X_{j}),\label{varcov}\end{equation}
 we have\begin{align}
\text{Var}(X) & =\text{Var}(a)+\text{Var}(a^{\dagger})+2\text{Cov}(a,a^{\dagger}),\\
\text{Var}(P) & =-\text{Var}(a)-\text{Var}(a^{\dagger})+2\text{Cov}(a,a^{\dagger}),\end{align}
 which leads to \begin{align}
\text{Var}(X)+\text{Var}(P) & =4\text{Cov}(a,a^{\dagger}),\label{varvarcov}\\
\text{Var}(X)-\text{Var}(P) & =2\left(\text{Var}(a)+\text{Var}(a^{\dagger})\right).\label{varvarvar}\end{align}
 Substituting Eqs.~(\ref{covaa}), (\ref{varvarcov}), and (\ref{varvarvar})
into Eq.~(\ref{xib}), we finally obtain \begin{equation}
\zeta_{B}^{2}=2\left\{ \text{Cov}(a,a^{\dagger})-\sqrt{\text{Var}(a)\text{Var}(a^{\dagger})}\right\} ,\end{equation}
 which is just Eq.~(\ref{xibb}).


\section{Variances of $J_{\vec{n}}$ for the CSSs\label{varn}}

We obtain the variance $\left(\Delta J_{\vec{n}}\right)^{2}=\left\langle J_{\vec{n}}^{2}\right\rangle -\left\langle J_{\vec{n}}\right\rangle ^{2}$
for the CSS$.$ The expectation values $\left\langle J_{\alpha}^{2}\right\rangle $
are given by \begin{align}
\left\langle J_{\alpha}^{2}\right\rangle  & =\frac{1}{4}\sum_{i,j=1}^{N}\left\langle \sigma_{i\alpha}\sigma_{j\alpha}\right\rangle \nonumber \\
 & =\frac{N}{4}+\frac{1}{4}\sum_{i\neq j=1}^{N}\left\langle \sigma_{i\alpha}\sigma_{j\alpha}\right\rangle \nonumber \\
 & =\frac{N}{4}+\frac{N^{2}-N}{4}\langle\sigma_{1\alpha}\rangle^{2}.\end{align}
 The last equality is valid since we are considering a product state
with exchange symmetry. Note that \begin{equation}
\left\langle J_{\alpha}\right\rangle =\frac{N}{2}\langle\sigma_{1\alpha}\rangle,\end{equation}
 the variance is found to be\begin{equation}
\left(\Delta J_{\alpha}\right)^{2}=\frac{N}{4}(1-\langle\sigma_{1\alpha}\rangle^{2}).\end{equation}

Now, we consider a more general operator $J_{\vec{n}}$ as\begin{equation}
J_{\vec{n}}~=~\vec{J}\cdot\vec{n}~=~\frac{1}{2}\sum_{i=1}^{N}\vec{\sigma}_{i}\cdot\vec{n}~=~\frac{1}{2}\sum_{i=1}^{N}\sigma_{i\vec{n}},\label{gl}\end{equation}
 from which, its variance is given by\begin{align}
\text{Var}(J_{\vec{n}}) & =\left(\Delta J_{\vec{n}}\right)^{2}\nonumber \\
 & =\frac{1}{4}\sum_{i=1}^{N}\text{Var(}\sigma_{i\vec{n}})\nonumber \\
 & =\frac{1}{4}\sum_{i=1}^{N}\Big(1-\langle\sigma_{i\vec{n}}\rangle^{2}\Big)\nonumber \\
 & =\frac{N}{4}\Big(1-\langle\sigma_{1\vec{n}}\rangle^{2}\Big).\label{eq:var}\end{align}
 Here, the second equality follows from Eq.~(\ref{varcov}) and the
fact that the CSS is a product state and contains no correlations
(the covariance Cov$(X_{i},X_{j})=\langle\lbrack X_{i},X_{j}]_{+}/2\rangle-\langle X_{i}\rangle\langle X_{j}\rangle=0$).
The third equality is from the result $\sigma_{i\vec{n}}^{2}=\mathbb{I}.$

From Eqs.~(\ref{gl}) and (\ref{var}), we obtain\begin{align}
\langle J_{\vec{n}}\rangle & =\frac{N}{2}\langle\sigma_{1\vec{n}}\rangle=\frac{N}{2}|\langle\vec{\sigma}_{1}\rangle|\vec{n}_{0}\cdot\vec{n},\nonumber \\
\left(\Delta J_{\vec{n}}\right)^{2} & =\frac{N}{4}\Big(1-\langle\sigma_{1\vec{n}}\rangle^{2}\Big)=\frac{N}{4}\Big[1-\left|\langle\vec{\sigma}_{1}\rangle\right|^{2}\left(\vec{n}_{0}\cdot\vec{n}\right)^{2}\Big],\label{expvar}\end{align}
 which is valid for any product state with exchange symmetry. The
second equality is obtained from the fact \begin{equation}
\langle\vec{\sigma}_{1}\rangle~=~|\langle\vec{\sigma}_{1}\rangle|\vec{n}_{0}~=~|\langle\vec{\sigma}_{1}\rangle|(\sin\theta\cos\phi,\sin\theta\sin\phi,\cos\theta).\end{equation}

From the definition of the CSS, the expectation values $\left\langle J_{\alpha}\right\rangle $
($\alpha=x,y,z$) are obtained as \begin{align}
\left\langle J_{x}\right\rangle  & =\frac{N}{2}\sin\theta\cos\phi,\nonumber \\
\left\langle J_{y}\right\rangle  & =\frac{N}{2}\sin\theta\sin\phi,\nonumber \\
\left\langle J_{z}\right\rangle  & =\frac{N}{2}\cos\theta.\label{expectation}\end{align}
 with $N=2j$, and $\theta$, $\phi$ are the polar and azimuth angles,
respectively. Substituting Eq.~(\ref{expectation}) into Eq.~(\ref{eq:var})
leads to the explicit form of the variances\begin{align}
\left(\Delta J_{x}\right)^{2} & =\frac{N}{4}(1-\sin^{2}\theta\cos^{2}\phi),\nonumber \\
\left(\Delta J_{y}\right)^{2} & =\frac{N}{4}(1-\sin^{2}\theta\sin^{2}\phi),\nonumber \\
\left(\Delta J_{z}\right)^{2} & =\frac{N}{4}\sin^{2}\theta.\label{variances}\end{align}

Using Eq.~(\ref{fig_spin_bloch}), one finds the variance along the
MSD, \begin{equation}
\text{Var}(J_{\vec{n}_{0}})=\frac{N}{4}\left(1-|\langle\vec{\sigma}_{1}\rangle|^{2}\right).\label{varmsd}\end{equation}
 For the CSS, only Dicke states are populated, and thus,\begin{equation}
\langle J_{x}^{2}+J_{y}^{2}+J_{z}^{2}\rangle=\frac{N}{2}\left(\frac{N}{2}+1\right).\end{equation}
 From the above equation and from the relations between the collective
operator $J_{\alpha}$ and the local operators (\ref{gl}), one finds\begin{equation}
\frac{1}{N^{2}-N}\sum_{i\neq j}\langle\vec{\sigma}_{i}\cdot\vec{\sigma}_{j}\rangle=1.\end{equation}
 As the CSS is a product symmetric state, the above equation reduces
to\begin{equation}
\langle\vec{\sigma}_{1}\rangle\cdot\langle\vec{\sigma}_{1}\rangle=1.\end{equation}
 Applying this result to Eq.~(\ref{varmsd}), one immediately has\begin{equation}
\text{Var}(J_{\vec{n}_{0}})=0.\end{equation}

For the variances along the direction $\vec{n}_{\perp}$ perpendicular
to the MSD, from Eq.~(\ref{var}), we have\begin{align}
\text{Var}(J_{\vec{n}\perp}) & =\frac{N}{4}\left(1-\langle\sigma_{1\vec{n}_{\perp}}\rangle^{2}\right)\nonumber \\
 & =\frac{N}{4}\left[1-\left(\langle\vec{\sigma}_{1}\rangle\cdot\vec{n}_{\perp}\right)^{2}\right]\nonumber \\
 & =\frac{N}{4}\left[1-|\langle\vec{\sigma}_{1}\rangle|^{2}\left(\vec{n}_{0}\cdot\vec{n}_{\perp}\right)^{2}\right]\nonumber \\
 & =\frac{N}{4}=j/2,\end{align}
 where the third equality follows from Eq.~(\ref{msd}).



\section{The minimum value of the quantity $\vec{n}^{T}\Gamma\vec{n}$}

\label{nGamman}

Here, we prove that the minimum value of $\vec{n}^{T}\Gamma\vec{n}$
is a minimum eigenvalue of $\Gamma$. The minimum value of $\vec{n}^{T}\Gamma\vec{n}$
is obtained by searching the vector $\vec{n}=(n_{1},n_{2},n_{3})^{T}$.
Since the matrix $\Gamma$ is real and symmetric, we obtain \begin{align}
 & \min_{\vec{n}}\Big(\vec{n}^{T}\Gamma\vec{n}\Big)=\min_{\vec{n}}\Big[\vec{n}^{T}O^{T}\left(O\Gamma O^{T}\right)O\vec{n}\Big]=\min_{\vec{n}^{\prime}}\Big(\vec{n}^{\prime T}\Gamma^{d}\vec{n}^{\prime}\Big),\end{align}
 where $O$ is an orthogonal $3\times3$ matrix which diagonalizes
$\Gamma$, and\begin{equation}
\Gamma^{d}=\text{diag}\left(\lambda_{1},\lambda_{2},\lambda_{3}\right),\end{equation}
 where we set $\lambda_{1}\geq\lambda_{2}\geq\lambda_{3}$. The rotated
direction is $\vec{n}^{\prime}=O\vec{n}=\left(n_{1}^{\prime},n_{2}^{\prime},n_{3}^{\prime}\right)^{T}$.
Now, the minimum value of $\vec{n}^{T}\Gamma\vec{n}$ is obtained
as\begin{align}
 & \min_{\vec{n}}\Big(\vec{n}^{T}\Gamma\vec{n}\Big)=\min_{\vec{n}^{\prime}}\Big(\vec{n}^{\prime T}\Gamma^{d}\vec{n}^{\prime}\Big)=\min_{\vec{n}^{\prime}}\Big(\lambda_{1}n_{1}^{\prime2}+\lambda_{2}n_{2}^{\prime2}+\lambda_{3}n_{3}^{\prime2}\Big)=\lambda_{3},\end{align}
 since $n_{1}^{\prime2}+n_{2}^{\prime2}+n_{3}^{\prime2}=1$. Thus,
the minimum value of $\vec{n}^{T}\Gamma\vec{n}$ is obtained by choosing
$n_{3}^{\prime}=1$, and $n_{1}^{\prime}=n_{2}^{\prime}=0$, therefore,
$\vec{n}=O^{T}\vec{n}^{\prime}$ is the direction where the minimum
is achieved.



\section{Expectations for the one-axis twisted state\label{oat}}

Here, we will use the Heisenberg picture to derive the relevant expectation
values for the one-axis twisted state. To determine the spin-squeezing
parameters $\xi_{S}^{2}$ and $\xi_{R}^{2}$, as seen from Table \ref{tab_comp},
one needs to know the expectation $\langle\sigma_{1z}\rangle$, and
correlations $\langle\sigma_{1+}\sigma_{2-}\rangle$ and $\langle\sigma_{1-}\sigma_{2-}\rangle$.
We first consider the expectation $\langle\sigma_{1z}\rangle$.

The evolution operator can be written as, \begin{equation}
U=\exp({-i\chi tJ_{x}^{2}})=\exp\left({-i\theta\sum_{k>l}j_{kx}j_{lx}}\right)\end{equation}
 up to a trivial phase, where $\theta=2\chi t$. From this form, the
evolution of $j_{1z}$ can be obtained as \[
U^{\dagger}j_{1z}U=j_{1z}\cos[\theta j_{x}^{(2)}]+j_{1y}\sin[\theta j_{x}^{(2)}],\]
 where \begin{equation}
j_{x}^{(k)}=\sum_{l=k}^{N}j_{lx}.\end{equation}
 Therefore, the expectations are \begin{equation}
\langle j_{1z}\rangle=-\frac{1}{2}\langle\mathbf{1^{\prime}}|\cos[\theta j_{x}^{(2)}]|\mathbf{1^{\prime}}\rangle\label{jz1}\end{equation}
 since $\langle1|j_{1y}|1\rangle=0.$ Here, $|\mathbf{1^{\prime}}\rangle=|1\rangle_{2}\otimes...\otimes|1\rangle_{N}.$
So, one can find the following form for the expectation values \begin{align}
\langle\mathbf{1}|\cos\left[\theta J_{x}\right]|\mathbf{1}\rangle & =\left(\langle\mathbf{1}|e^{i\theta J_{x}}|\mathbf{1}\rangle+\mathrm{c.c.}\right)/2\nonumber \\
 & =\left(\Pi_{k=1}^{N}\langle1|e^{i\theta j_{kx}}|1\rangle+\mathrm{c.c.}\right)/2\nonumber \\
 & =\cos^{N}({\theta^{\prime}}),\label{jz2}\end{align}
 where $\theta^{\prime}=\theta/2$ and $|\mathbf{1}\rangle=|1\rangle^{\otimes N}$.
By using Eqs.~(\ref{jz1}) and (\ref{jz2}), one gets \begin{equation}
\langle\sigma_{z}\rangle=-\cos^{N-1}\left({\theta^{\prime}}\right).\label{sigmaz}\end{equation}

{\bigskip{}
 }Since the operator $\sigma_{1x}\sigma_{2x}$ commutes with the
unitary operator $U,$ we easily obtain \begin{equation}
\langle\sigma_{1x}\sigma_{2x}\rangle=0.\label{xx}\end{equation}
 We now compute the correlations $\langle\sigma_{1z}\sigma_{2z}\rangle.$
From the unitary operator, \begin{align*}
U^{\dagger}j_{1z}j_{2z}U & =\left[j_{1z}\cos(\theta j_{x}^{(2)})+j_{1y}\sin(\theta j_{x}^{(2)})\right]\left[j_{2z}\cos[\theta(j_{1x}+j_{x}^{(3)})]+j_{2y}\sin[\theta(j_{1x}+j_{x}^{(3)})]\right]\\
 & =\left[j_{1z}\cos(\theta j_{2x})\cos(\theta j_{x}^{(3)})-j_{1z}\sin(\theta j_{2x})\sin(\theta j_{x}^{(3)})\right.\\
 & \left.+j_{1y}\sin(\theta j_{2x})\cos(\theta j_{x}^{(3)})+j_{1y}\cos(\theta j_{2x})\sin(\theta j_{x}^{(3)})\right]\\
 & \times\left[j_{2z}\cos(\theta j_{1x})\cos(\theta j_{x}^{(3)})-j_{2z}\sin(\theta j_{1x})\sin(\theta j_{x}^{(3)})\right.\\
 & \left.+j_{2y}\sin(\theta j_{1x})\cos(\theta j_{x}^{(3)})+j_{2y}\cos(\theta j_{1x})\sin(\theta j_{x}^{(3)})\right].\end{align*}
 Although there are 16 terms after expanding the above equation, only
four terms survive when calculating $\langle j_{1z}j_{2z}\rangle.$
We then have \begin{align}
\langle j_{1z}j_{2z}\rangle & =\langle\mathbf{1}|j_{1z}j_{2z}\cos^{2}(\theta/2)\cos^{2}(\theta j_{x}^{(3)})-j_{1z}j_{2x}j_{2y}\sin(\theta)\sin^{2}(\theta j_{x}^{(3)})\nonumber \\
 & +4j_{1y}j_{1x}j_{2x}j_{2y}\sin^{2}(\theta/2)\cos^{2}(\theta j_{x}^{(3)})-j_{1y}j_{1x}j_{2z}\sin(\theta)\sin^{2}(\theta j_{x}^{(3)})|\mathbf{1}\rangle\nonumber \\
 & =\frac{1}{4}\langle\mathbf{1}^{\prime}|\cos^{2}(\theta j_{x}^{(3)})|\mathbf{1}^{\prime}\rangle\nonumber \\
 & =\frac{1}{8}\langle\mathbf{1}^{\prime}|\left[1+\cos(2\theta j_{x}^{(3)})\right]|\mathbf{1}^{\prime}\rangle\nonumber \\
 & =\frac{1}{8}\left[1+\cos^{N-2}(\theta)\right],\label{eef}\end{align}
 where $|\mathbf{1}^{\prime}\rangle=|1\rangle_{3}\otimes...\otimes|1\rangle_{N}$.
The second equality in Eq.~(\ref{eef}) is due to the property $j_{x}j_{y}=-j_{y}j_{x}={i}$
${j_{z}}/{2}$, and the last equality is from Eq.~(\ref{jz2}). Finally,
from the above equation, one finds \begin{equation}
\langle\sigma_{1z}\sigma_{2z}\rangle=\frac{1}{2}\left(1+\cos^{N-2}\theta\right).\label{zz}\end{equation}
 Due to the relation $\langle\sigma_{1x}\sigma_{2x}+\sigma_{1y}\sigma_{2y}+\sigma_{1z}\sigma_{2z}\rangle=1$
for the initial state, the correlation $\langle\sigma_{1y}\sigma_{2y}\rangle$
is obtained from Eqs.~(\ref{xx}) and (\ref{zz}) as \begin{equation}
\langle\sigma_{1y}\sigma_{2y}\rangle=\frac{1}{2}\left(1-\cos^{N-2}\theta\right).\label{yy}\end{equation}
 Substituting Eqs.~(\ref{xx}) and (\ref{yy}) into the following
relations \begin{equation}
\sigma_{1x}\sigma_{2x}+\sigma_{1y}\sigma_{2y}=2\left(\sigma_{1+}\sigma_{2-}+\sigma_{1-}\sigma_{2+}\right)\end{equation}
 leads to \begin{equation}
\langle\sigma_{1+}\sigma_{2-}\rangle=\frac{1}{8}\left(1-\cos^{N-2}\theta\right),\label{y0}\end{equation}
 where the relation $\langle\sigma_{1+}\sigma_{2-}\rangle=\langle\sigma_{1-}\sigma_{2+}\rangle$
is used due to the exchange symmetry.

To calculate the correlation $\langle\sigma_{1-}\sigma_{2-}\rangle,$
due to the following relations\begin{align}
\sigma_{1x}\sigma_{2x}-\sigma_{1y}\sigma_{2y} & =2\left(\sigma_{1+}\sigma_{2+}+\sigma_{1-}\sigma_{2-}\right),\label{sigma1}\\
i\left(\sigma_{1x}\sigma_{2y}+\sigma_{1y}\sigma_{2x}\right) & =2\left(\sigma_{1+}\sigma_{2+}-\sigma_{1-}\sigma_{2-}\right),\quad\label{sigma2}\end{align}
 we need to know the expectations $\langle j_{1x}j_{2y}\rangle.$
The evolution of $j_{1x}j_{2y}$ is given by \begin{align}
U^{\dagger}s_{1x}s_{2y}U & =j_{1x}\left\{ j_{2y}\cos\left[\theta(j_{1x}+j_{x}^{(3)})\right]-j_{2z}\sin\left[\theta(j_{1x}+j_{x}^{(3)})\right]\right\} ,\end{align}
 and the expectation is obtained as \begin{align}
\langle j_{1x}j_{2y}\rangle & =\frac{1}{2}\langle\mathbf{1^{\prime}}|j_{1x}\sin\left[\theta(j_{1x}+j_{x}^{(3)})\right]|\mathbf{1^{\prime}}\rangle\nonumber \\
 & =\frac{{1}}{4i}\langle\mathbf{1^{\prime}}|j_{1x}e^{i\theta j_{1x}}\Pi_{k=3}^{N}e^{i\theta j_{kx}}-j_{1x}e^{-i\theta j_{1x}}\Pi_{k=3}^{N}e^{-i\theta j_{kx}}|\mathbf{1^{\prime}}\rangle\nonumber \\
 & =\frac{{1}}{4i}{\cos^{N-2}\left({\theta^{\prime}}{}\right)}\langle1|j_{1x}e^{i\theta j_{1x}}-j_{1x}e^{-i\theta j_{1x}}|1\rangle\nonumber \\
 & =\frac{1}{2}{\cos^{N-2}\left({\theta^{\prime}}\right)}\langle1|j_{1x}\sin(\theta j_{1x})|1\rangle\nonumber \\
 & =\frac{1}{4}{\sin\left({\theta^{\prime}}{}\right)\cos^{N-2}\left(\theta^{\prime}\right)}.\end{align}
 Here, $|\mathbf{1^{\prime}}\rangle=|1\rangle_{1}\otimes|1\rangle_{3}\otimes...\otimes|1\rangle_{N}$,
where $|1\rangle_{2}$ is absent. Moreover, $\langle j_{1y}j_{2x}\rangle=\langle j_{1x}j_{2y}\rangle$
due to the exchange symmetry, and thus, \begin{equation}
\langle j_{1x}j_{2y}+j_{1y}j_{2x}\rangle=\frac{1}{2}{\sin\left({\theta^{\prime}}{}\right)\cos^{N-2}\left({\theta^{\prime}}{}\right)}.\end{equation}
 For the state (\ref{initial-1}), we obtain the following expectations
\begin{equation}
\langle\sigma_{1x}\sigma_{2y}+\sigma_{1y}\sigma_{2x}\rangle=2\sin\left({\theta^{\prime}}{}\right)\cos^{N-2}\left({\theta^{\prime}}{}\right).\label{b12}\end{equation}
 The combination of Eqs.~(\ref{xx}), (\ref{yy}), (\ref{sigma1}),
(\ref{sigma2}), and (\ref{b12}) leads to the correlation \begin{align}
\langle\sigma_{1-}\sigma_{2-}\rangle & =-\frac{{1}}{8}\left(1-\cos^{N-2}\theta\right)-\frac{{i}}{2}\sin\left({\theta^{\prime}}{}\right)\cos^{N-2}\left({\theta^{\prime}}\right).\label{u0}\end{align}
 Substituting Eqs.~(\ref{y0}) and (\ref{u0}) to the expression
$\xi_{S}^{2}$ in terms of local expectation values (Table \ref{tab_comp})
leads to the desirable expression of the squeezing parameter for the
one-axis twisted state.

\section{Photon polarization and Stokes operators}

Here, we briefly introduce the Stokes operators for photons. The light
field has two orthogonal polarization directions. If the light field
propagates along the $z$-axis, we use the bases given by $\hat{e}_{x}$
and $\hat{e}_{y}$, which denote the vertical and horizontal polarizations,
respectively. The other two sets of bases can be expressed with $\hat{e}_{x}^{\prime}$
and $\hat{e}_{y}^{\prime}$. If the phase difference between the $\hat{e}_{x}$-
and $\hat{e}_{y}$-polarized components is $\pm\pi/2$, then the light
is $45^{\circ}$ polarized, with new bases expressed as\begin{equation}
\hat{e}_{x}^{\prime}=\frac{1}{\sqrt{2}}\left(\hat{e}_{x}+\hat{e}_{y}\right),\text{ \ }\hat{e}_{y}^{\prime}=\frac{1}{\sqrt{2}}\left(-\hat{e}_{x}+\hat{e}_{y}\right).\label{Stokes1}\end{equation}
 If the phase difference between the $\hat{e}_{x}$ and $\hat{e}_{y}$
components is $\pm\pi/4$, then the light is circularly polarized,
with right- and left-circular bases\begin{equation}
\hat{e}_{+}=-\frac{1}{\sqrt{2}}\left(\hat{e}_{x}+i\hat{e}_{y}\right),\text{ \ }\hat{e}_{-}=\frac{1}{\sqrt{2}}\left(\hat{e}_{x}-i\hat{e}_{y}\right).\label{Stokes2}\end{equation}
 The two polarized components can build up an angular momentum operator
analogous to the Schwinger representation (\ref{Schwinger_Rep}).
Denote the $x$- and $y$-polarized field operators as $a_{x}$ and
$a_{y}$, respectively. Then, the Stokes operators are defined as\begin{align}
S_{x} & =\frac{1}{2}\left(a_{x}^{\dagger}a_{x}-a_{y}^{\dagger}a_{y}\right),\notag\\
S_{y} & =\frac{1}{2}\left(a_{x}^{\prime\dagger}a_{x}^{\prime}-a_{y}^{\prime\dagger}a_{y}^{\prime}\right),\notag\\
S_{z} & =\frac{1}{2}\left(a_{+}^{\dagger}a_{+}-a_{-}^{\dagger}a_{-}\right).\label{Stokes_Op}\end{align}
 The three Stokes operators ($S_{x}$, $S_{y}$ and $S_{z}$) shown
in (\ref{Stokes_Op}) measure the photon number differences of the
$x$- to $y$-polarization, $\pi/4-$ to $-\pi/4$-polarization, and
right- to left-polarization, respectively. The polarization
squeezing of light field was studied in
Refs.~\cite{Bowen2002,Heersink2003,Corney2006,Luis2006,Marquardt2007,Shalm2009}.
\end{appendix}

\end{document}